\documentclass[a4paper,11pt]{article}
\pdfoutput=1 

\usepackage{jcappub} 

\usepackage[T1]{fontenc} 
\usepackage{mathrsfs}
\usepackage{amsmath}
\usepackage{upgreek}

\title{\boldmath Novel inner shadows of the Kerr black hole with a tilted thin accretion disk}

\author[a,1]{Shiyang Hu,\note{Corresponding author.}}
\author[a]{Dan Li,}
\author[b]{and Chen Deng}

\affiliation[a]{School of Mathematics and Physics, University of South China, \\ Hengyang, 421001 People's Republic of China}
\affiliation[b]{School of Astronomy and Space Science, Nanjing University, \\ Nanjing, 210023 People's Republic of China}

\emailAdd{husy\_arcturus@163.com}
\emailAdd{danli@usc.edu.cn}
\emailAdd{dengchen@smail.nju.edu.cn}

\abstract{The inner shadow of a black hole, as a projection of the event horizon, is regarded as a potential tool for testing gravitational theories and constraining system parameters. Whether this holds in the case of a tilted accretion disk warrants further investigation. In this paper, we employ a ray-tracing algorithm to simulate images of the Kerr black hole with a tilted thin accretion disk, with particular attention to the relationship between the inner shadow and system parameters. Our findings reveal that in the case of an equatorial accretion disk, the Kerr black hole exhibits a minimum inner shadow size of $S_{\textrm{min}} = 13.075$ M$^{2}$, where M denotes the black hole mass. This minimum is achieved when the viewing angle is $0^{\circ}$ and the spin parameter approaches $1$. However, with a non-zero disk tilt, the inner shadow exhibits novel configurations--taking on petal, crescent, or eyebrow shapes--significantly smaller than $S_{\textrm{min}}$ across various parameter spaces. This indicates that the inner shadow is highly sensitive to the accretion environment, suggesting that caution is needed when using it as a diagnostic tool for black holes. Notably, an observed inner shadow smaller than $S_{\textrm{min}}$ would either indicate the presence of a tilted accretion disk or support the viability of modified gravity. Moreover, in certain parameter spaces, we identify the emergence of a dual-shadow structure, which could also serve as a probe for the tilted accretion disk.}

\begin{document}
\maketitle
\flushbottom

\section{Introduction}
Since the advent of general relativity, black holes have remained a subject of profound fascination within the astronomical community. In recent years, the study of black holes has evolved from theoretical speculation to direct observation. Notably, the Event Horizon Telescope (EHT) collaboration, utilizing a Very Long Baseline Interferometry (VLBI) array, successfully captured and released the first images of the supermassive black holes at the centers of the M87 galaxy and the Milky Way \cite{Akiyama et al. (2019),Akiyama et al. (2022)}. These landmark observations not only conclusively verified the existence of black holes but also opened a novel channel for their exploration. Subsequently, a substantial body of research on black hole shadows \cite{Narayan et al. (2019),Gralla et al. (2019),Bronzwaer and Falcke (2021),Chael et al. (2021),Lin et al. (2022),Vincent et al. (2022),Hou et al. (2022),Heydari-Fard et al. (2023a),Heydari-Fard et al. (2023b),Zhang et al. (2023),Hu et al. (2023),Hu et al. (2024),Wang et al. (2024),Li et al. (2024b),Chen et al. (2024a)}, black hole-jet system images \cite{Chatterjee et al. (2020),Kawashima et al. (2021),Lu et al. (2023),Papoutsis et al. (2023),Zhang et al. (2024a)}, hot-spot images \cite{Tamm and Rosa (2024),Rosa et al. (2024),Huang et al. (2024),Chen et al. (2024b),Zhou et al. (2024),Wu et al. (2024)}, and polarized images \cite{Chen et al. (2020),Zhang et al. (2021),Gelles et al. (2021),Qin et al. (2022),Hu et al. (2022b),Chen et al. (2022a),Chen et al. (2022b),Zhang et al. (2024b),Shi et al. (2024),Guo et al. (2024)}, within the frameworks of general relativity and modified gravity theories, has emerged, injecting unprecedented vitality into this field.

Theoretically, a black hole can capture light rays emitted by surrounding sources, resulting in a distinct region of brightness depression on the observation plane from Earth. This region is referred to as the `shadow', first introduced by Falcke et al. \cite{Falcke et al. (2000)}, and its boundary is the 2D projection of the black hole's unstable critical photon orbits. Notably, the black hole shadow is considered the `fingerprint' of spacetime, unaffected by the accretion environment, making it a powerful tool for studying black hole physics and gravity theories. Consequently, many pioneering studies have focused on constraining system parameters through the shadow's profile \cite{Hioki and Maeda (2009),Johannsen (2013),Tsukamoto et al. (2014),Johannsen et al. (2016),Tsukamoto (2018),Wei et al. (2019),Gralla et al. (2020),Li et al. (2020),Farah et al. (2020),Hou et al. (2021),Afrin and Ghosh (2022),Cao et al. (2023),Allahyari et al. (2020),Vagnozzi et al. (2023),Khodadi et al. (2024)}. However, accurately extracting the geometric information of the black hole shadow requires extremely high angular resolution, which is currently beyond reach with the technology of the EHT \cite{Gralla et al. (2019),Gralla (2021)}. As a result, despite the groundbreaking achievements of the EHT, there remains ample room for discussion regarding modifications to gravity theories and the accretion environment.

On the other hand, it is interesting to note that for some supermassive black holes at the centers of galaxies, such as Sagittarius A$^{*}$, the accreting material near the event horizon is optically thin at $230$ GHz \cite{Akiyama et al. (2022)}. In other words, following the backward ray-tracing approach, the accretion flow within the innermost stable circular orbit (ISCO) can intercept light rays that would otherwise pass through the black hole's event horizon, allowing the observer to discern a dim region smaller than the traditional shadow. Numerous studies have confirmed that such a dim region can appear when the inner edge of an equatorial, optically and geometrically thin accretion disk coincides with the event horizon, and its boundary is clearly separated from the higher-order bright rings \cite{Gralla et al. (2019),Hou et al. (2022),Hu et al. (2024),Li et al. (2024b),Peng et al. (2021),Li and He (2021),Hu et al. (2022a),Zeng et al. (2022),Guo et al. (2022),Gao et al. (2023),Meng et al. (2023),Wang et al. (2023),Yang et al. (2023),Li et al. (2024a),Sui et al. (2024)}. More importantly, Chael et al. identified a similar region in general relativistic magnetohydrodynamic (GRMHD) simulations in Kerr spacetime, which they named the `inner shadow' to distinguish it from the traditional shadow \cite{Chael et al. (2021)}. Furthermore, researchers have pointed out that the boundary of the inner shadow is a projection of the black hole's event horizon, and its geometric informations have the potential to constrain system parameters such as the viewing angle and spin. Building on this, several authors have explored the geometric properties of the inner shadow of Kerr-Melvin black holes illuminated by an equatorial accretion disk across different parameter spaces, finding that the evolution of the inner shadow with varying parameters aligns with the changes in higher-order photon rings. This has led them to propose that the black hole's inner shadow could serve as a tool for identifying black holes and studying the magnetic field around them \cite{Hou et al. (2022)}.

In astrophysics, the angular momentum of accreting matter is not always aligned with the angular momentum of the black hole, which leads to the formation of tilted accretion disk. There is substantial evidence showing that tilted accretion disks are commonly present in X-ray binary systems \cite{Maccarone (2002),Caproni et al. (2006),van den Eijnden et al. (2017)}, and the tilted accretion disk-jet system of the supermassive black hole M87$^{*}$ has also been confirmed \cite{Cui et al. (2023)}. Research on the structure evolution and energy spectrum of tilted accretion disks has garnered considerable attention in the scientific community \cite{Mckinney et al. (2013),Liska et al. (2019),Liska et al. (2021),Musoke et al. (2023),Liska et al. (2023)}. Inspired by this, Hu and colleagues conducted a simulation of the images of spherically symmetric hairy black holes illuminated by a geometrically thin, tilted accretion disk within the framework of Horndeski theory \cite{Hu et al. (2024)}. They found that the inner shadow of the black hole tends to be obscured by the tilted accretion disk\footnote{The tilted accretion disk acts as a light source, making its obscuration of the black hole's inner shadow seemingly paradoxical. However, this is physically reasonable. From the perspective of backward ray-tracing, the tilted disk can intercept some photons that would otherwise fall into the black hole (In the case of an equatorial accretion disk, these photons would contribute to the inner shadow), resulting in non-zero luminosities for the corresponding observational pixels. In other words, regions that should appear dark (the inner shadow) are illuminated by the tilted accretion disk, while the inner shadow itself becomes obscured.}, and it also rotates along with the precession of the disk. Therefore, they argued that, given the difficulties in constraining the accretion environment, using the inner shadow to test gravity theories requires particular caution. However, the scientific community has reached a consensus on the existence of rotation in astrophysical black holes, indicating that Hu et al. did not tell the whole story. In the scenario of a rotating black hole, does a tilted accretion disk obscure the inner shadow? What insights can the inner shadow provide for black hole physics and high-energy physics? To address these issues, this study defines a tilted plane from a purely geometric perspective to emulate an inclined accretion disk surrounding a Kerr black hole. Employing a ray-tracing algorithm, we compute trajectories for a large number of light rays, identifying those intersecting either the tilted accretion disk or the black hole's event horizon. This approach reveals intriguing features of the inner shadow induced by the disk inclination in rotating black holes.

The remainder of this paper is organized as follows. In section 2, we briefly review the Kerr black hole and the canonical equation of photon motion, which form the foundation of the ray-tracing algorithm. Then, in section 3, we provide a detailed introduction to the backward ray-tracing method and the model of the tilted accretion disk. Before the conclusions and discussion in section 5, we present in section 4 the novel inner shadows cast by the Kerr black hole surrounded by a tilted thin accretion disk across different parameter spaces, and carefully elaborate on their astrophysical implications. Throughout this paper, we adopt geometric units, where the black hole mass $M$, the speed of light $c$, and the gravitational constant $G$ are set to unity.
\section{Kerr metric and null-like geodesics}
Astrophysical black holes are usually described by the Kerr metric, which can be expressed in the Boyer-Lindquist coordinates $x = (t,r,\theta,\varphi)$ with the spacelike signature $(-,+,+,+)$ as \cite{Papoutsis et al. (2023),Hioki and Maeda (2009),Cunha et al. (2016),Pu et al. (2016),Cadavid et al. (2022)}
\begin{eqnarray}\label{1}
\textrm{d}s^{2} &=& -\left(1-\frac{2r}{\Sigma}\right)\textrm{d}t^{2}-\frac{4ar\sin^{2}\theta}{\Sigma}\textrm{d}t\textrm{d}\varphi+\frac{\Sigma}{\Delta}\textrm{d}r^{2} +\Sigma \textrm{d}\theta^{2} \nonumber \\
&& +\sin^{2}\theta\left(r^{2}+a^{2}+\frac{2a^{2}r\sin^{2}\theta}{\Sigma}\right)\textrm{d}\varphi^{2},
\end{eqnarray}
where $a$ denotes the black hole spin parameter, and $\Sigma$ and $\Delta$ are given by
\begin{eqnarray}\label{2}
\Sigma = r^{2}+a^{2}\cos^{2}\theta,
\end{eqnarray}
\begin{equation}\label{3}
\Delta = r^{2}-2r+a^{2}.
\end{equation}
By solving the equation $\Delta = 0$, we obtain the radius of the event horizon $r_{\textrm{e}}$, expressed as
\begin{equation}\label{4}
r_{\textrm{e}} = 1+\sqrt{1-a^{2}}.
\end{equation}
It is important to emphasize that the size of the inner shadow of a black hole is associated with the event horizon \cite{Chael et al. (2021)}. Consequently, it is anticipated that the inner shadow of a Kerr black hole shrinks as the spin parameter increases. Given that there is an upper limit to the spin parameter for a Kerr black hole solution, the size of the inner shadow theoretically has a minimum value. We will delve into this topic in detail in section 4.

The motion of photons in the Kerr spacetime satisfies the Lagrangian formula
\begin{equation}\label{5}
\mathscr{L} = \frac{1}{2}g_{\mu\nu}\dot{x^{\mu}}\dot{x^{\nu}},
\end{equation}
where $\dot{x^{\mu}}$ represents the derivative of the generalized coordinates with respect to an affine parameter $\lambda$, and $g_{\mu\nu}$ denotes the covariant metric tensor, which can be read from spacetime line elements. Based on the Legendre transformation, we further obtain the Hamiltonian function governing the motion of the photon:
\begin{equation}\label{6}
\mathscr{H} = p_{\mu}\dot{x^{\mu}} - \mathscr{L} = \frac{1}{2}g^{\mu\nu}p_{\mu}p_{\nu},
\end{equation}
where $g^{\mu\nu}$ is the contravariant metric tensor, and $p_{\mu}$ denotes the covariant four-momentum, which relate to the four-velocity as
\begin{eqnarray}\label{7}
p_{t} = -\left(1-\frac{2r}{\Sigma}\right)\dot{t}-\frac{2ar\sin^{2}\theta}{\Sigma}\dot{\varphi},
\end{eqnarray}
\begin{equation}\label{8}
p_{r} = \frac{\Sigma}{\Delta}\dot{r},
\end{equation}
\begin{equation}\label{9}
p_{\theta} = \Sigma \dot{\theta},
\end{equation}
\begin{equation}\label{10}
p_{\varphi} = \sin^{2}\theta\left(r^{2}+a^{2}+\frac{2a^{2}r\sin^{2}\theta}{\Sigma}\right)\dot{\varphi}-\frac{2ar\sin^{2}\theta}{\Sigma}\dot{t}.
\end{equation}

It is necessary to point out that since the Kerr spacetime is of Petrov type D, the metric is independent of the coordinates $t$ and $\varphi$. Consequently, $p_{t}$ and $p_{\varphi}$ are conserved quantities in photon motion, corresponding to the specific energy $E=-p_{t}$ and specific angular momentum $L=p_{\varphi}$, respectively. Hence, we obtain the specific expression of equation \eqref{6} as
\begin{eqnarray}\label{11}
\mathscr{H} =\frac{1}{2}\left[-\frac{\left(\Delta + 2r\right)^{2}-\Delta a^{2}\sin^{2}\theta}{\Delta\Sigma}(-E)^{2}+\frac{\Delta}{\Sigma}p_{r}^{2}+\frac{p_{\theta}^{2}}{\Sigma}+\frac{\Sigma-2r}{\Delta \Sigma \sin^{2}\theta}L^{2}-\frac{4ar}{\Delta \Sigma}(-E)L\right].
\end{eqnarray}
By utilizing the canonical equations
\begin{equation}\label{12}
\dot{x^{\mu}}=\frac{\partial \mathscr{H}}{\partial p_{\mu}}, \quad \dot{p_{\mu}} = -\frac{\partial \mathscr{H}}{\partial x^{\mu}},
\end{equation}
one can conveniently compute the propagation of light rays in curved spacetime, thereby circumventing the turning point issue that arises when solving $\dot{\theta}^{2}$ and $\dot{r}^{2}$.
\section{Ray-tracing method}
To simulate black hole images, it is essential to trace the trajectory of each photon within the local frame of the black hole. In the context of the backward ray-tracing algorithm, the initial conditions for the photons are typically specified in the observer's local coordinate system. Therefore, the primary task in ray-tracing is to transform initial position and four-momentum into the local coordinate system of the black hole.
\begin{figure*}
\center{
\includegraphics[width=8cm]{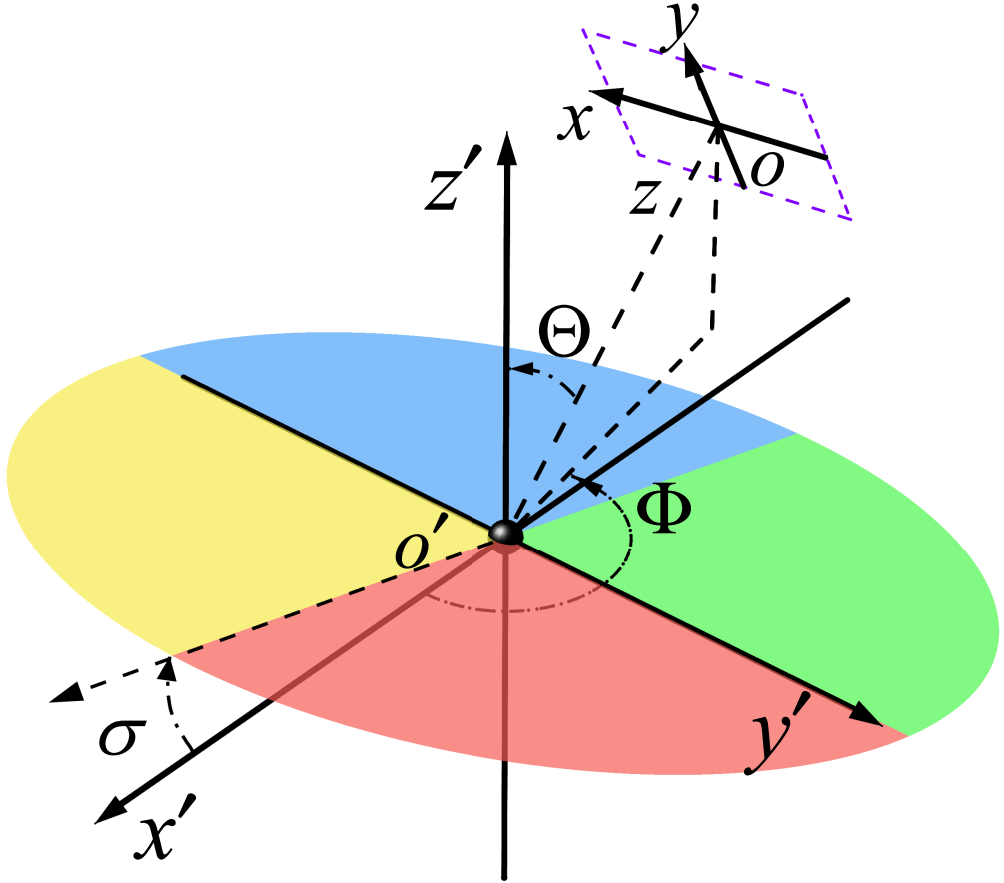}
\caption{Diagram of the ray-tracing method. The local reference frames of the black hole and the observer are denoted by $(x^{\prime}, y^{\prime}, z^{\prime})$ and $(x, y, z)$, respectively, with the observer's screen located in the $\overline{xoy}$ plane. $\Theta$ and $\Phi$ represent the observer's inclination angle and azimuthal angle, respectively. The disk, idealized as a tilted thin accretion disk, is composed of four distinct color-coded quadrants, with its inner boundary connected to the event horizon of the black hole. The tilt angle $\sigma$ is defined as the angle between the black hole's equatorial plane $\overline{x^{\prime} o^{\prime} y^{\prime}}$ and the tilted accretion disk. Following the positive direction of $\Phi$, the quadrants of the disk are colored red, green, blue, and yellow, respectively. By employing numerical integration algorithms, the trajectories of light rays originating from each pixel $(x, y, 0)$ on the observer's screen can be tracked, with those terminating at the event horizon forming the inner shadow.}}\label{fig1}
\end{figure*}

As shown in figure 1, the local frames of the observer and the Kerr black hole are represented by quasi-Cartesian coordinate systems $(x,y,z)$ and $(x^{\prime}, y^{\prime}, z^{\prime})$, respectively. Here, the $z$-axis coincides with $\overline{oo^{\prime}}$ and points towards the point $o^{\prime}$, with an angle $\Theta$ relative to the black hole's rotation axis ($z^{\prime}$-axis), which defined as the viewing angle. The observer's screen, depicted as a purple block, lies in $\overline{xoy}$ plane, thus $(x,y,0)$ naturally denotes the initial position of the photon. The solid disk inclined at an angle $\sigma$ to the black hole's equatorial plane represents a tilted, geometrically thin accretion disk. It is divided into four regions of different colors by quadrants to vividly illustrate the black hole's inner shadow and the spacetime distortion. Consequently, for each set of system parameters, the observation plane will exhibit regions colored red, yellow, blue, and green contributed by rays hitting the tilted plane, along with black regions corresponding to rays that fall into the black hole. It is important to note that for the equatorial accretion disk scenario $(\sigma = 0^{\circ})$, the black hole image is independent of the observer's azimuthal angle. However, this property is expected to not hold when a non-zero $\sigma$ is introduced. Therefore, an additional parameter, the observation azimuth $\Phi$, is included in our model.

Next, we derive the initial position of light rays in the black hole's local coordinate system. From figure 1, it can be observed that under the condition of asymptotically flatness of spacetime, the observer's local coordinate system $(x, y, z)$ can be aligned with the black hole's local coordinate system $(x^{\prime}, y^{\prime}, z^{\prime})$ through rotation and translation. Thus, the light ray emitted from a pixel $(x, y, z)$ on the observer's plane can be expressed in the black hole's local coordinate system as \cite{Lin et al. (2022),Pu et al. (2016),Younsi et al. (2016)}
\begin{eqnarray}\label{13}
x^{\prime} = \mathscr{T}\cos\Phi - x\sin\Phi,
\end{eqnarray}
\begin{equation}\label{14}
y^{\prime} = \mathscr{T}\sin\Phi + x\cos\Phi,
\end{equation}
\begin{equation}\label{15}
z^{\prime} = \left(r_{\textrm{obs}} - z\right)\cos\Theta + y\sin\Theta,
\end{equation}
where $r_{\textrm{obs}}$ denotes the observation distance(the length of $\overline{oo^{\prime}}$), and $\mathscr{T}$ takes the form
\begin{equation}\label{16}
\mathscr{T} = \left(\sqrt{r_{\textrm{obs}}^{2}+a^{2}}-z\right)\sin\Theta-y\cos\Theta.
\end{equation}
Further, we obtain the initial position of the light ray described in Boyer-Lindquist coordinates, given by
\begin{eqnarray}\label{17}
r = \sqrt{\frac{\mathscr{D}+\sqrt{\mathscr{D}^{2}+4a^{2}z^{\prime2}}}{2}},
\end{eqnarray}
\begin{equation}\label{18}
\theta = \arccos\left(\frac{z^{\prime}}{r}\right),
\end{equation}
\begin{equation}\label{19}
\varphi = \textrm{atan}2\left(y^{\prime},x^{\prime}\right),
\end{equation}
where $\mathscr{D}$ is expressed as
\begin{equation}\label{20}
\mathscr{D} = x^{\prime2} + y^{\prime2} + z^{\prime2} - a^{2}.
\end{equation}

To integrate the geodesic equations, in addition to determining the initial position of the photon, it is also necessary to specify its covariant four-momentum. Considering that the observer is situated in asymptotically flat spacetime and that the photon is emitted perpendicular to the observation plane, we can safely assume that the initial velocity of the photon, as measured by the observer, is $(\dot{x}, \dot{y}, \dot{z}) = (0, 0, 1)$. Substituting this into the differential forms of equations \eqref{13}-\eqref{15}, we obtain
\begin{eqnarray}\label{21}
\dot{x}^{\prime} = -\sin\Theta\cos\Phi,
\end{eqnarray}
\begin{equation}\label{22}
\dot{y}^{\prime} = -\sin\Theta\sin\Phi,
\end{equation}
\begin{equation}\label{23}
\dot{z}^{\prime} = -\cos\Theta.
\end{equation}
Subsequently, by inserting $(\dot{x}^{\prime}, \dot{y}^{\prime}, \dot{z}^{\prime})$ into the differential forms of equations \eqref{17}-\eqref{19}, the components of the photon's velocity in Boyer-Lindquist coordinates are given by \cite{Lin et al. (2022),Pu et al. (2016),Younsi et al. (2016)}
\begin{eqnarray}\label{24}
\dot{r} = -\frac{r\mathscr{R}\sin\theta\sin\Theta\cos\Psi+\mathscr{R}^{2}\cos\theta\cos\Theta}{\Sigma},
\end{eqnarray}
\begin{equation}\label{25}
\dot{\theta} = \frac{r\sin\theta\cos\Theta-\mathscr{R}\cos\theta\sin\Theta\cos\Psi}{\Sigma},
\end{equation}
\begin{equation}\label{26}
\dot{\varphi} = \frac{\sin\Theta\sin\Psi}{\mathscr{R}\sin\theta},
\end{equation}
where $\mathscr{R}$ and $\Psi$ are defined as $\sqrt{r^{2}+a^{2}}$ and $(\varphi-\Phi)$, respectively.

Based on equations \eqref{8} and \eqref{9}, it is straightforward to compute $p_{r}$ and $p_{\theta}$ using equations \eqref{24} and \eqref{25}. However, due to the presence of the coupling term $g_{t\varphi}$, the calculation of $p_{t}$ and $p_{\varphi}$ is not as intuitive. Recalling equation \eqref{7}, we have
\begin{equation}\label{27}
\dot{t} =\left(\frac{\Sigma}{\Sigma-2r}\right)\left(-p_{t} - \frac{2ar\sin^{2}\theta}{\Sigma}\dot{\varphi}\right).
\end{equation}
Substituting $\dot{t}$ into the Lagrangian formula \eqref{5} and applying the photon constraint $\mathscr{L} = 0$, we obtain
\begin{equation}\label{28}
(-p_{t})^{2} = \left(\frac{\Sigma-2r}{\Sigma\Delta}\right)\left(\Sigma\dot{r}^{2}+\Sigma\Delta\dot{\theta}^{2}\right)+\Delta\dot{\varphi}^{2}\sin^{2}\theta.
\end{equation}
Next, $p_{\varphi}$ can be written as
\begin{equation}\label{29}
p_{\varphi} = \frac{\left(\Sigma\Delta\dot{\varphi}+2arp_{t}\right)\sin^{2}\theta}{\Sigma-2r}.
\end{equation}
Now, we have obtained the initial conditions of the photon in the local frame of the black hole as $(t, r, \theta, \phi, p_{t}, p_{r}, p_{\theta}, p_{\phi})$. Using a fifth- and sixth-order Runge-Kutta-Fehlberg integrator with adaptive step size (RKF56), we can trace the propagation of the photon corresponding to each pixel on the observation plane, thereby simulating the black hole's inner shadow. It is crucial to emphasize that when determining whether a photon crosses the black hole's event horizon, achieving a photon radial coordinate exactly satisfying $r=r_{\textrm{e}}$ is numerically improbable, as this constitutes a coordinate singularity. To address this, we implement a numerical criterion: once a photon's position meets $r\leq r_{\textrm{e}}+10^{-4}$, the photon is deemed to have entered the black hole, and the simulation for that particular ray is terminated.

Additionally, to improve the efficiency of the ray-tracing method, we supplemented the RKF56 with an orbit-adaptive step size adjustment, which significantly enhances the convergence speed of the algorithm. Specifically, the iteration step size $\tau$ in the algorithm is dynamically adjusted based on the distance between the photon and the black hole's event horizon, $\tau = \tau_{0}(r/r_{\textrm{e}})^{\kappa}$. In this paper, we set $\tau_{0}$ and $\kappa$ to $10^{-5}$ and $1.8$, respectively.
\section{Results}
In this section, we fix the observation distance at $r_{\textrm{obs}} = 1000$ M and the field of view at $16 \times 16$ M with a resolution of $400 \times 400$ pixels to numerically simulate the inner shadow of a Kerr black hole with a tilted thin accretion disk across different parameter spaces. It is worth noting that, in the case of an equatorial accretion disk, the black hole image remains independent of the observer's azimuth $\Phi$. However, this invariance does not hold for an inclined accretion disk scenario. Therefore, we present the results for $\Phi = 0^{\circ}$ and $\Phi = 90^{\circ}$ separately and analyze the impact of the accretion environment on the inner shadow.
\subsection{$\Phi=0^{\circ}$ case}
When the observer's azimuth is fixed at zero, we numerically simulate the inner shadows of the Kerr black holes with a tilted thin accretion disk with varying inclinations for observation angle of $17^{\circ}$, $50^{\circ}$, and $85^{\circ}$. The results are presented in figures 2-4, respectively.
\begin{figure*}
\center{
\includegraphics[width=2.8cm]{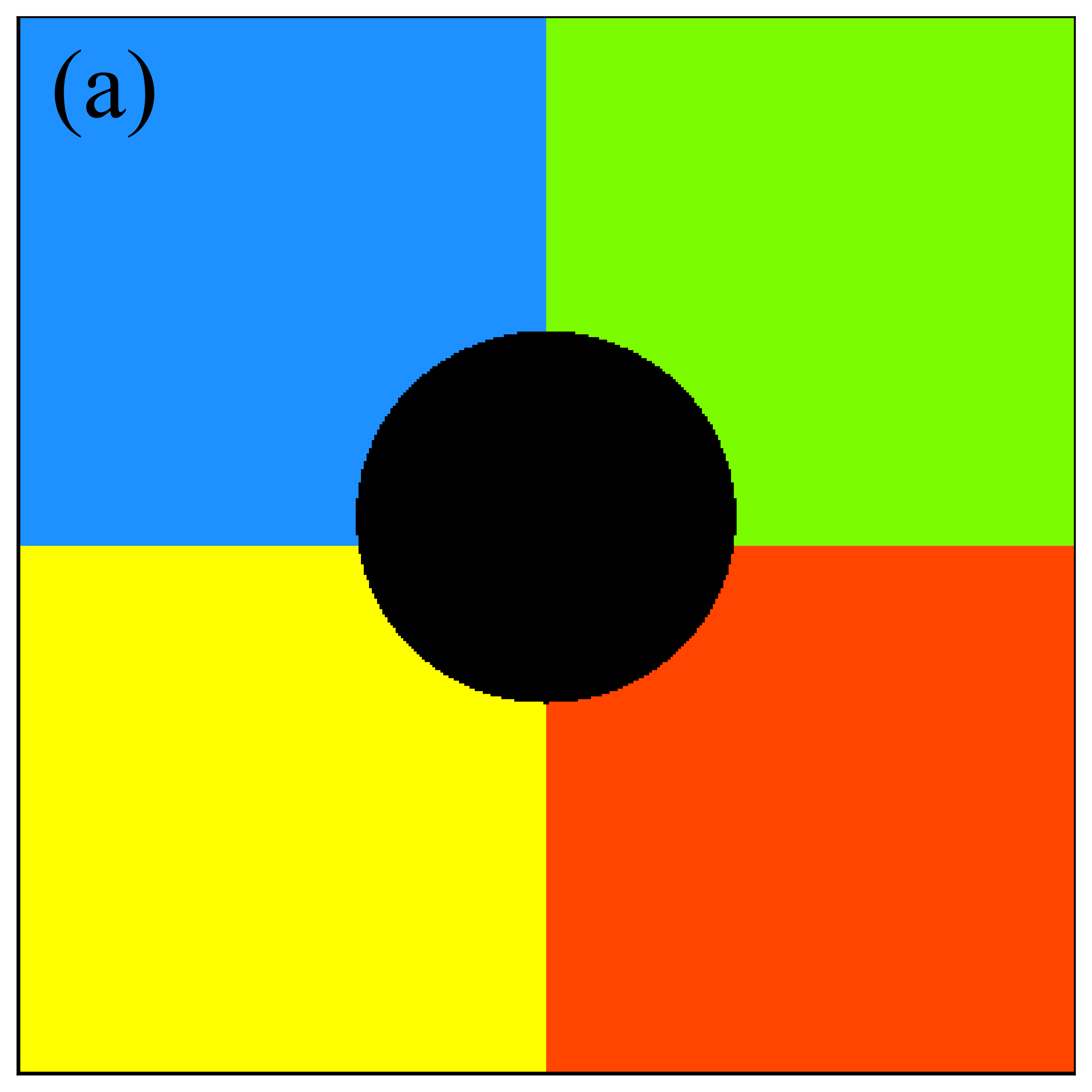}
\includegraphics[width=2.8cm]{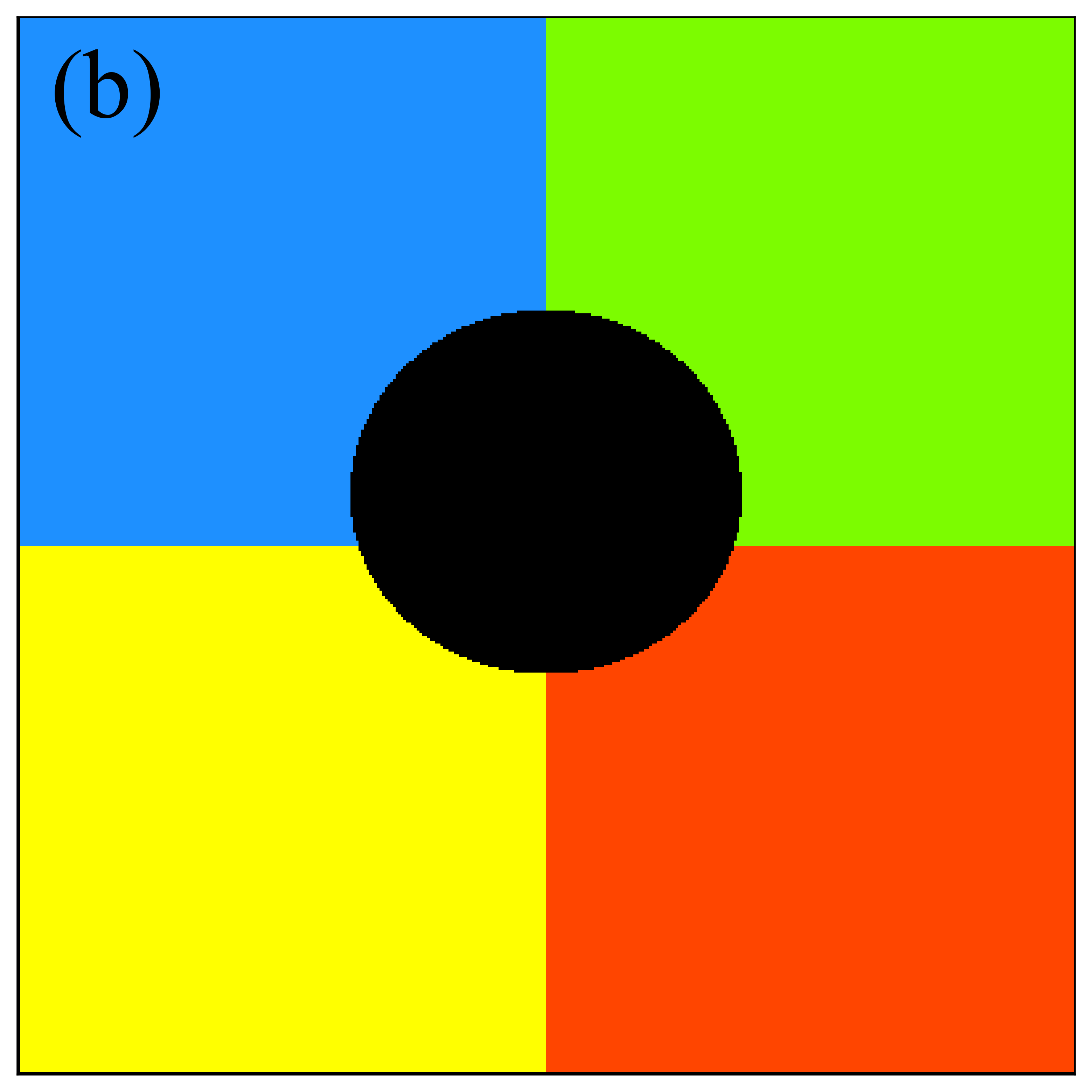}
\includegraphics[width=2.8cm]{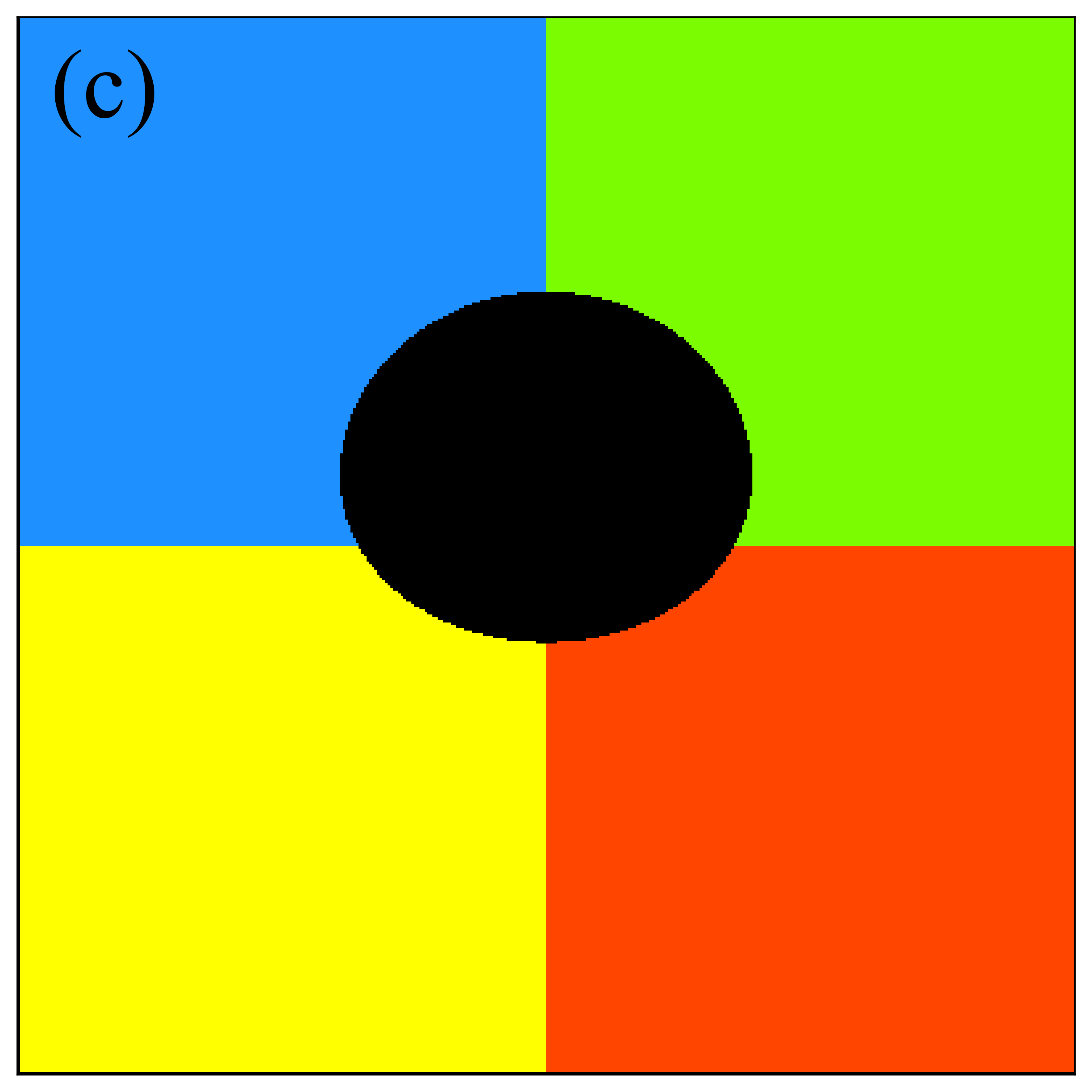}
\includegraphics[width=2.8cm]{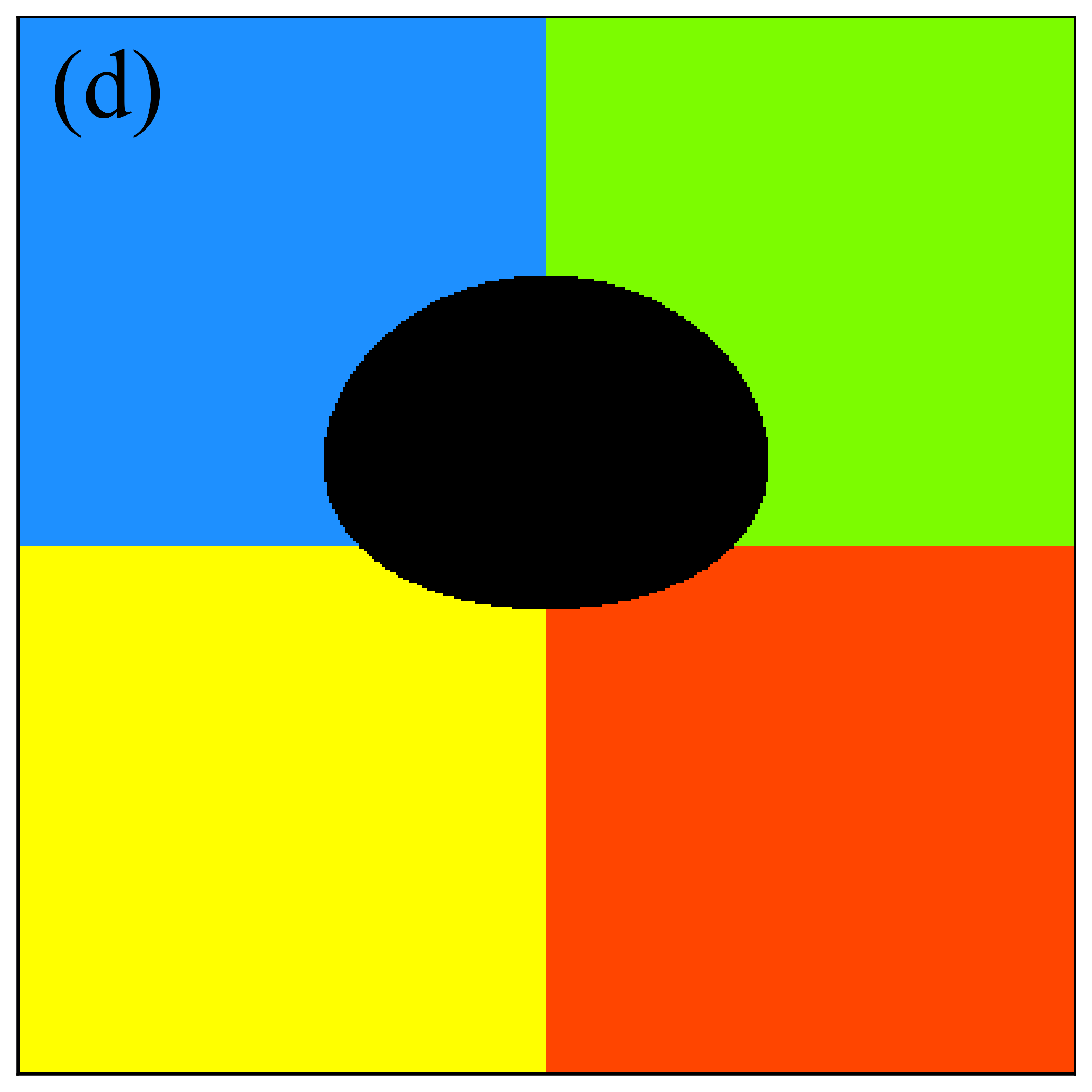}
\includegraphics[width=2.8cm]{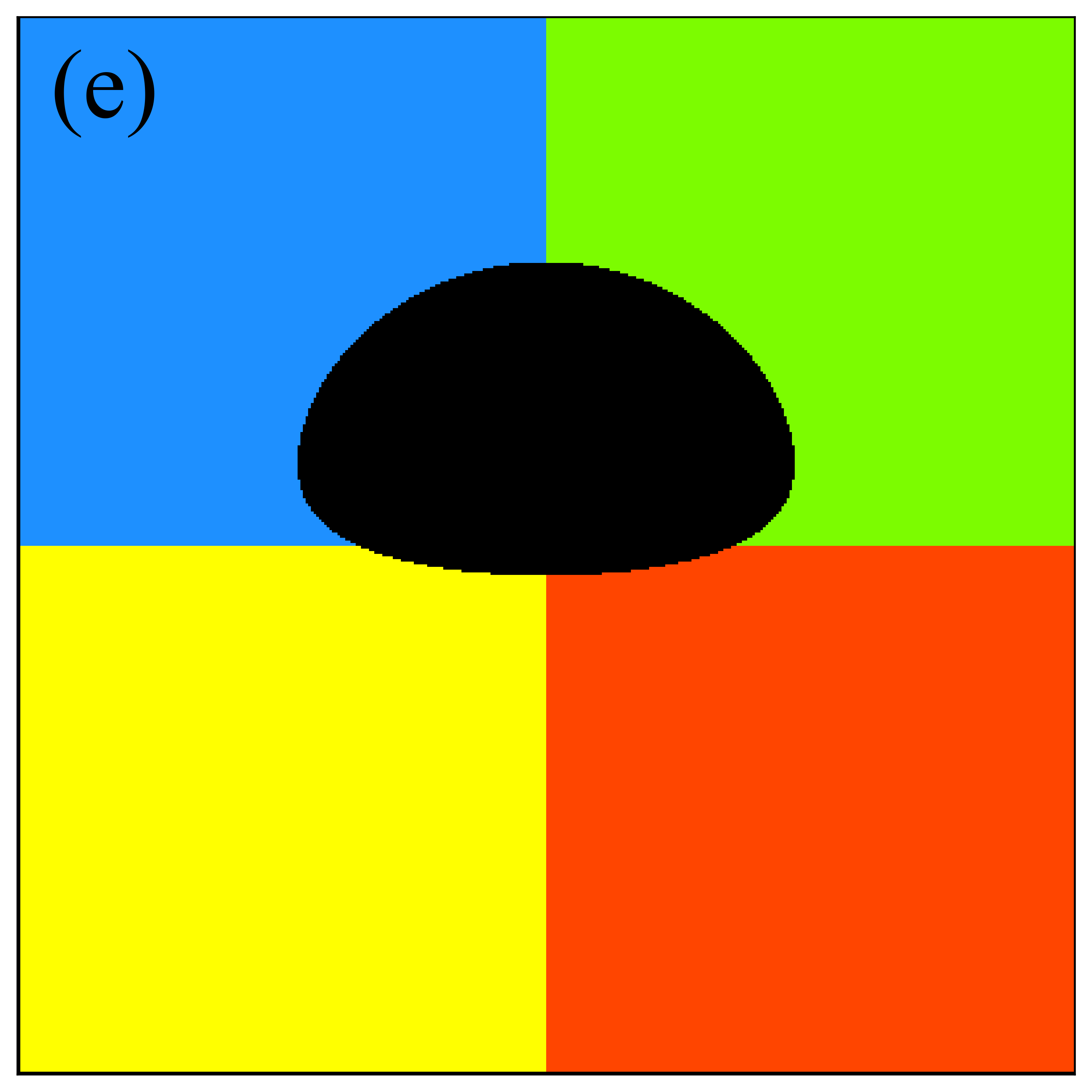}
\includegraphics[width=2.8cm]{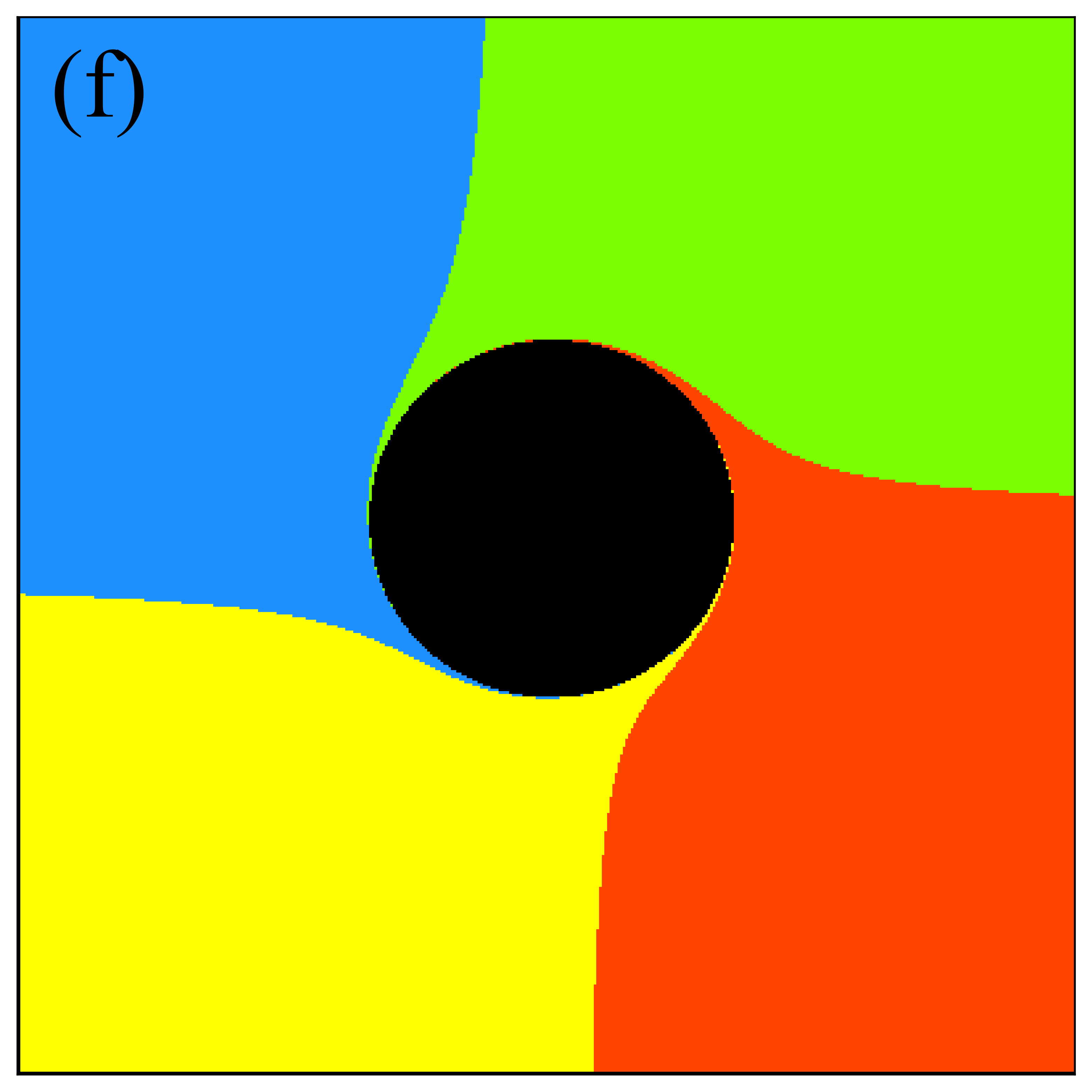}
\includegraphics[width=2.8cm]{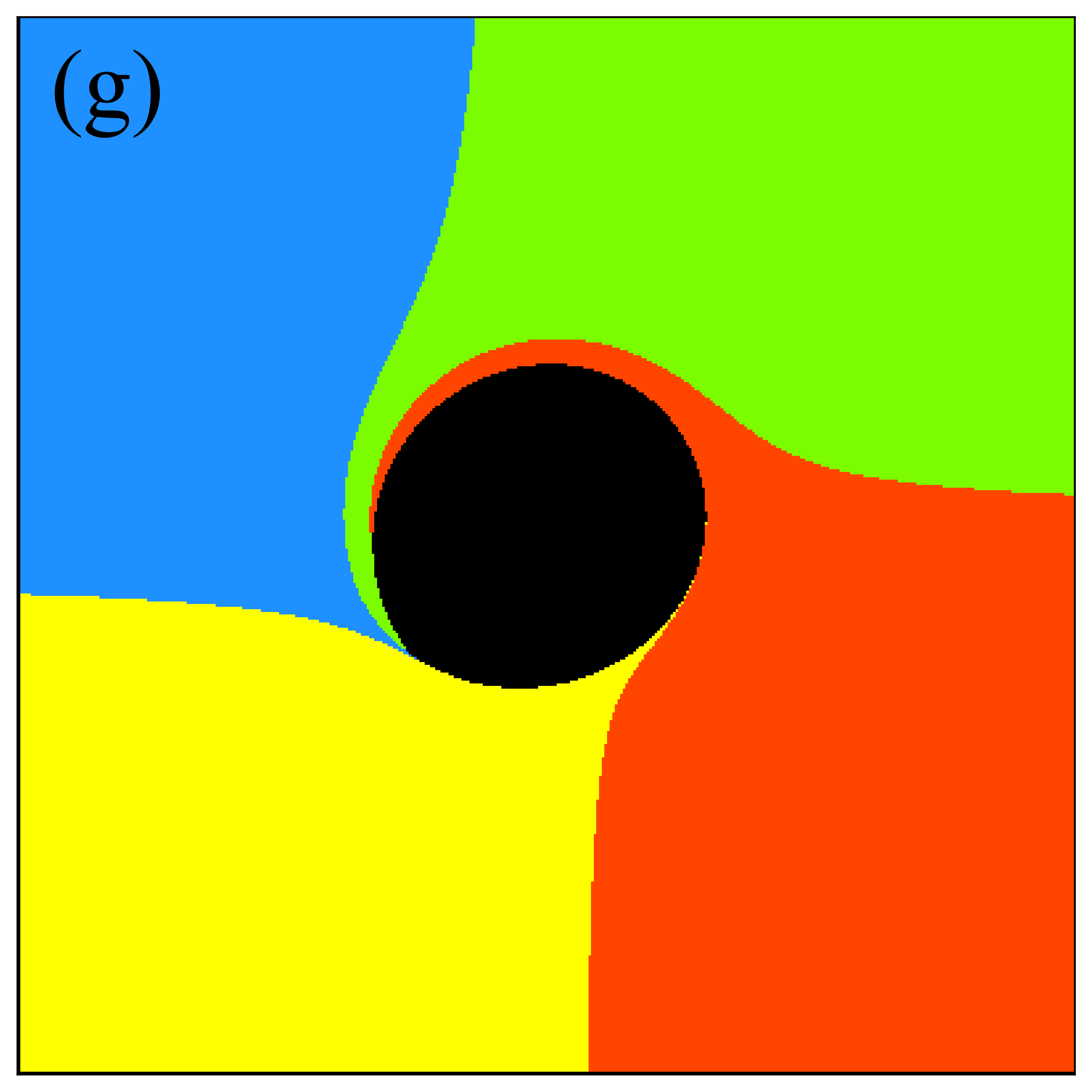}
\includegraphics[width=2.8cm]{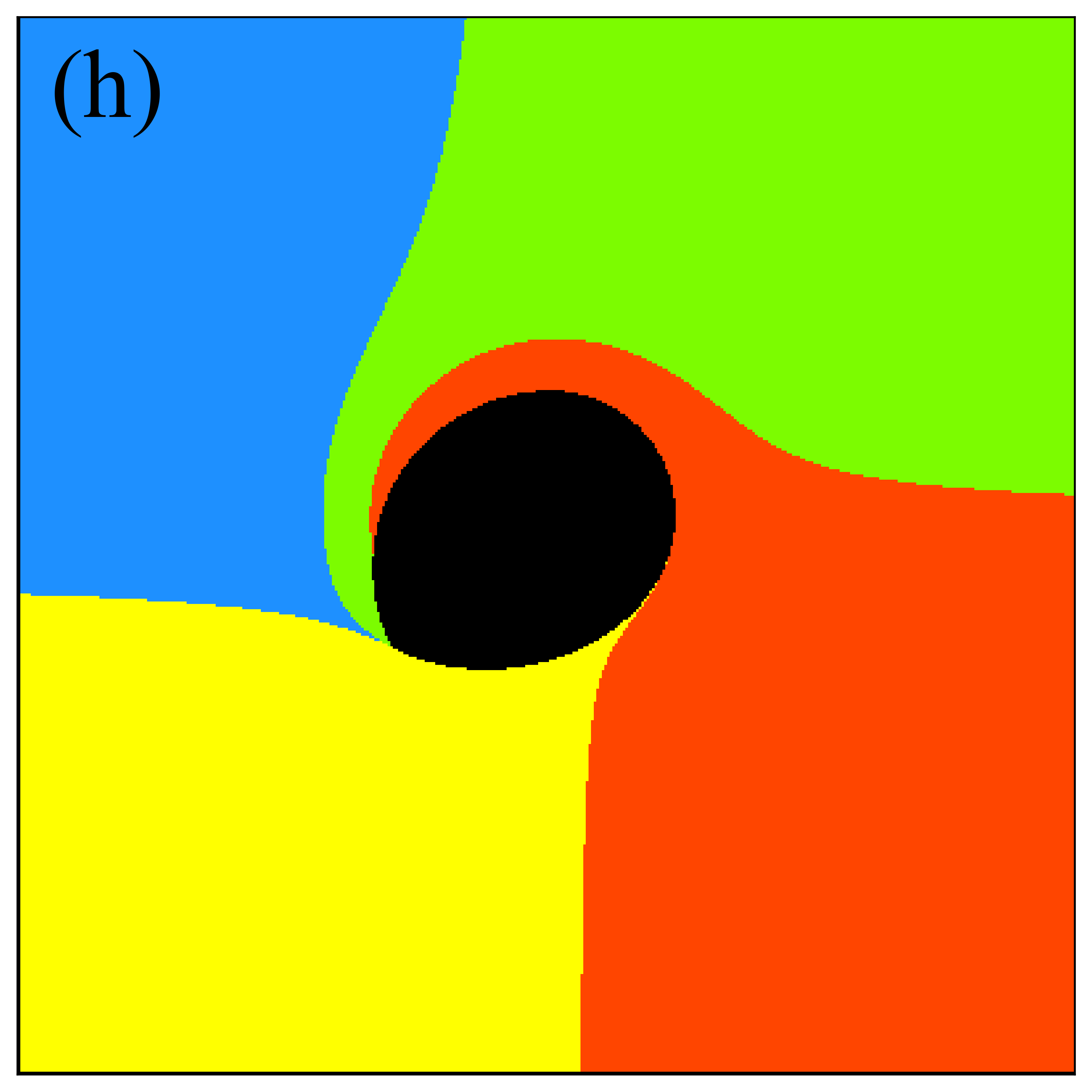}
\includegraphics[width=2.8cm]{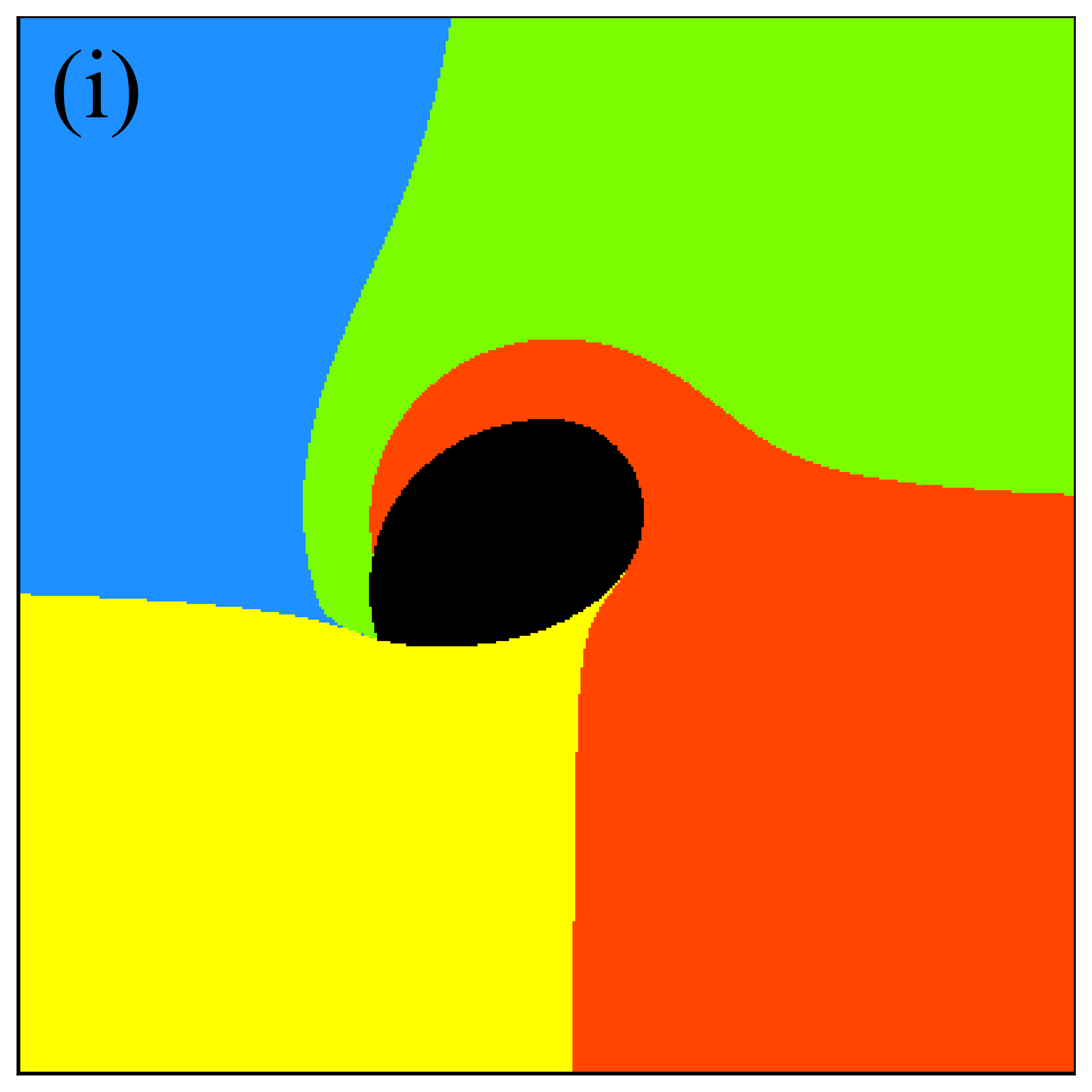}
\includegraphics[width=2.8cm]{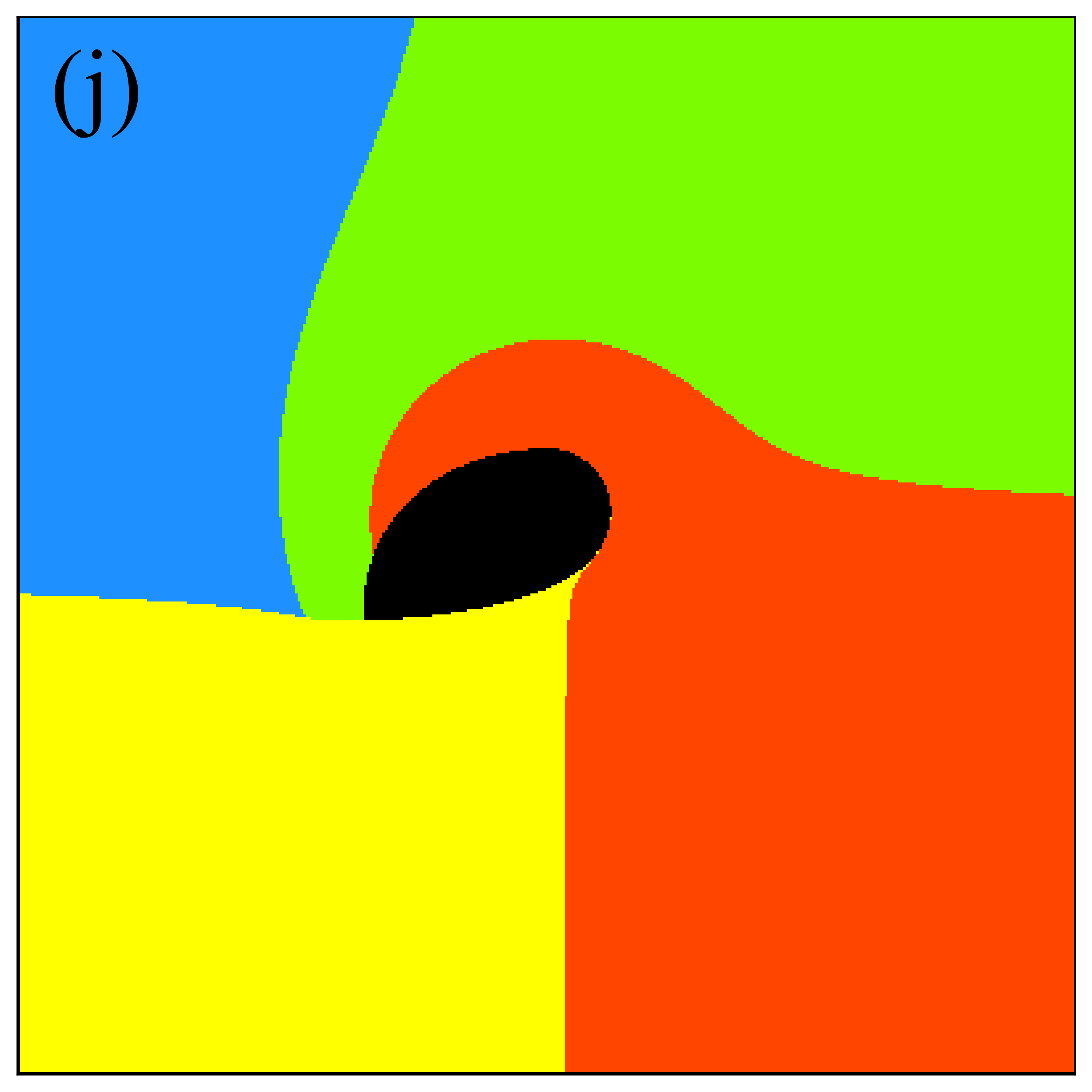}
\includegraphics[width=2.8cm]{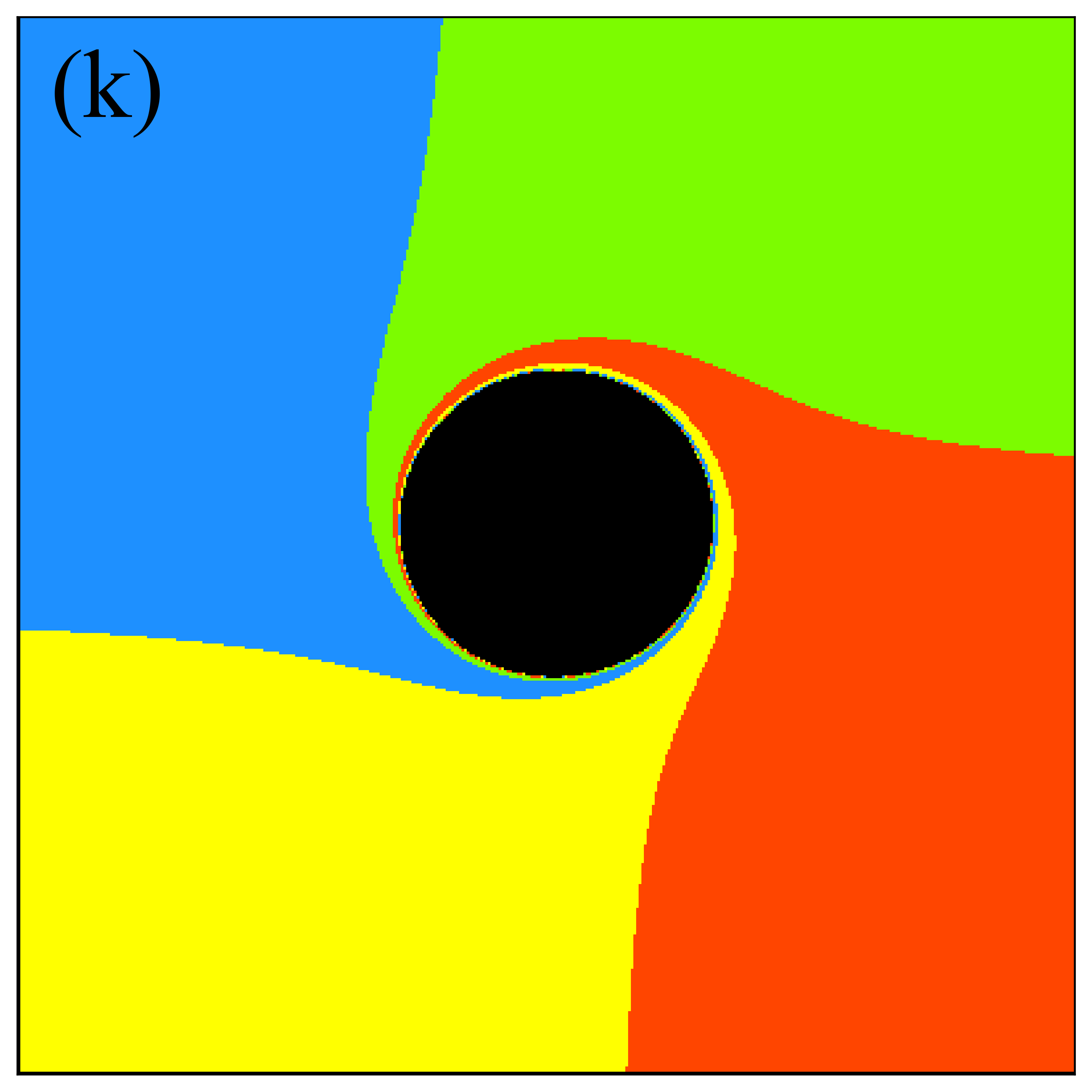}
\includegraphics[width=2.8cm]{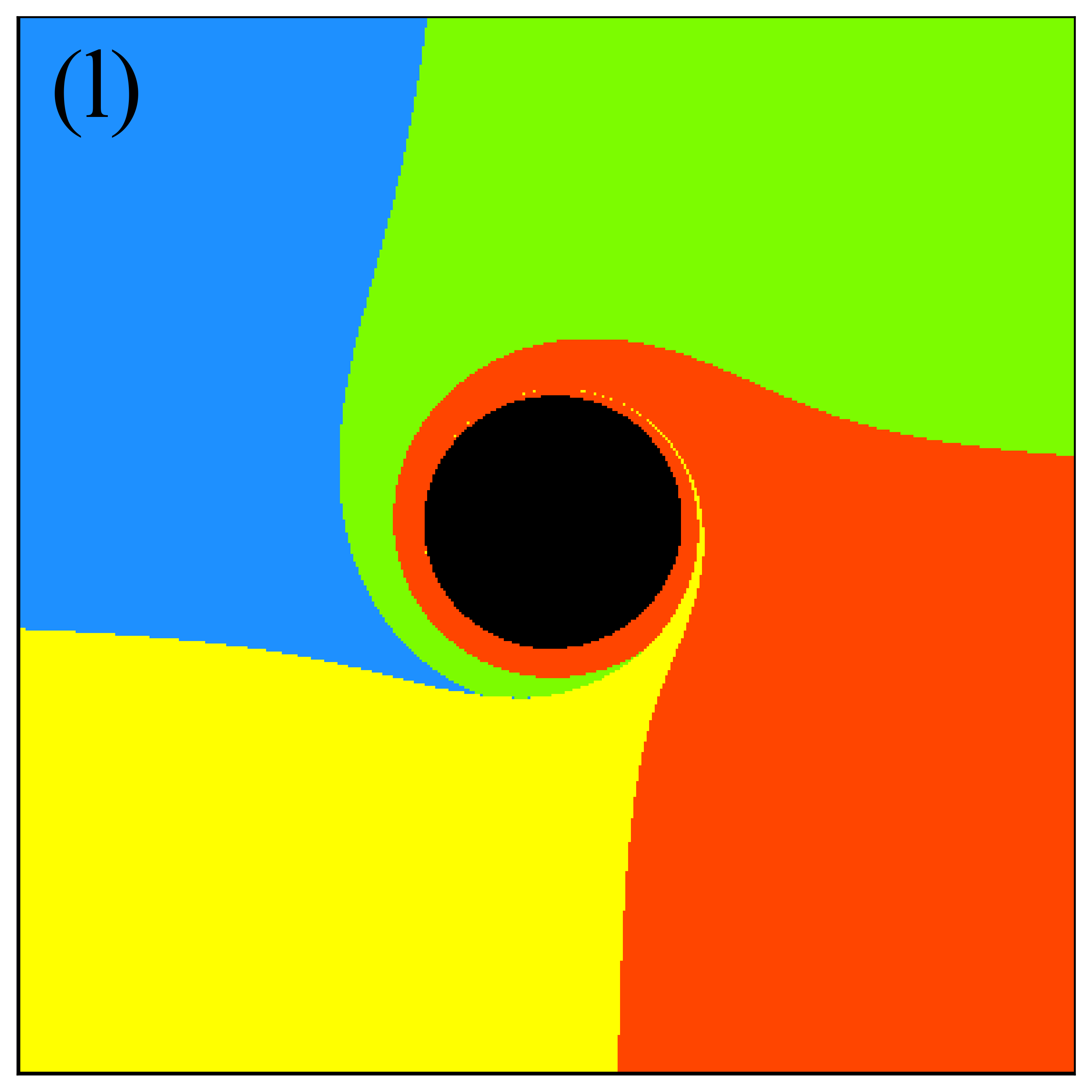}
\includegraphics[width=2.8cm]{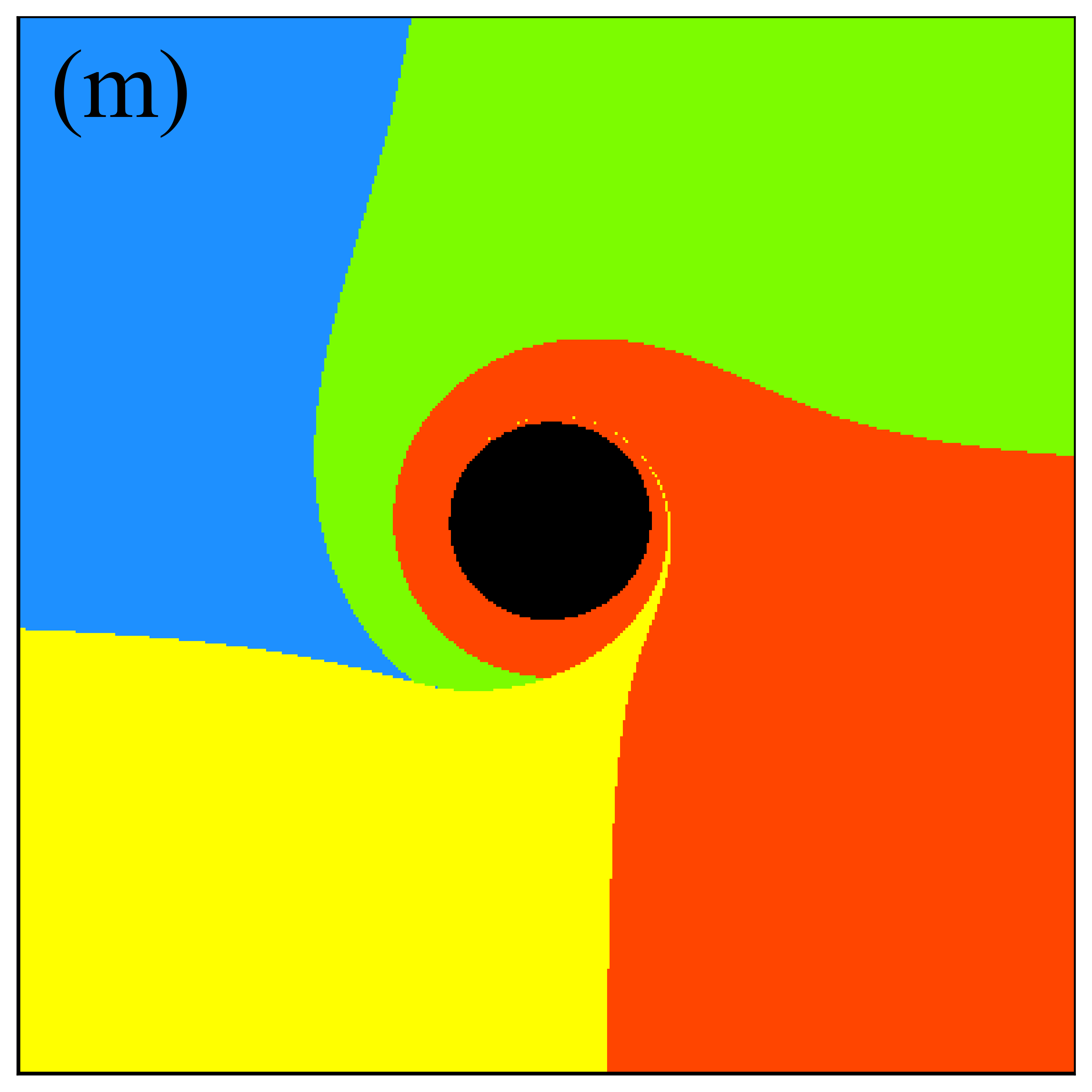}
\includegraphics[width=2.8cm]{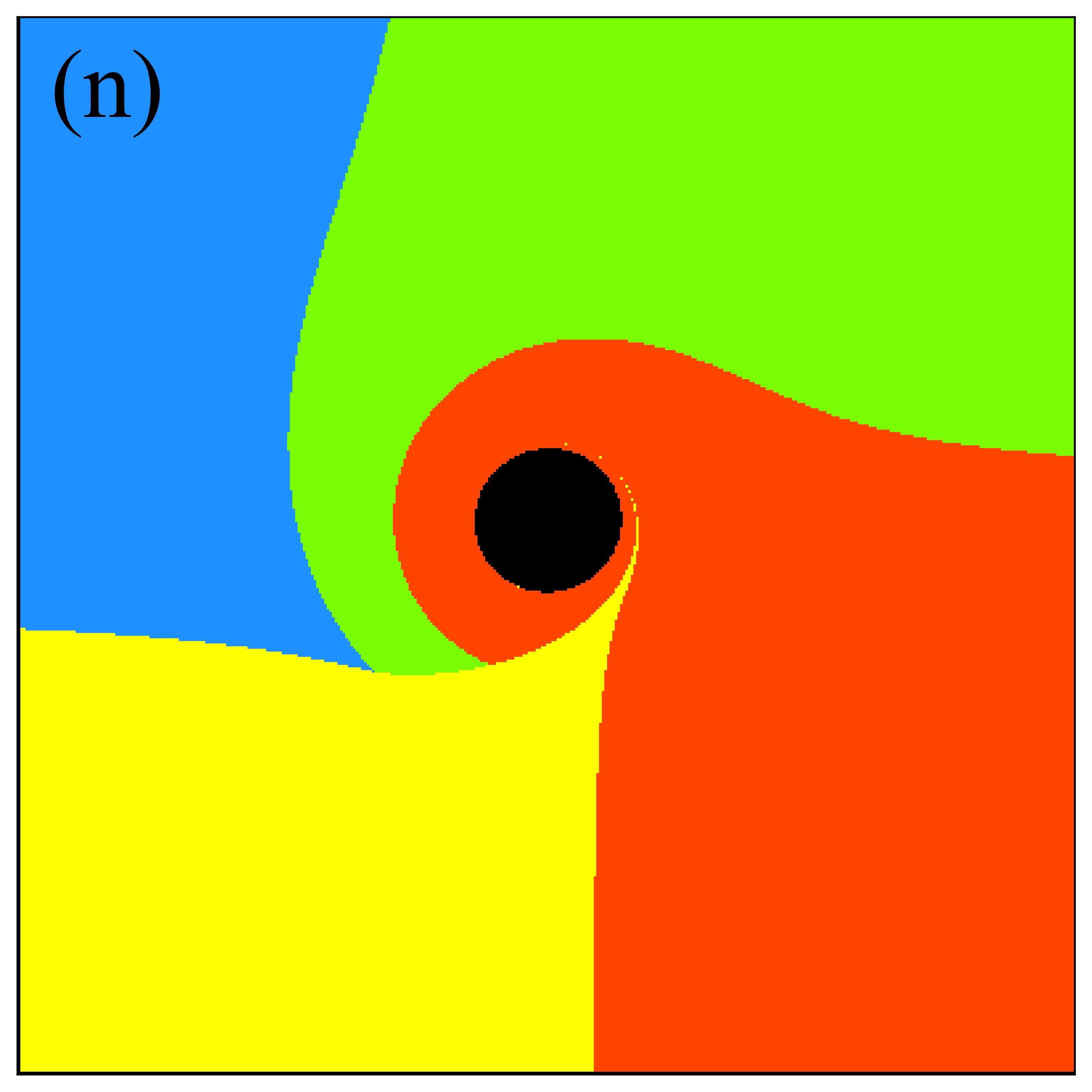}
\includegraphics[width=2.8cm]{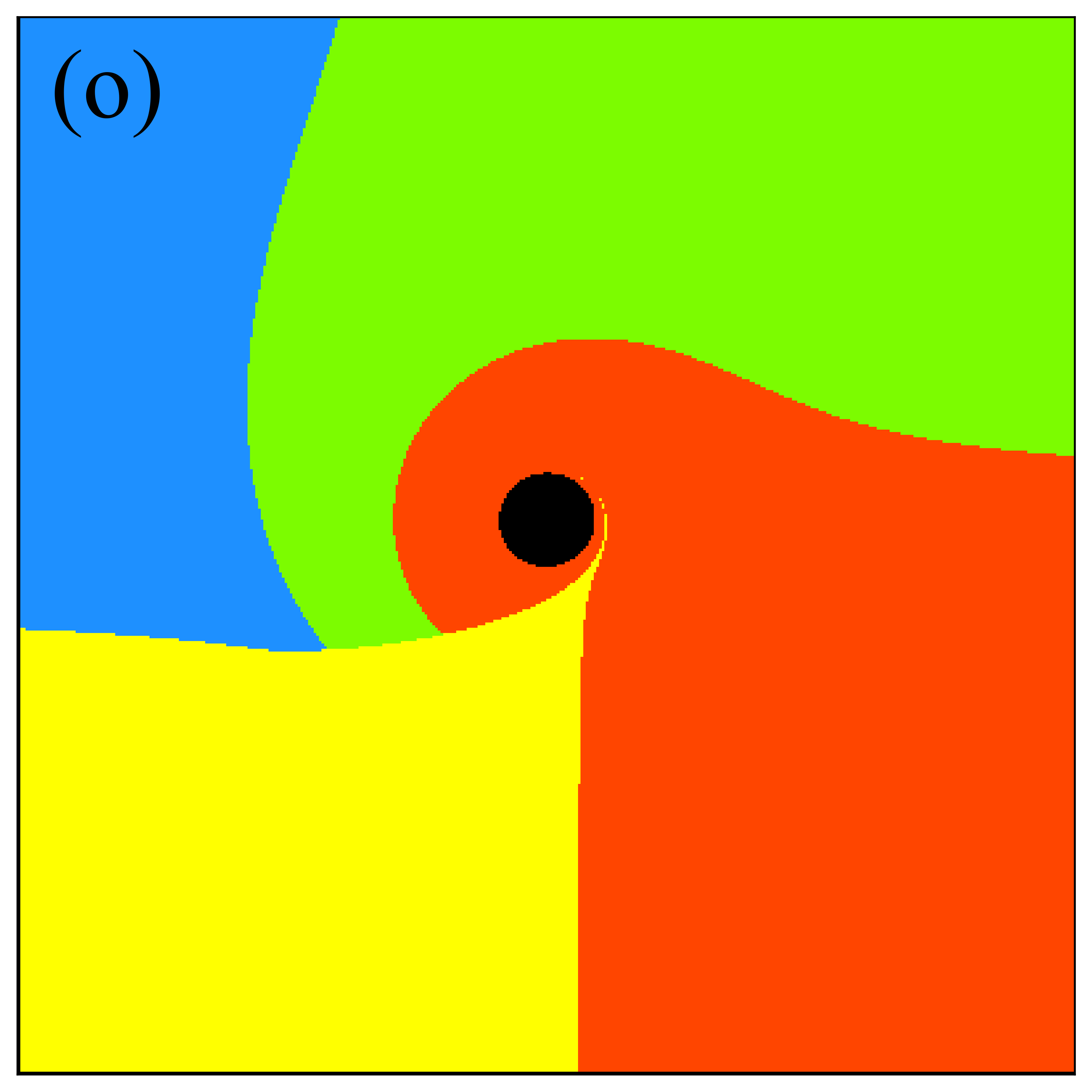}
\includegraphics[width=2.8cm]{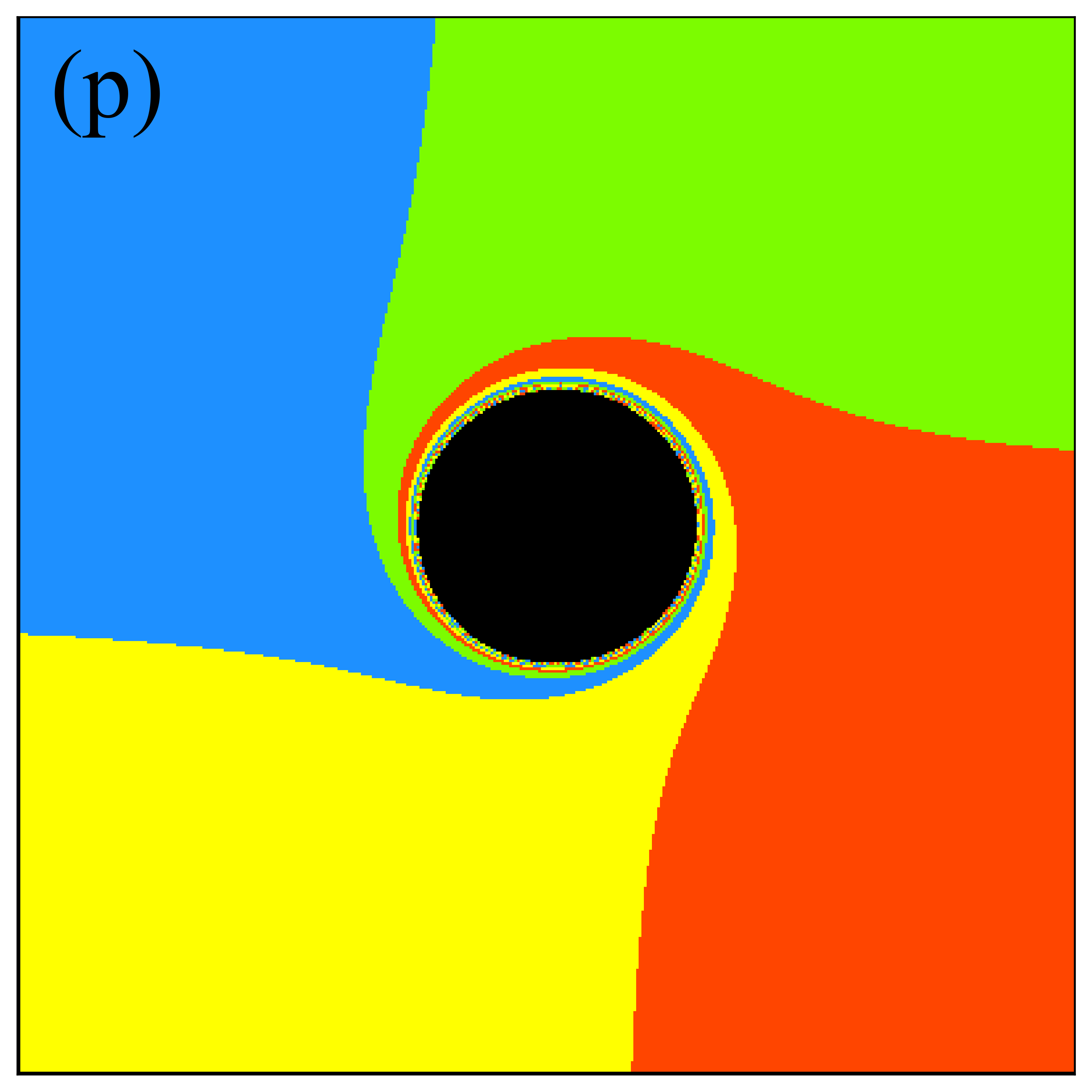}
\includegraphics[width=2.8cm]{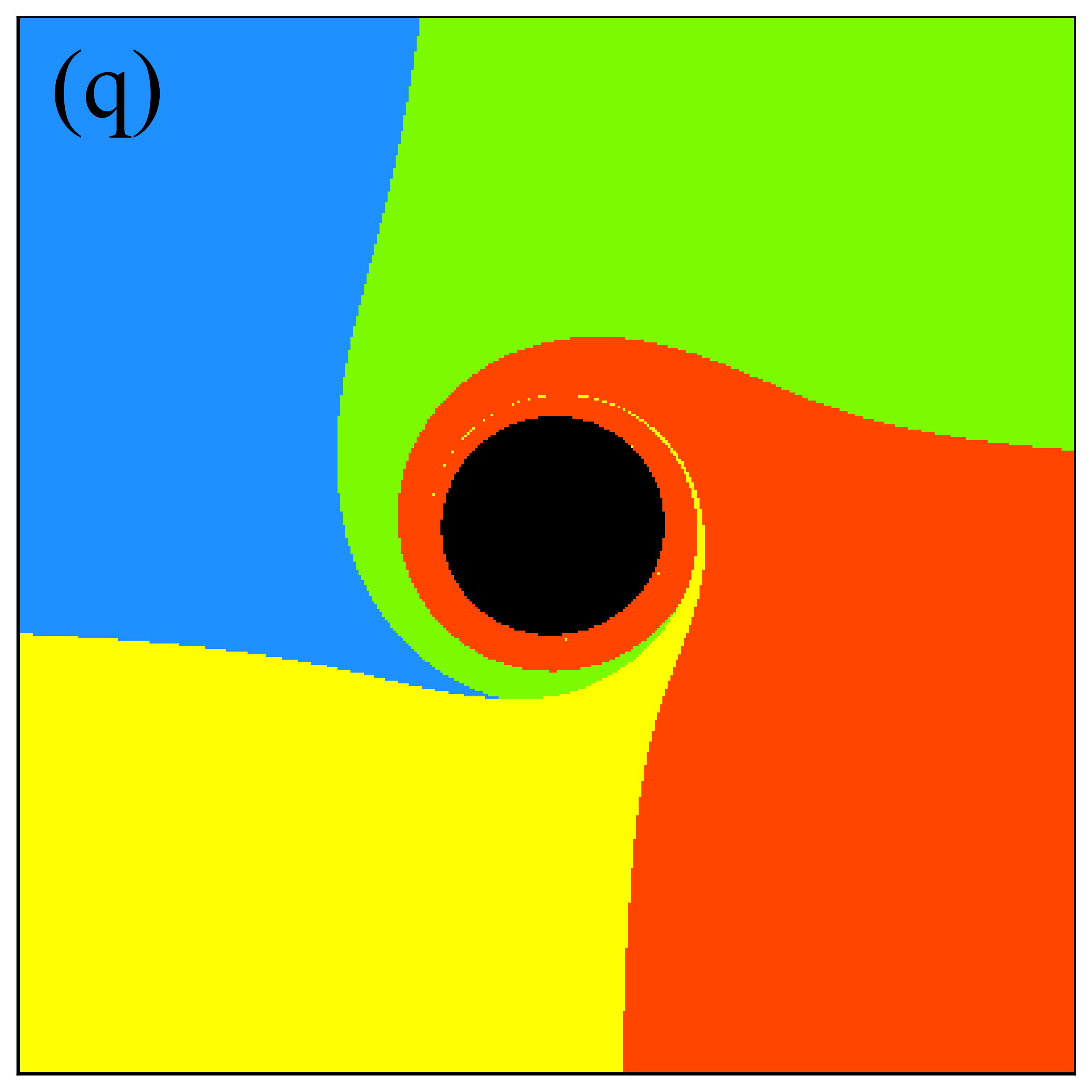}
\includegraphics[width=2.8cm]{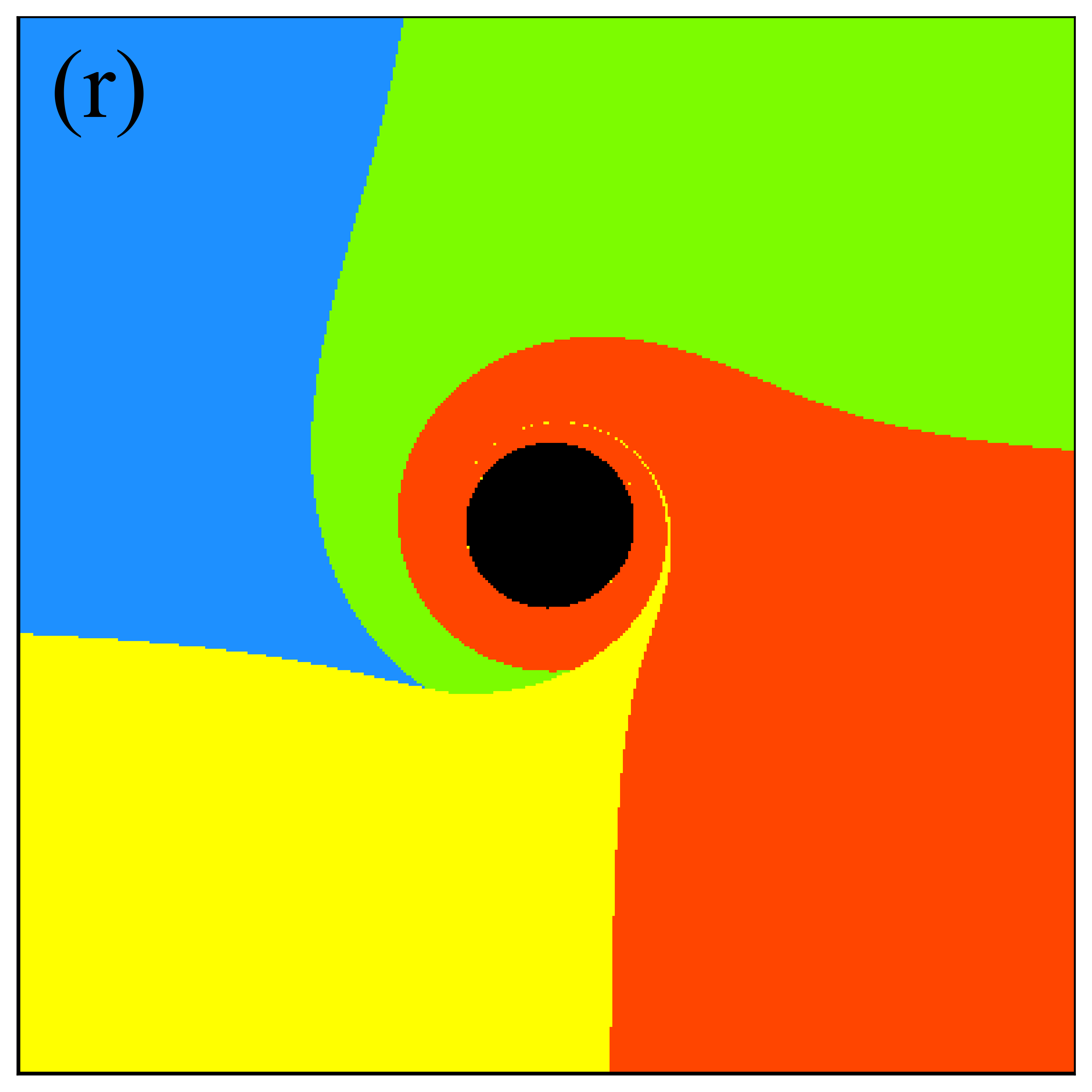}
\includegraphics[width=2.8cm]{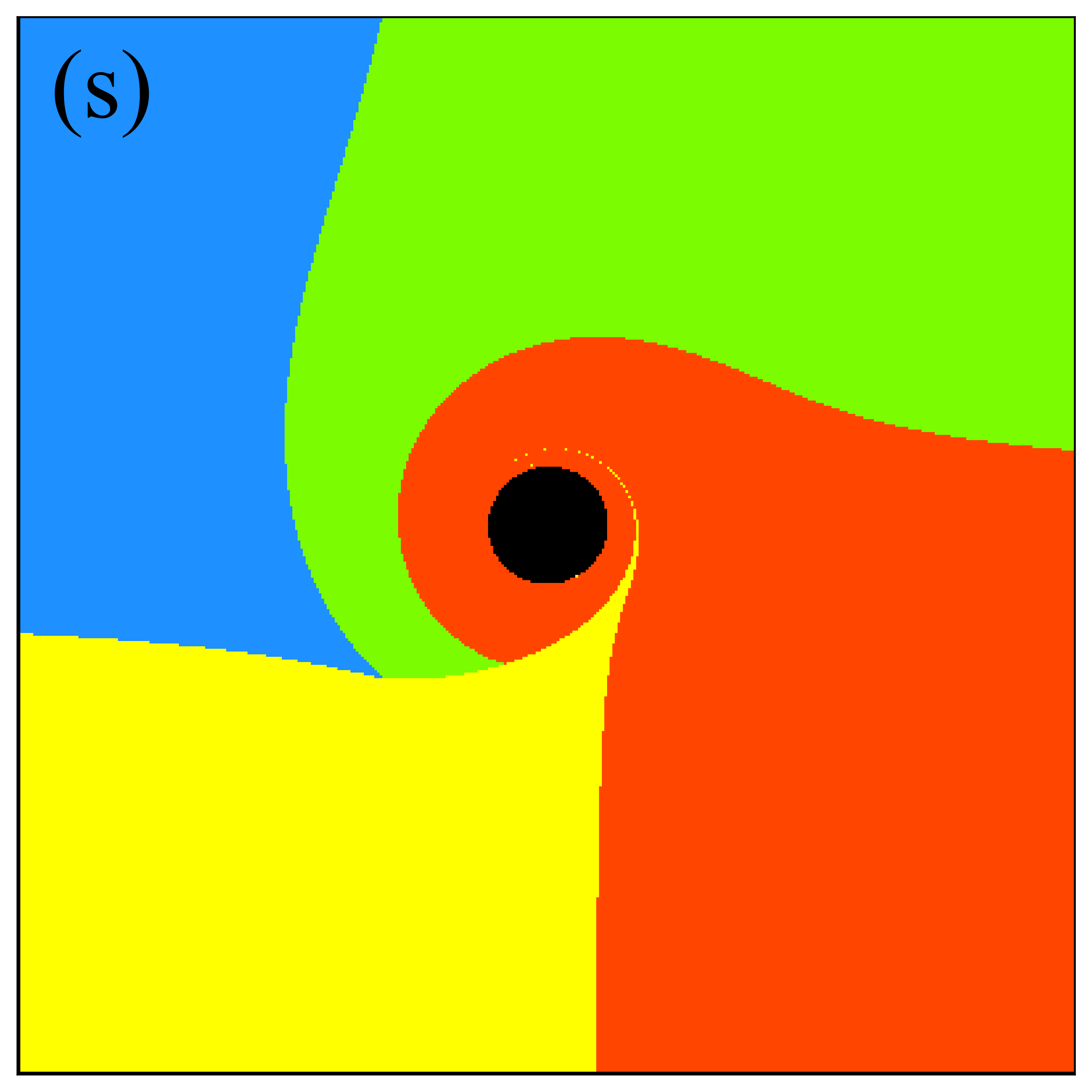}
\includegraphics[width=2.8cm]{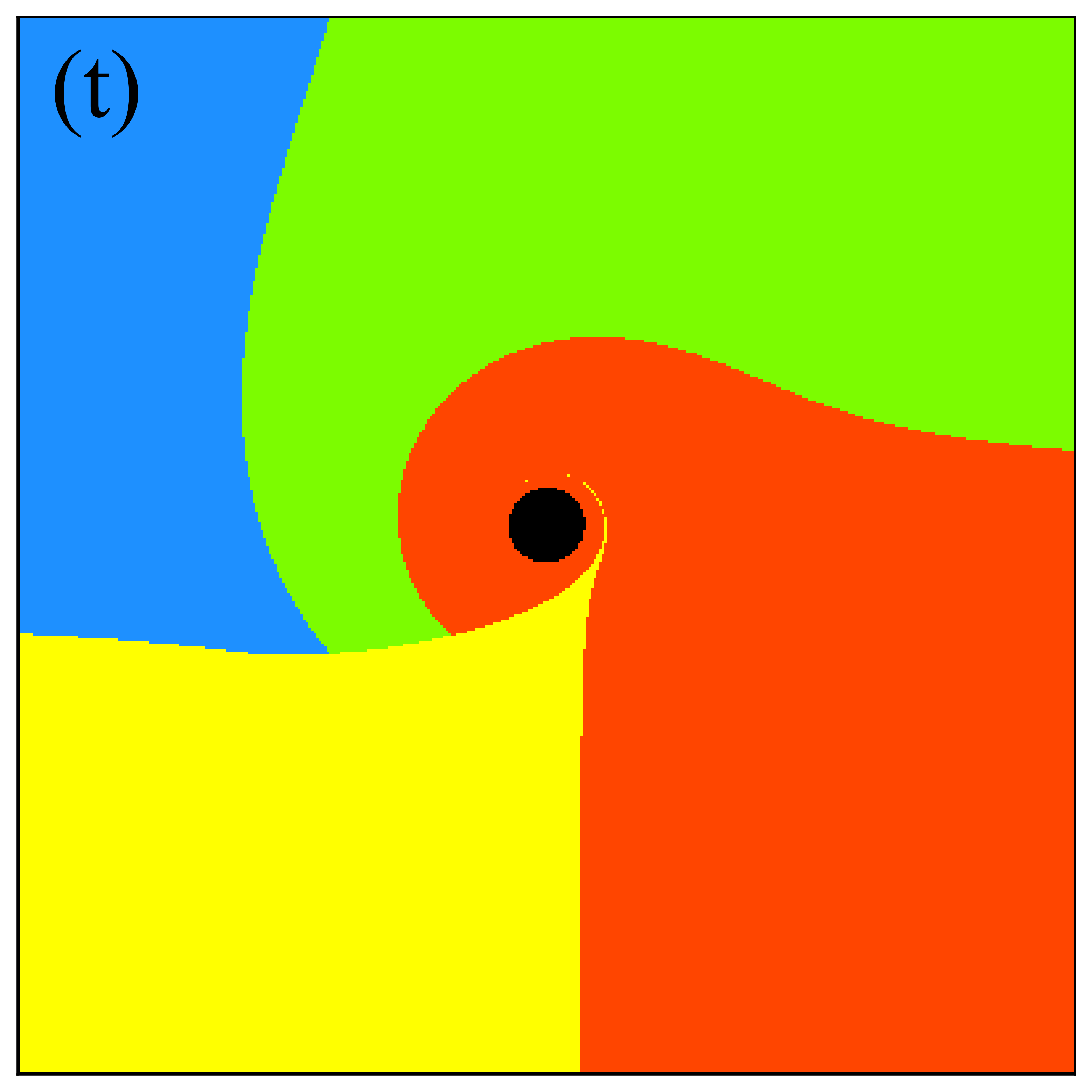}
\caption{Evolution of the inner shadow of a Kerr black hole with a tilted thin accretion disk with respect to disk inclination $\sigma$ and spin parameter $a$. From left to right, the values of disk inclination are $0^{\circ}$, $15^{\circ}$, $30^{\circ}$, $45^{\circ}$, and $60^{\circ}$; from top to bottom, the spin parameters are $0$, $0.54$, $0.94$, and $0.9985$. Here, the observation angle and azimuth are fixed at $17^{\circ}$ and $0^{\circ}$, respectively, and the field of view is set to $16 \times 16$ M, with a resolution of $400 \times 400$ pixels. It is evident that the tilted accretion disk introduces a significant obscuration effect on the inner shadow.}}\label{fig2}
\end{figure*}

First, we examine the images corresponding to the case of equatorial disk, specifically the leftmost column of each figure. It is evident that the rotational effect of background light source become more pronounced with increasing spin parameter at any observation angle, as a higher black hole rotation enhances the frame-dragging effect. Furthermore, the shape of the inner shadow is predominantly influenced by the observation angle: as the observation angle increases, the inner shadow elongates horizontally and compresses vertically, approaching a closed arc shape.

Additionally, the size of the inner shadow demonstrates a positive correlation with the observation angle and a negative correlation with the spin parameter. In other words, for a Kerr black hole endowed with an equatorial accretion disk, there exists a minimum area of the inner shadow, denoted as $S^{\textrm{min}}_{\Theta}$, for each observation angle. It is anticipated that as the spin parameter approaches $1$, $S^{\textrm{min}}_{\Theta=0^{\circ}}$ reaches its minimum value, corresponding to the smallest inner shadow area, $S_{\textrm{min}}$, that a Kerr black hole surrounded by an equatorial accretion disk can exhibit. We find that $S_{\textrm{min}} = 13.075$ M$^{2}$.

\begin{figure*}
\center{
\includegraphics[width=2.8cm]{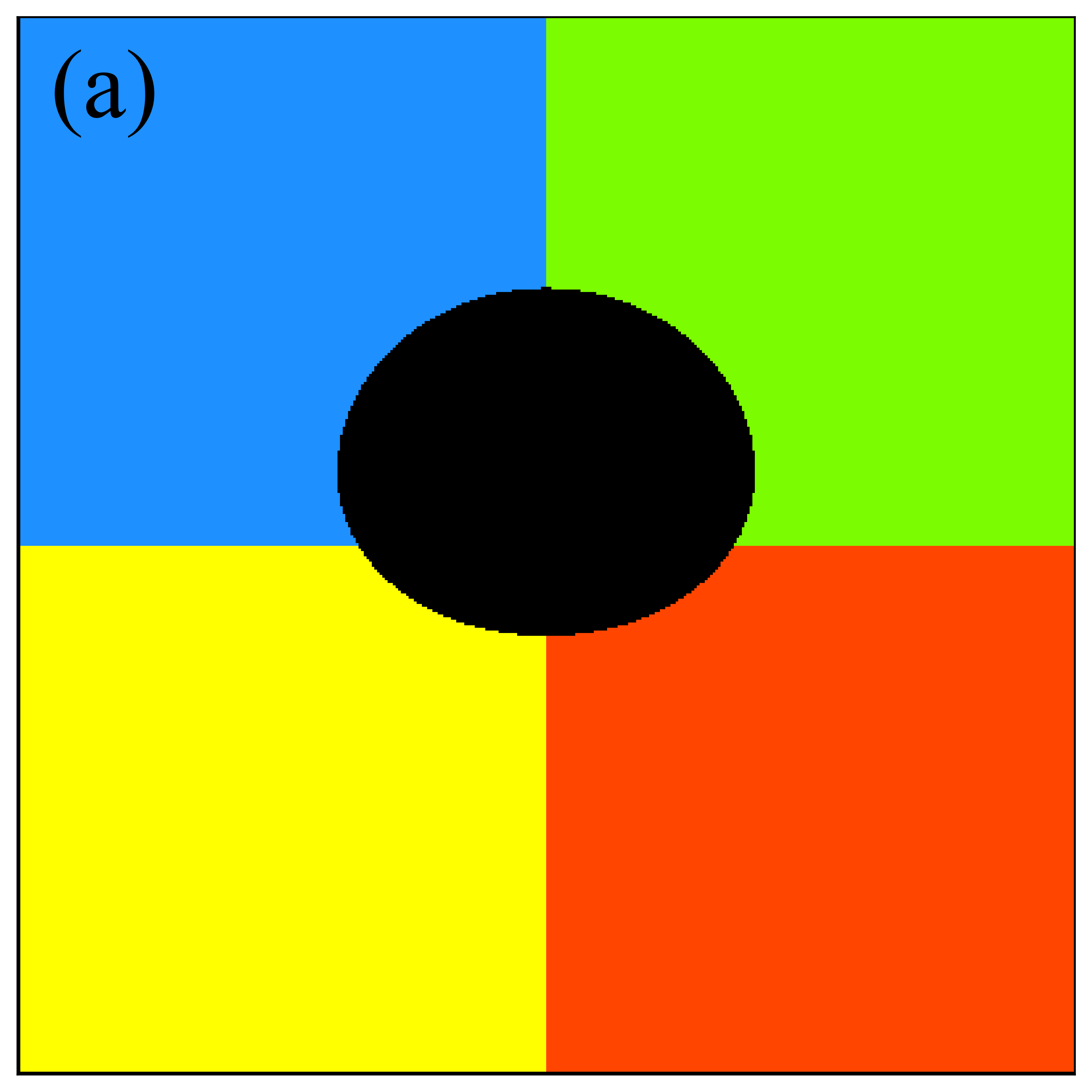}
\includegraphics[width=2.8cm]{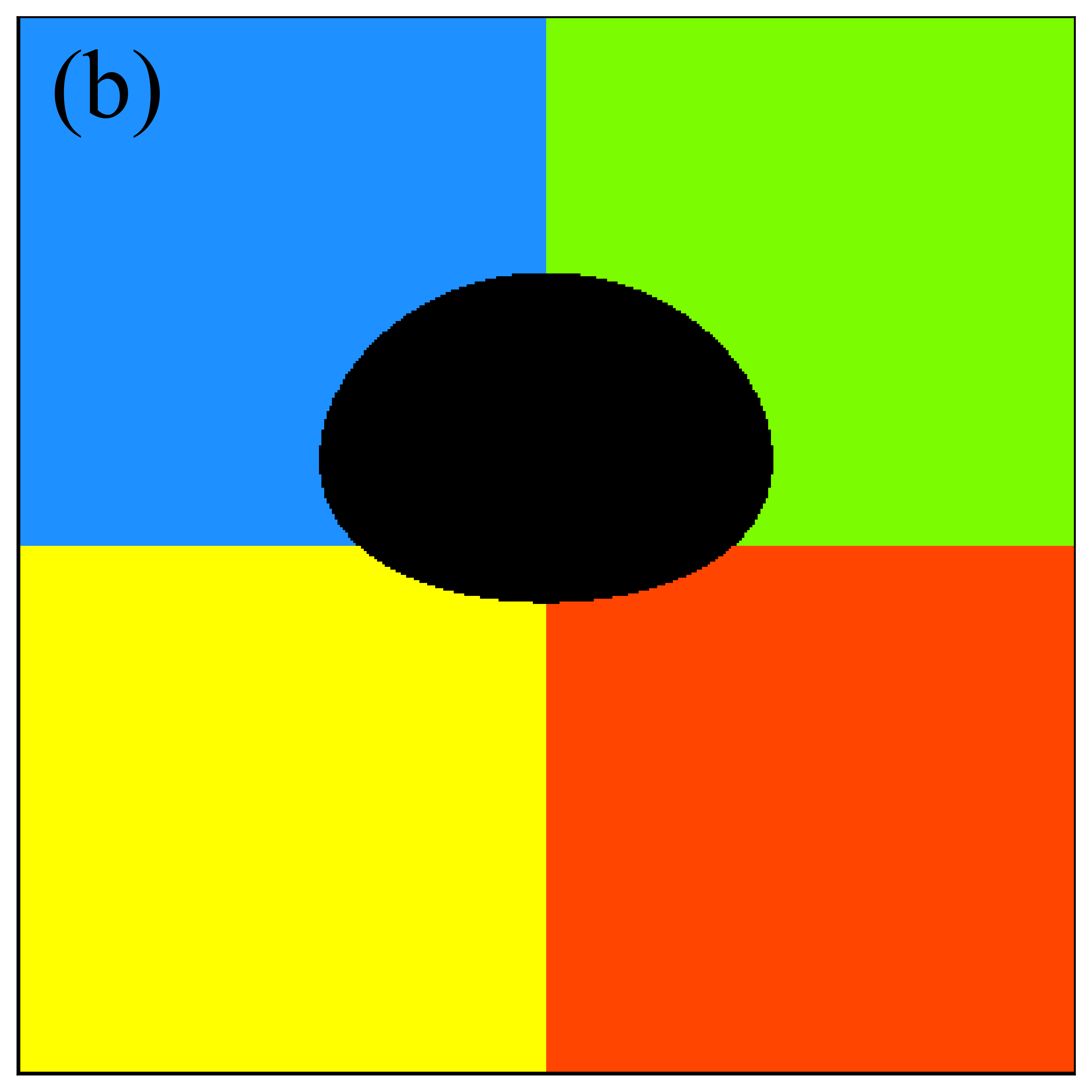}
\includegraphics[width=2.8cm]{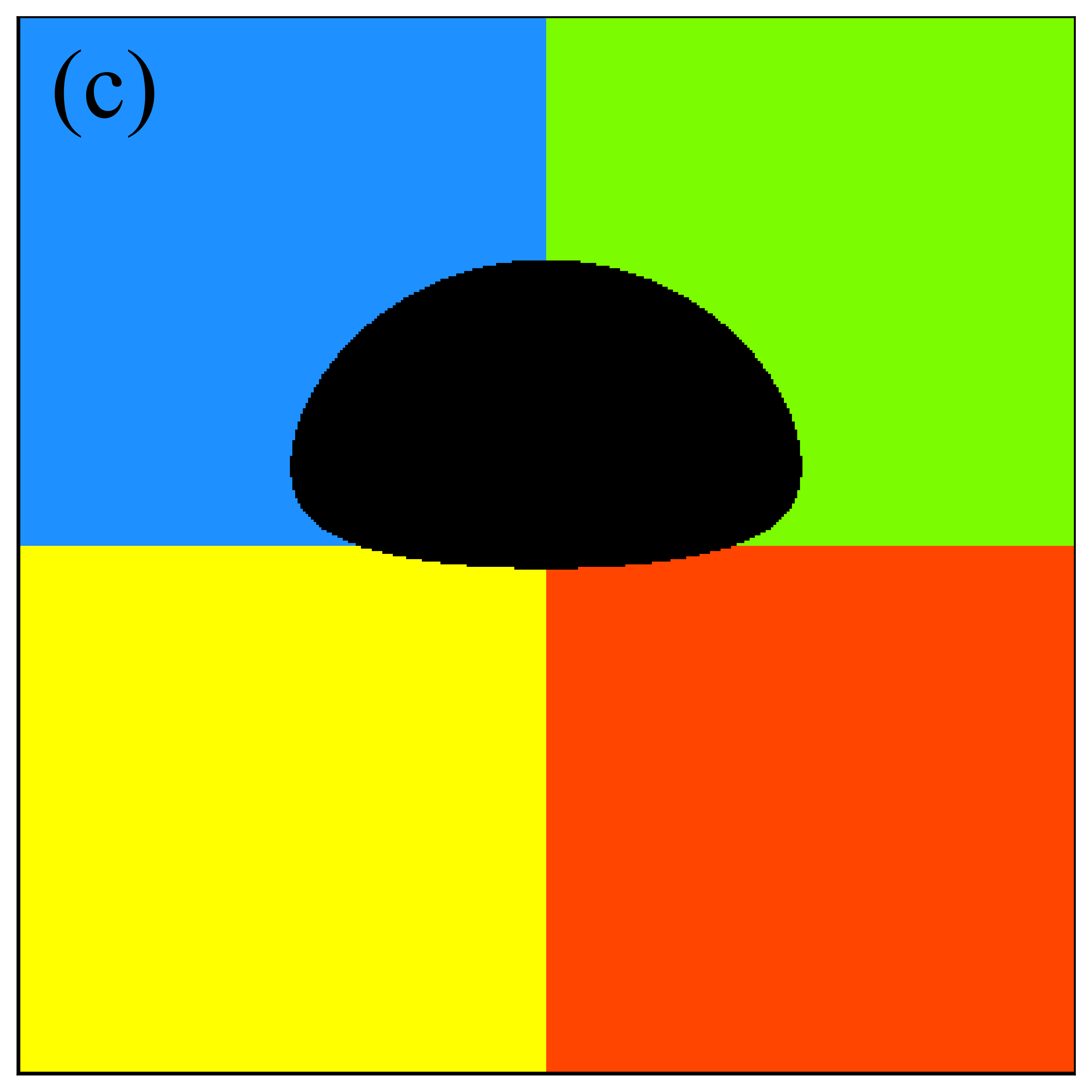}
\includegraphics[width=2.8cm]{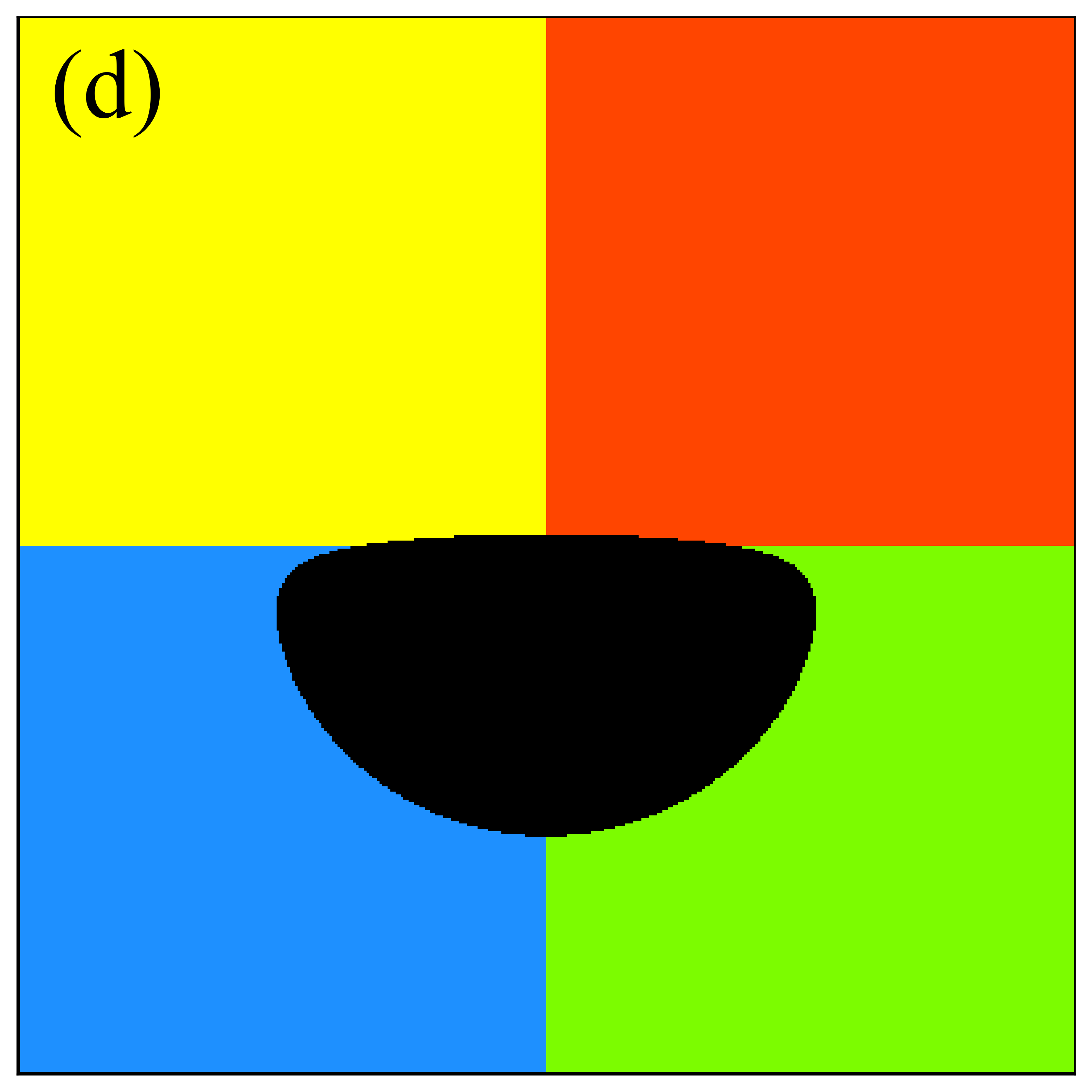}
\includegraphics[width=2.8cm]{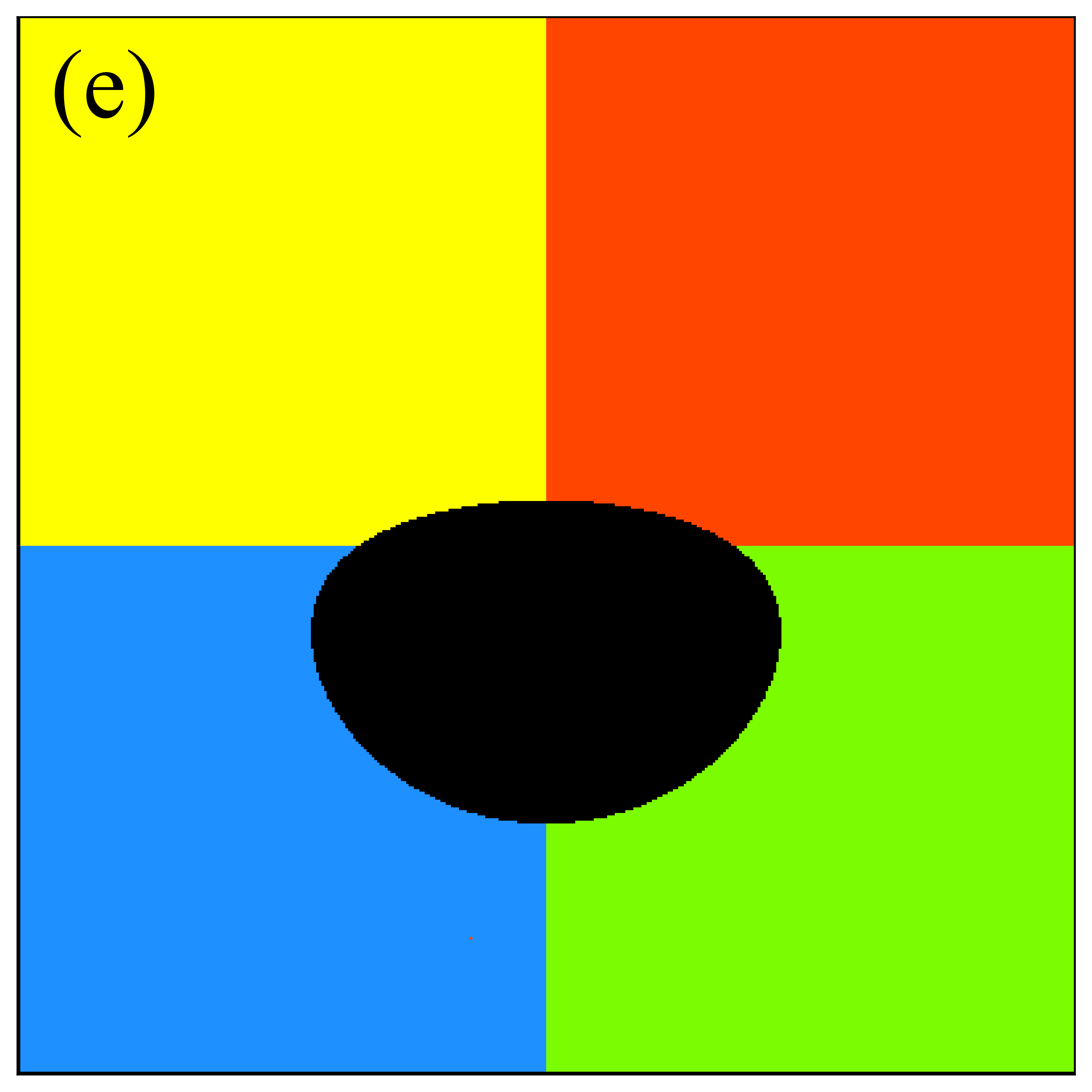}
\includegraphics[width=2.8cm]{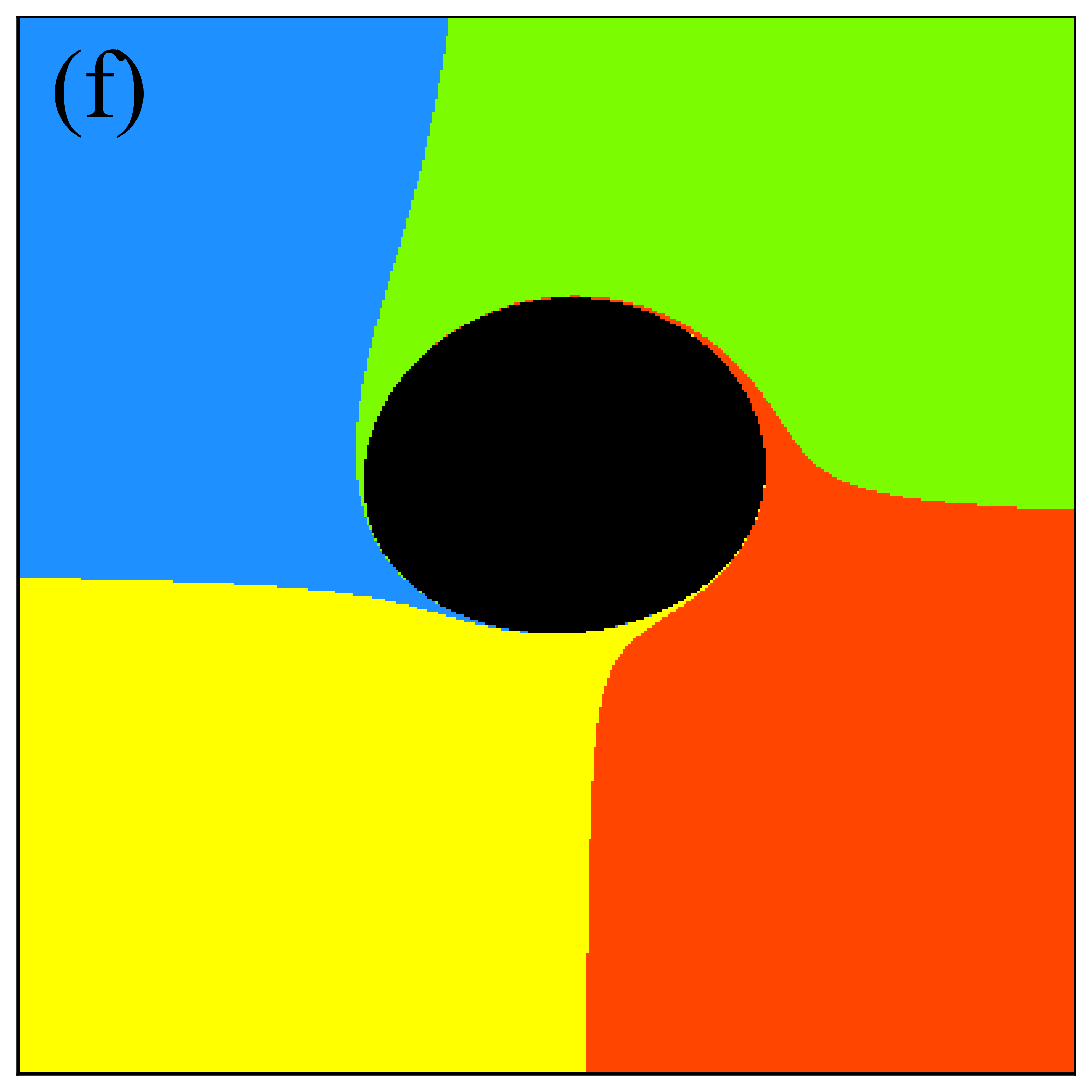}
\includegraphics[width=2.8cm]{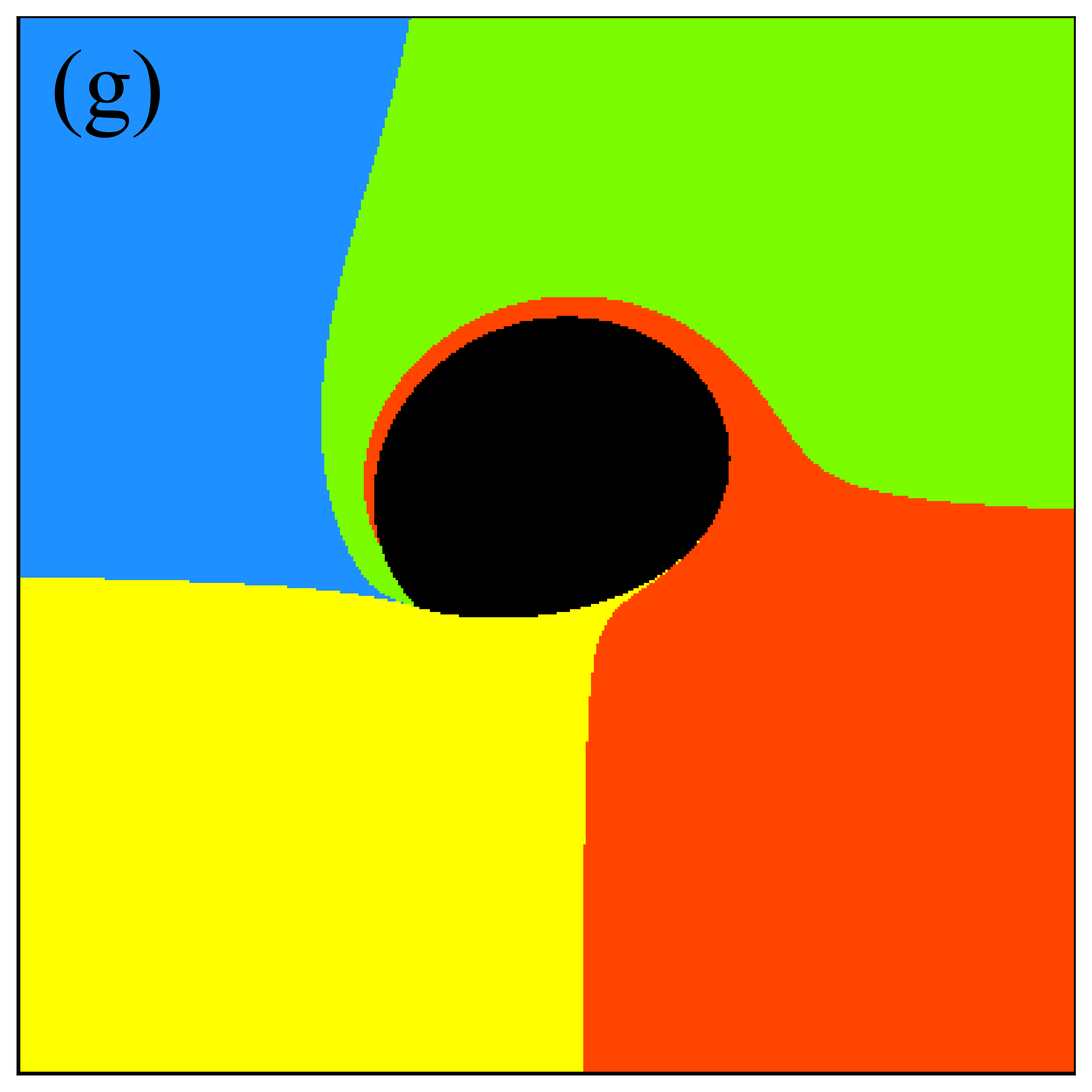}
\includegraphics[width=2.8cm]{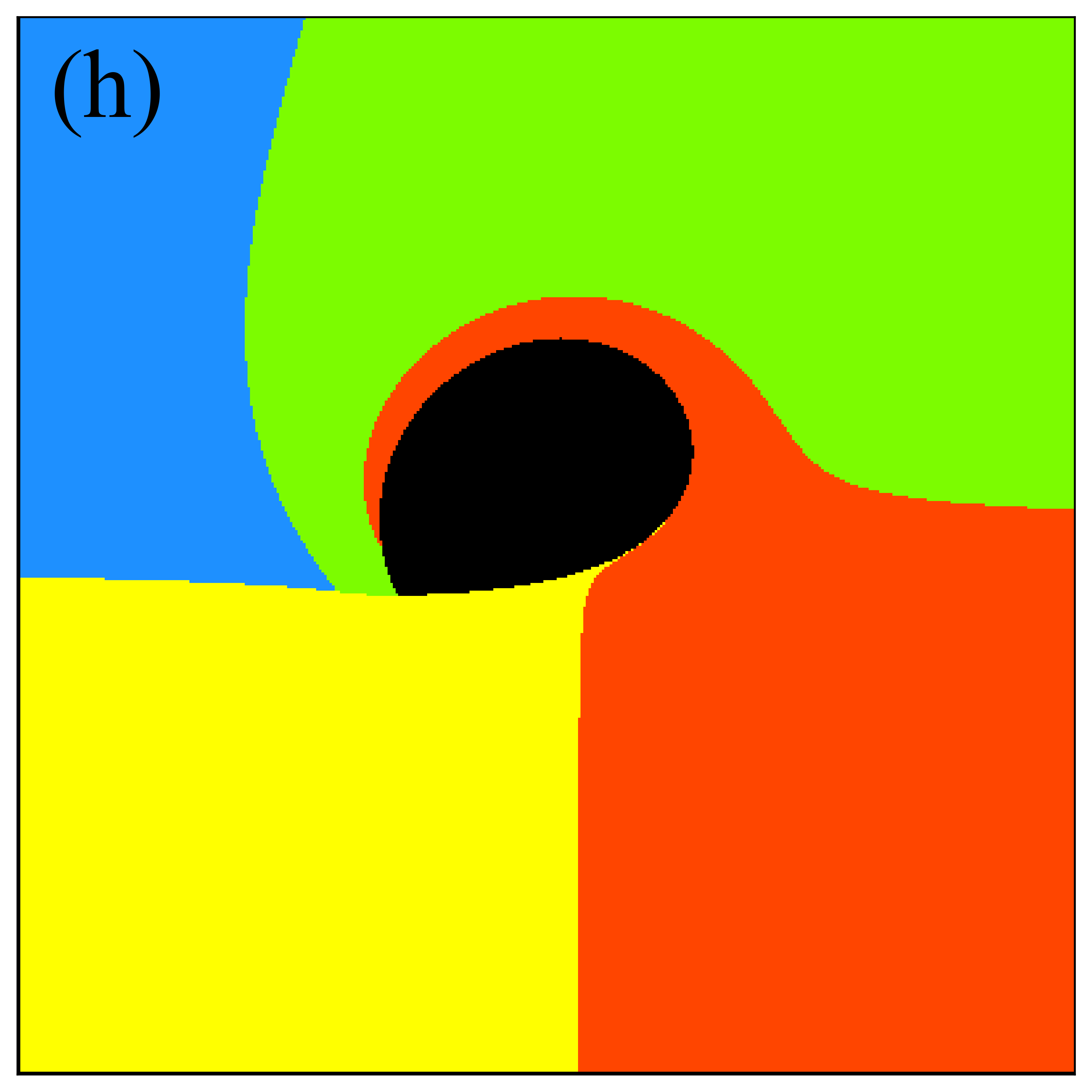}
\includegraphics[width=2.8cm]{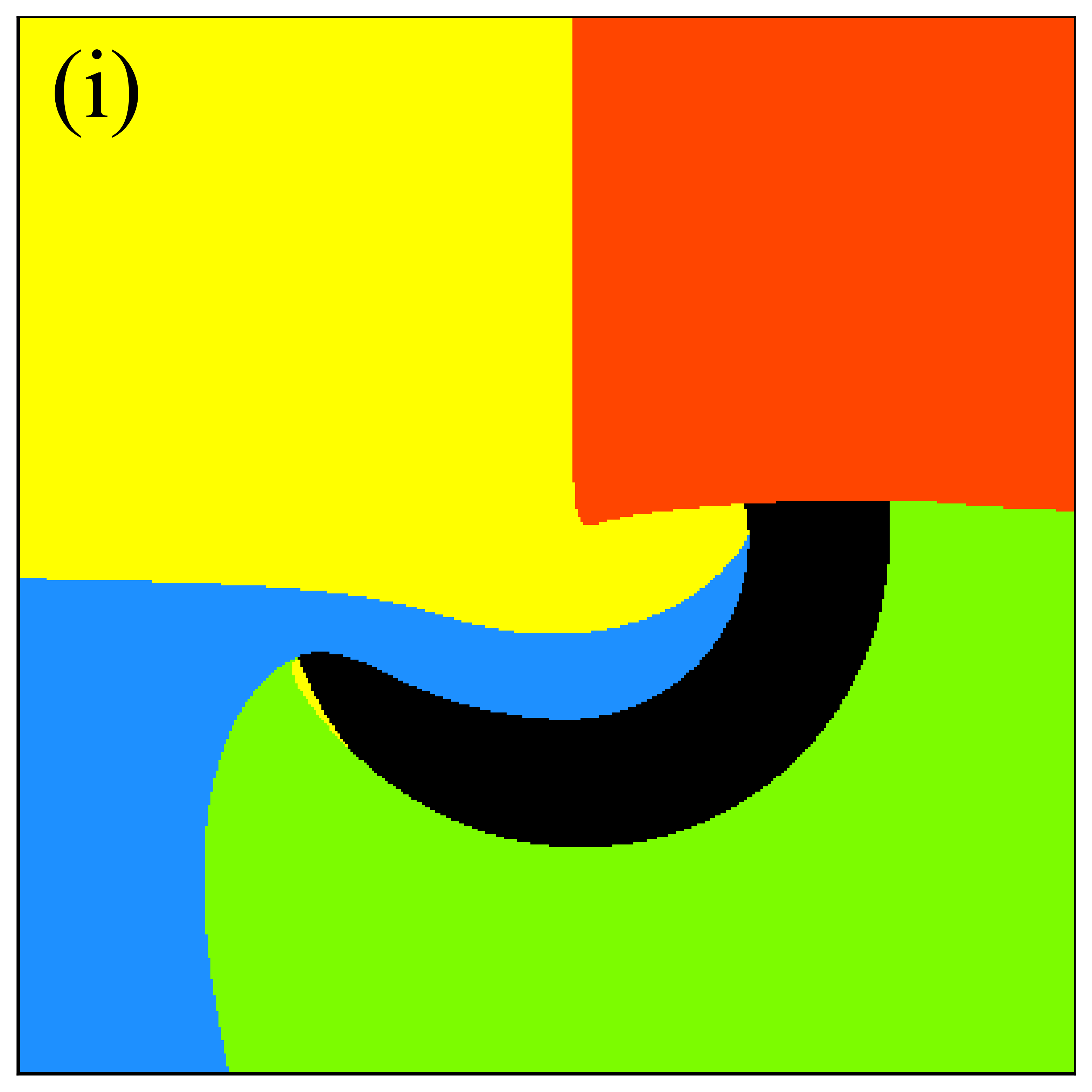}
\includegraphics[width=2.8cm]{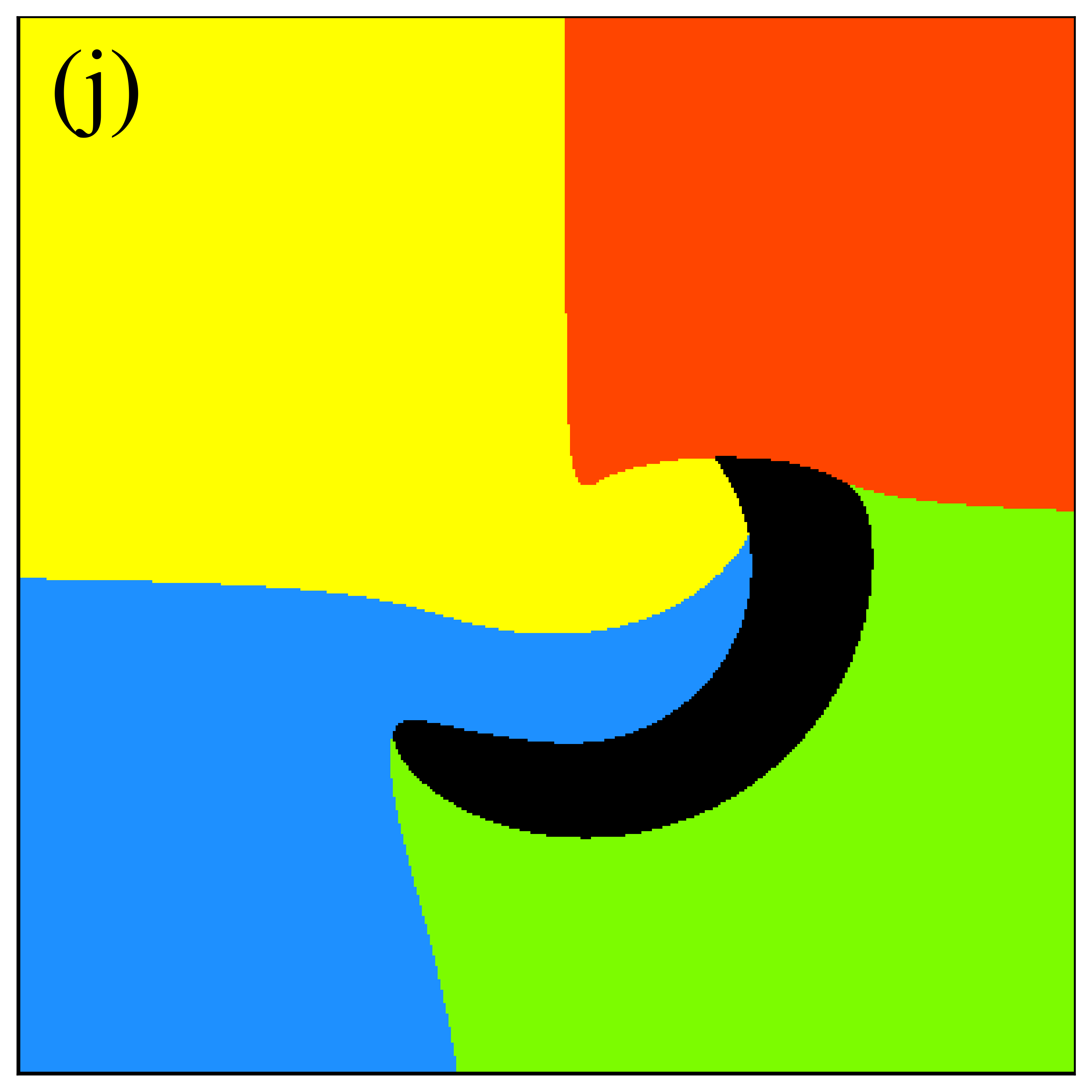}
\includegraphics[width=2.8cm]{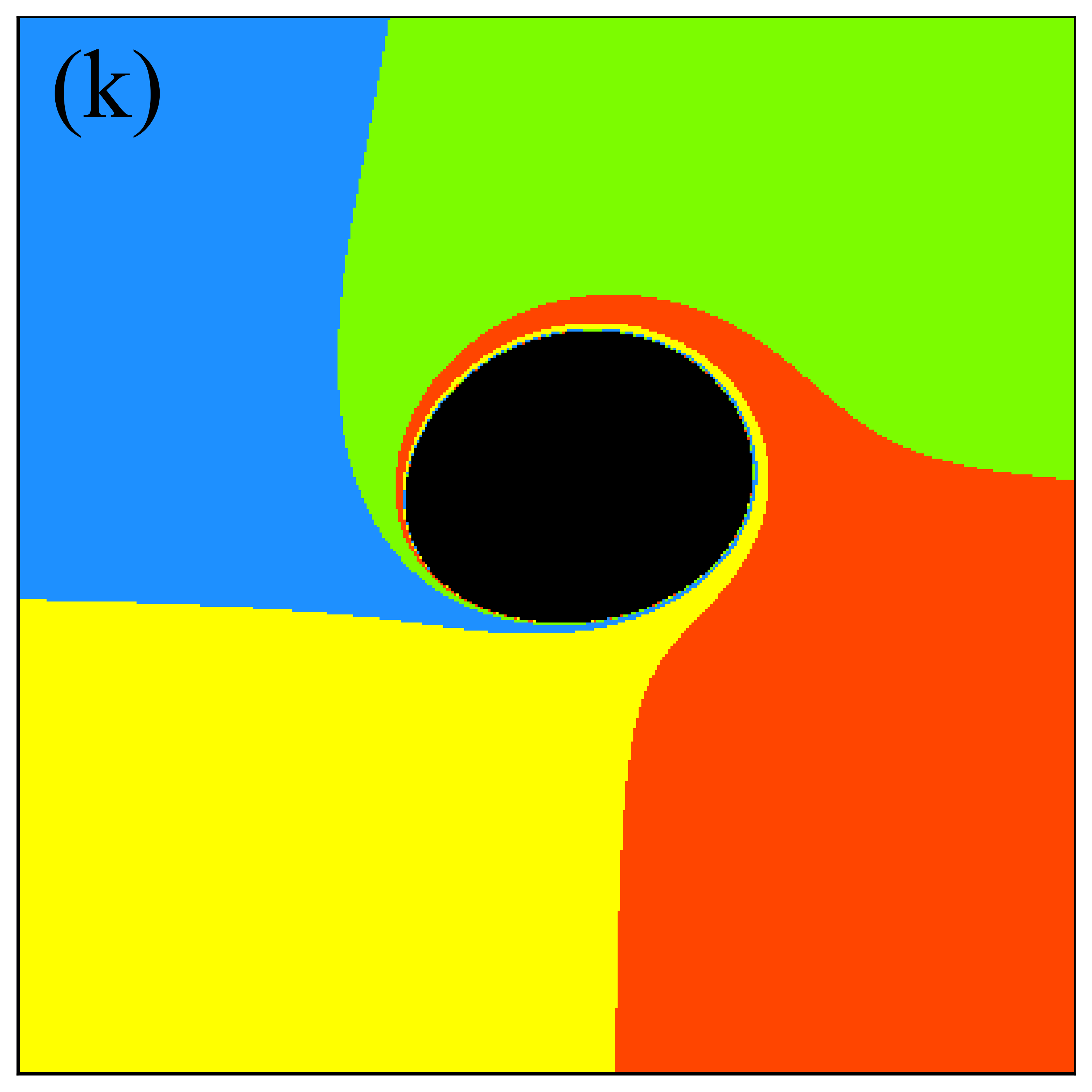}
\includegraphics[width=2.8cm]{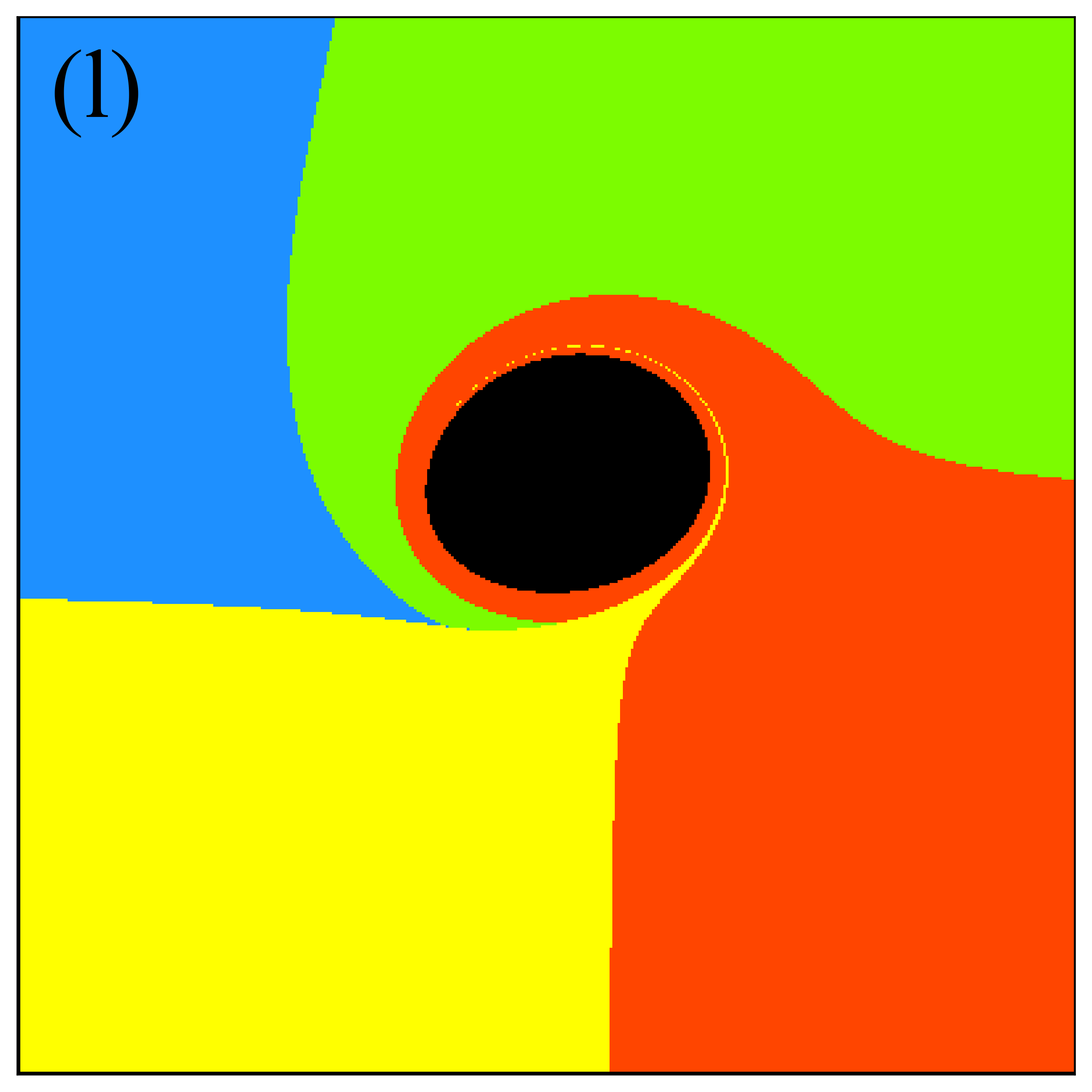}
\includegraphics[width=2.8cm]{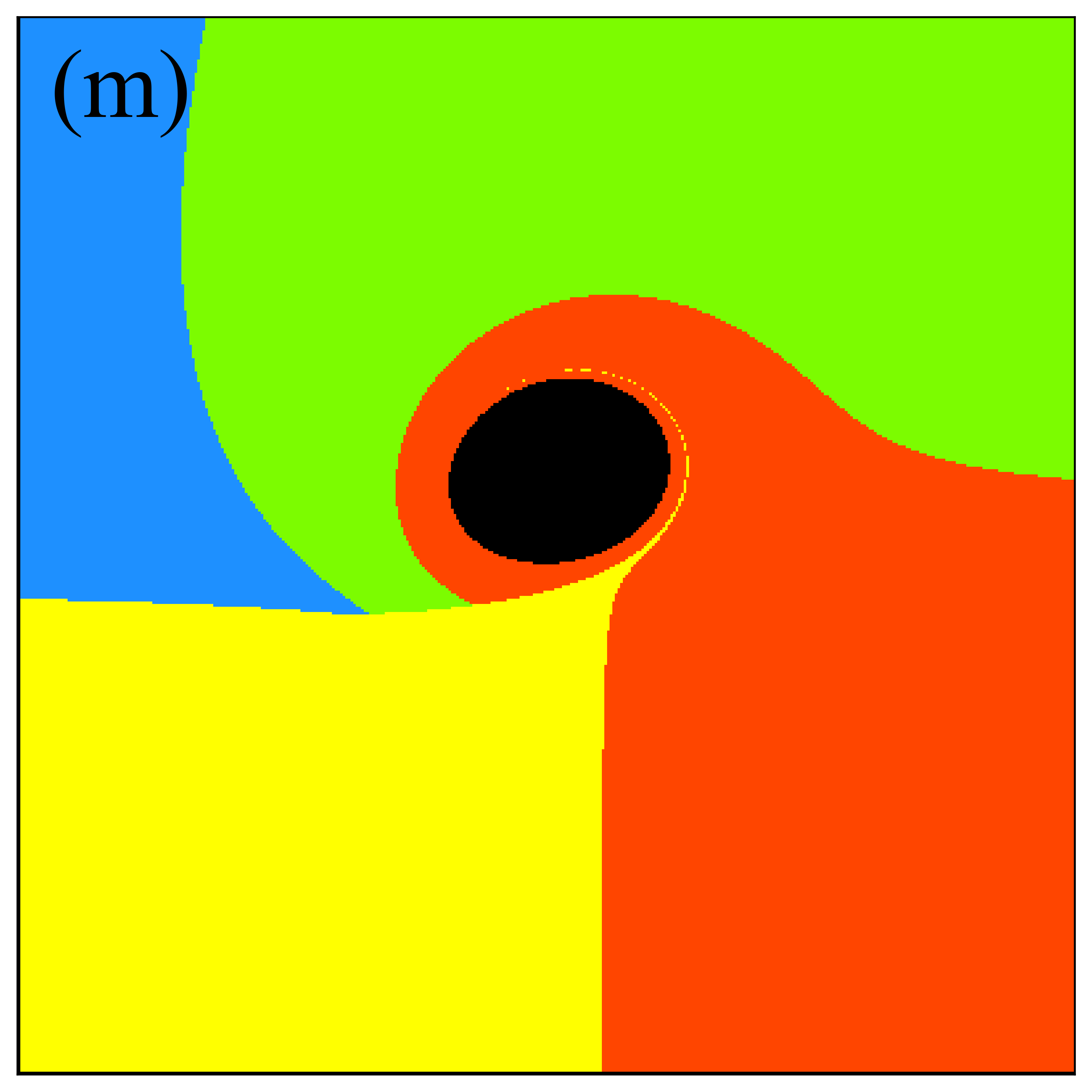}
\includegraphics[width=2.8cm]{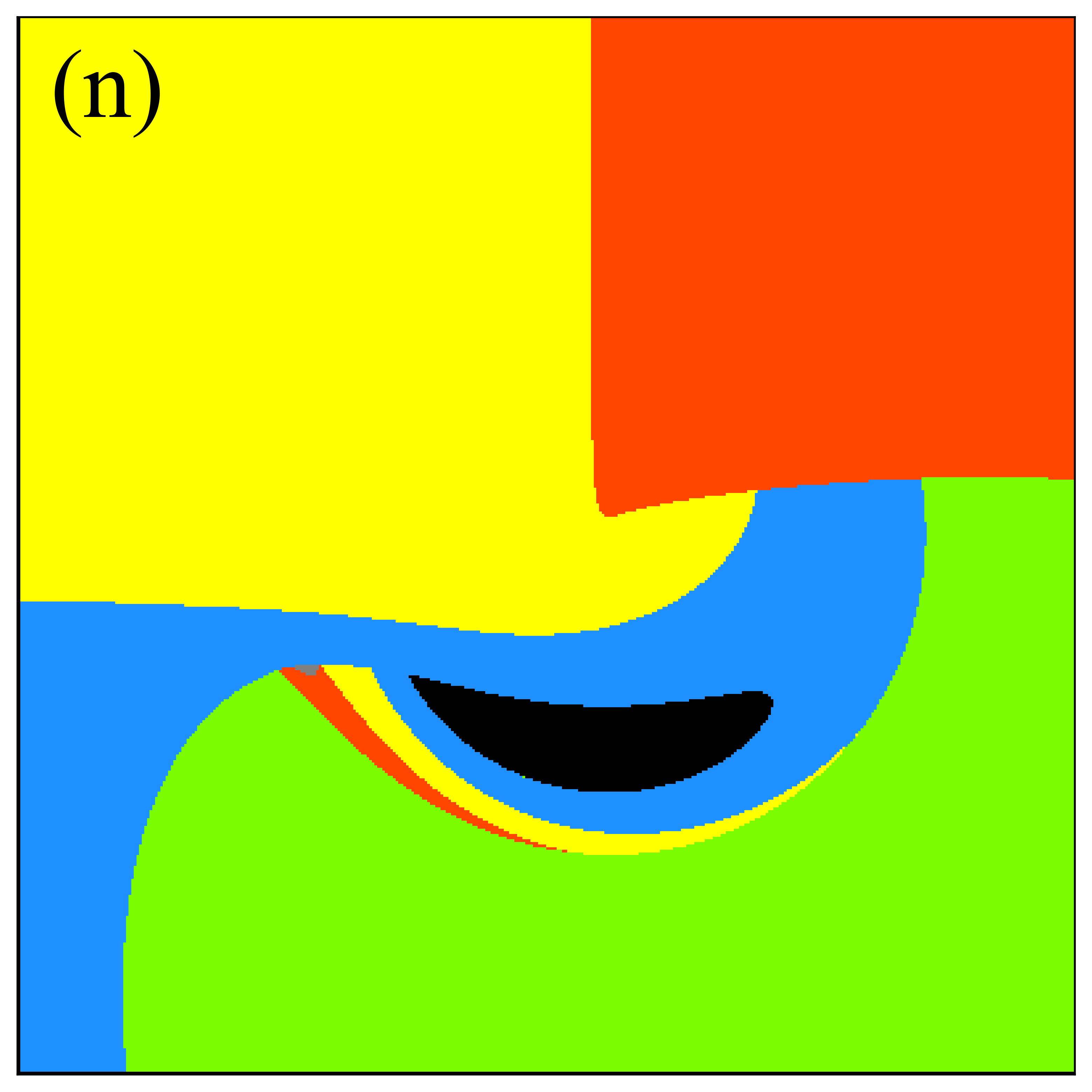}
\includegraphics[width=2.8cm]{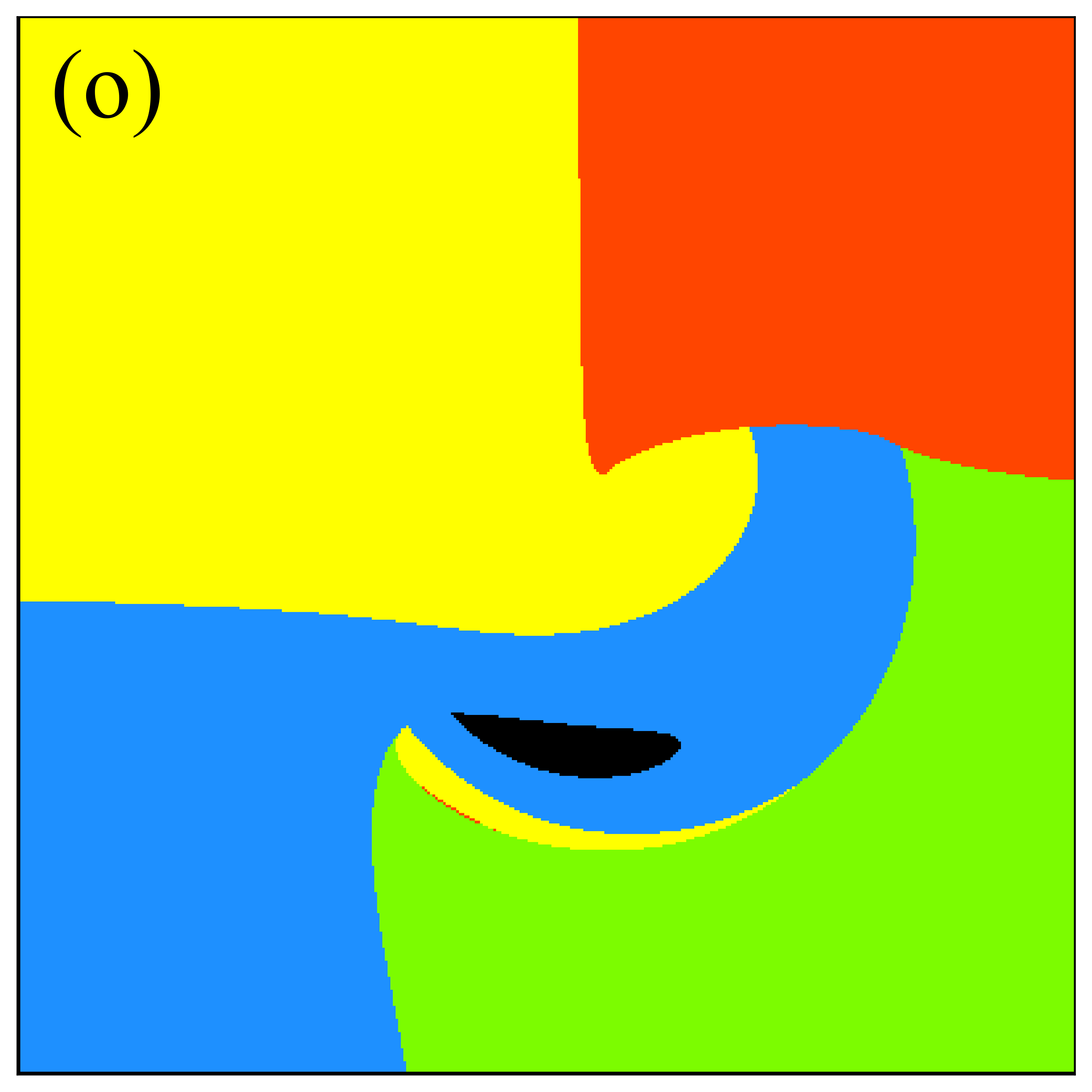}
\includegraphics[width=2.8cm]{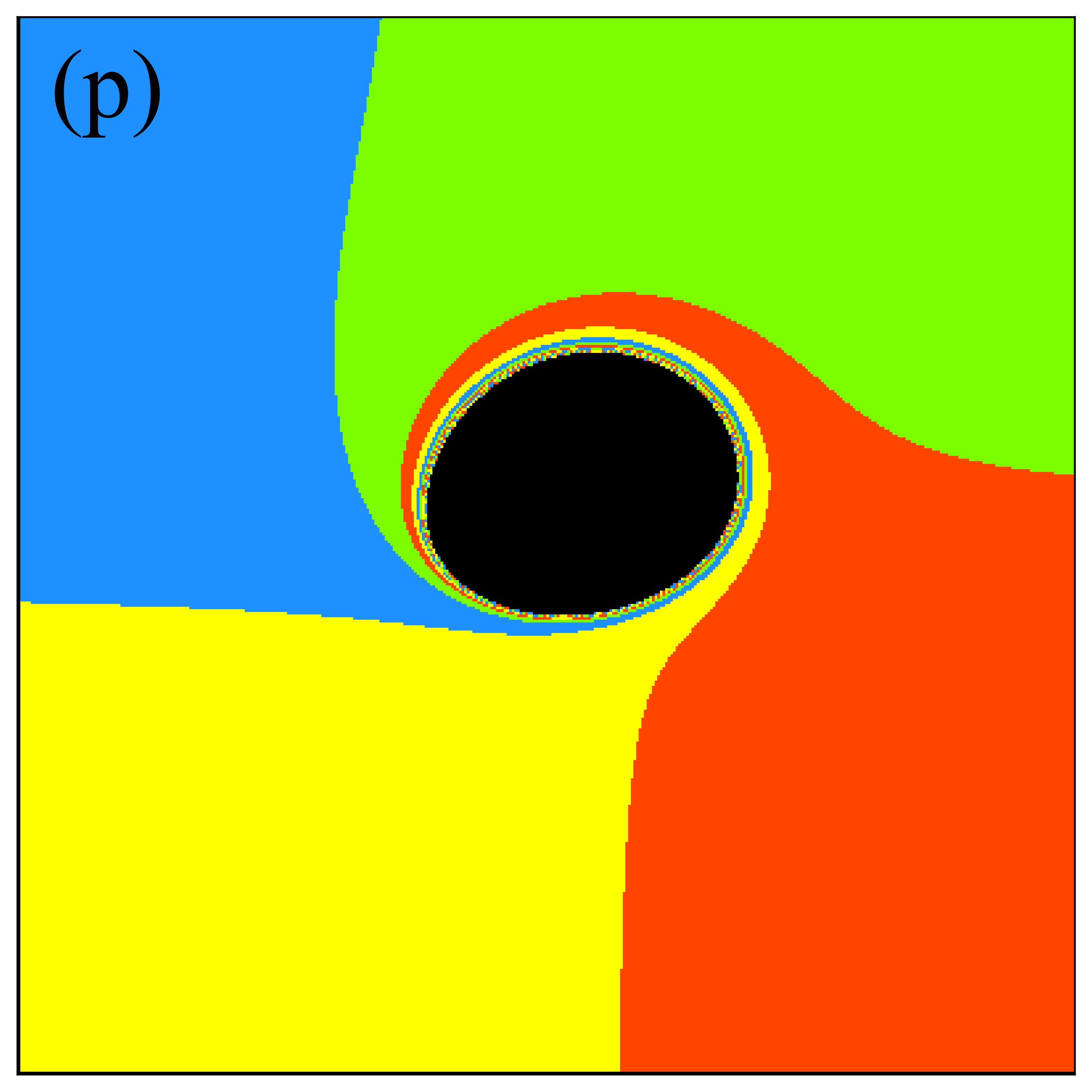}
\includegraphics[width=2.8cm]{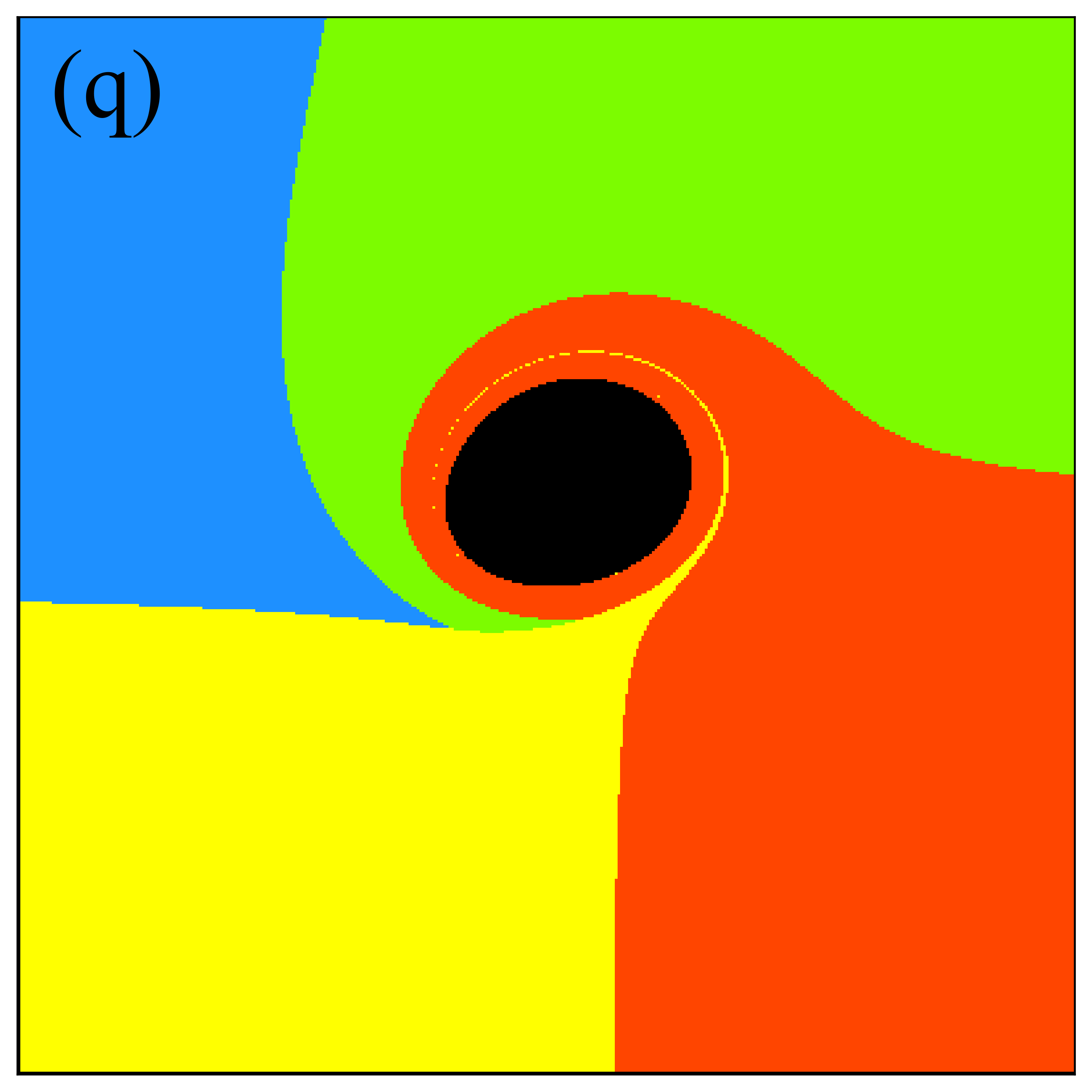}
\includegraphics[width=2.8cm]{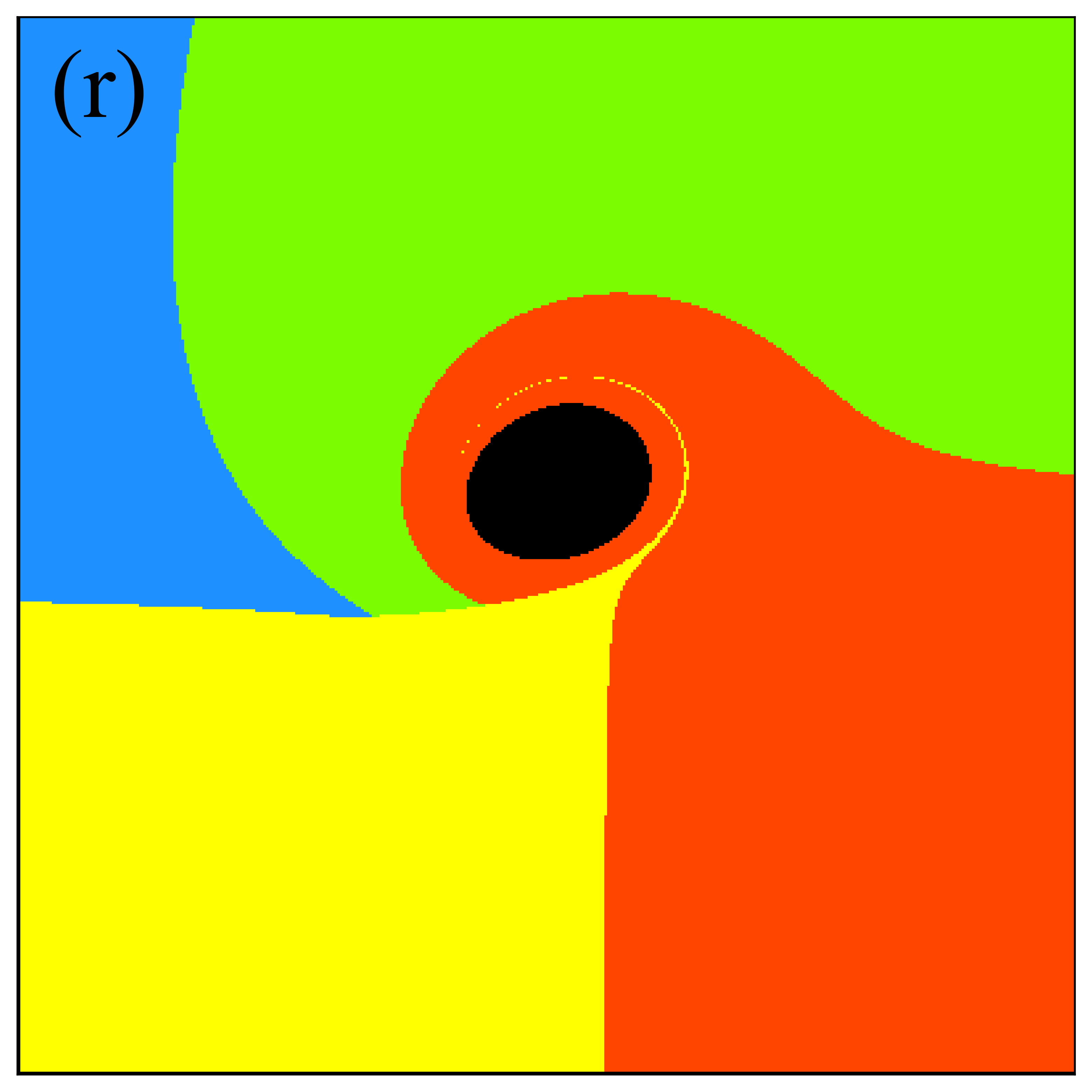}
\includegraphics[width=2.8cm]{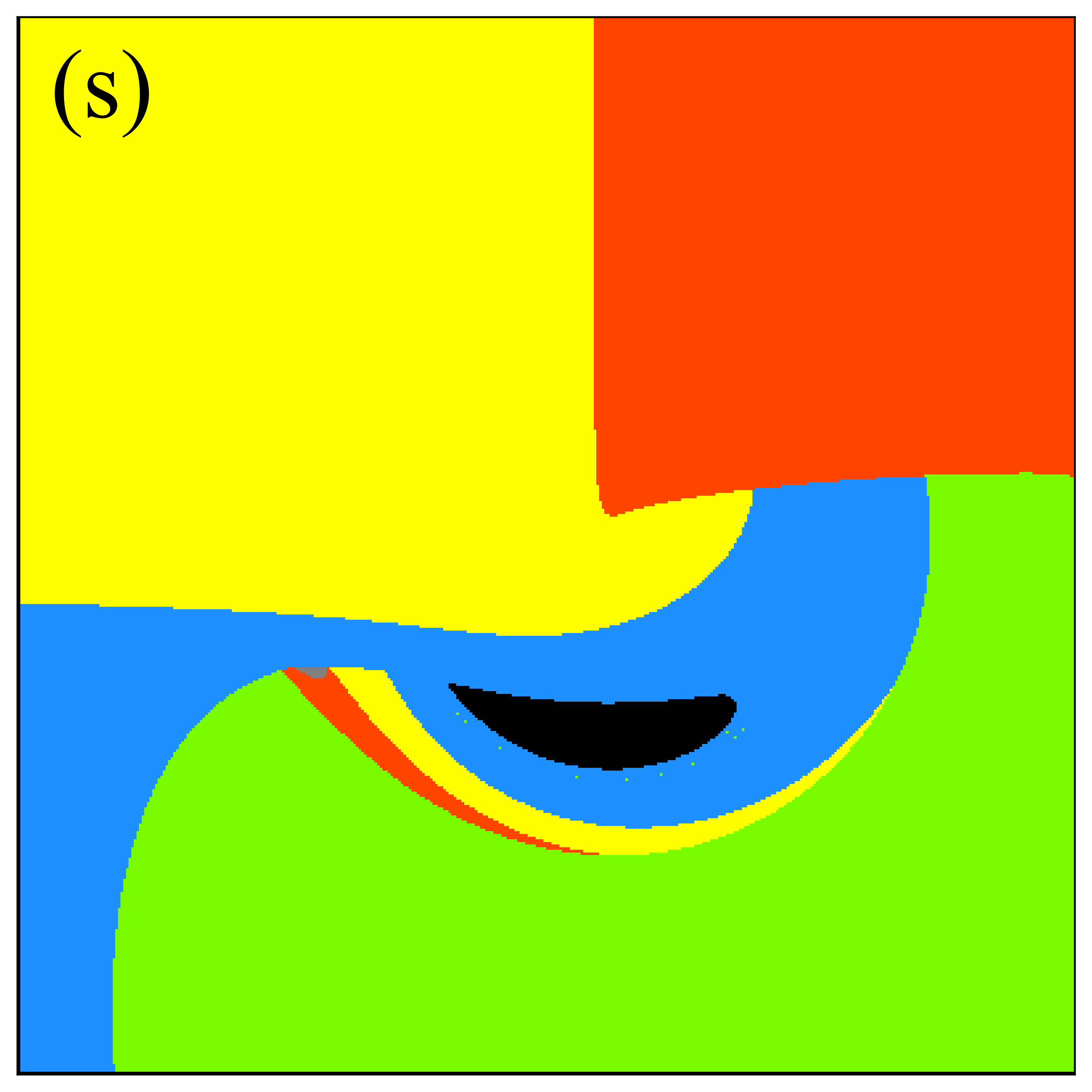}
\includegraphics[width=2.8cm]{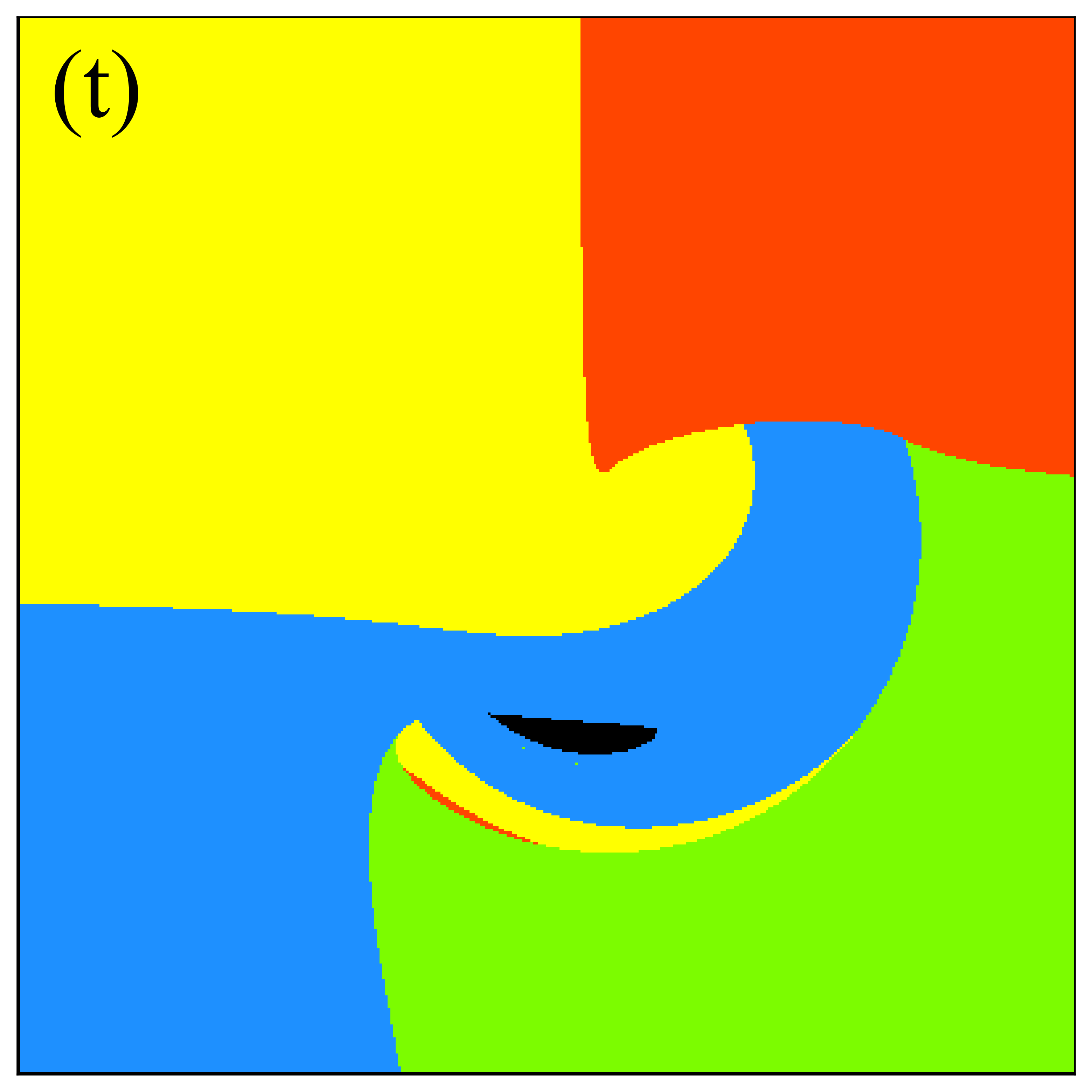}
\caption{Similar to figure 2, but for the observation angle of $50^{\circ}$.}}\label{fig3}
\end{figure*}
Interestingly, we find that when the accretion disk deviates from the equatorial plane, the inner shadow of the black hole can shrink further from $S^{\textrm{min}}_{\Theta}$: the size of the inner shadow decreases with an increase in the tilt of the accretion disk, and this reduction is positively correlated with the black hole's spin parameter, as vividly illustrated in the last row of figure 2. This suggests that if a significantly smaller than expected inner shadow is observed, it may indicate the presence of a tilted accretion disk surrounding the black hole. In other words, the size of the inner shadow can serve as a probe for identifying tilted accretion disks.

Moreover, we note that the introduction of the accretion disk inclination can alter the shape of the black hole's inner shadow. When the spin parameter is zero, changes in the inner shadow due to the disk inclination can be fully attributed to a specific observation angle, a degeneracy that was identified in our previous work \cite{Hu et al. (2024)}. However, for non-zero spin parameters, we do not observe a degeneracy between the observation angle and the disk tilt; instead, the presence of the tilted disk leads to a richer variety of inner shadow profiles, including crescent, eyebrow, and petal shapes. This phenomenon, along with the inner shadow shrinkage reported earlier, occurs because the tilted accretion disk blocks light rays that would otherwise cross the event horizon, thus reducing the inner shadow's visibility. To better visualize this characteristic, we employ ray propagation to demonstrate how the tilted accretion disk erodes the black hole's inner shadow. As shown in figure 5(a), a Kerr black hole with spin parameter $0.94$ is represented by a black sphere, while the accretion disk with inclination $\sigma=30^{\circ}$ is depicted as a four-colored tilted plane, and the equatorial disk appears as a gray disk. The purple, blue, and yellow curves correspond to three light rays originating from observational coordinates $(2,1)$, $(-2,0)$, and $(1,0.2)$, respectively. Clearly, in the equatorial disk scenario, all three light rays can plunge into the black hole without intersecting the accretion disk. However, when the disk is tilted to $30^{\circ}$, it intercepts the blue and purple rays before they reach the black hole. From an astrophysical standpoint, this implies that while these rays would contribute to the dark regions of the observation plane for an equatorial disk, in the tilted configuration they instead carry disk radiation, resulting in non-zero luminosities at the observational coordinates $(2,1)$ and $(-2,0)$. By contrast, the yellow ray consistently contributes to the black hole's inner shadow, as it directly intersects the event horizon in both disk configurations. Figures 5(b) and (c) display the projections of the three light rays from panel (a) in the $\overline{x^{\prime}o^{\prime}y^{\prime}}$ and $\overline{x^{\prime}o^{\prime}z^{\prime}}$ planes, respectively. A clear distinction emerges: when the accretion disk lies in the equatorial plane (green dashed line in the right panel), all three rays bypass the disk and plunge directly into the black hole, thereby contributing to the inner shadow. However, when the disk rotates to the red dashed position, it intercepts both the blue and purple rays, inducing deformation of the inner shadow.

\begin{figure*}
\center{
\includegraphics[width=2.8cm]{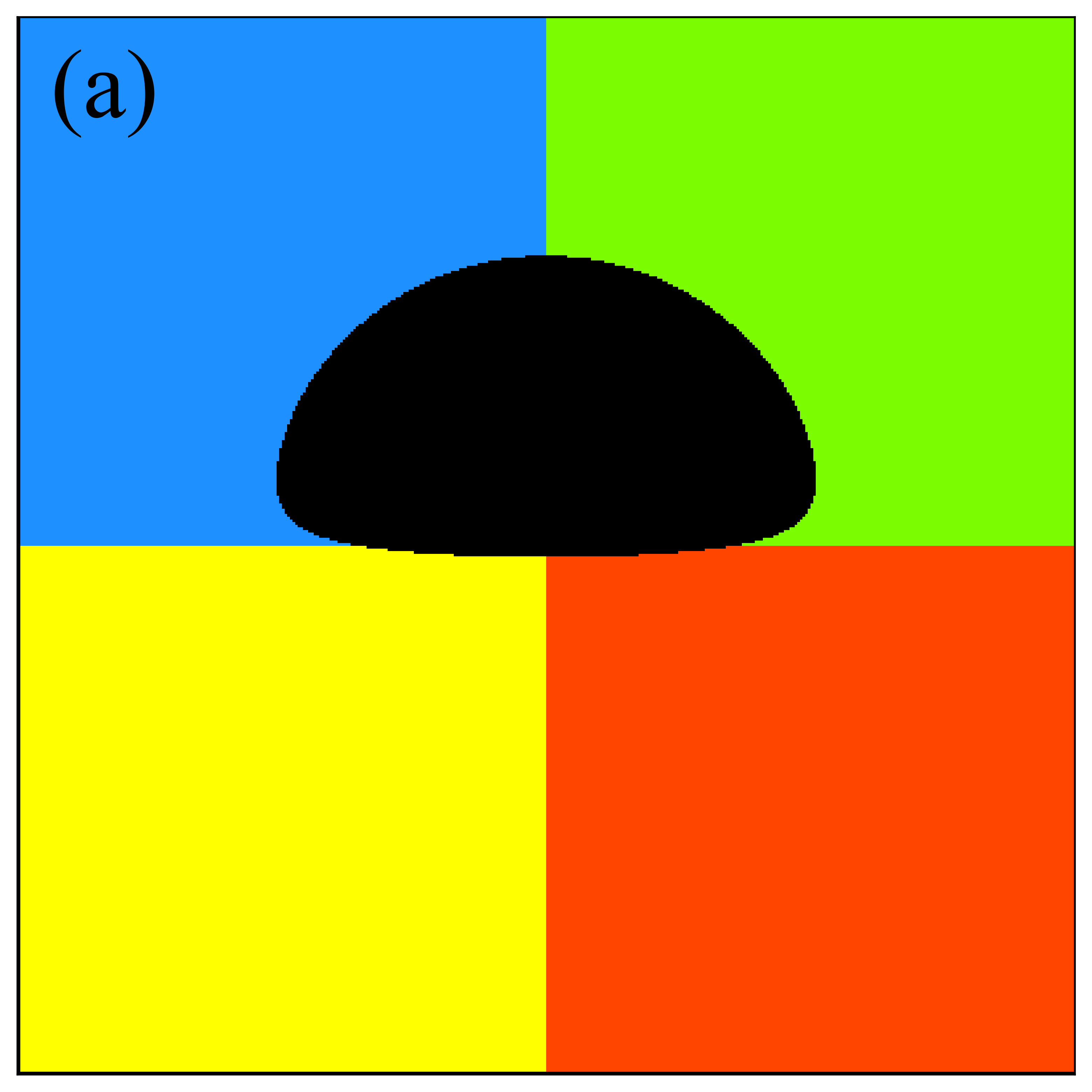}
\includegraphics[width=2.8cm]{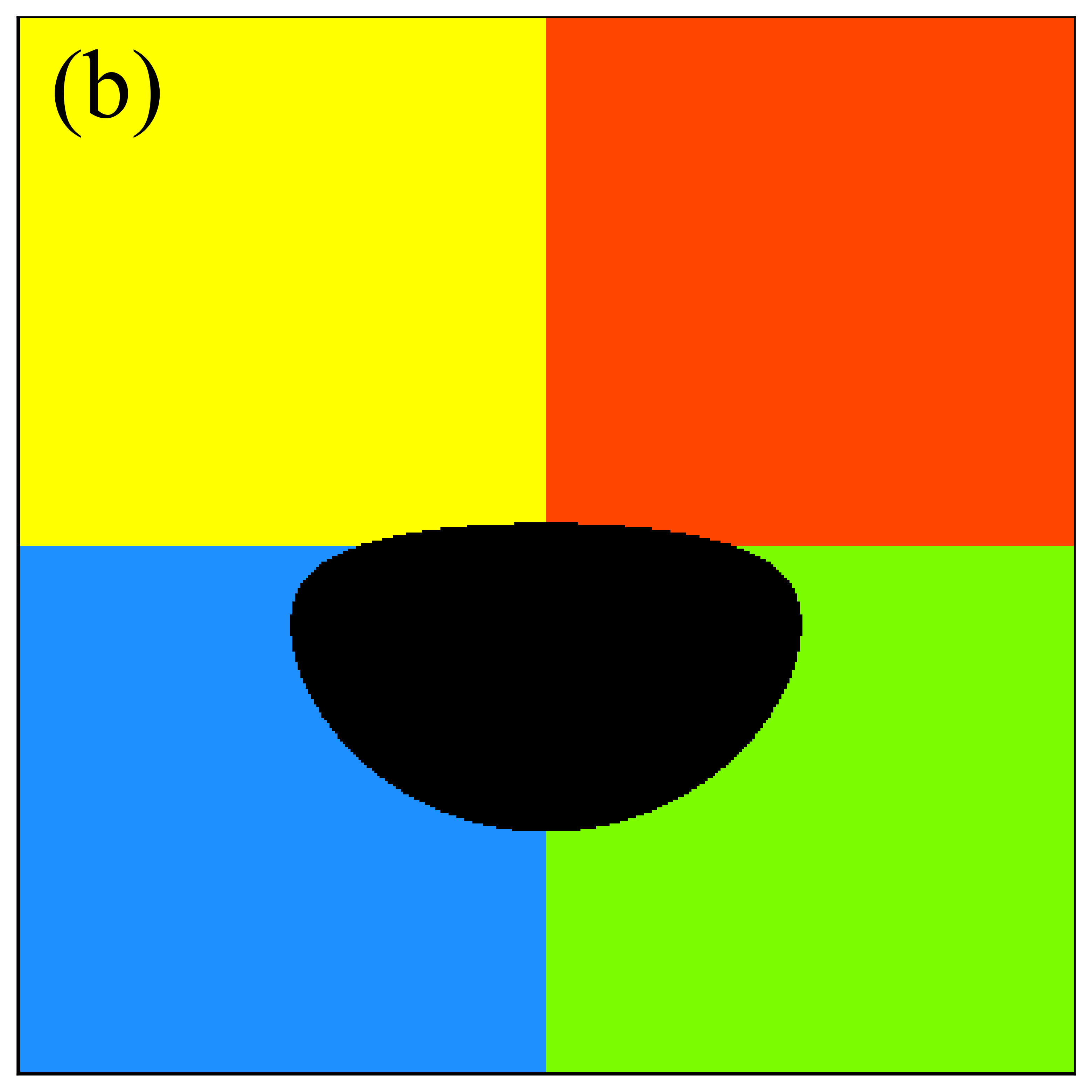}
\includegraphics[width=2.8cm]{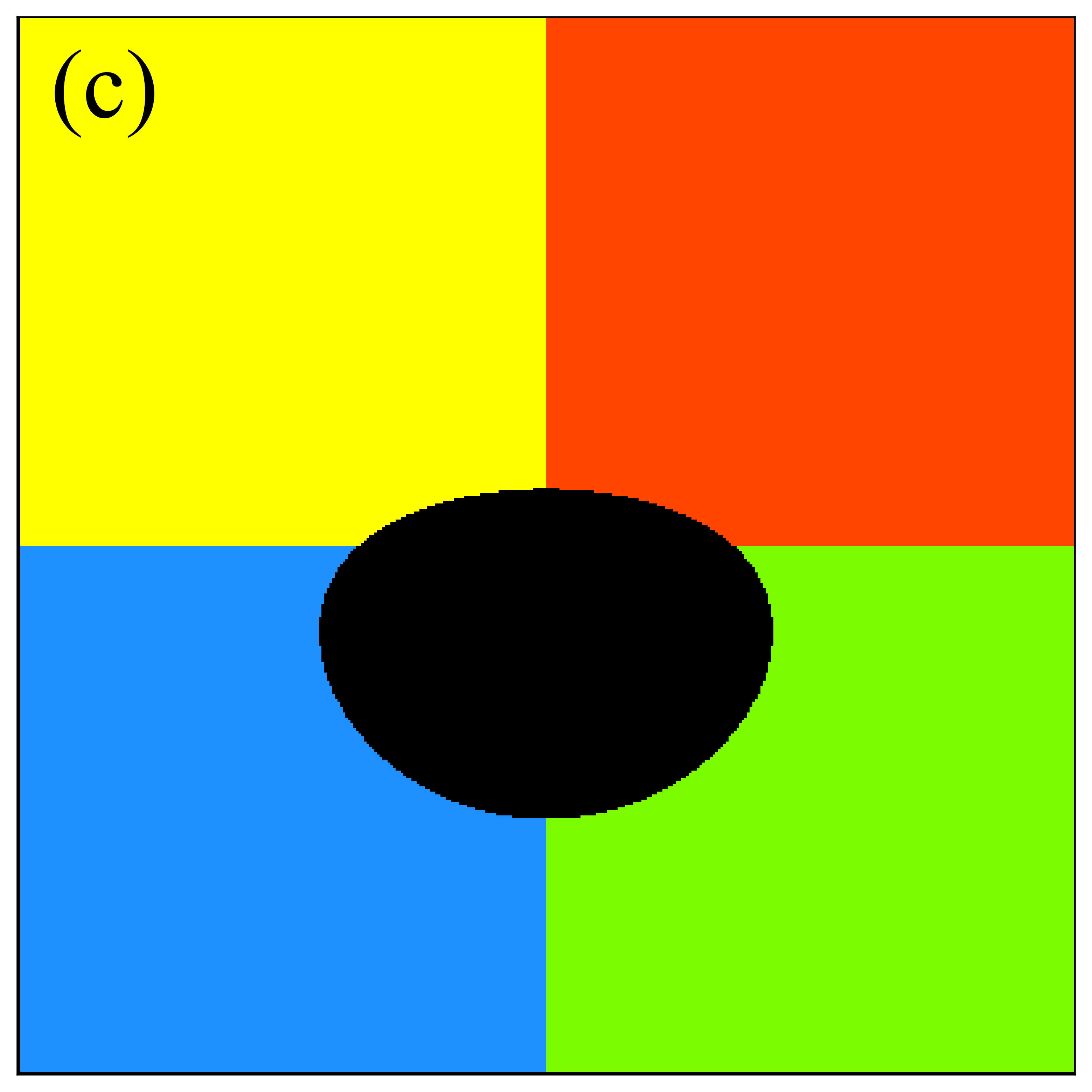}
\includegraphics[width=2.8cm]{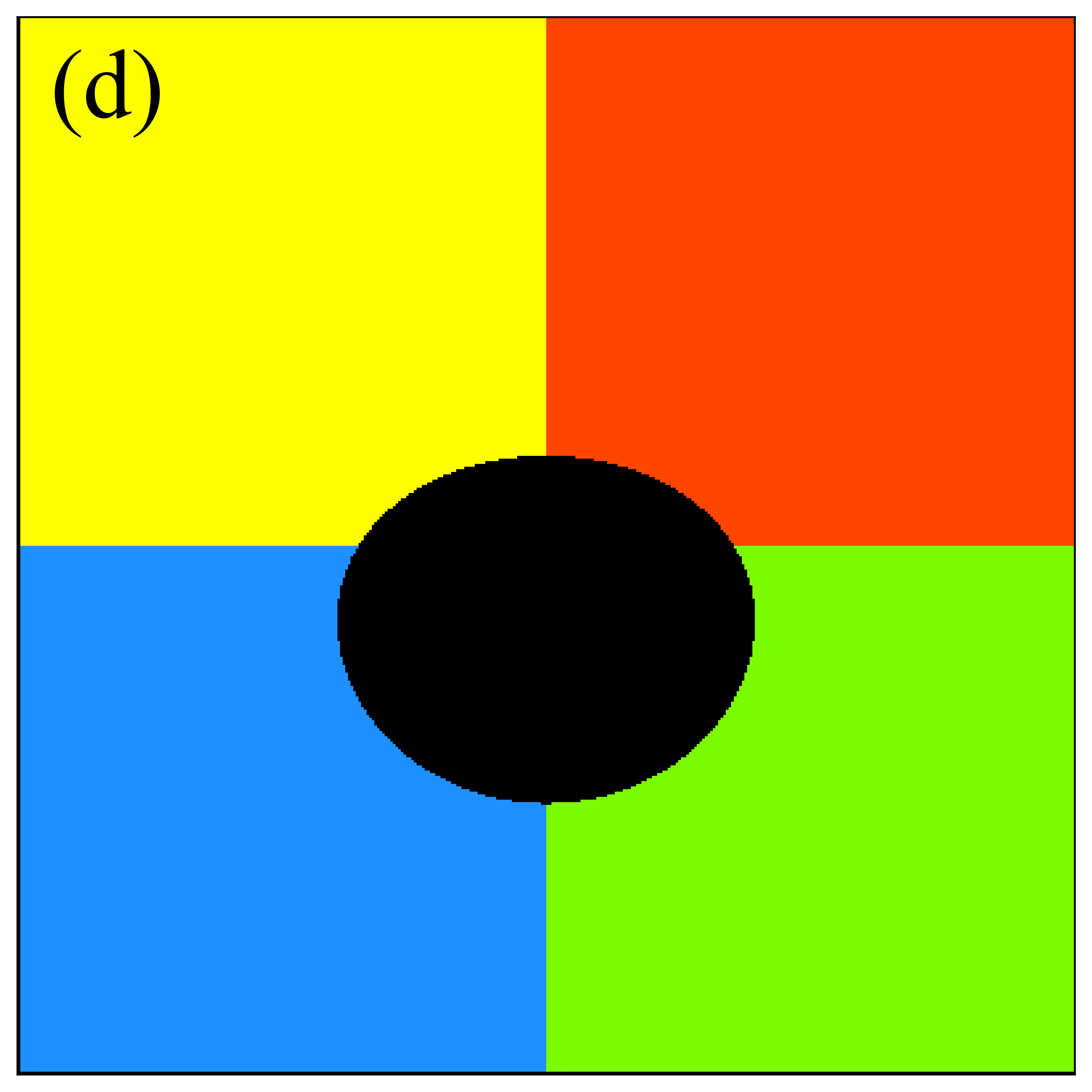}
\includegraphics[width=2.8cm]{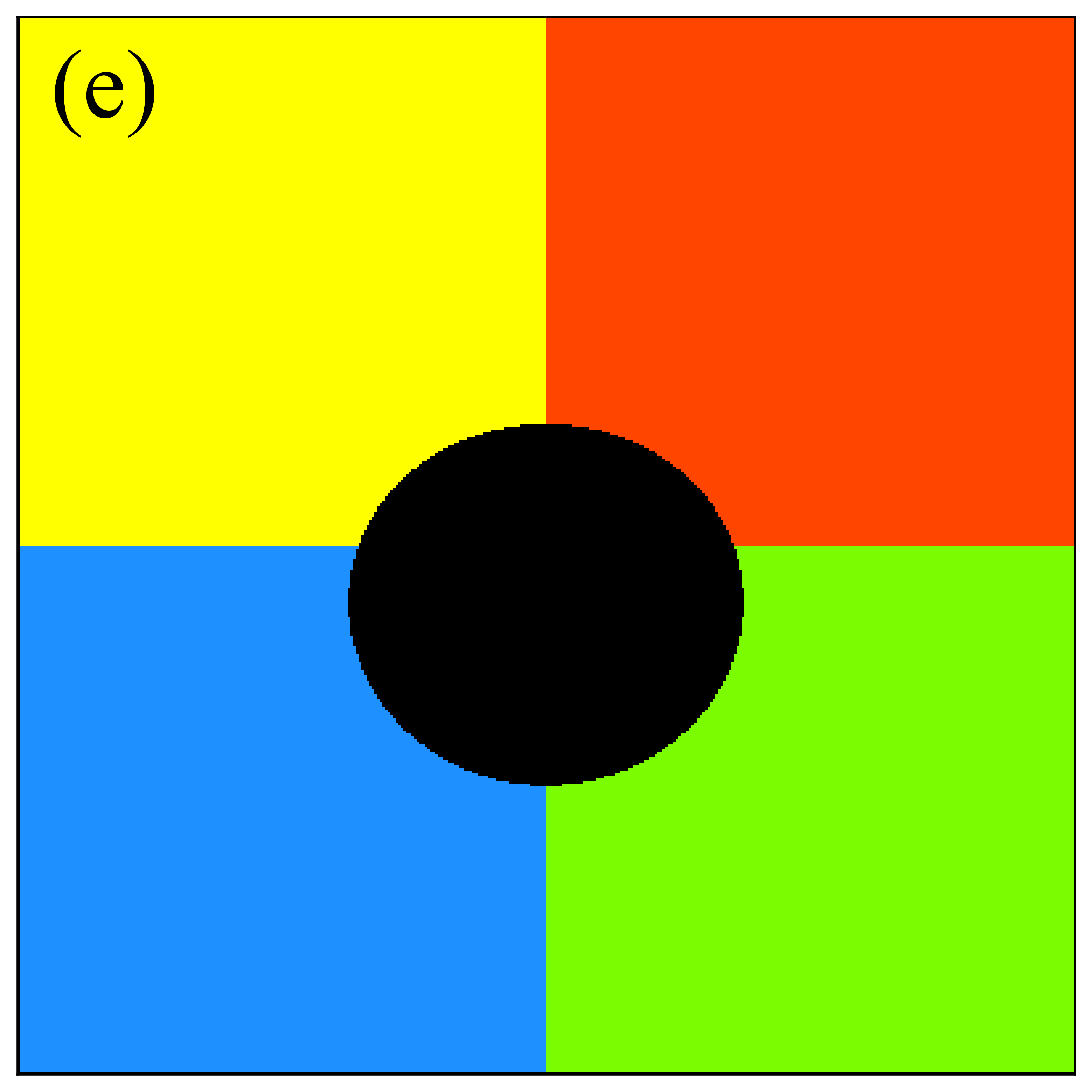}
\includegraphics[width=2.8cm]{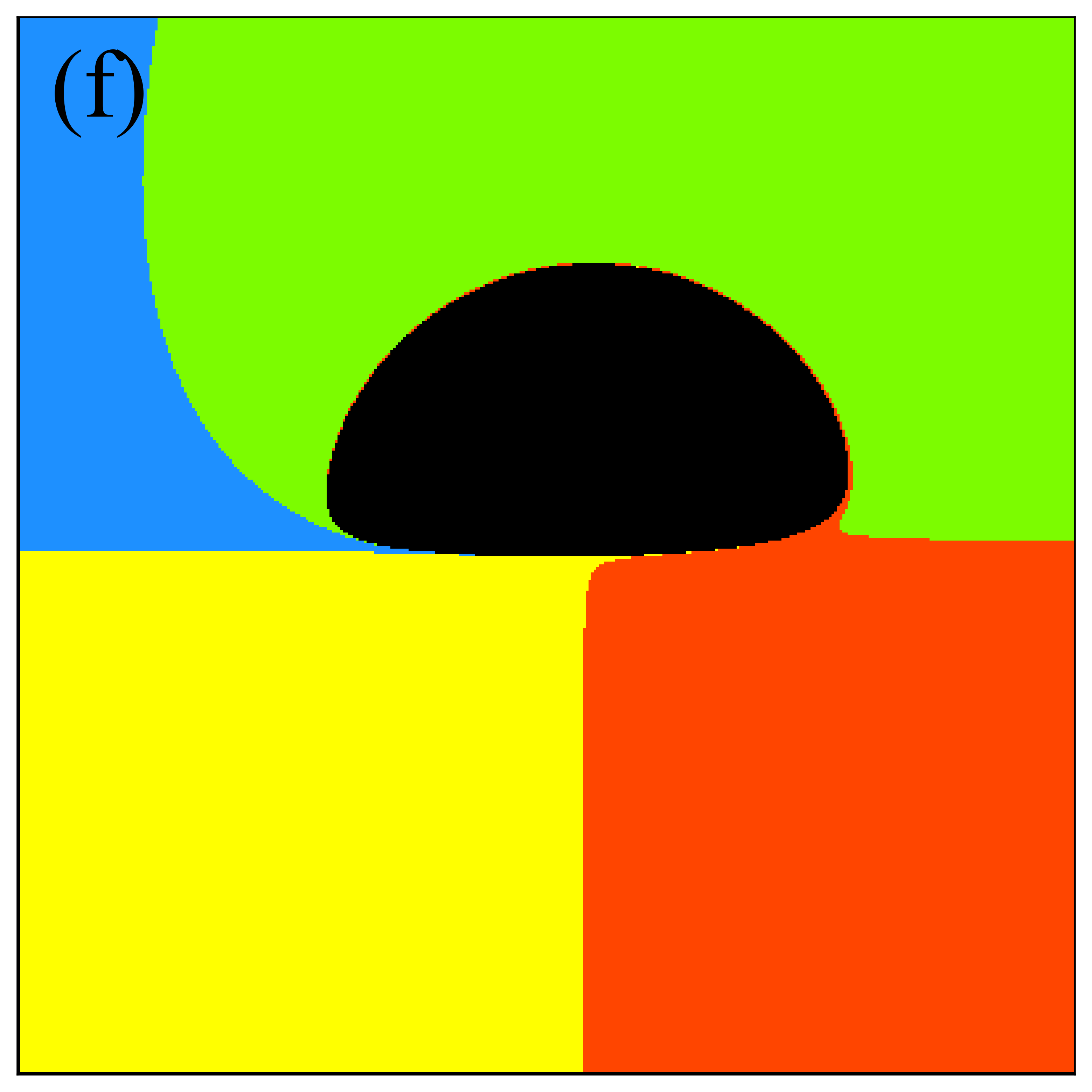}
\includegraphics[width=2.8cm]{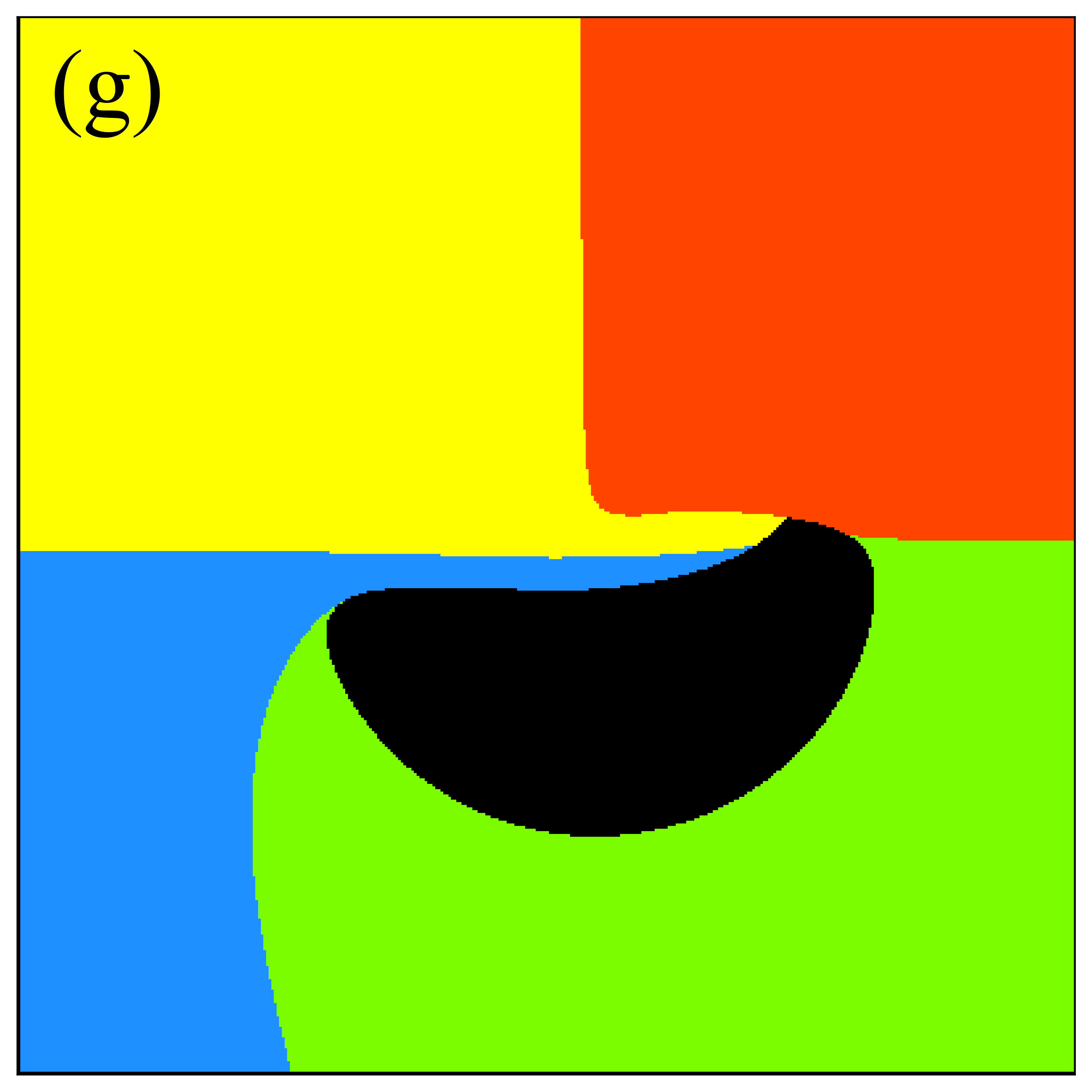}
\includegraphics[width=2.8cm]{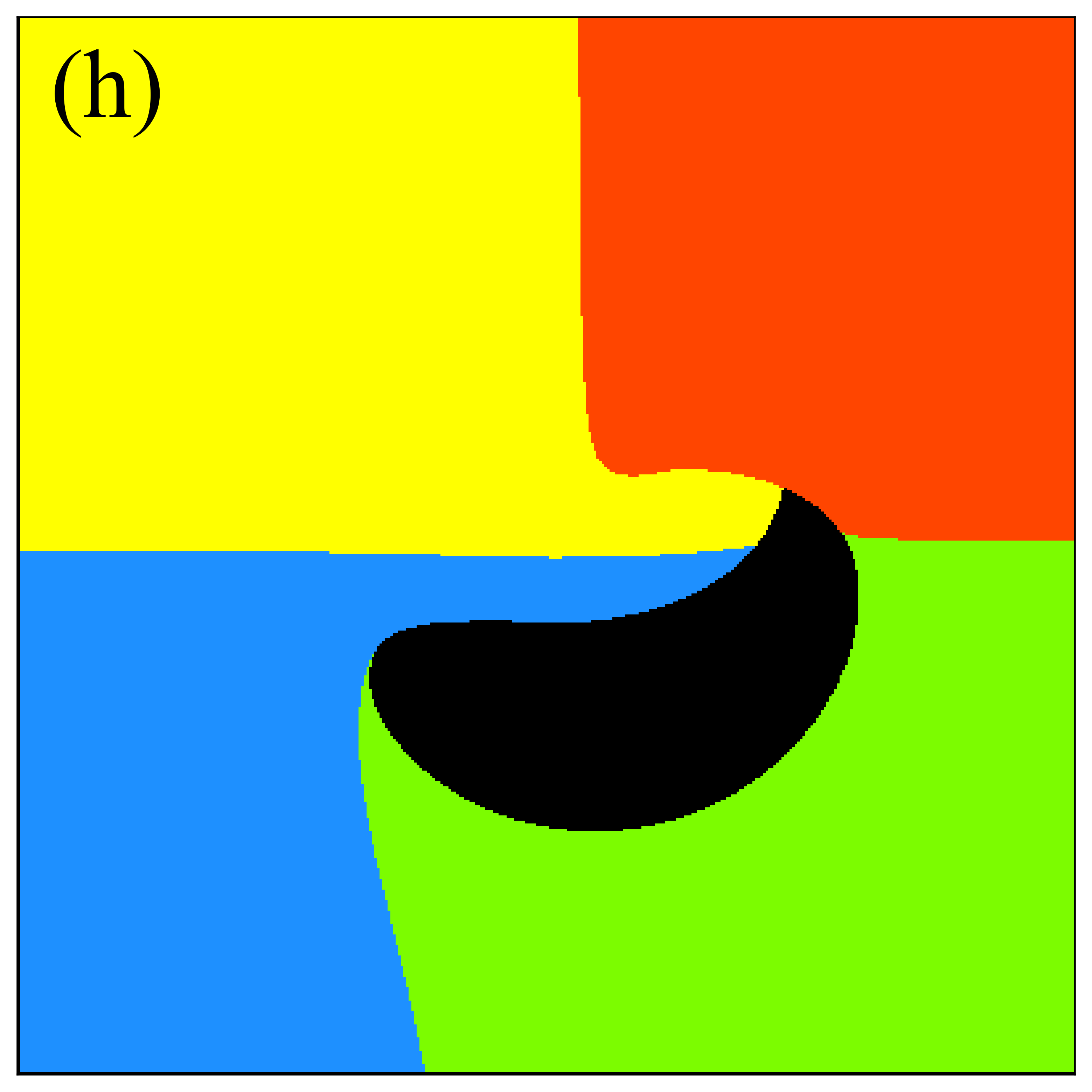}
\includegraphics[width=2.8cm]{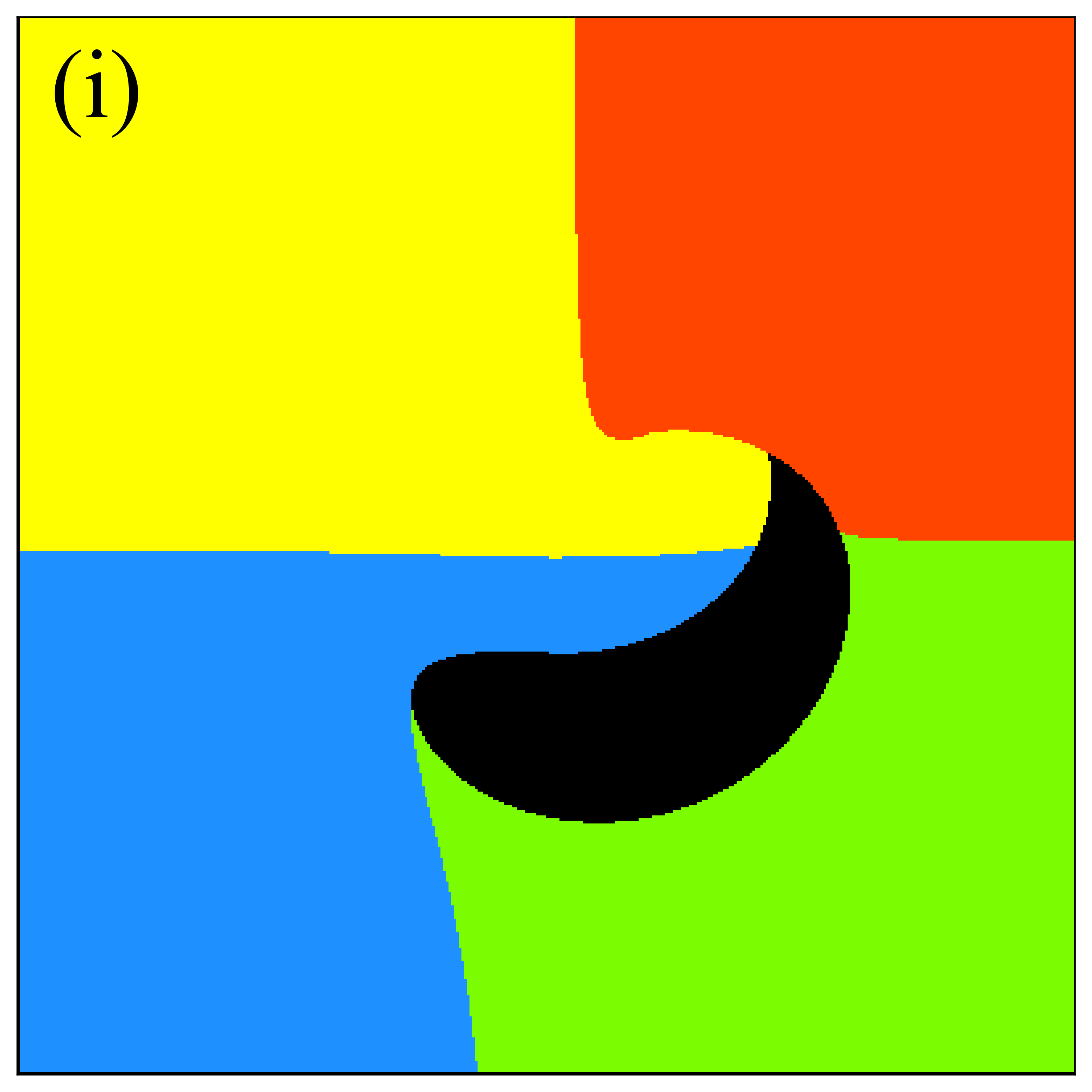}
\includegraphics[width=2.8cm]{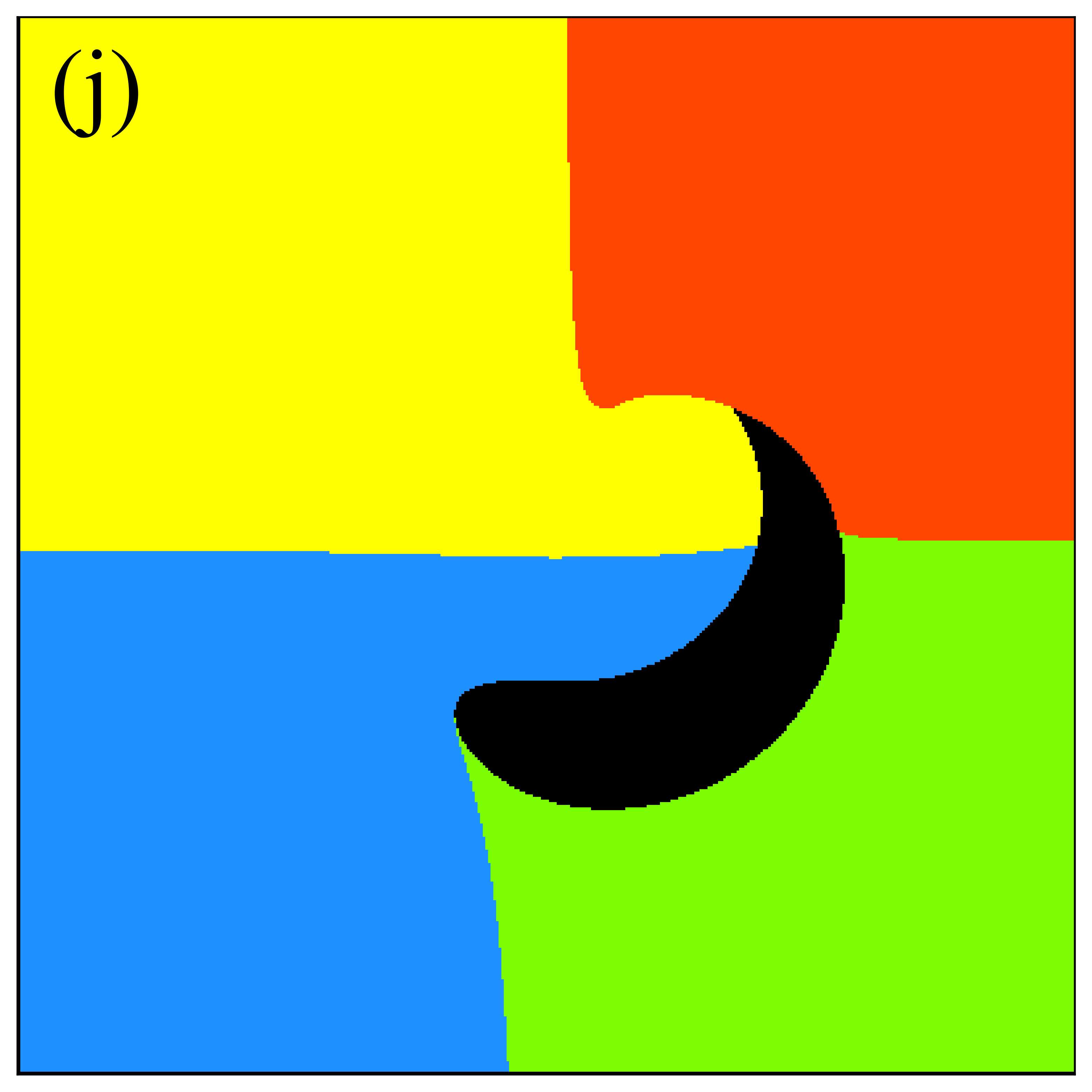}
\includegraphics[width=2.8cm]{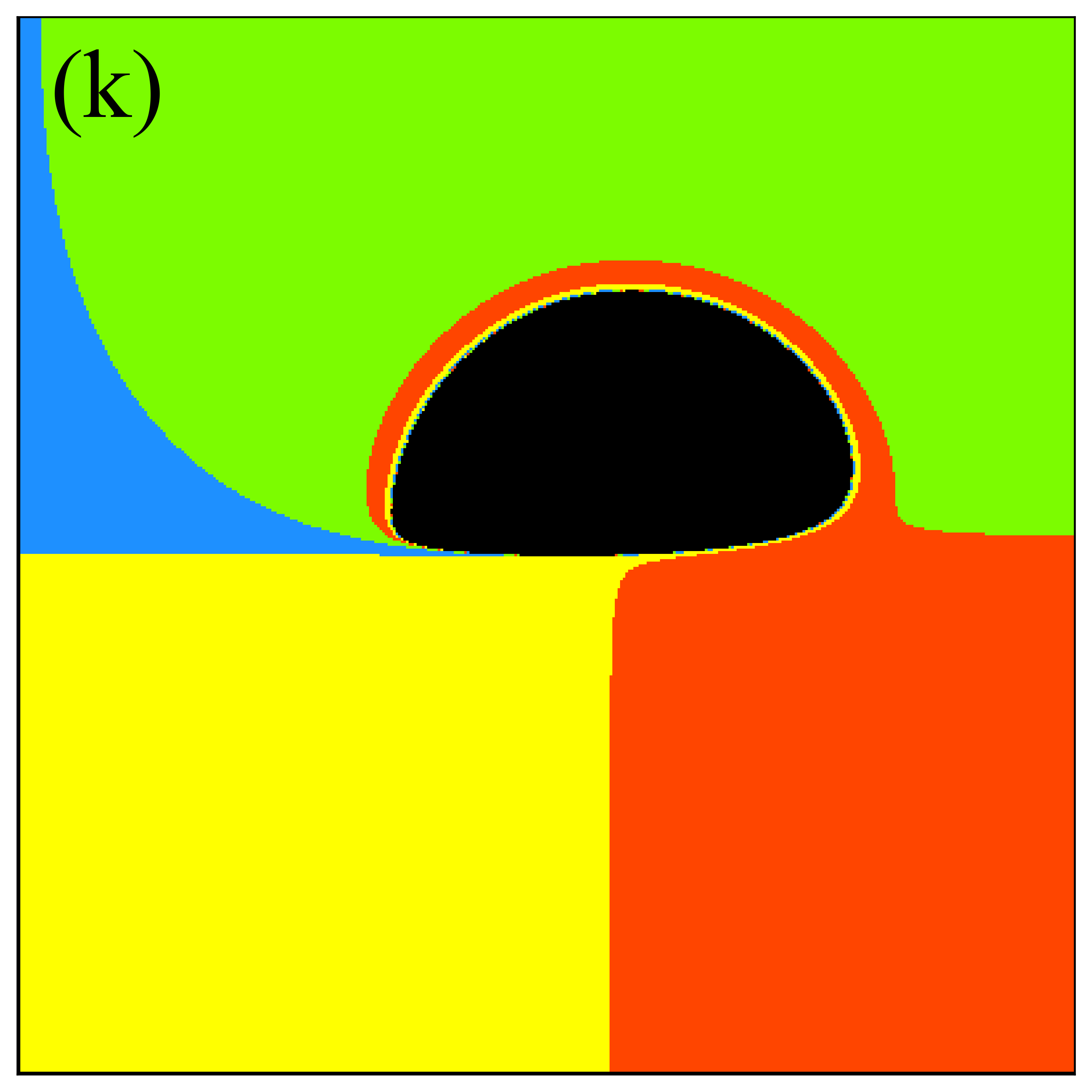}
\includegraphics[width=2.8cm]{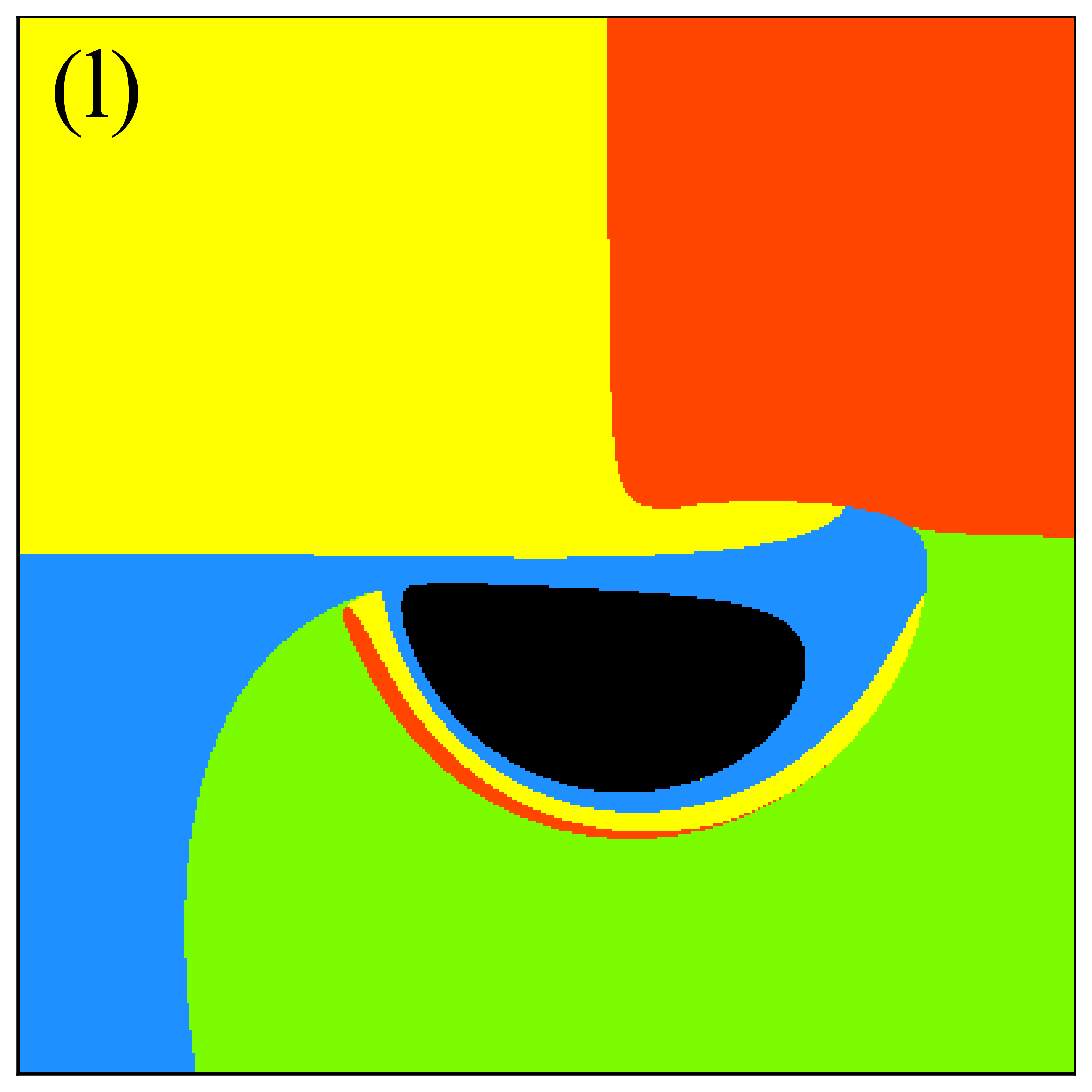}
\includegraphics[width=2.8cm]{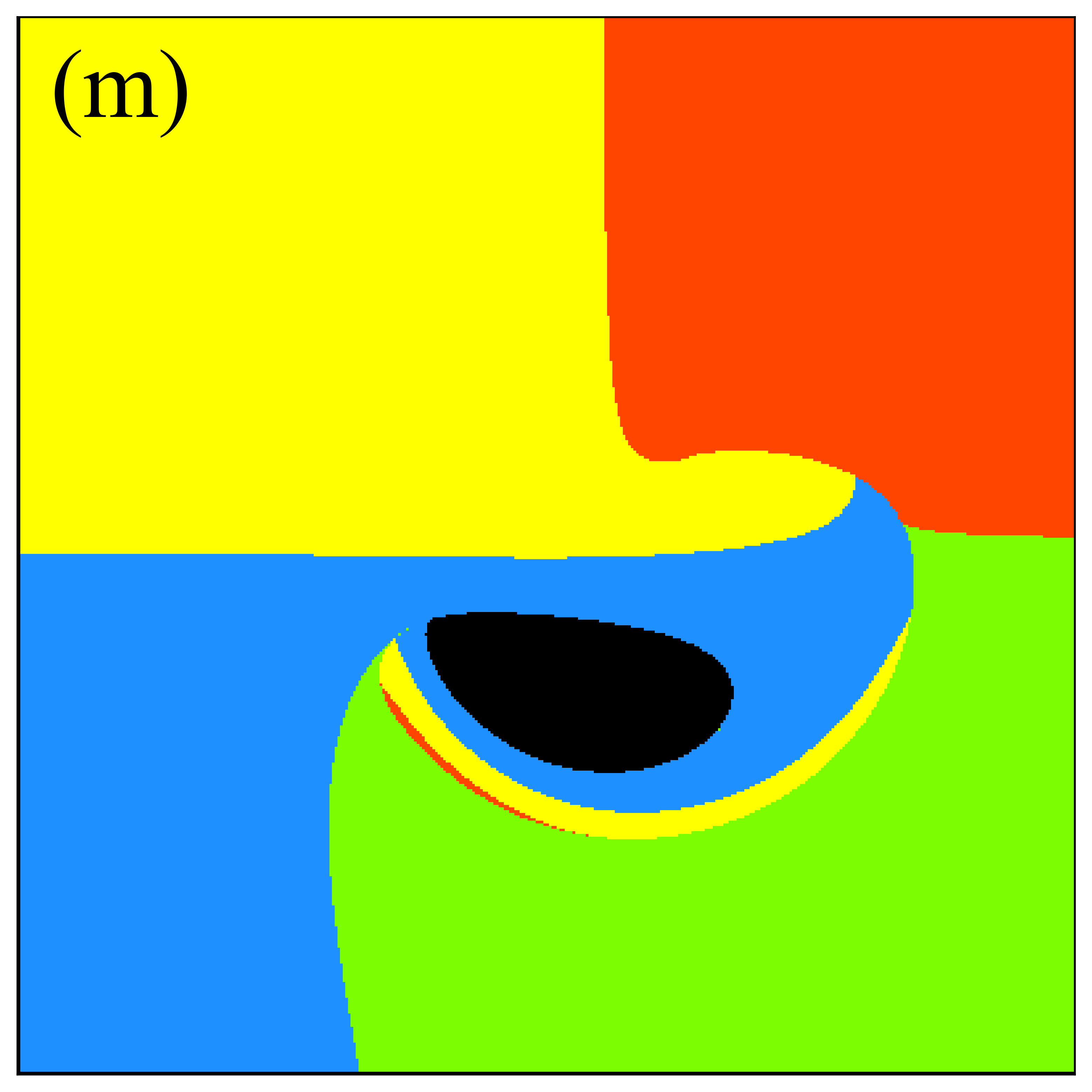}
\includegraphics[width=2.8cm]{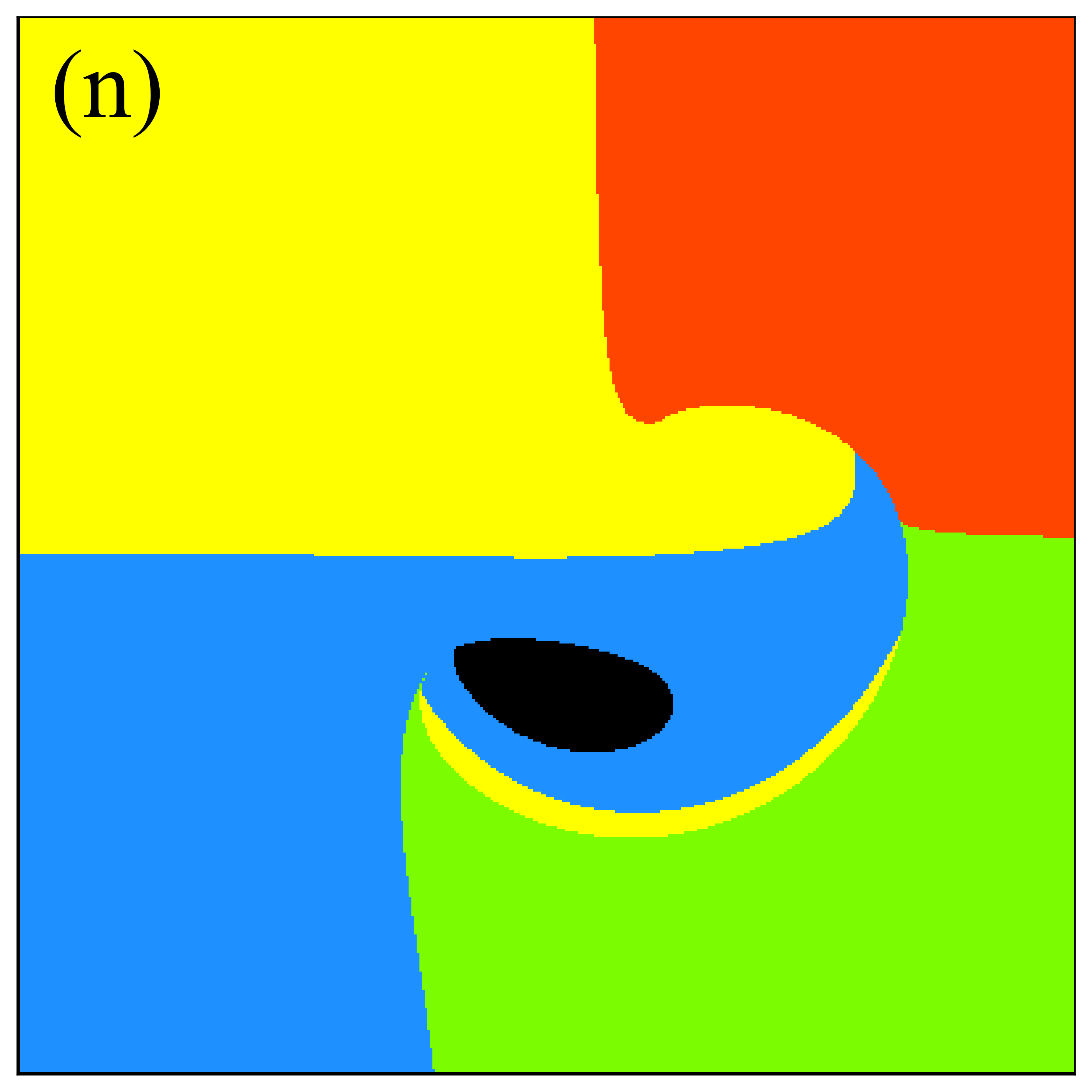}
\includegraphics[width=2.8cm]{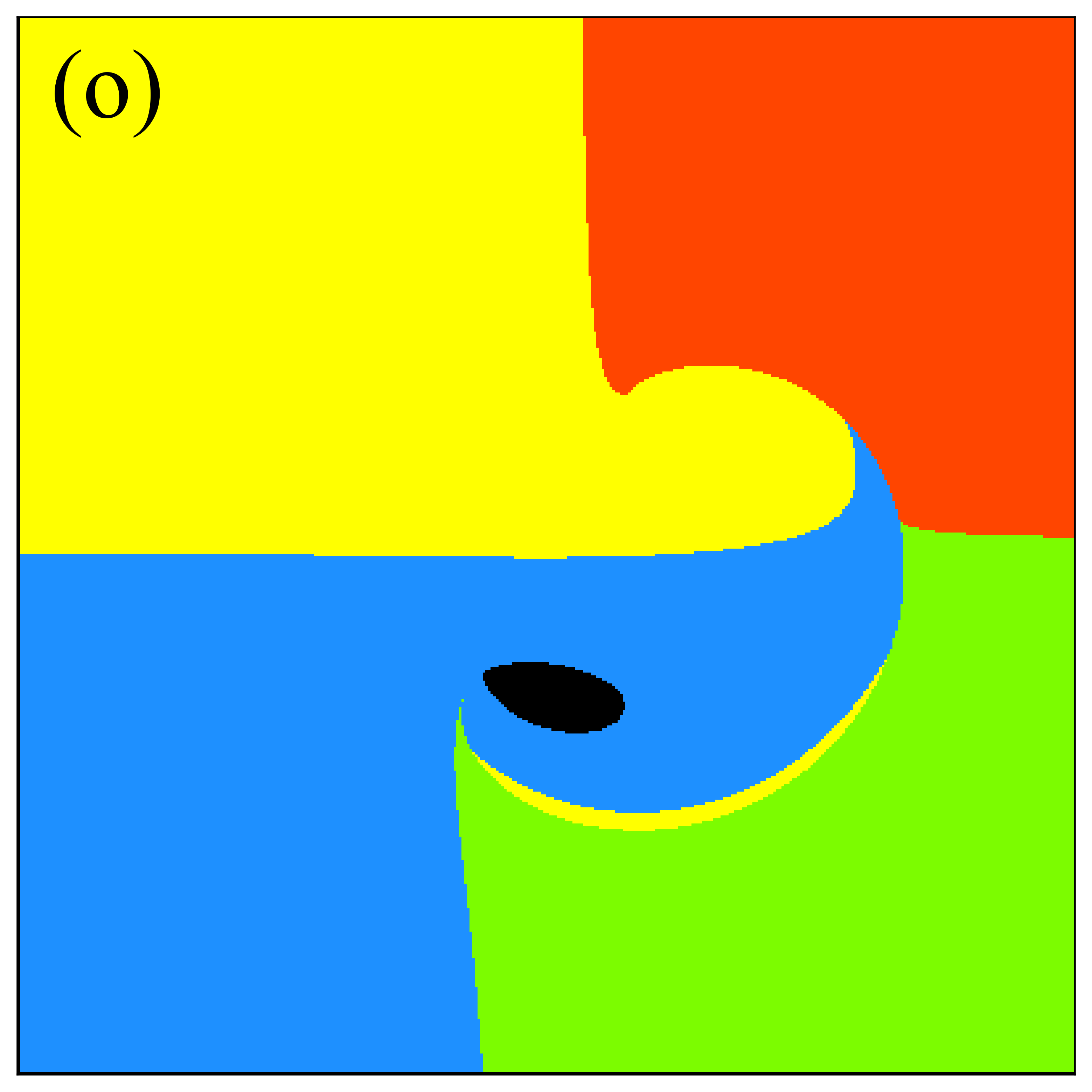}
\includegraphics[width=2.8cm]{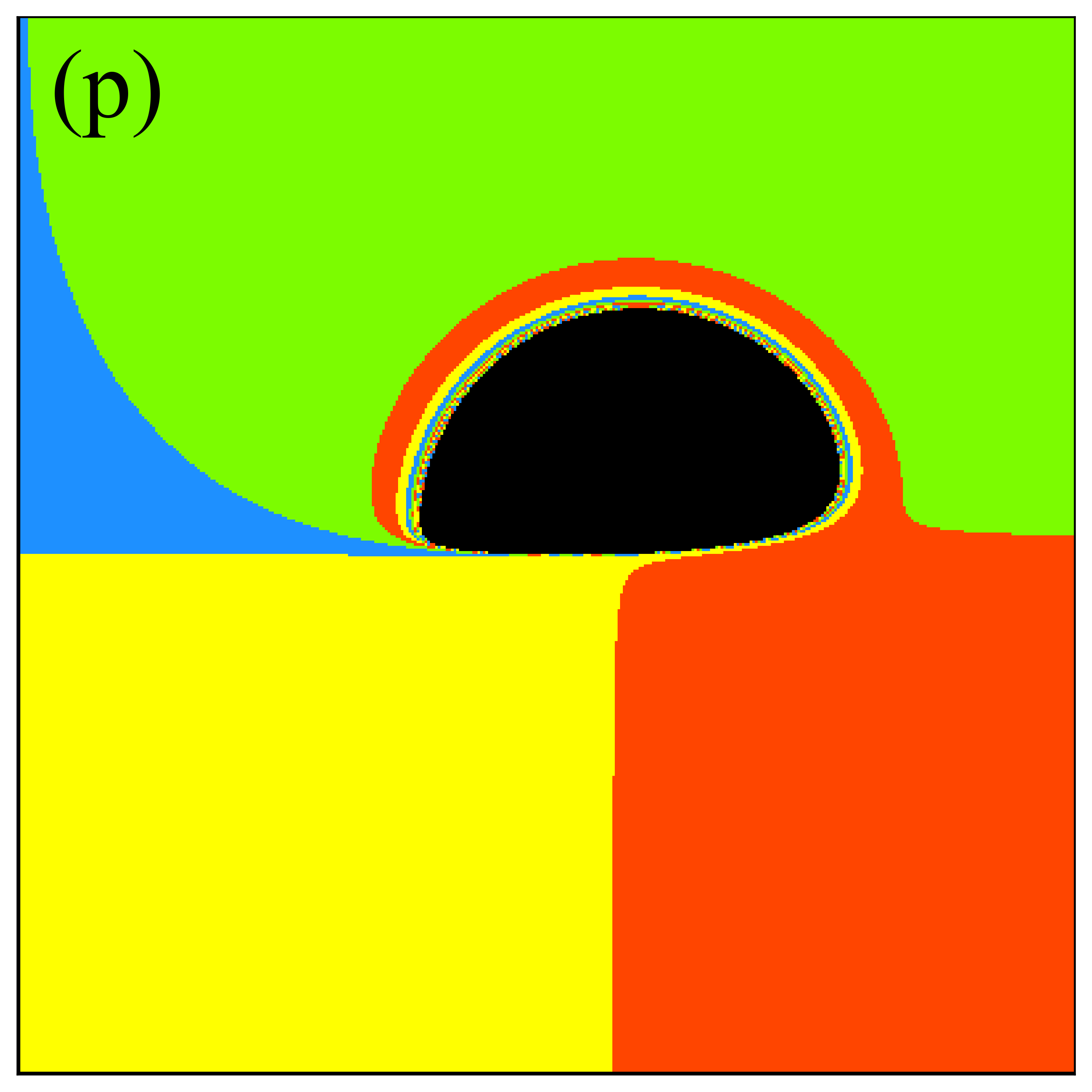}
\includegraphics[width=2.8cm]{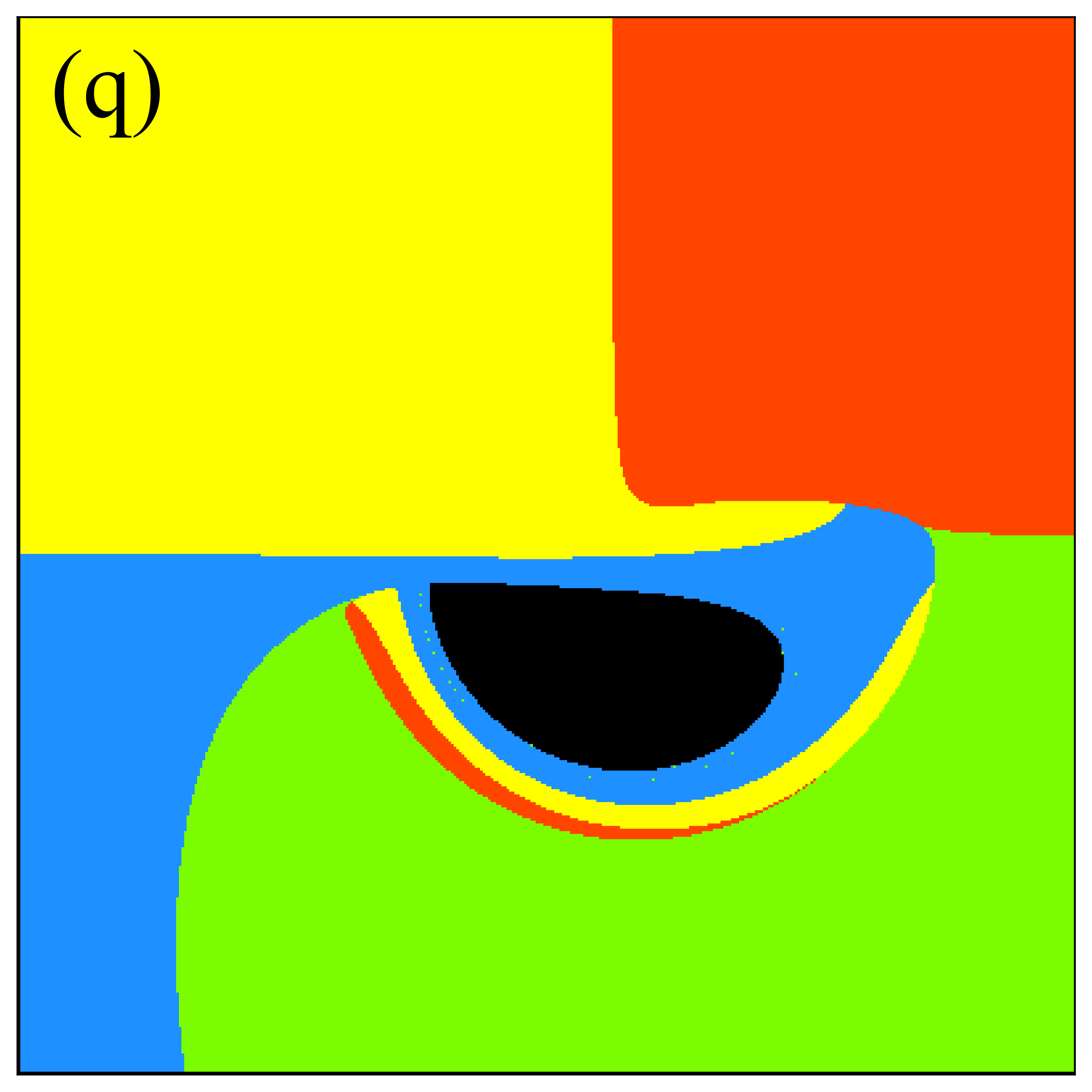}
\includegraphics[width=2.8cm]{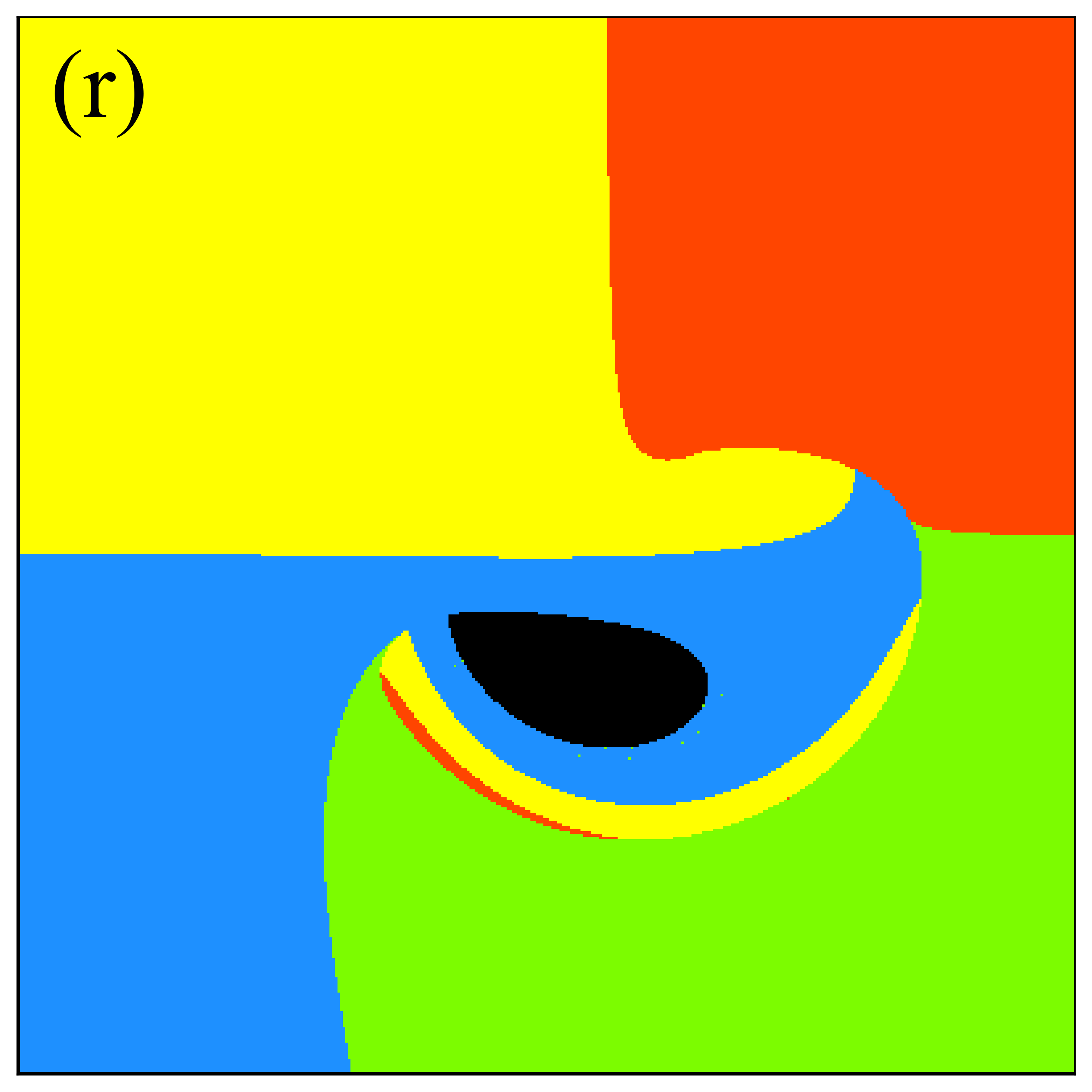}
\includegraphics[width=2.8cm]{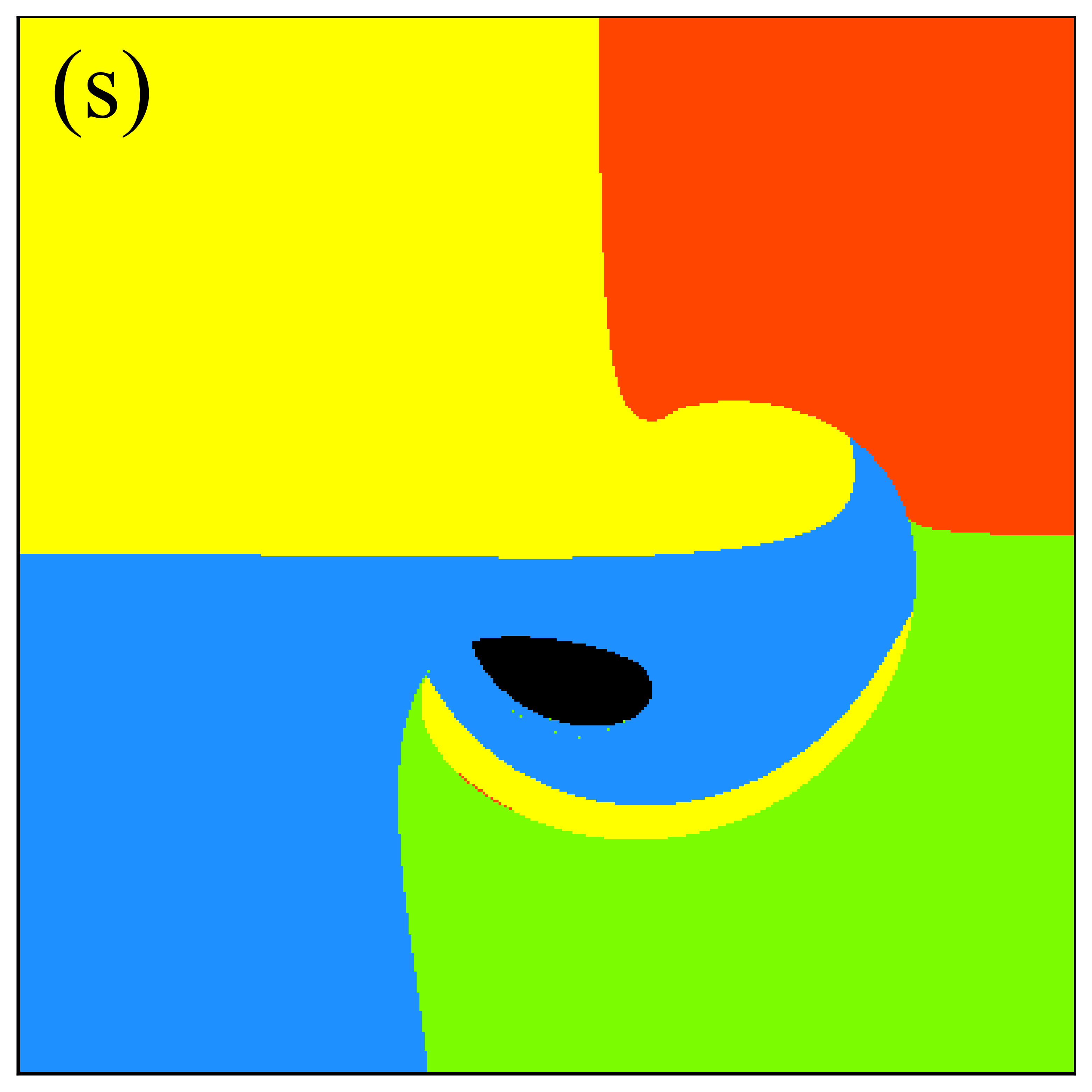}
\includegraphics[width=2.8cm]{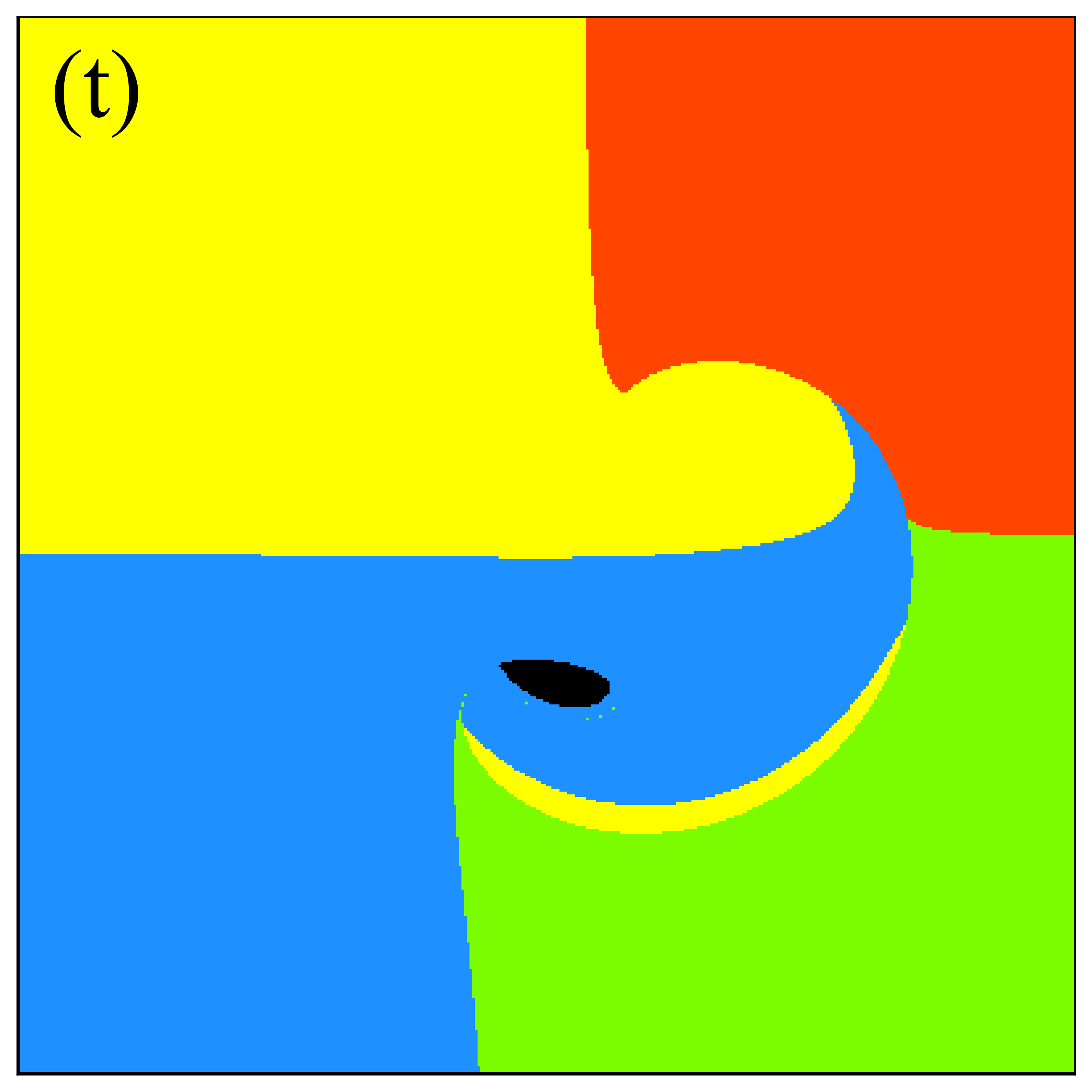}
\caption{Similar to figure 2, but for the observation angle of $85^{\circ}$.}}\label{fig4}
\end{figure*}
Additionally, we observe inverted inner shadows in certain parameter spaces, as shown in the right two columns of figure 3 and the right four columns of figure 4. This inversion arises because the observer's line of sight is situated beneath the accretion disk. The inverted inner shadow can provide insights into the relative positioning between the accretion disk and the observer.
\begin{figure*}
\center{
\includegraphics[width=5cm]{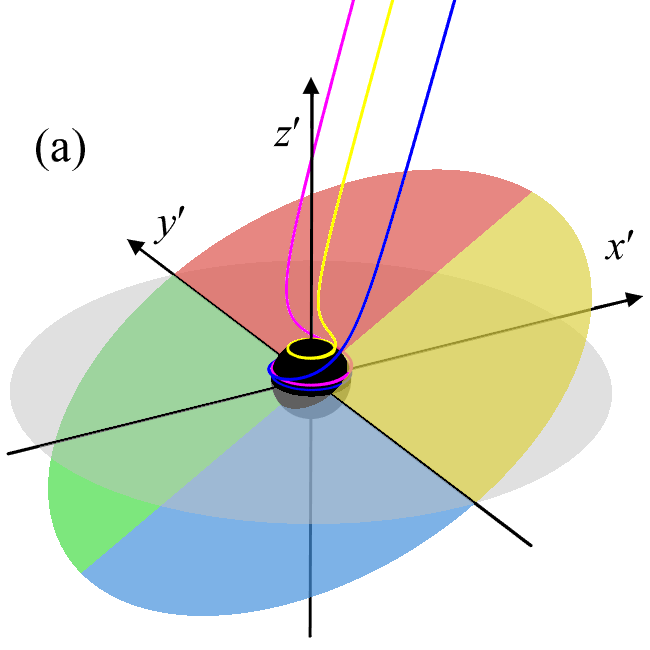}
\includegraphics[width=5cm]{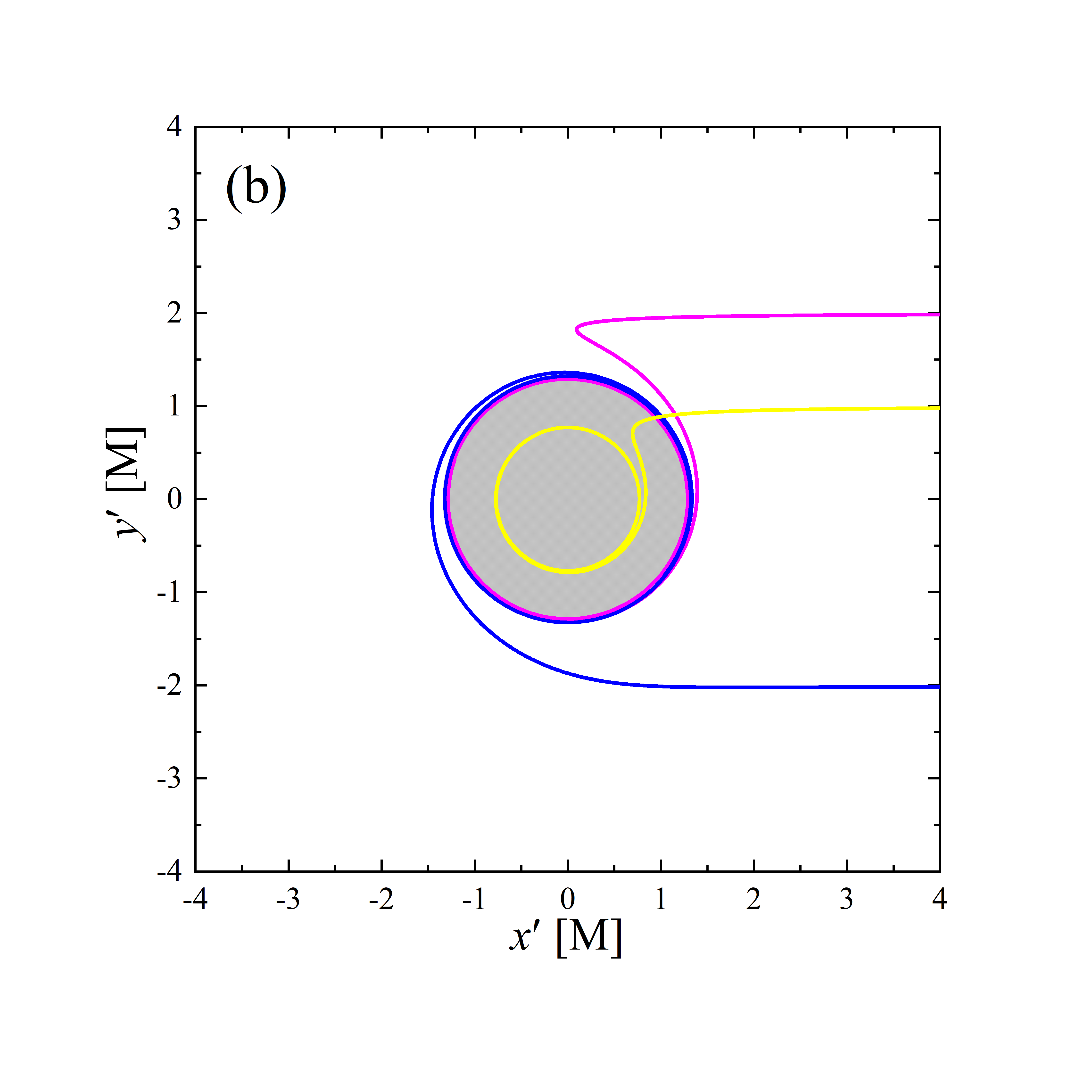}
\includegraphics[width=5cm]{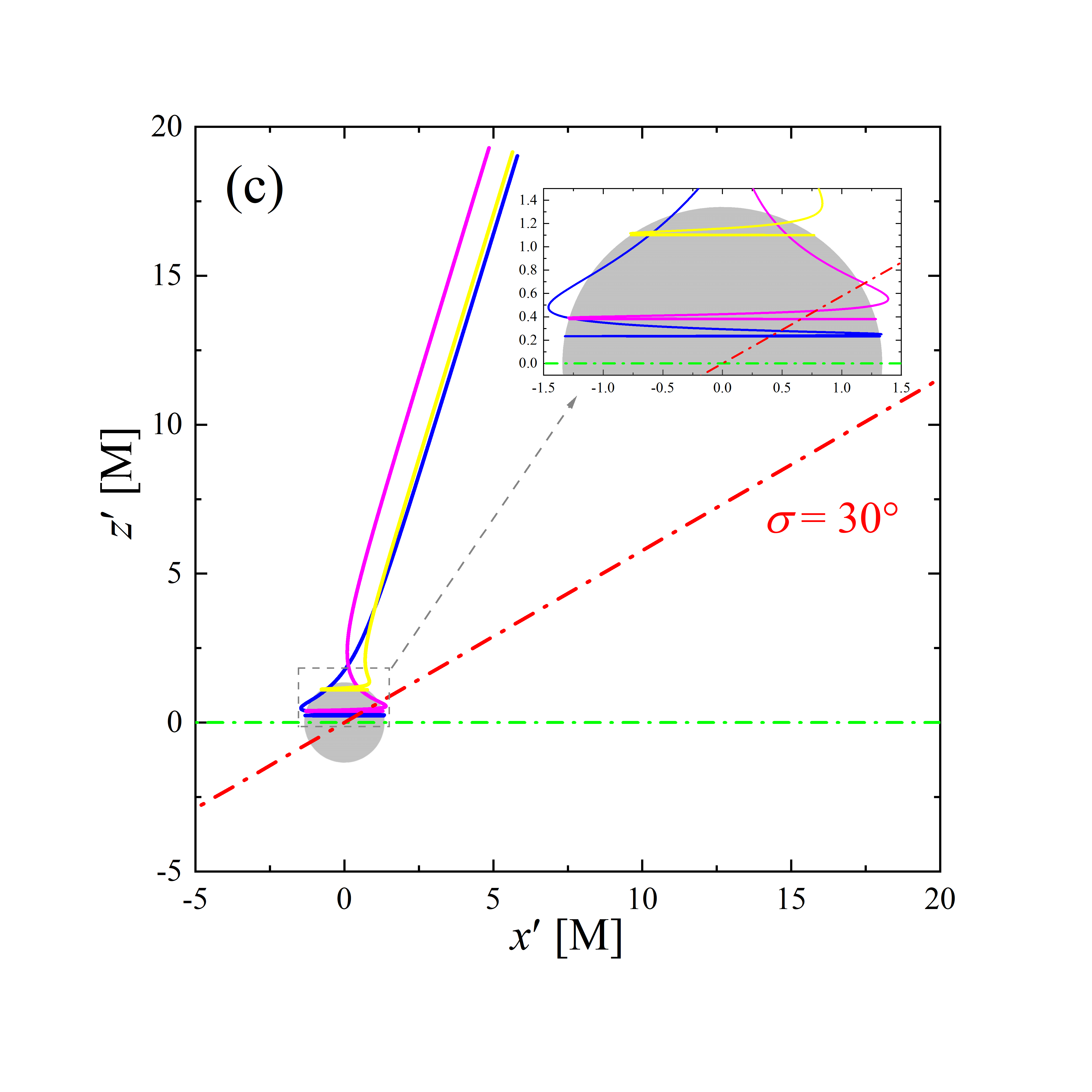}
\caption{Trajectories of three photons in the black hole's local reference frame (a) and their projections onto the (b) $\overline{x^{\prime}o^{\prime}y^{\prime}}$ and $\overline{x^{\prime}o^{\prime}z^{\prime}}$ planes. In panel (a), the equatorial accretion disk is represented by a gray disk, while the black hole's event horizon is shown as a black sphere. Panels (b) and (c) display the event horizon as a gray disk, with the $30^{\circ}$-tilted accretion disk and equatorial disk marked by red and green dashed lines, respectively. The results demonstrate that under the equatorial disk configuration, all three rays (blue, purple, and yellow) contribute to the inner shadow, as they reach the event horizon without intersecting the disk. As evident in panel (c), these photons maintain trajectories entirely above the equatorial plane. However, when the disk is tilted to $30^{\circ}$, the blue and purple rays intersect the inclined plane before entering the black hole, indicating that their corresponding regions on the observational screen will exhibit non-zero luminosity. This reveals that the tilted accretion disk effectively illuminates portions of the black hole's inner shadow. Here, we have the viewing angle of $17^{\circ}$, and the spin parameter of $0.94$.}}\label{fig5}
\end{figure*}

In figures 3(n) and (s), we identify escaping photons (depicted in gray) that, upon emission from the observation plane, neither get absorbed by the black hole nor collide with the accretion disk; instead, they deflect near the black hole and travel to infinity. Figure 6 illustrates the trajectory of the photon corresponding to the observational coordinate $(-3.46867, -1.82456)$ from figure 3(n) within the black hole's local reference frame (a), along with its projections onto the $\overline{x^{\prime}o^{\prime}z^{\prime}}$ plane (b) and $\overline{x^{\prime}o^{\prime}y^{\prime}}$ plane (c). It is evident that the photon consistently remains below the tilted accretion disk throughout its trajectory, without making contact with the disk, as demonstrated in the zoomed-in view in (b). Notably, in the case of the equatorial plane, the occurrence of escaping photons is considerably less likely, as evidenced by the intersection of the photon trajectory with the green dashed line in (b). For a tilted accretion disk of finite size, escaping photons can only manifest in specific parameter spaces, a phenomenon resulting from the combined effects of frame dragging and the tilted disk.
\begin{figure*}
\center{
\includegraphics[width=5cm]{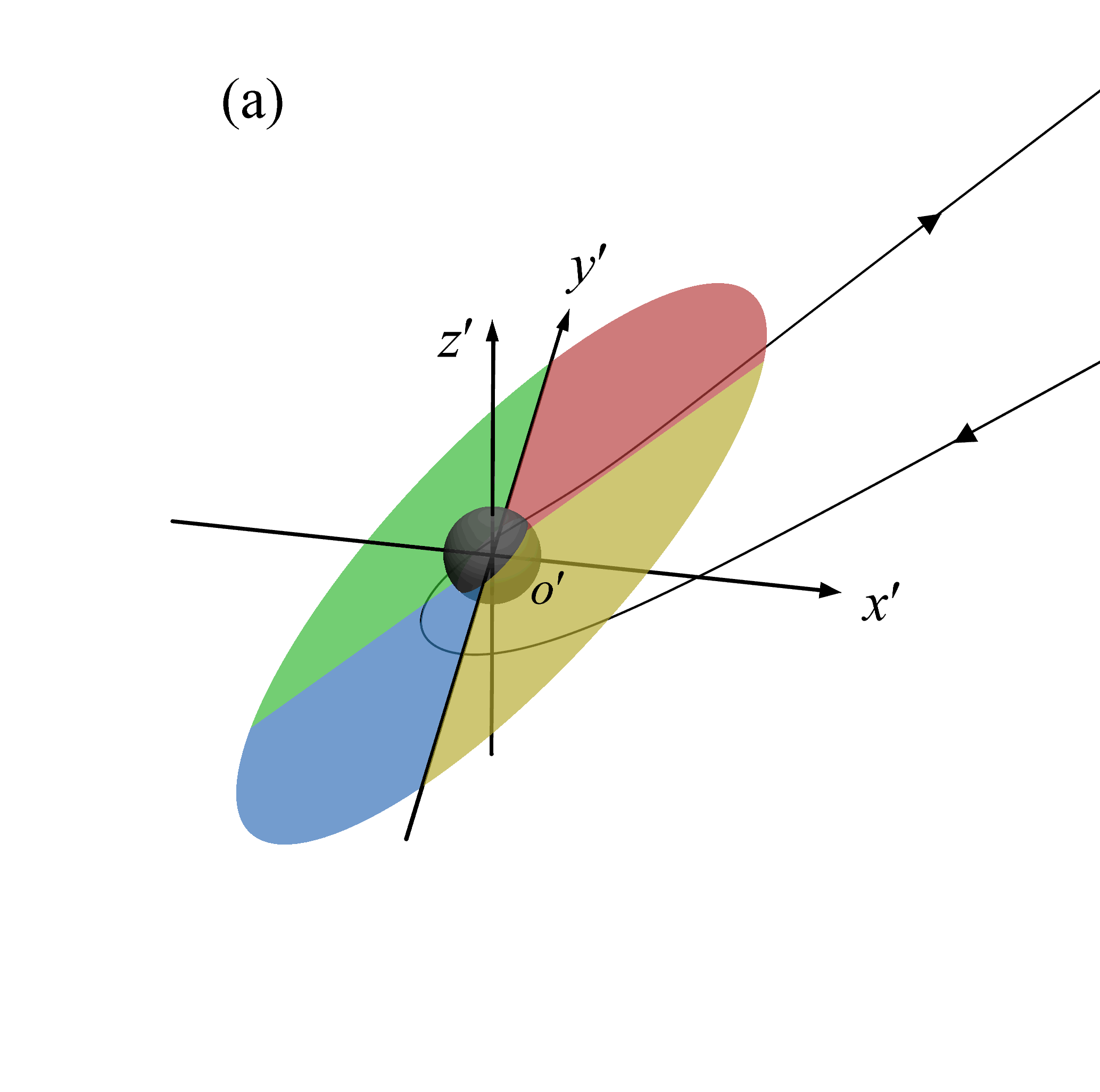}
\includegraphics[width=5cm]{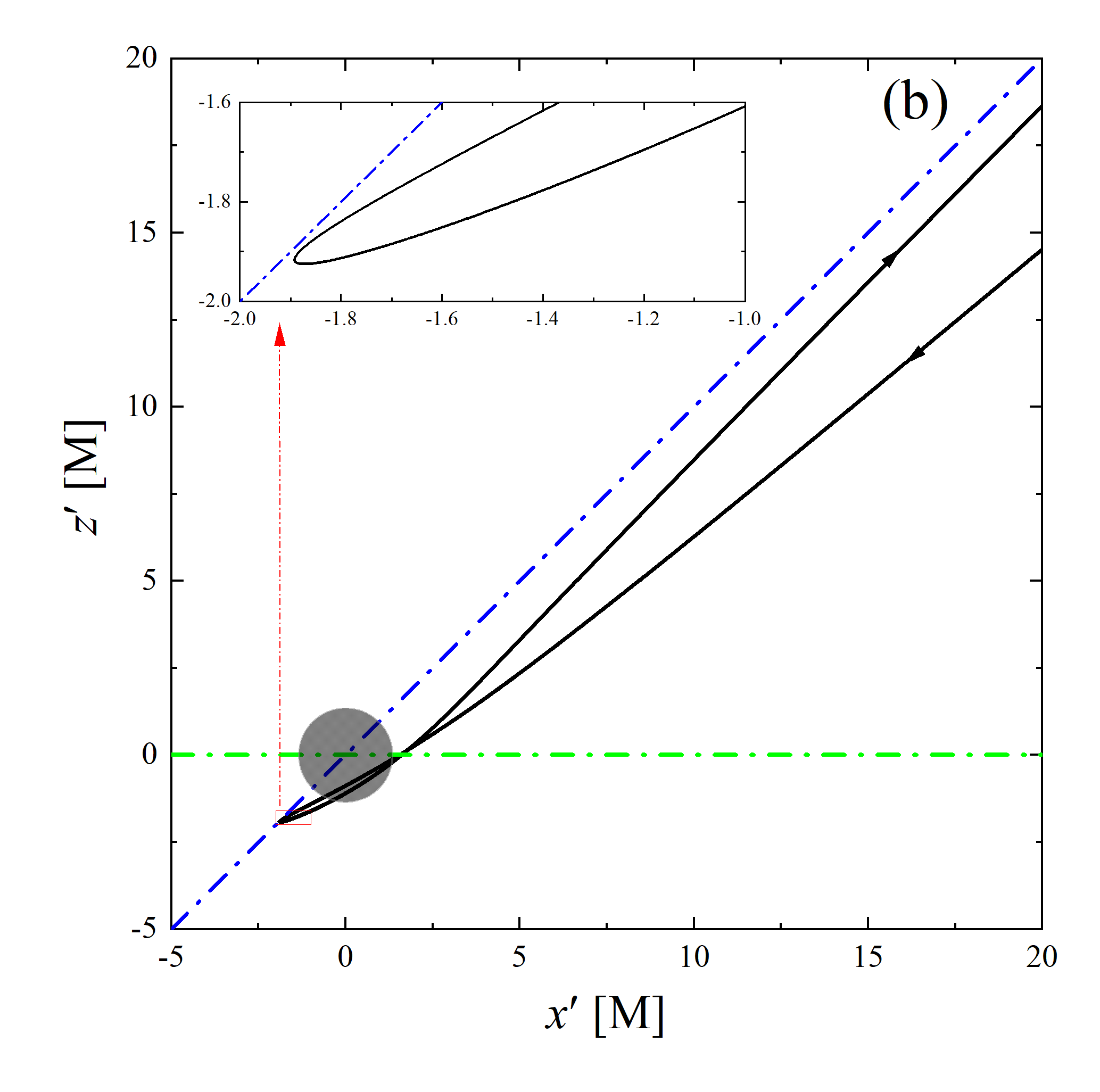}
\includegraphics[width=5cm]{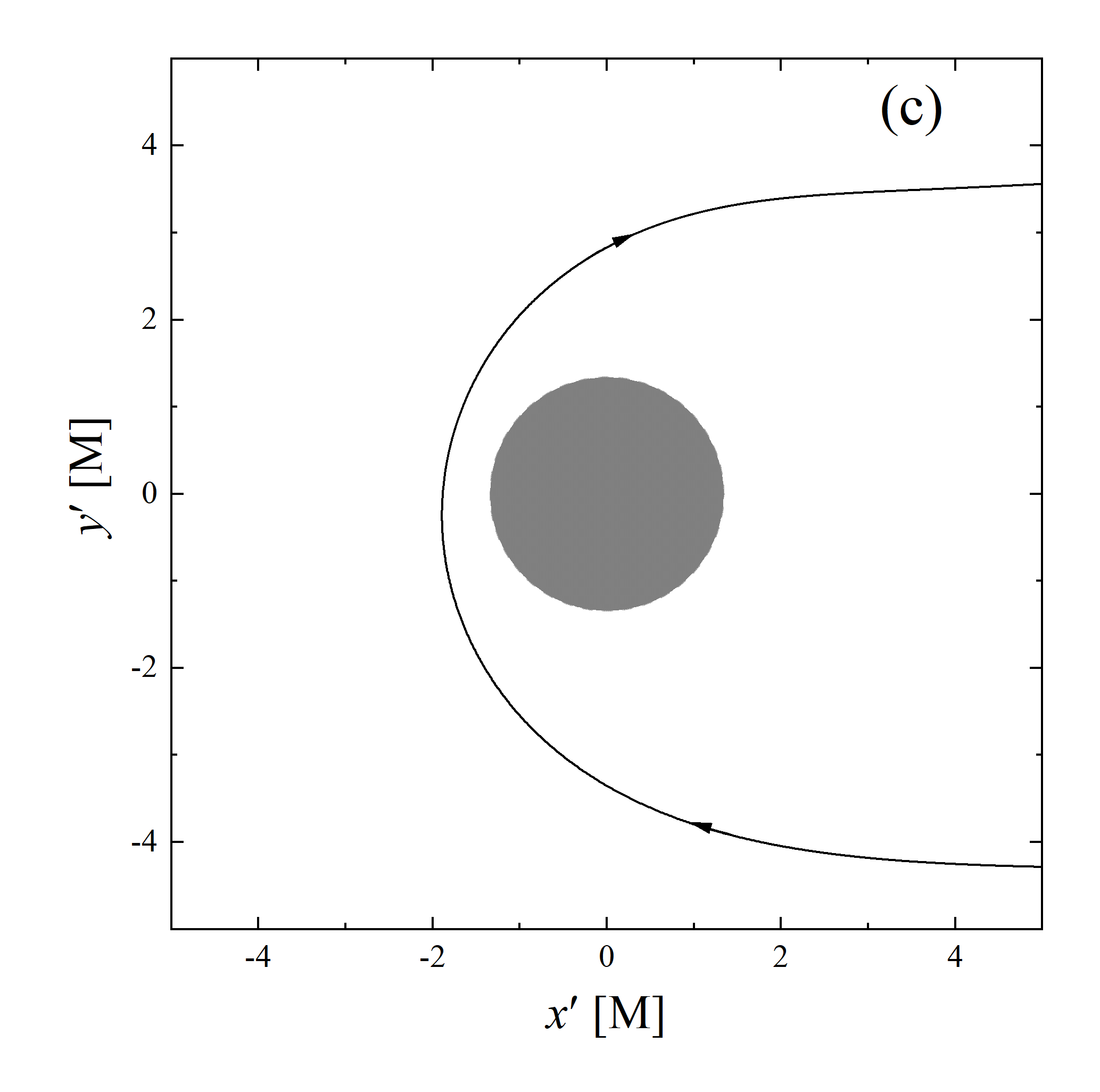}
\caption{Trajectory of an escaping photon in the black hole's local reference frame (a) and its projection onto the $\overline{x^{\prime}o^{\prime}z^{\prime}}$ plane (b) and the $\overline{x^{\prime}o^{\prime}y^{\prime}}$ plane (c). In panel (b), the tilted accretion disk ($\sigma = 45^{\circ}$) and the equatorial plane are represented by the blue and green dashed lines, respectively. It can be observed that the escaping photon neither crosses the black hole's event horizon nor collides with the accretion disk. Instead, it escapes to infinity. Notably, this photon can intersect the equatorial accretion disk, as indicated by the intersection of the photon trajectory and the green dashed line in (b). Here, the viewing angle is $50^{\circ}$, and the spin parameter is $0.94$.}}\label{fig6}
\end{figure*}

It should be emphasized that the inner shadow under tilted disk conditions studied in this work constitutes a rigorous generalization of the black hole's inner shadow in the equatorial disk scenario \cite{Chael et al. (2021)}. Specifically, the inner shadow boundaries in figures 2-4 represent the direct image of the intersection between the event horizon and the tilted plane of the accretion disk.
In other words, the inner shadow studied in this work is defined purely geometrically: the boundary corresponds to the projection onto the observation screen of the intersection curve between an arbitrarily inclined plane and the black hole's event horizon. The formation of such an inner shadow requires two conditions to be met. First, the accretion disk's inner boundary must extend to the black hole's event horizon. Second, from the perspective of backward ray tracing, photons contributing to the inner shadow must intersect the event horizon without crossing the accretion disk. Figures 7 and 8 display the number of intersections between light rays and the tilted accretion disk for each pixel on the observation plane across a broad parameter space. Here, the black color indicates zero intersections, meaning the light rays bypass the accretion disk and directly cross the event horizon, thereby contributing to the inner shadow. Indeed, one can clearly observe that the black regions in figures 7 and 8 exhibit coincidence with the inner shadows shown in the corresponding panels of figures 2 and 4. In the equatorial accretion disk scenario (left columns of figures 7 and 8), rays intersecting the disk multiple times contribute to the lensing band (yellow regions). Notably, these lensing bands progressively approach the critical curve \cite{Gralla et al. (2019)} as the intersection count increases, as demonstrated by the green curves within the bands. By contrast, such higher-order image convergence behavior is rarely observed in the tilted disk configuration. We find that for smaller black hole spins and disk inclination angles (i.e., panels (g) and (h) in figure 7), multiply disk-crossing rays primarily concentrate near the critical curve, though their signatures remain identifiable near the inner shadow. However, as both $\sigma$ and $a$ increase, these higher-order images become distributed across multiple distinct regions without convergence behavior, as evidenced by the green, blue, and gray patterns in panels (l)-(o) and (q)-(t) of figure 7. This phenomenon occurs because a tilted accretion disk effectively introduces additional photon-intercepting planes on both sides of the equatorial plane, substantially increasing the probability of ray-disk interactions. This effect is particularly pronounced for photons orbiting near the equatorial plane under frame-dragging effects, which can rapidly accumulate multiple disk intersections. However, such intersection multiplicity does not provide meaningful guidance for identifying the black hole's characteristic geometry or lensing bands. Nevertheless, from an astrophysical perspective, the multiple intersections between light rays and the tilted accretion disk can enhance the specific intensities of these photons, producing bright features in the image. This phenomenon warrants further investigation in future studies.
\begin{figure*}
\center{
\includegraphics[width=2.8cm]{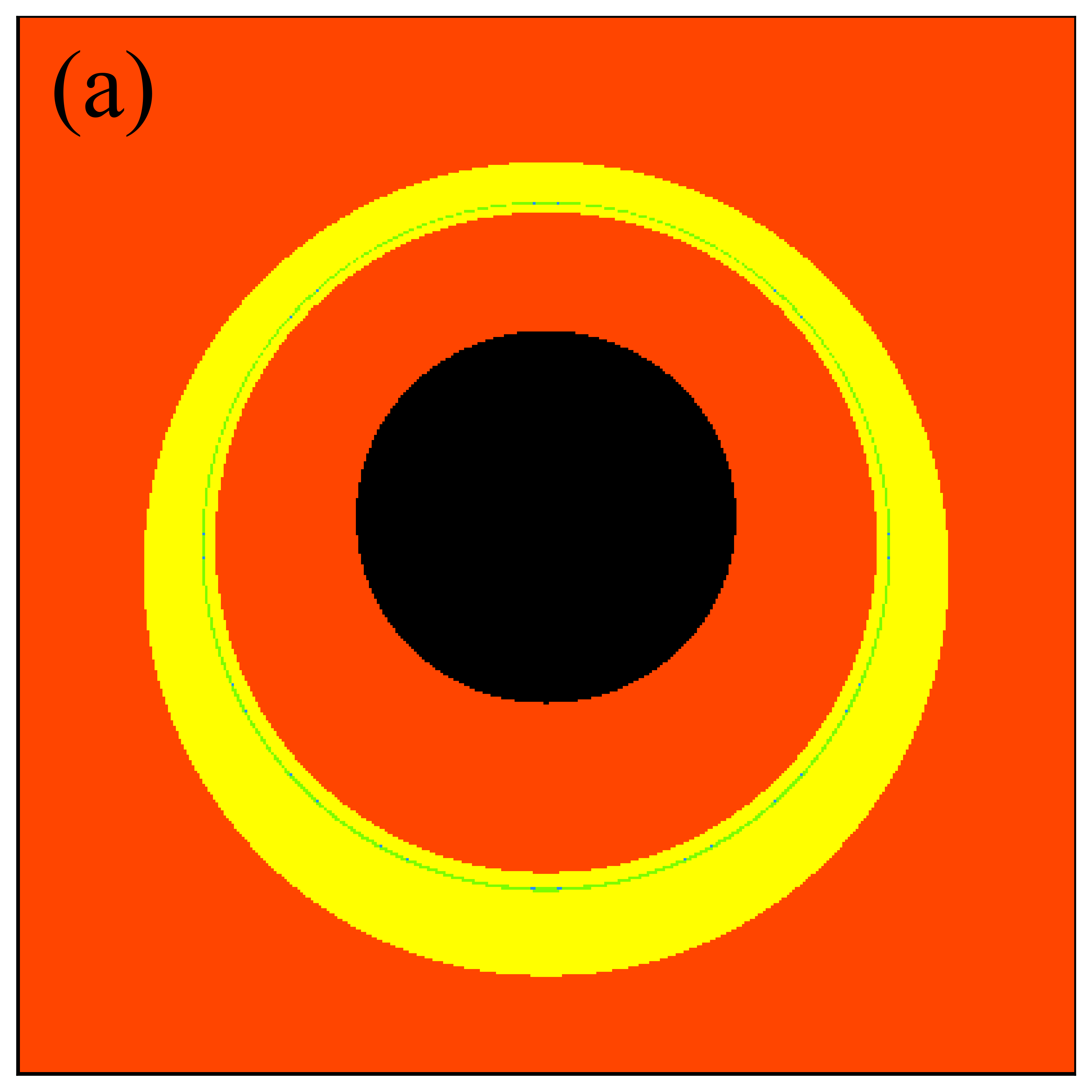}
\includegraphics[width=2.8cm]{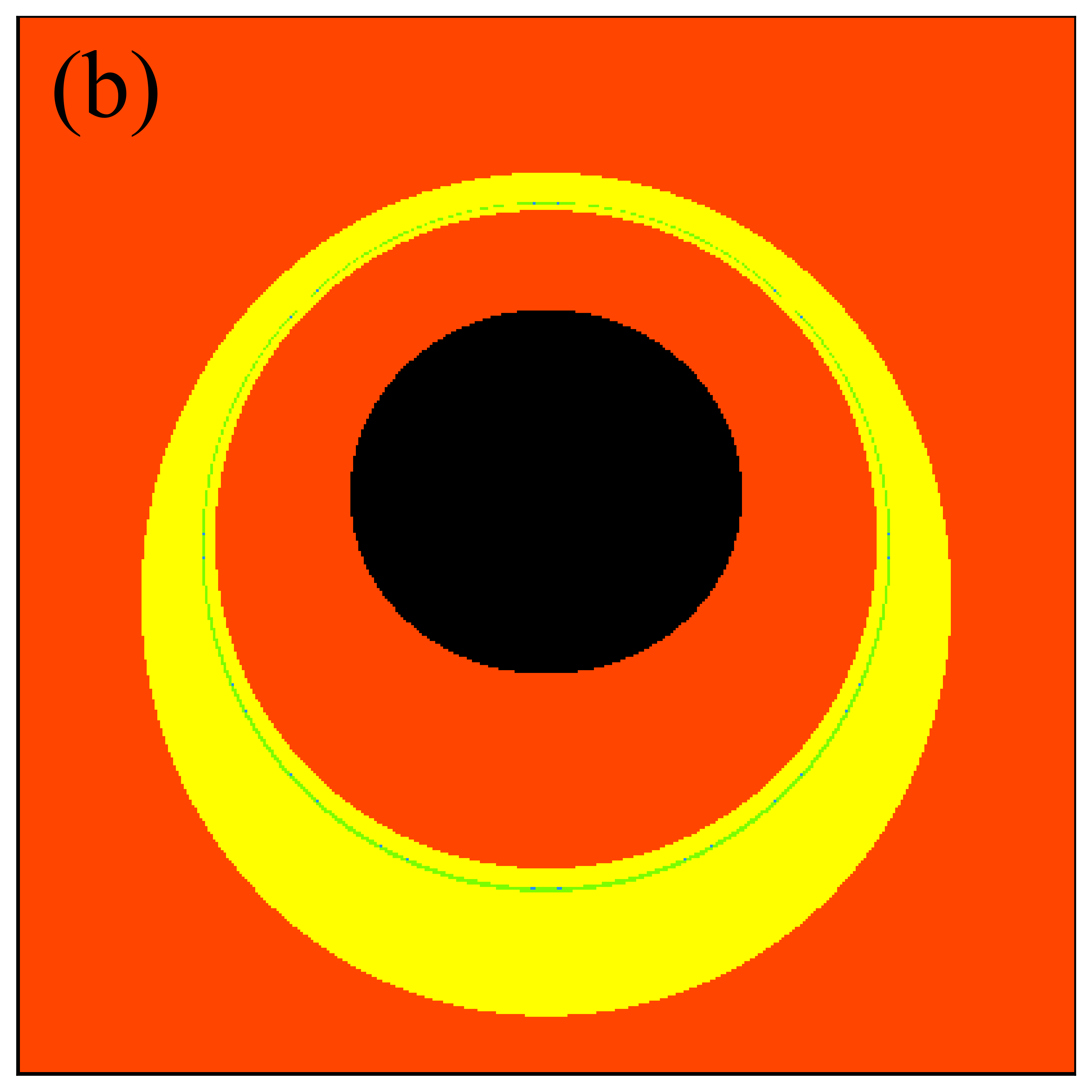}
\includegraphics[width=2.8cm]{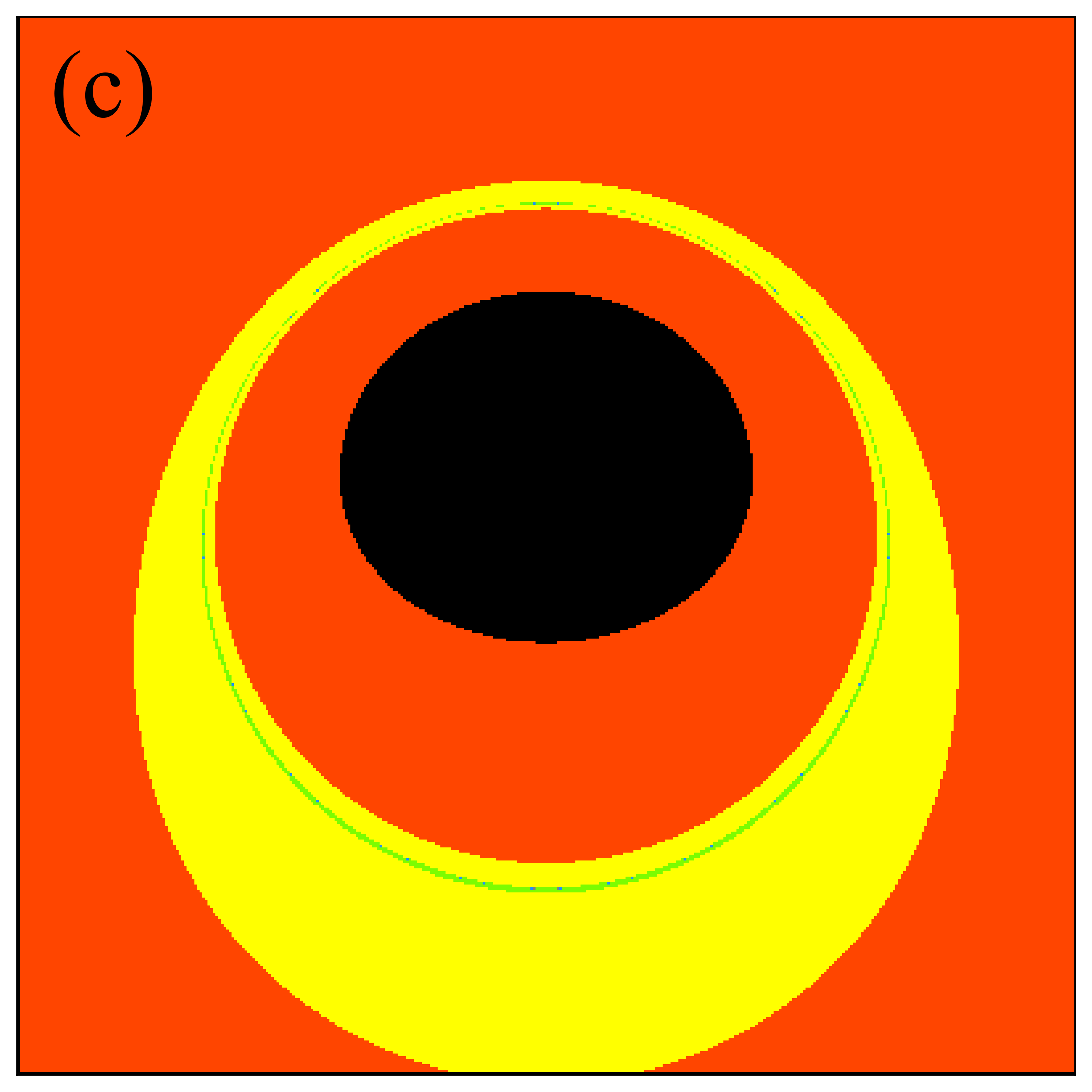}
\includegraphics[width=2.8cm]{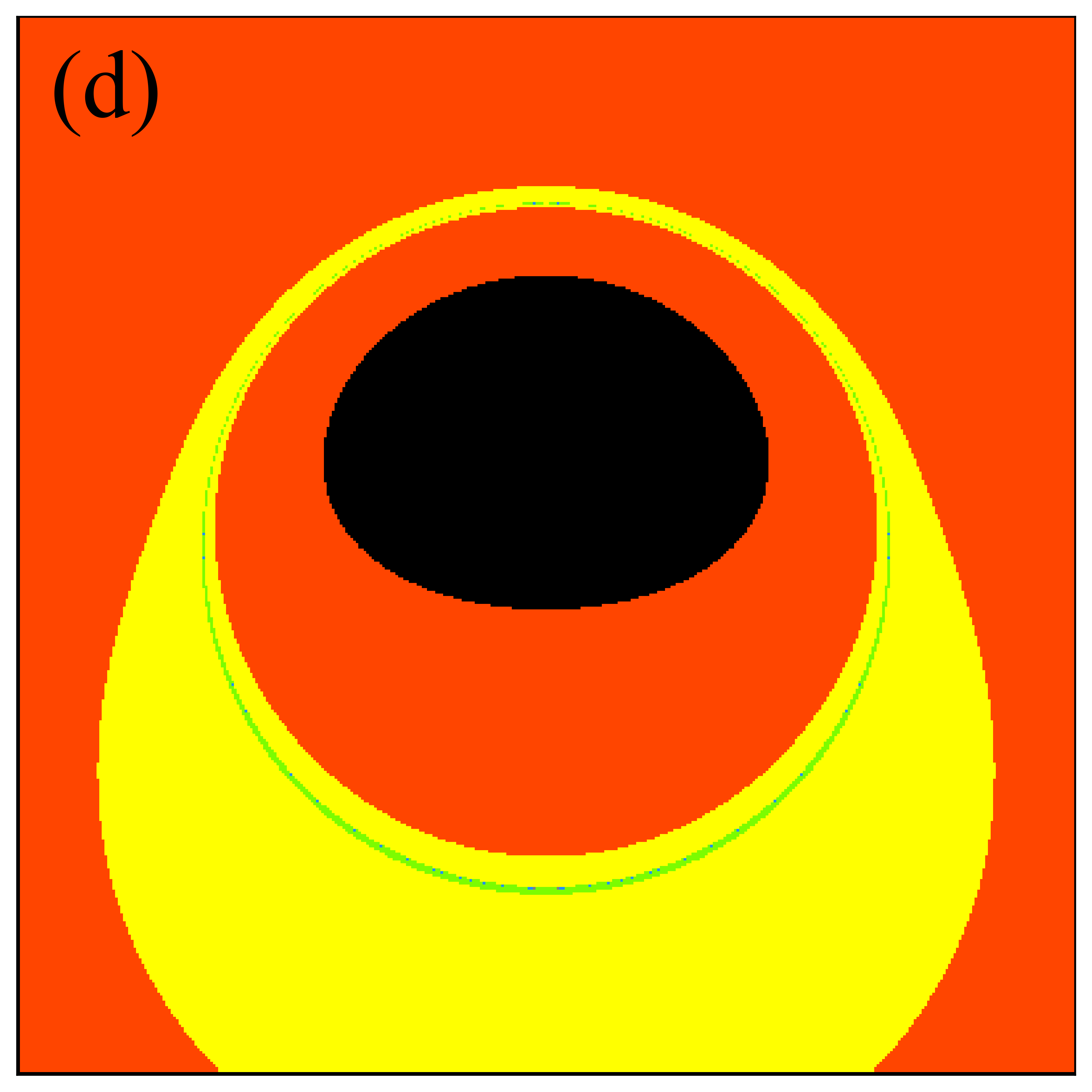}
\includegraphics[width=2.8cm]{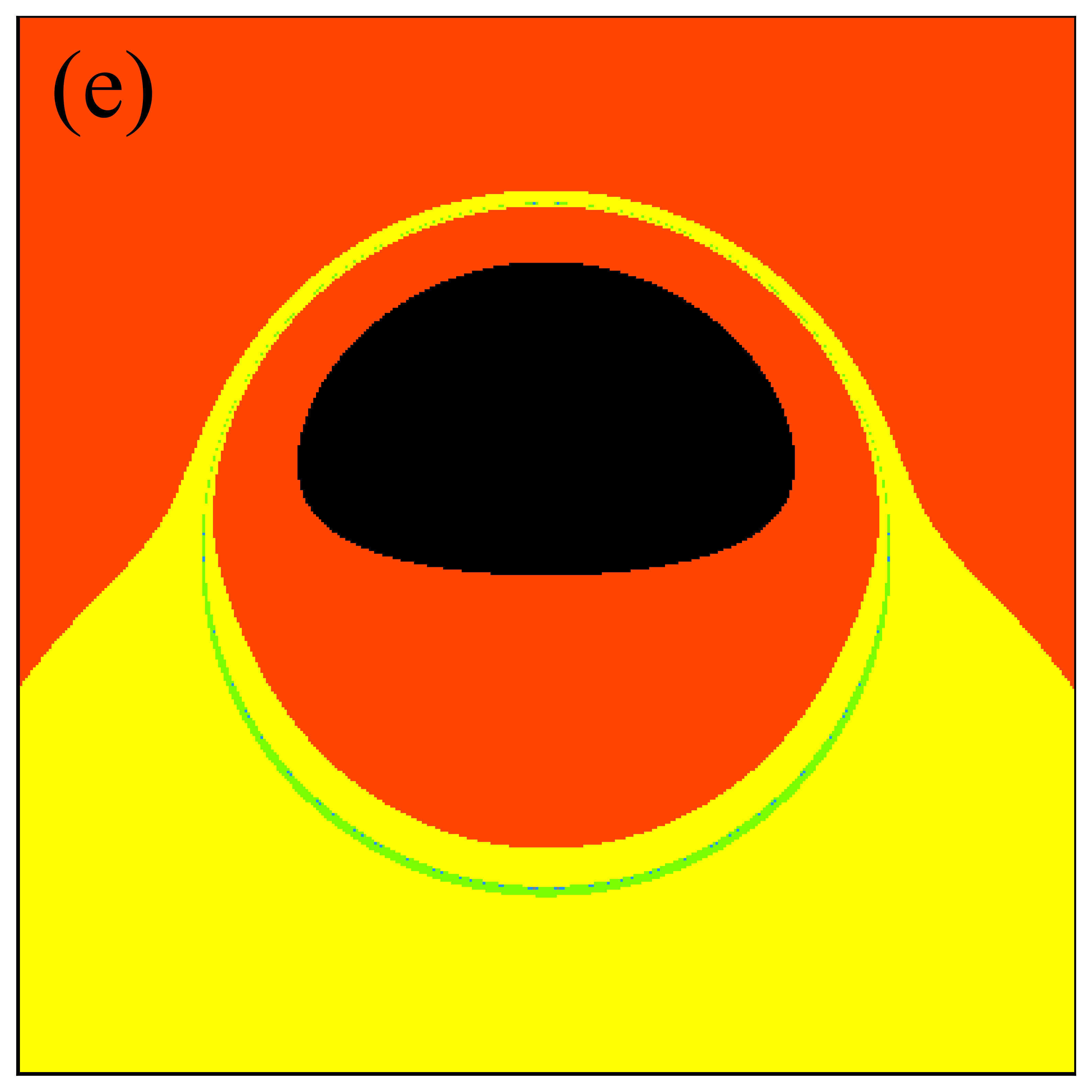}
\includegraphics[width=2.8cm]{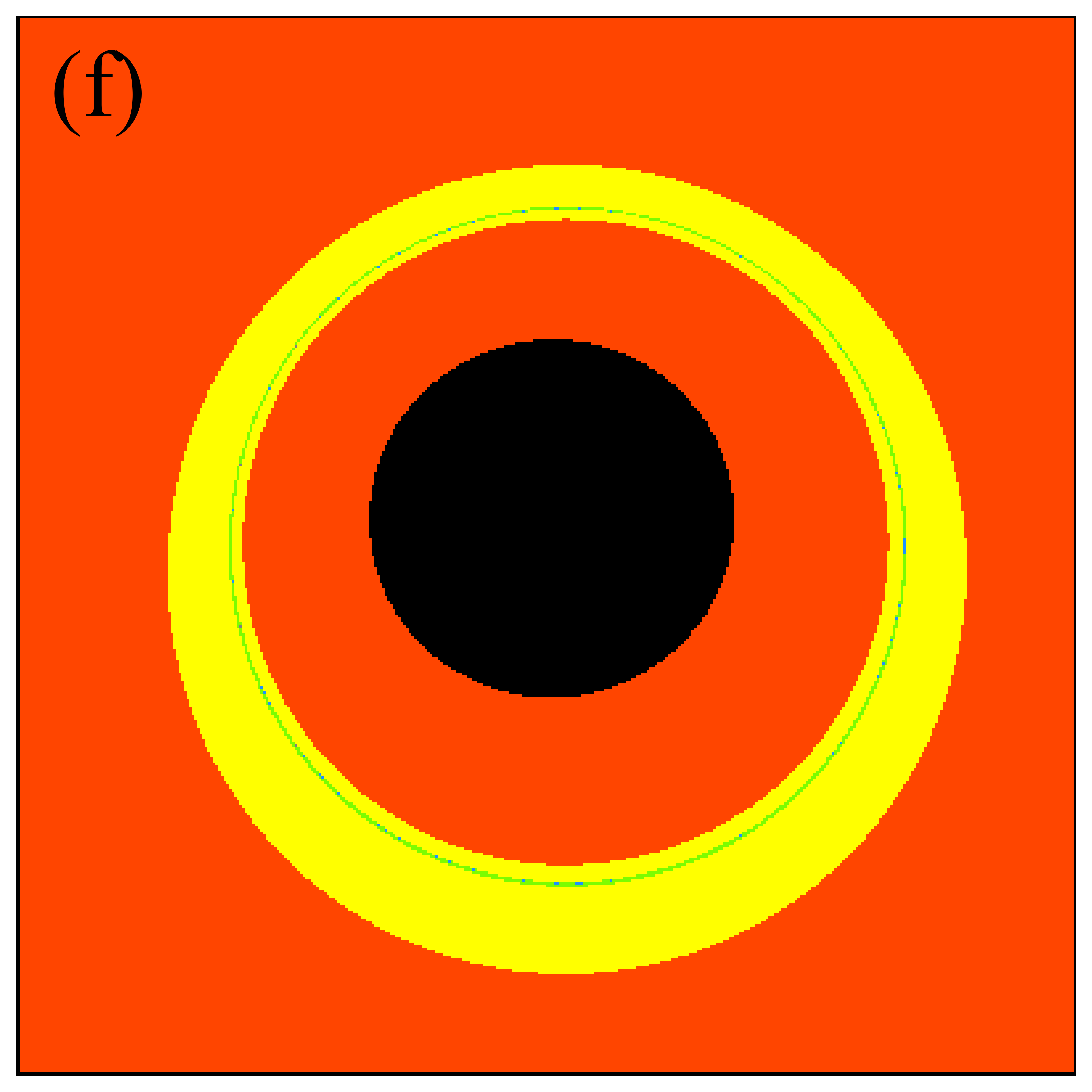}
\includegraphics[width=2.8cm]{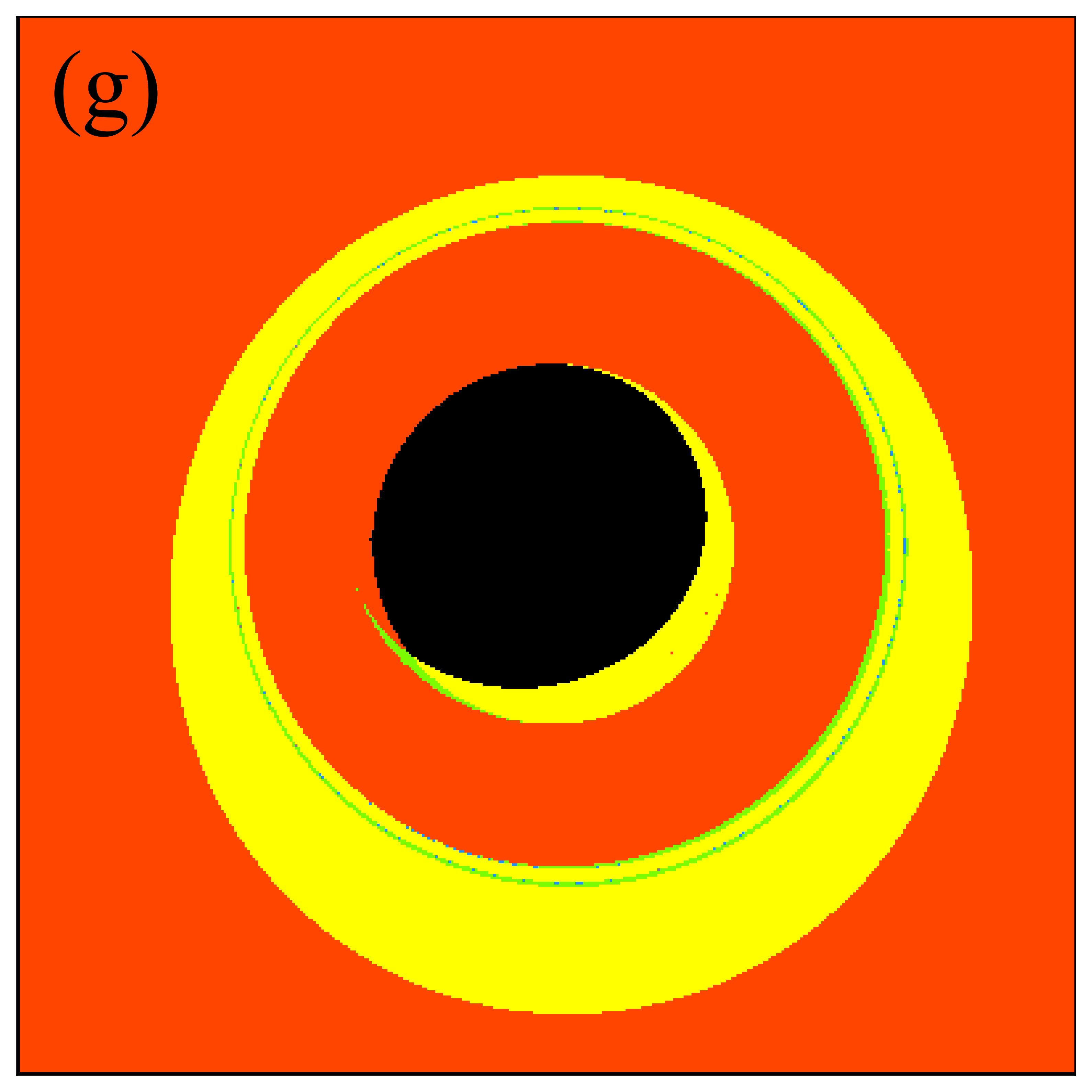}
\includegraphics[width=2.8cm]{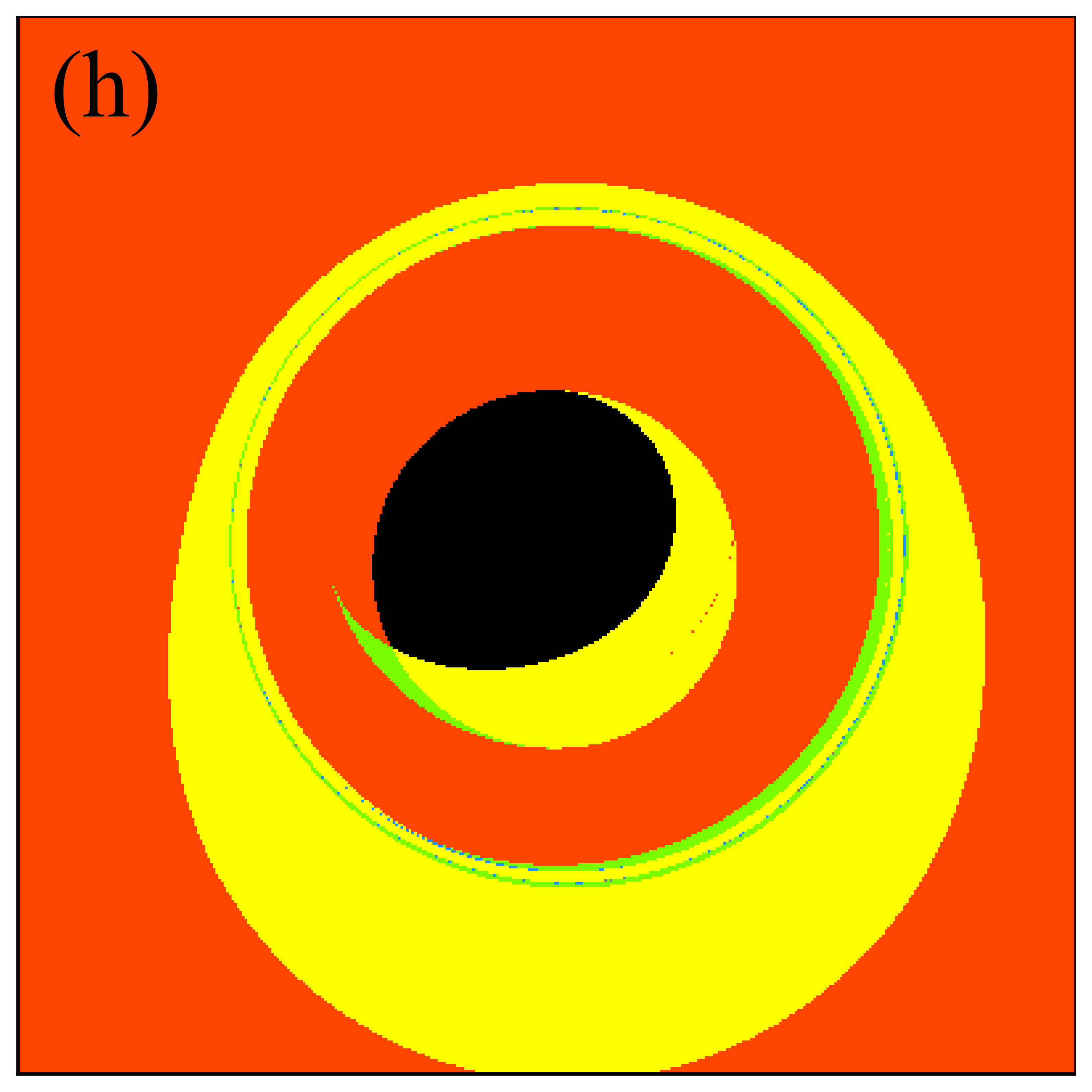}
\includegraphics[width=2.8cm]{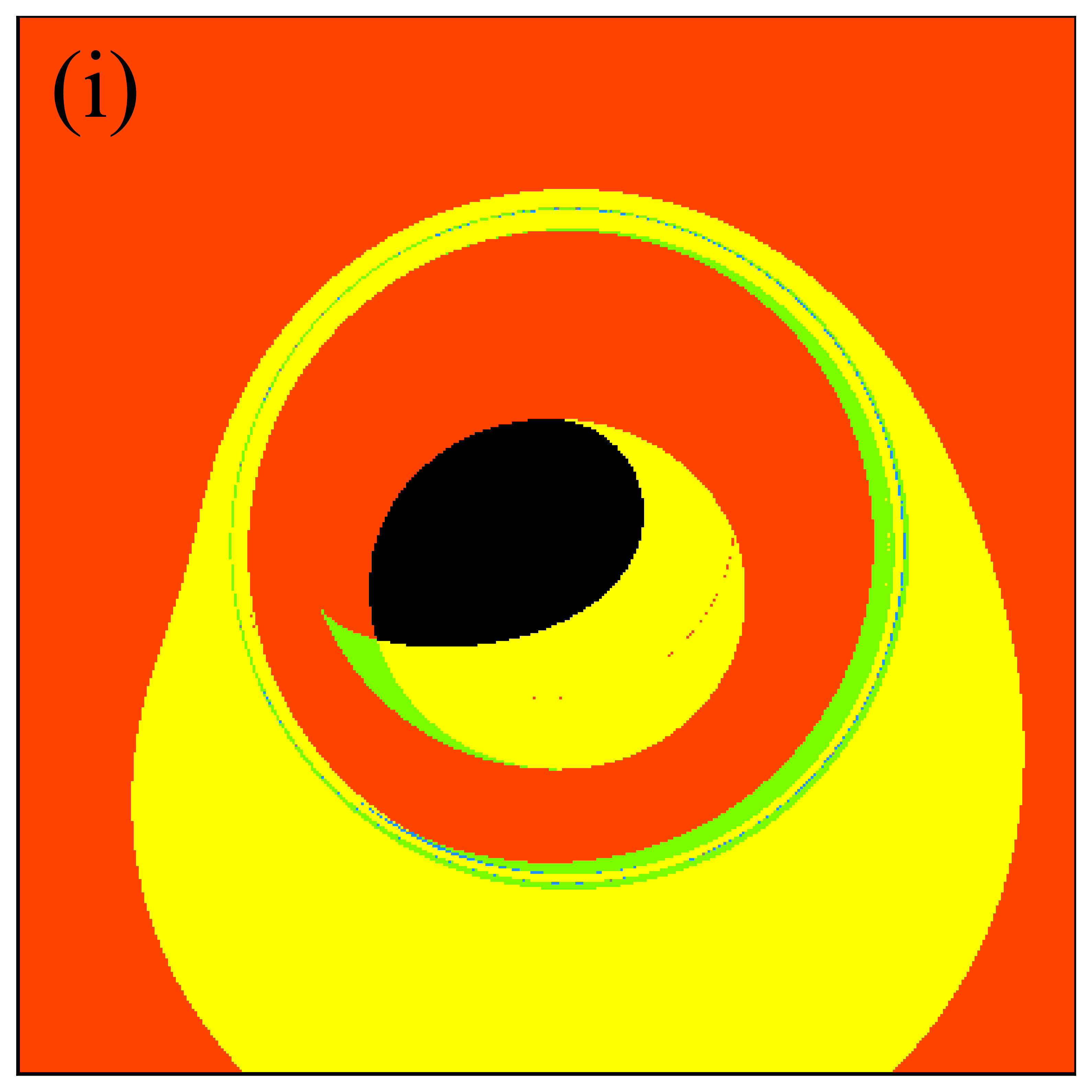}
\includegraphics[width=2.8cm]{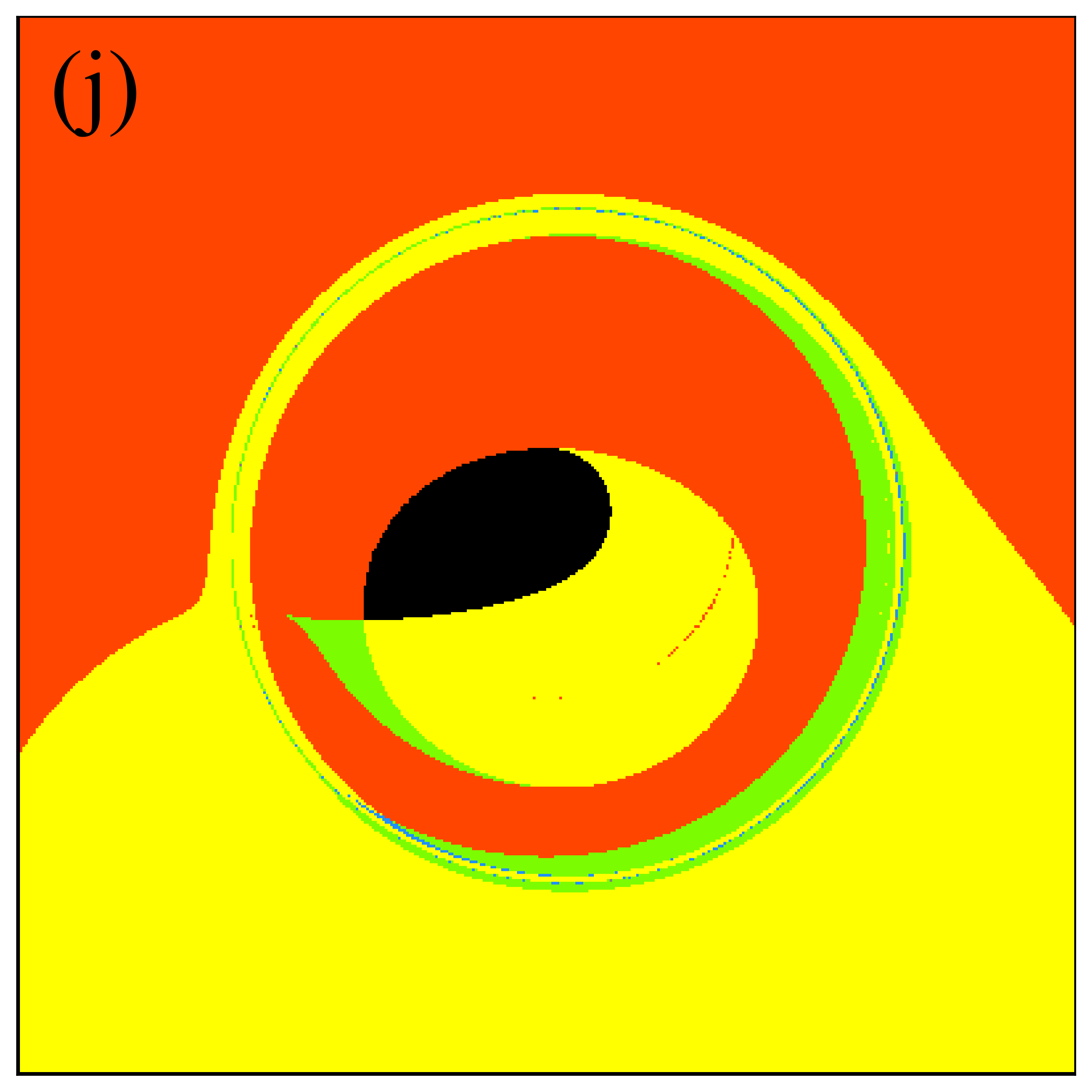}
\includegraphics[width=2.8cm]{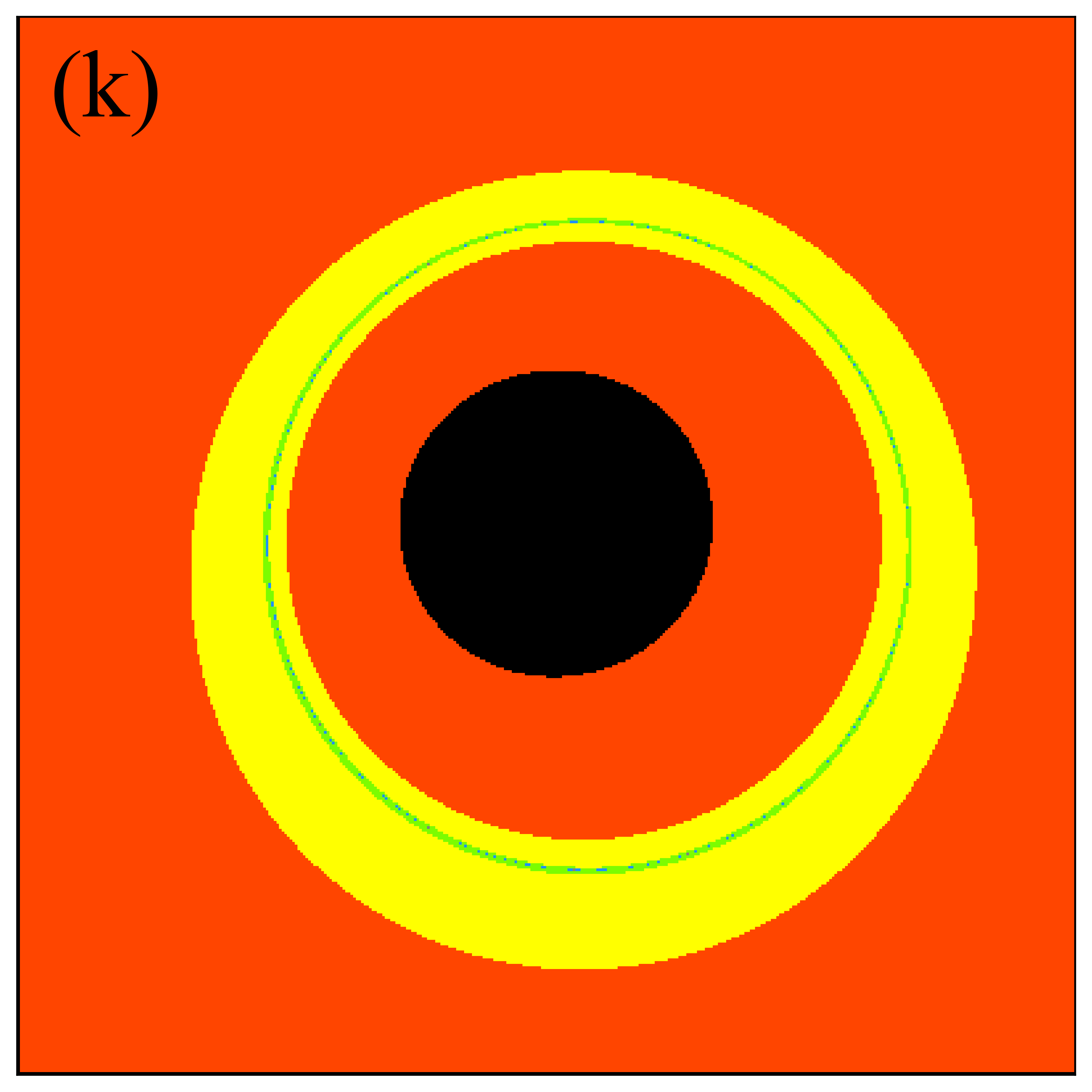}
\includegraphics[width=2.8cm]{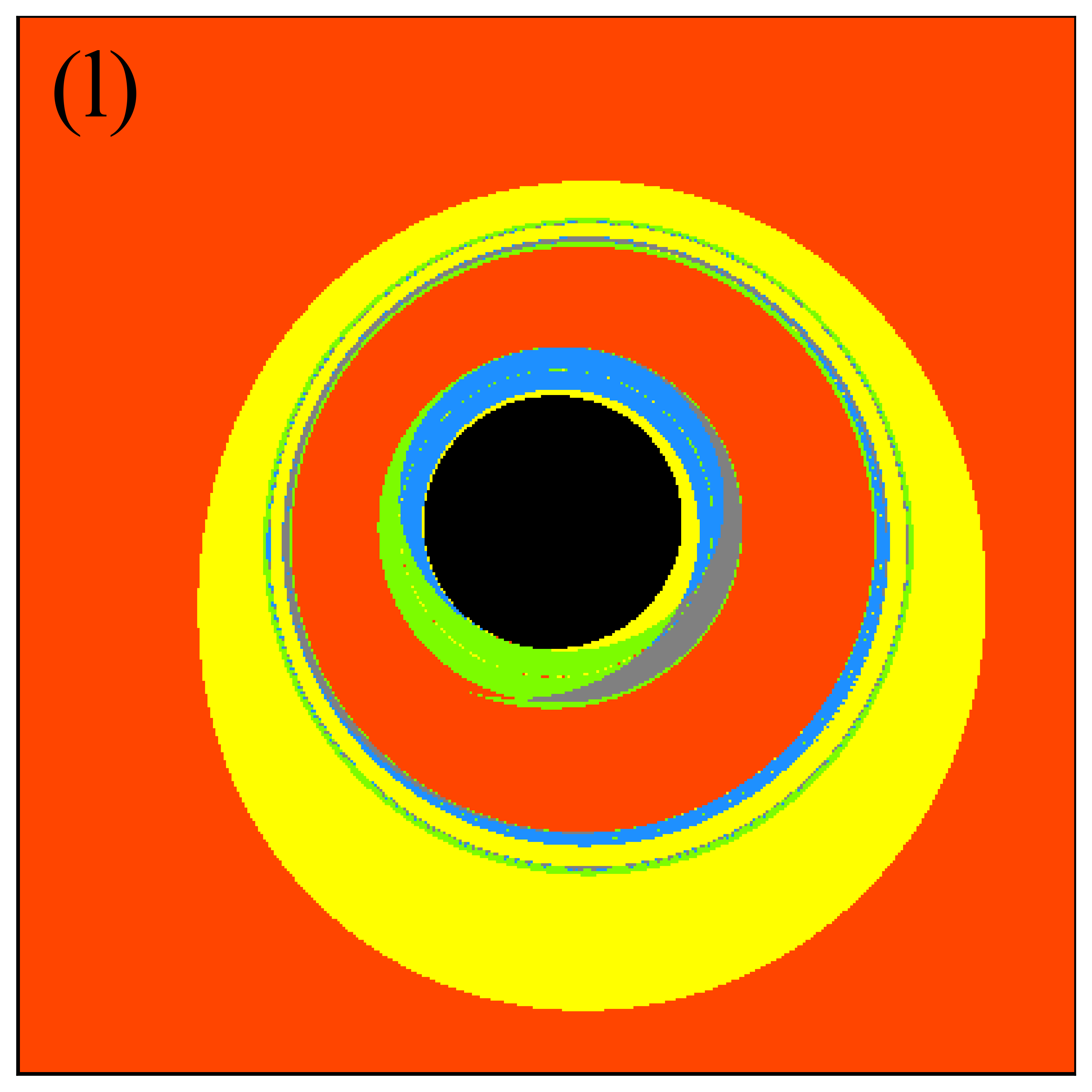}
\includegraphics[width=2.8cm]{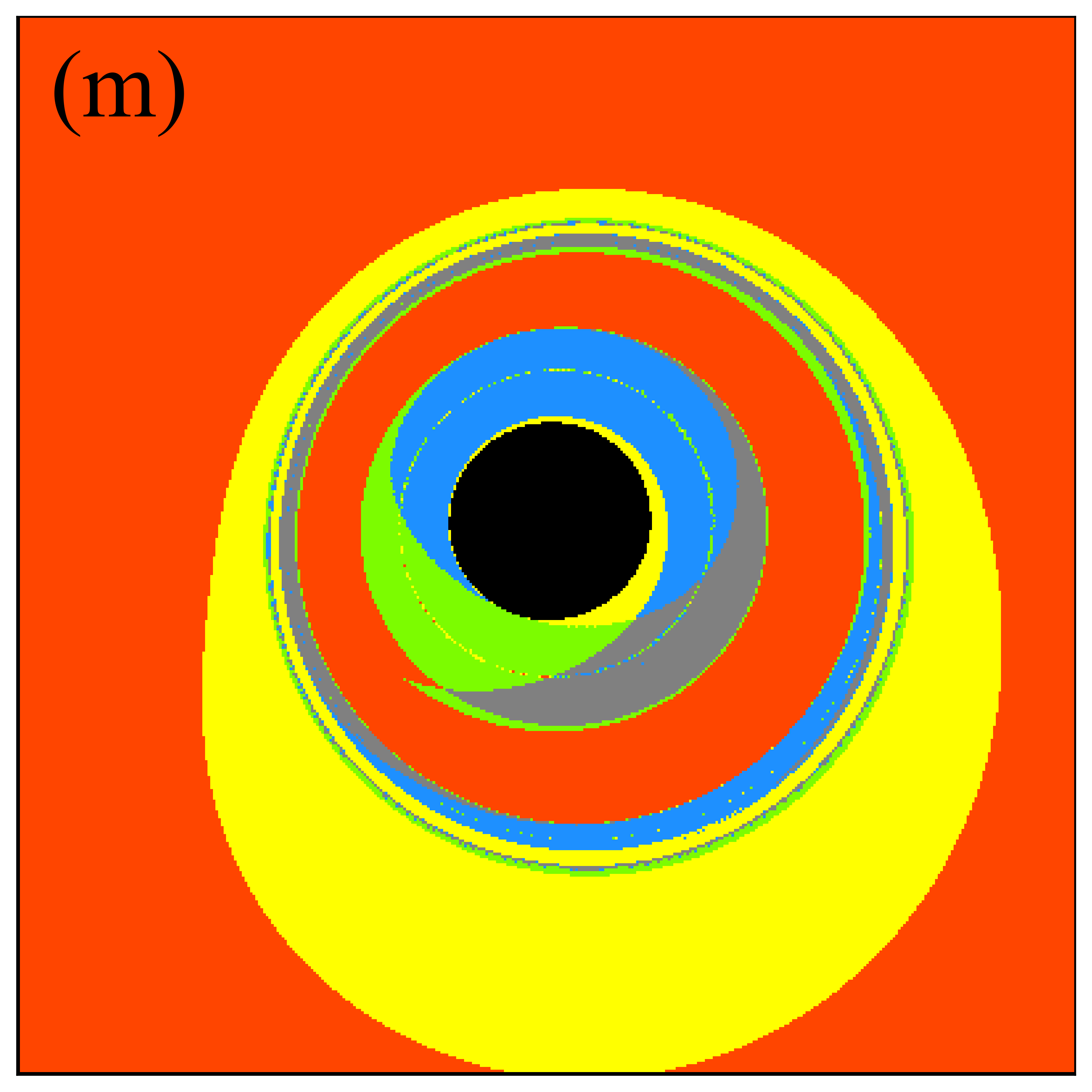}
\includegraphics[width=2.8cm]{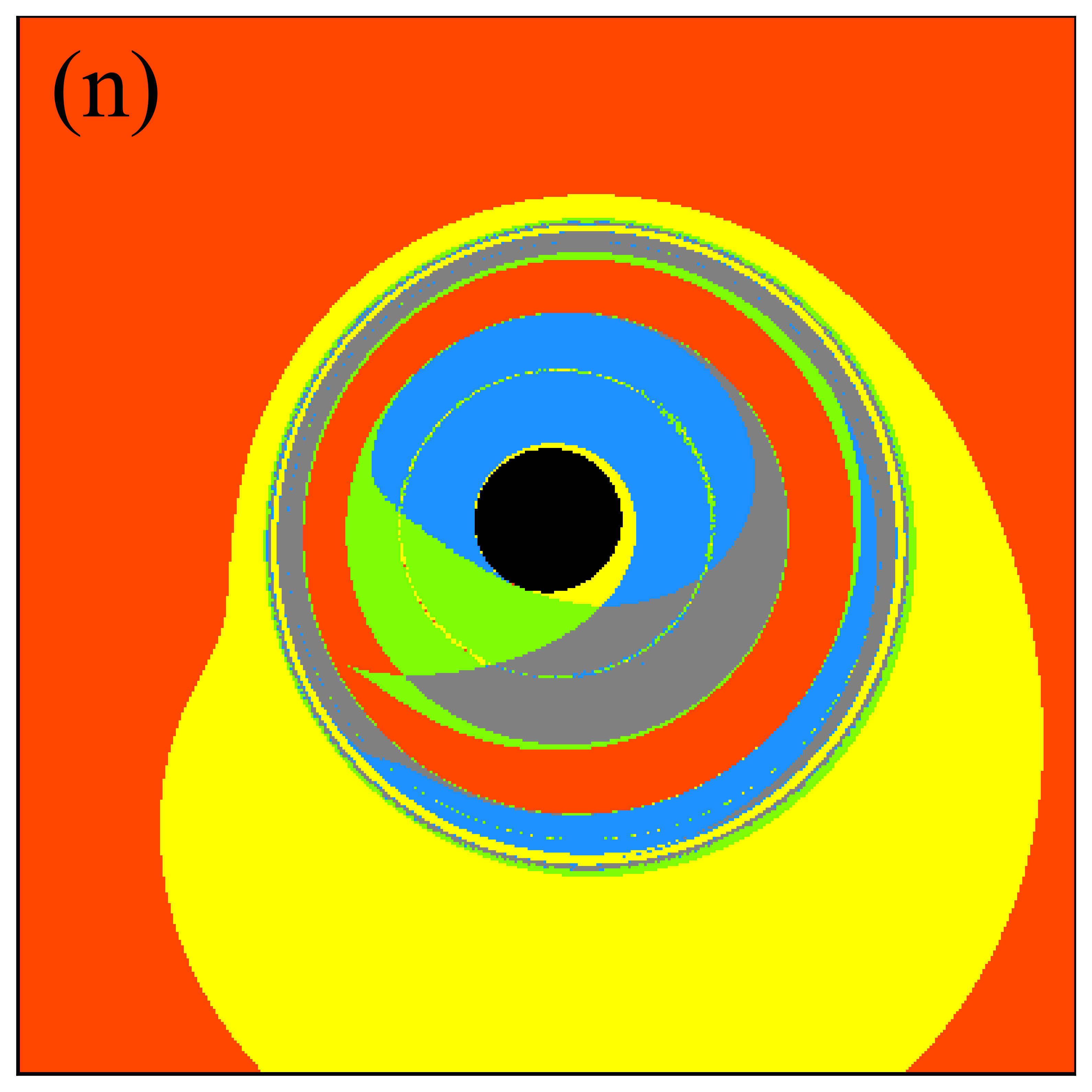}
\includegraphics[width=2.8cm]{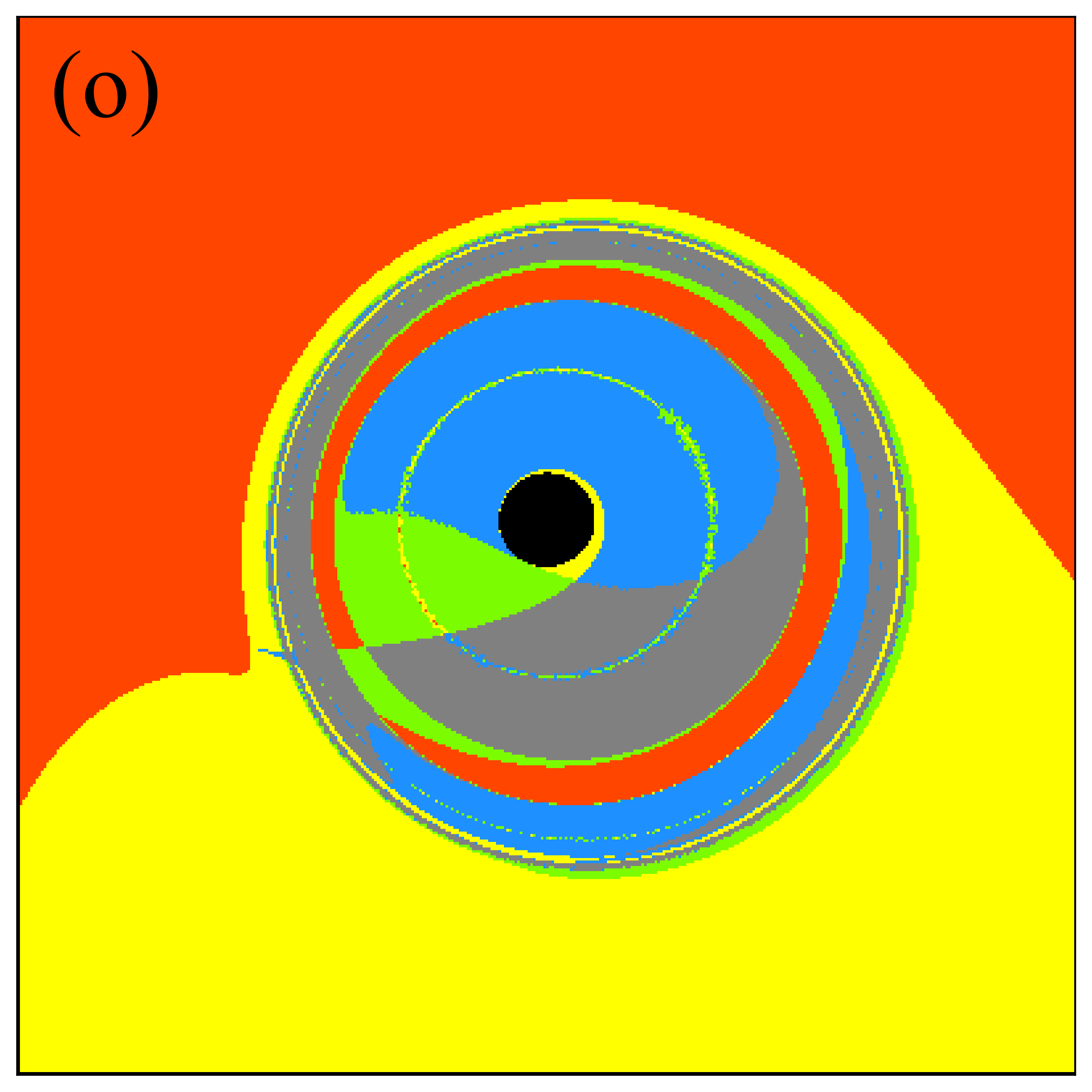}
\includegraphics[width=2.8cm]{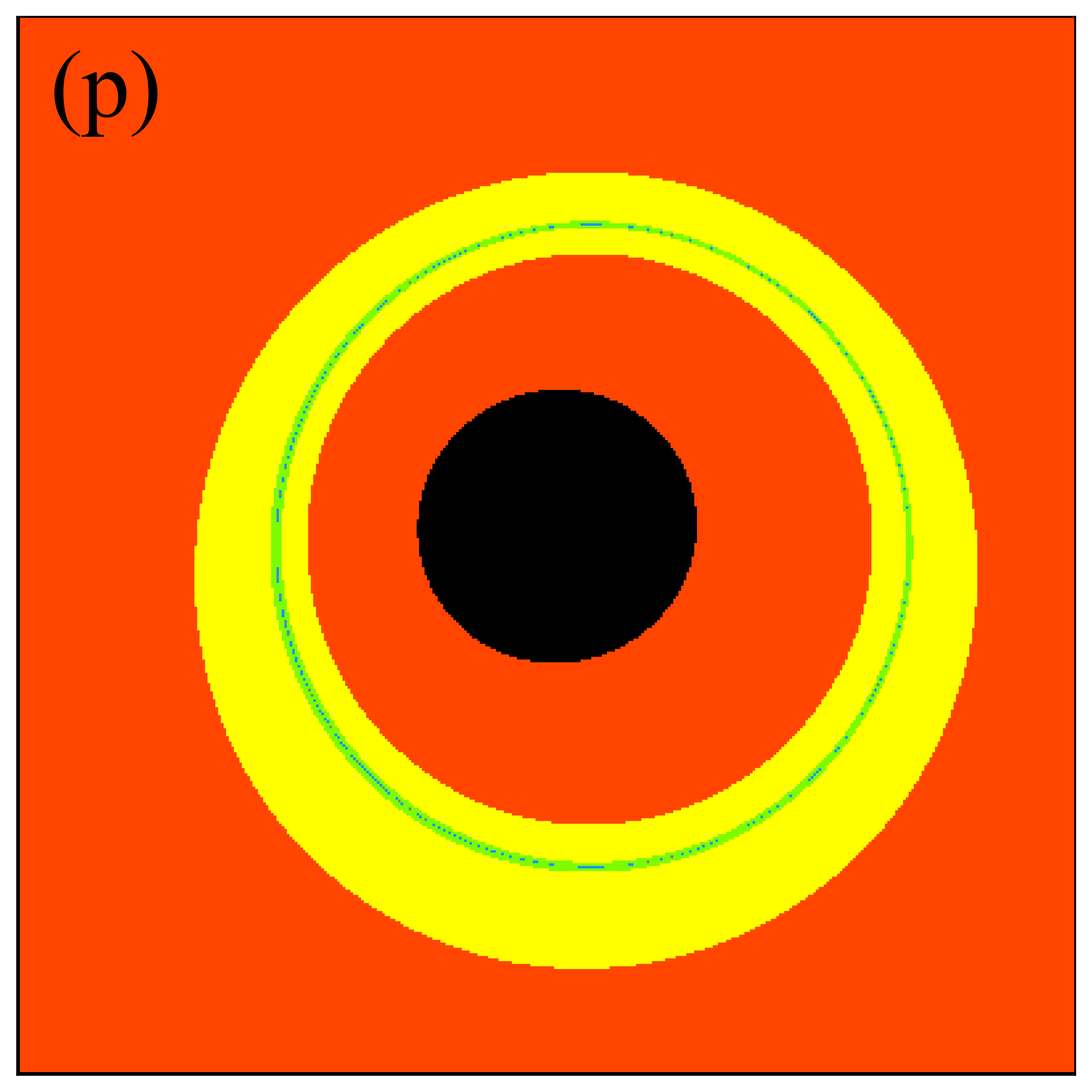}
\includegraphics[width=2.8cm]{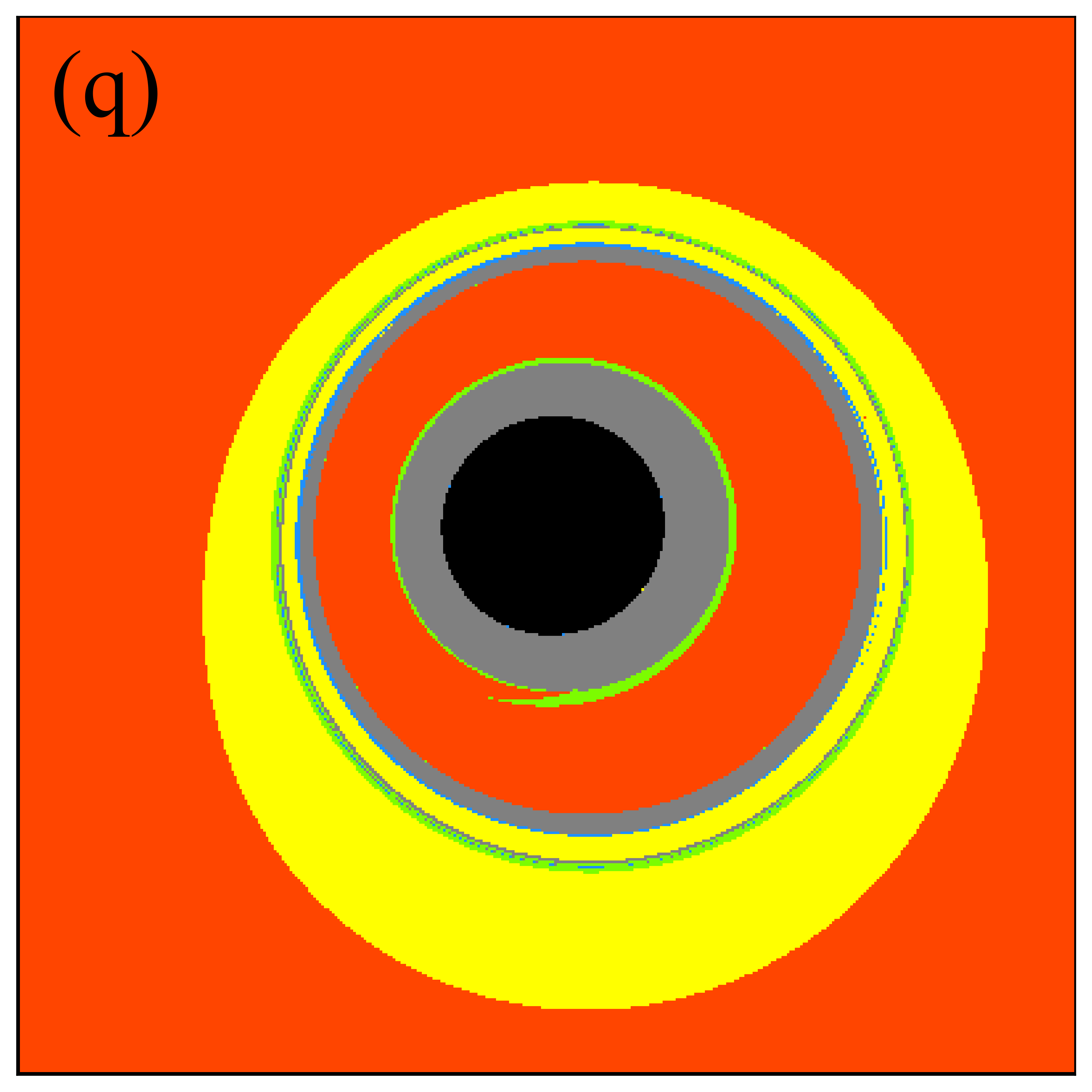}
\includegraphics[width=2.8cm]{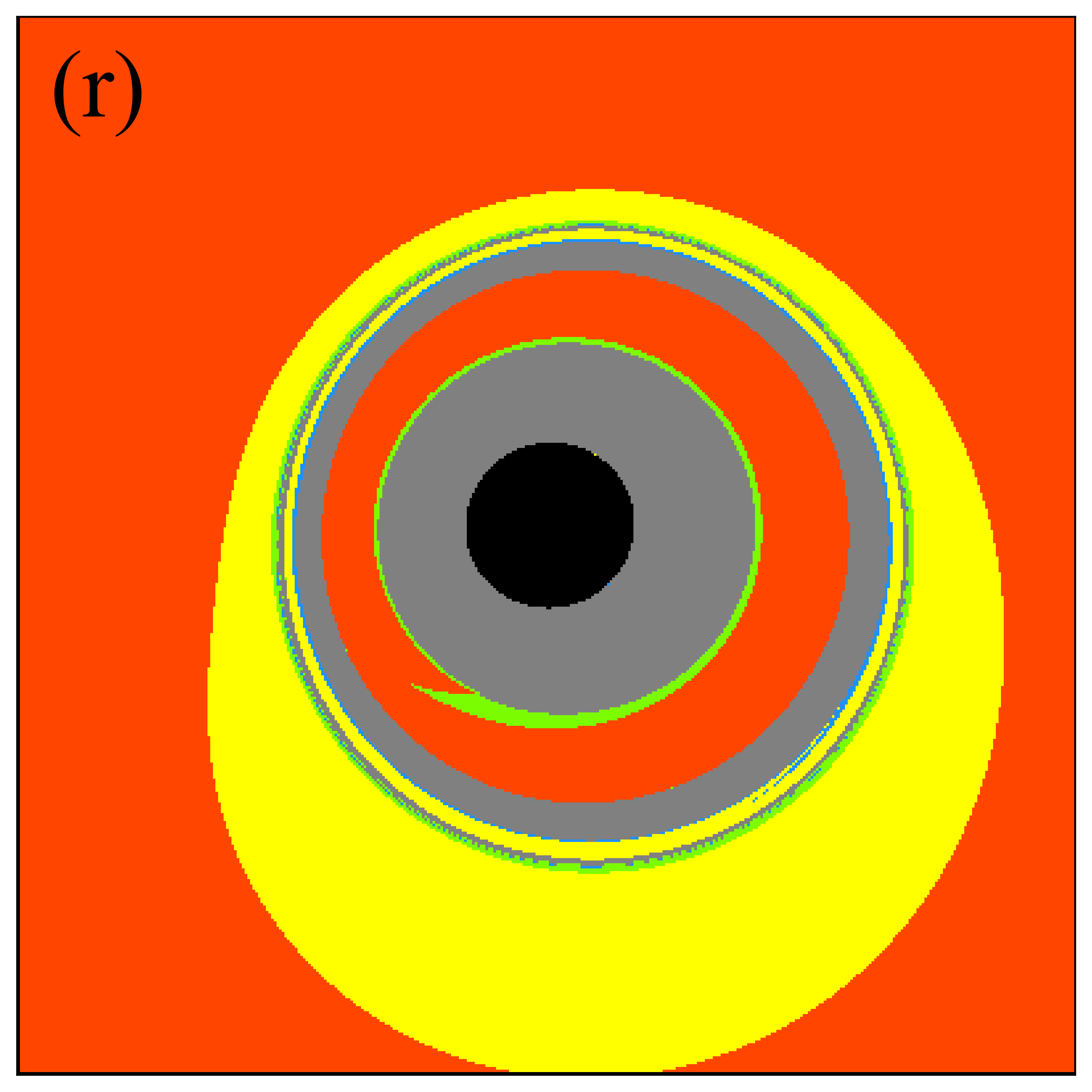}
\includegraphics[width=2.8cm]{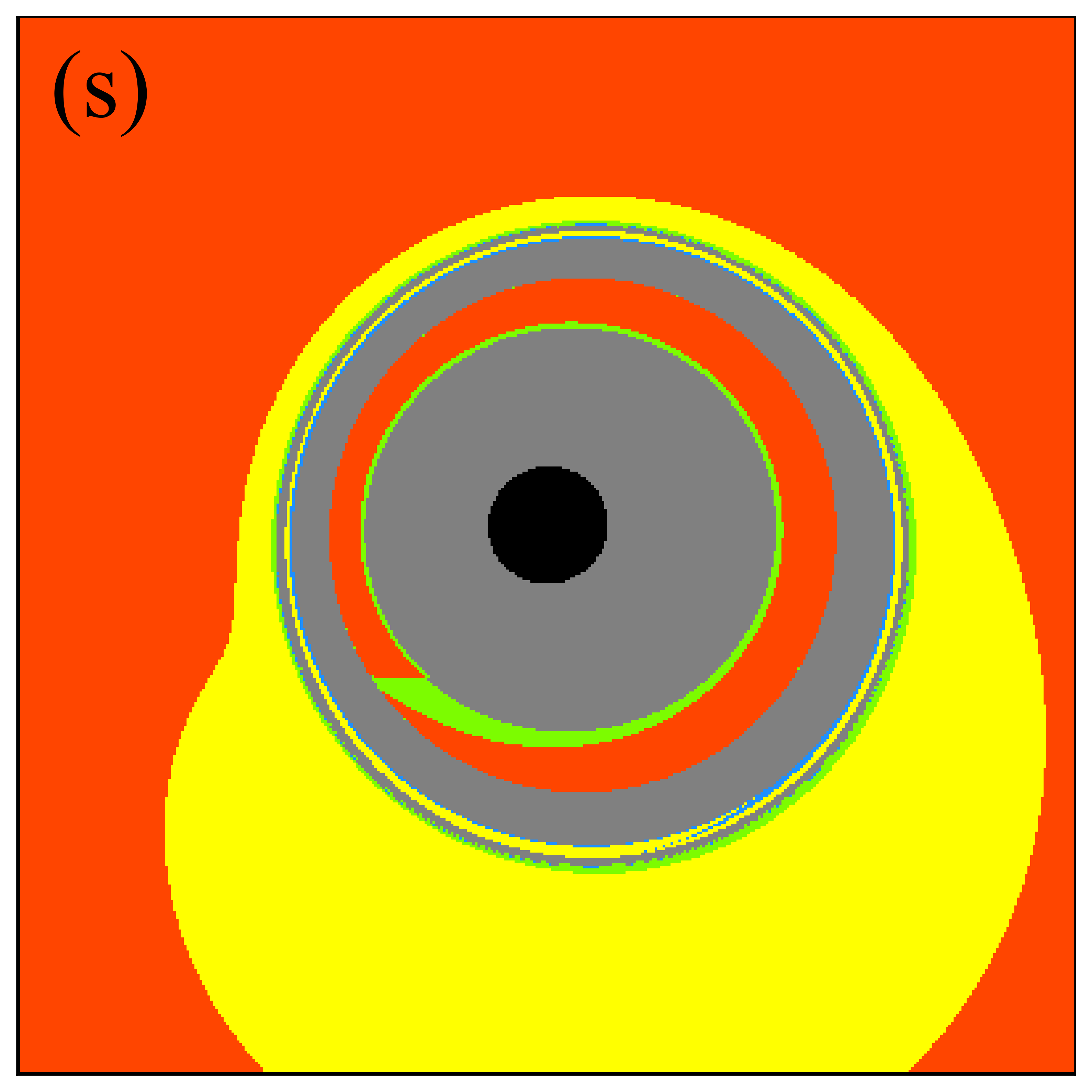}
\includegraphics[width=2.8cm]{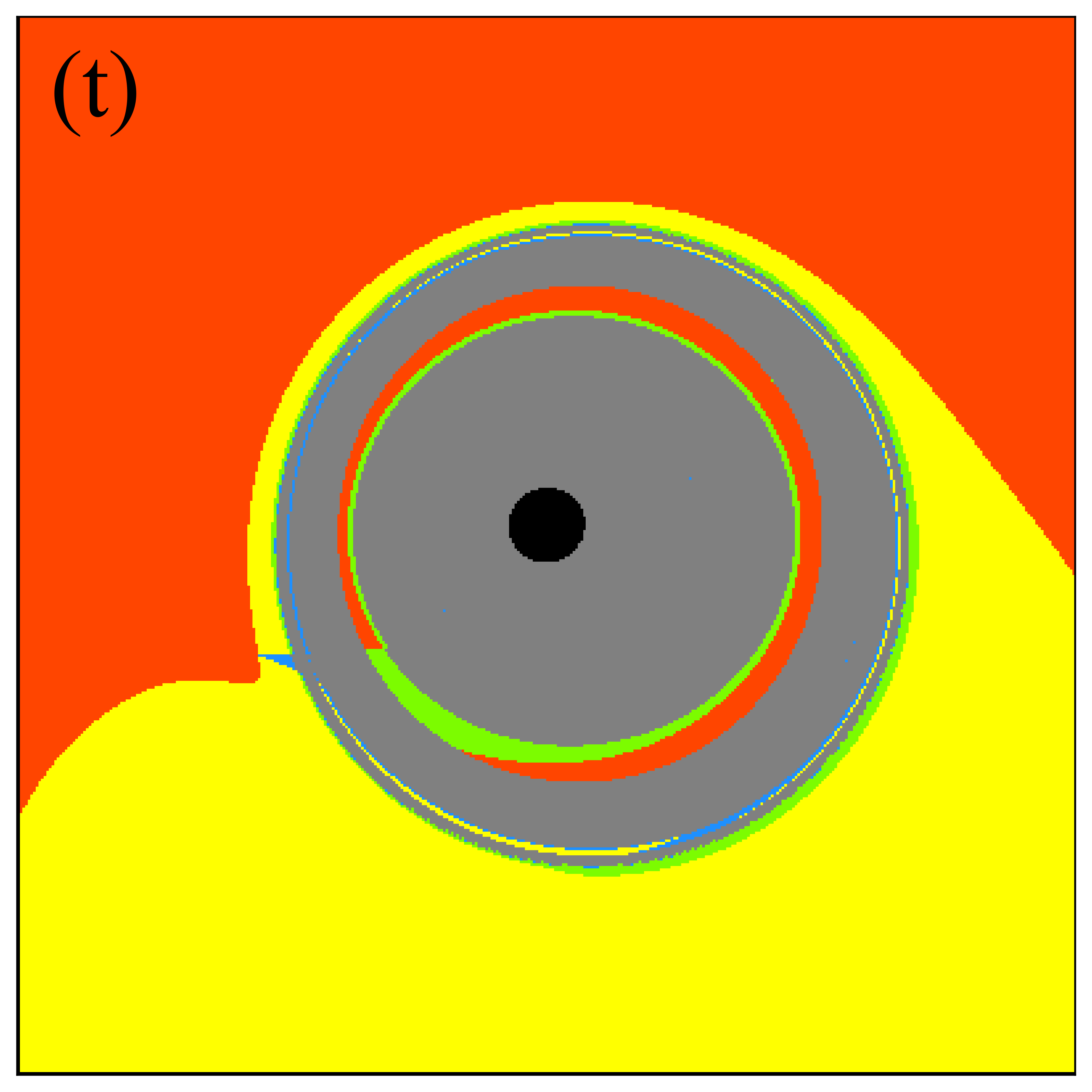}
\caption{Distribution of light ray-disk intersection counts on the observation plane. From left to right, the values of disk inclination are $0^{\circ}$, $15^{\circ}$, $30^{\circ}$, $45^{\circ}$, and $60^{\circ}$; from top to bottom, the spin parameters are $0$, $0.54$, $0.94$, and $0.9985$. Here, the observation angle and azimuth are fixed at $17^{\circ}$ and $0^{\circ}$, respectively. Light rays intersecting the accretion disk $0$, $1$, $2$, $3$, $4$, and $5$ times are represented by black, red, yellow, green, blue, and gray, respectively. Notably, in the tilted disk scenario, photons contributing to the black hole's inner shadow must have zero intersections with the disk. Moreover, compared to the equatorial case (left column), images formed by multiply disk-crossing rays under tilted disk conditions generally lack asymptotic convergence.}}\label{fig7}
\end{figure*}
\begin{figure*}
\center{
\includegraphics[width=2.8cm]{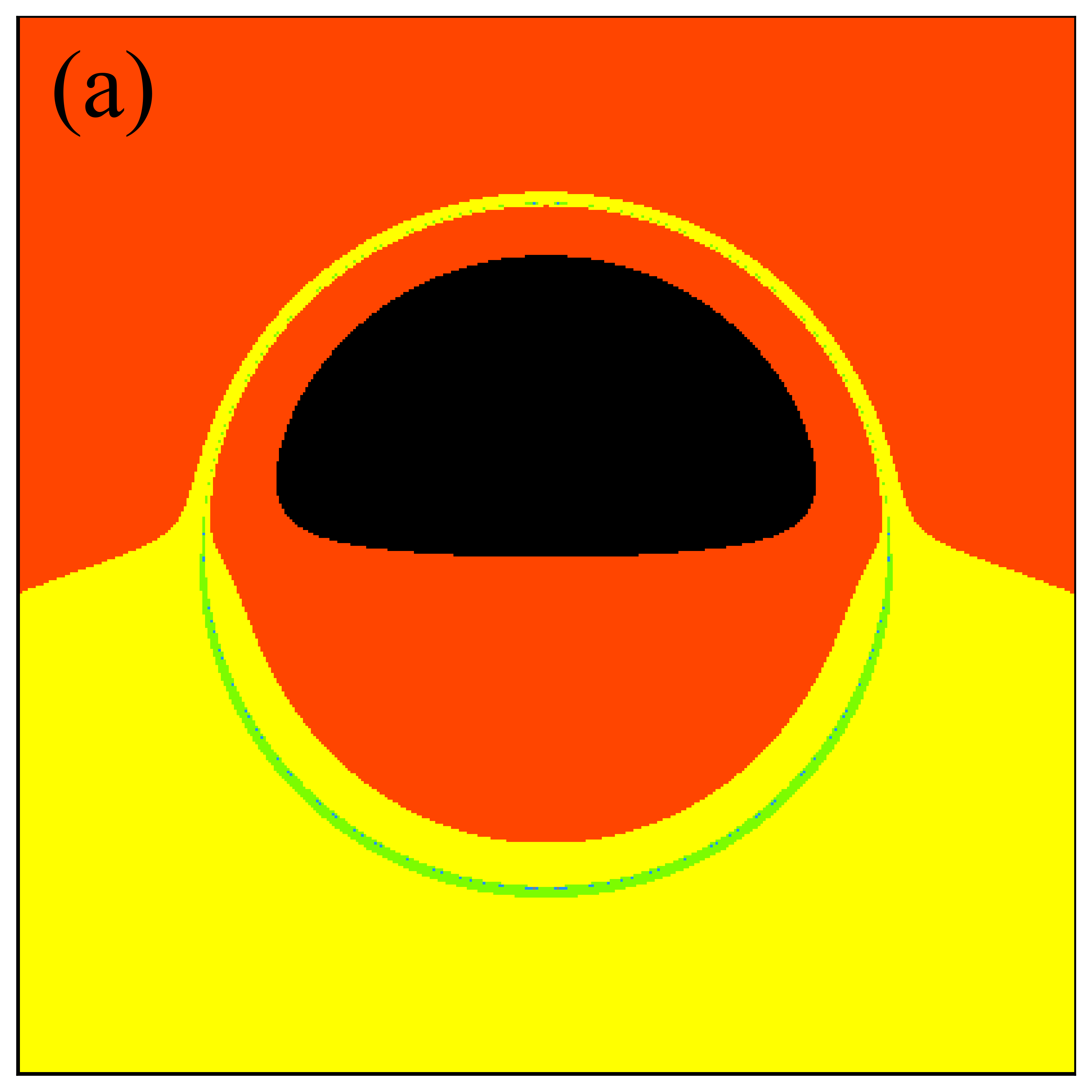}
\includegraphics[width=2.8cm]{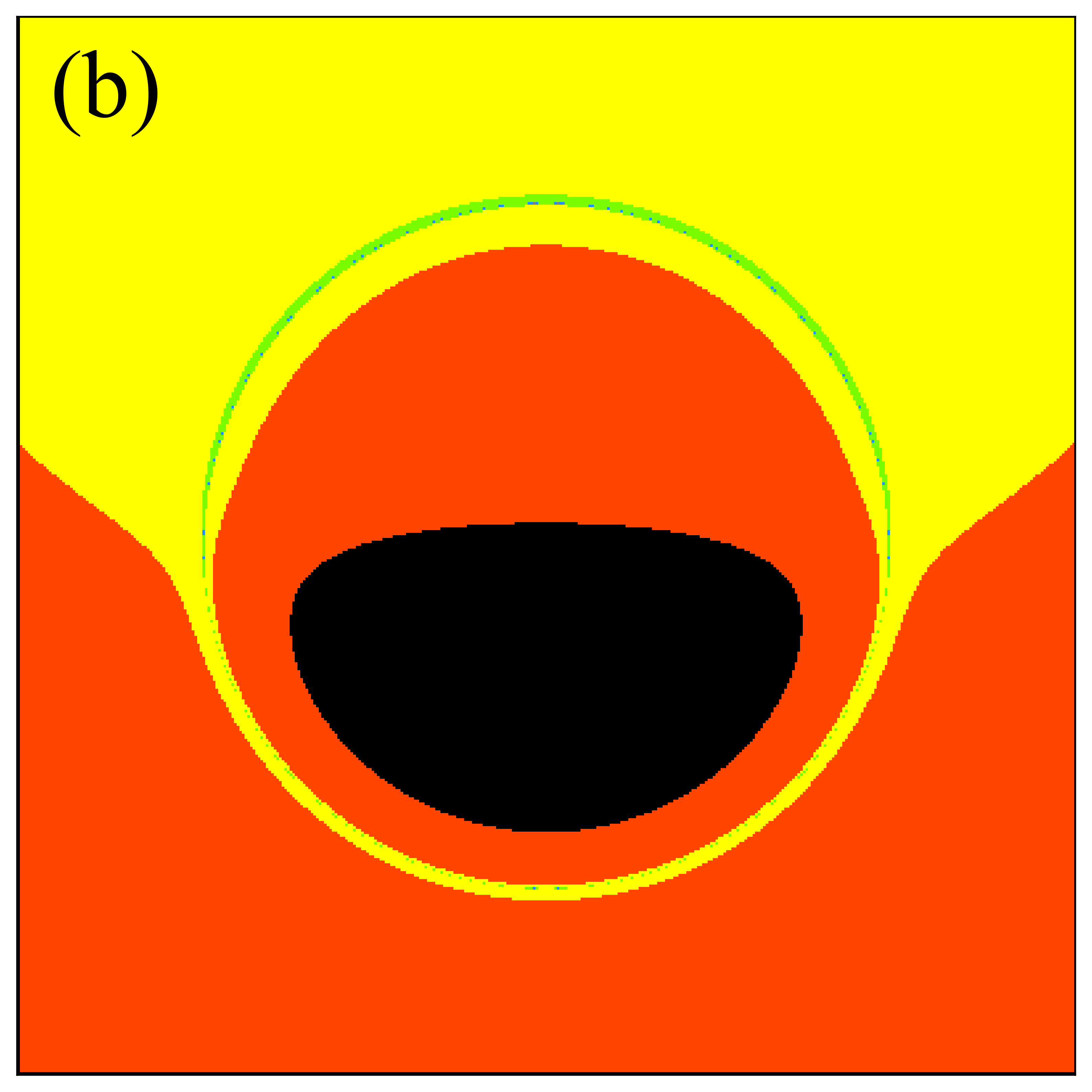}
\includegraphics[width=2.8cm]{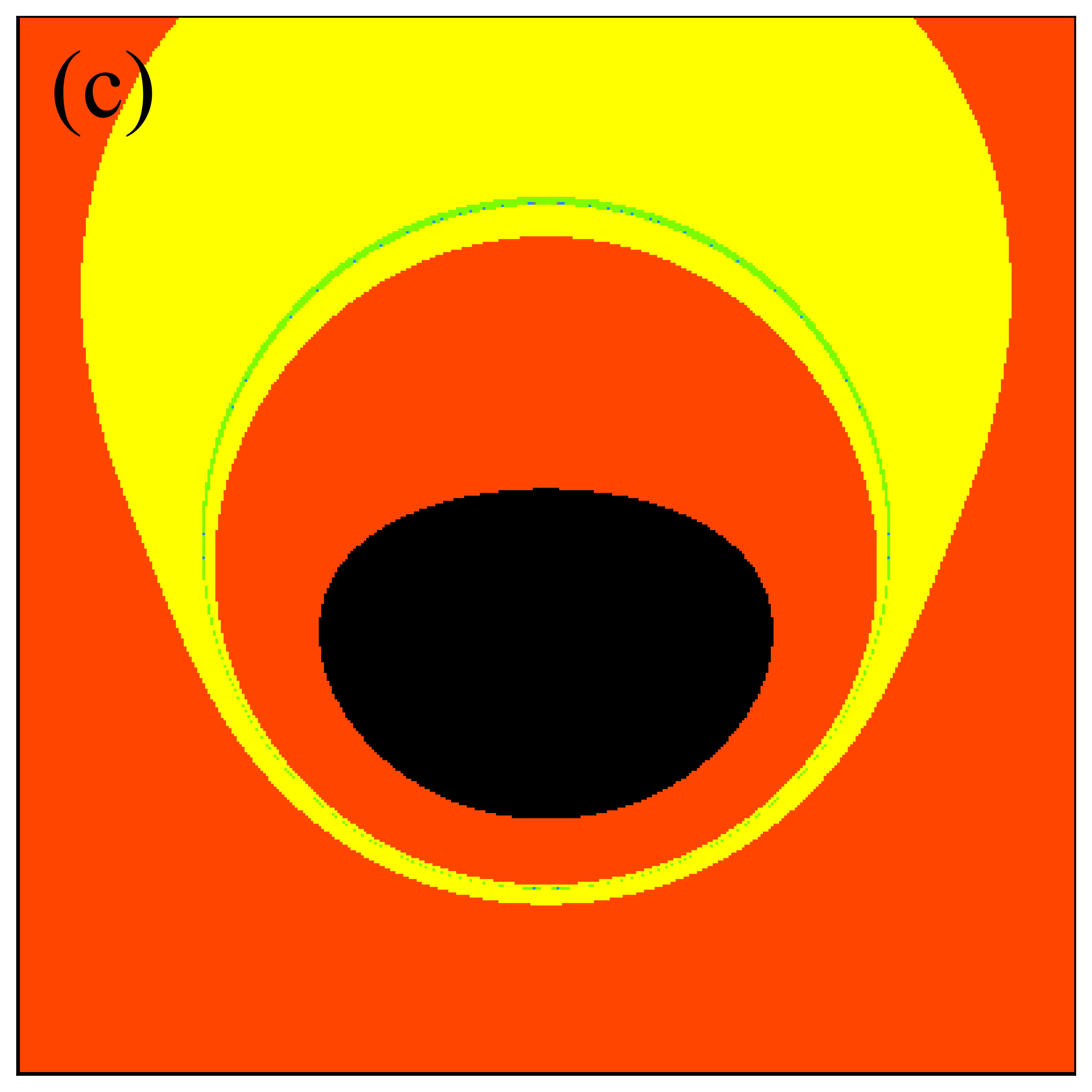}
\includegraphics[width=2.8cm]{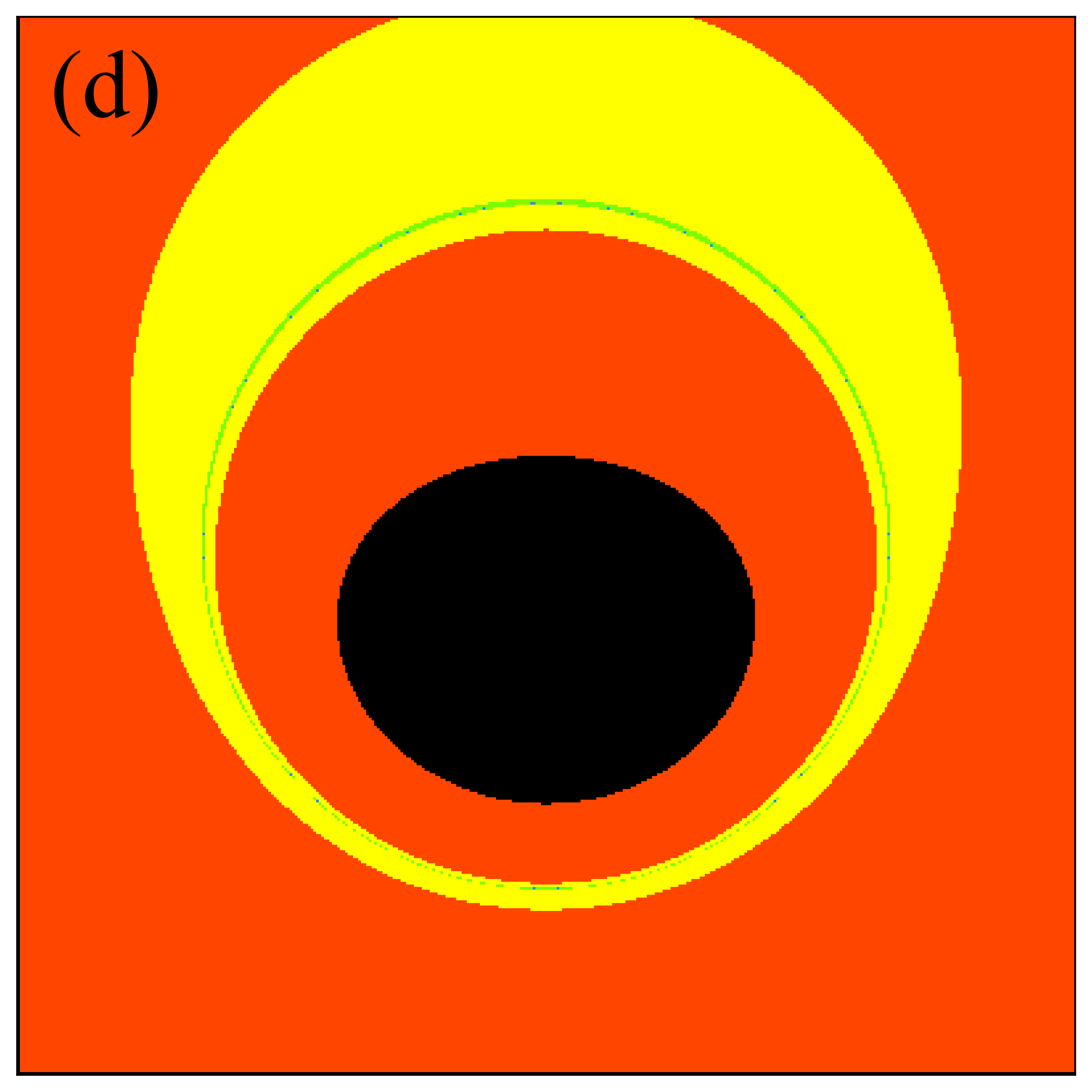}
\includegraphics[width=2.8cm]{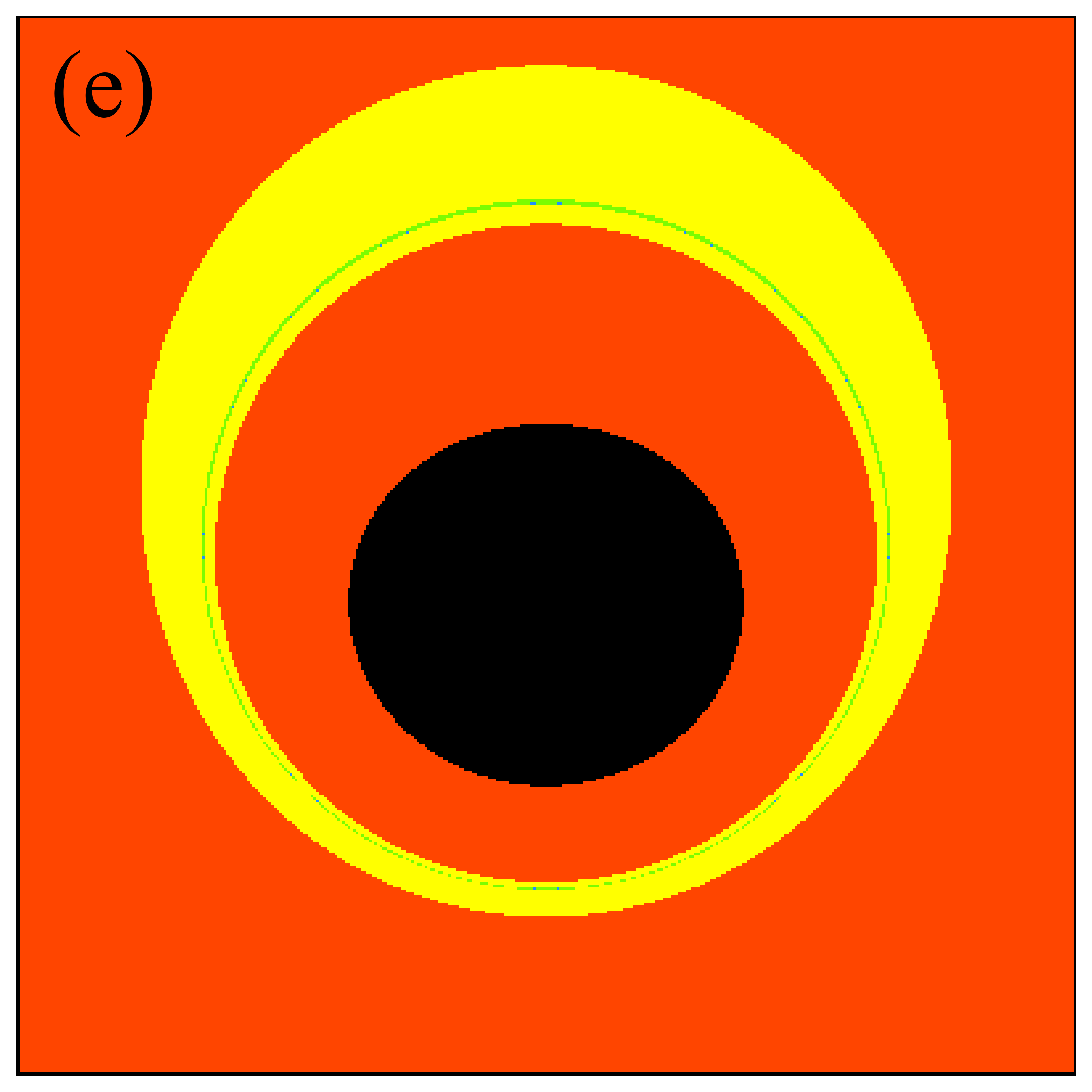}
\includegraphics[width=2.8cm]{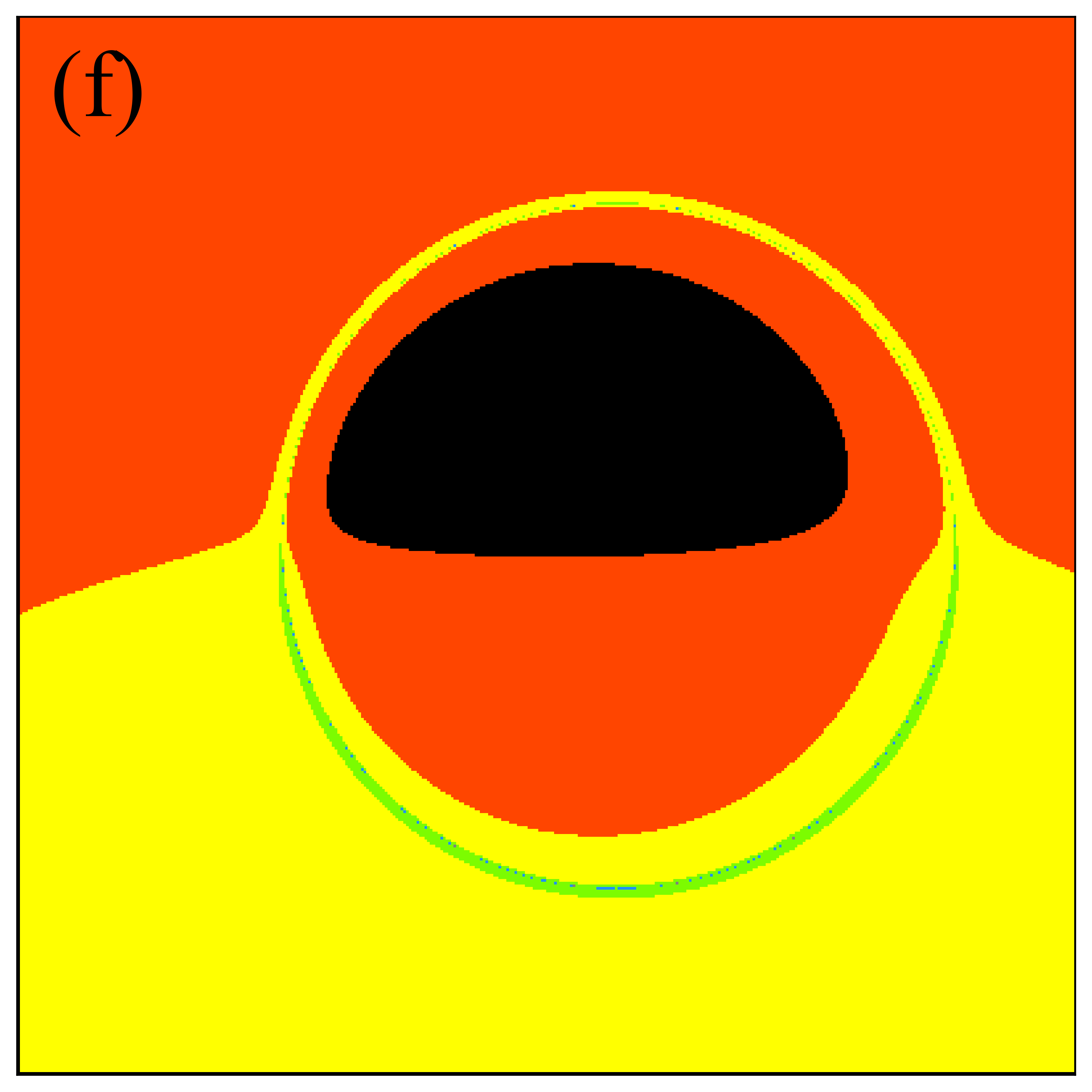}
\includegraphics[width=2.8cm]{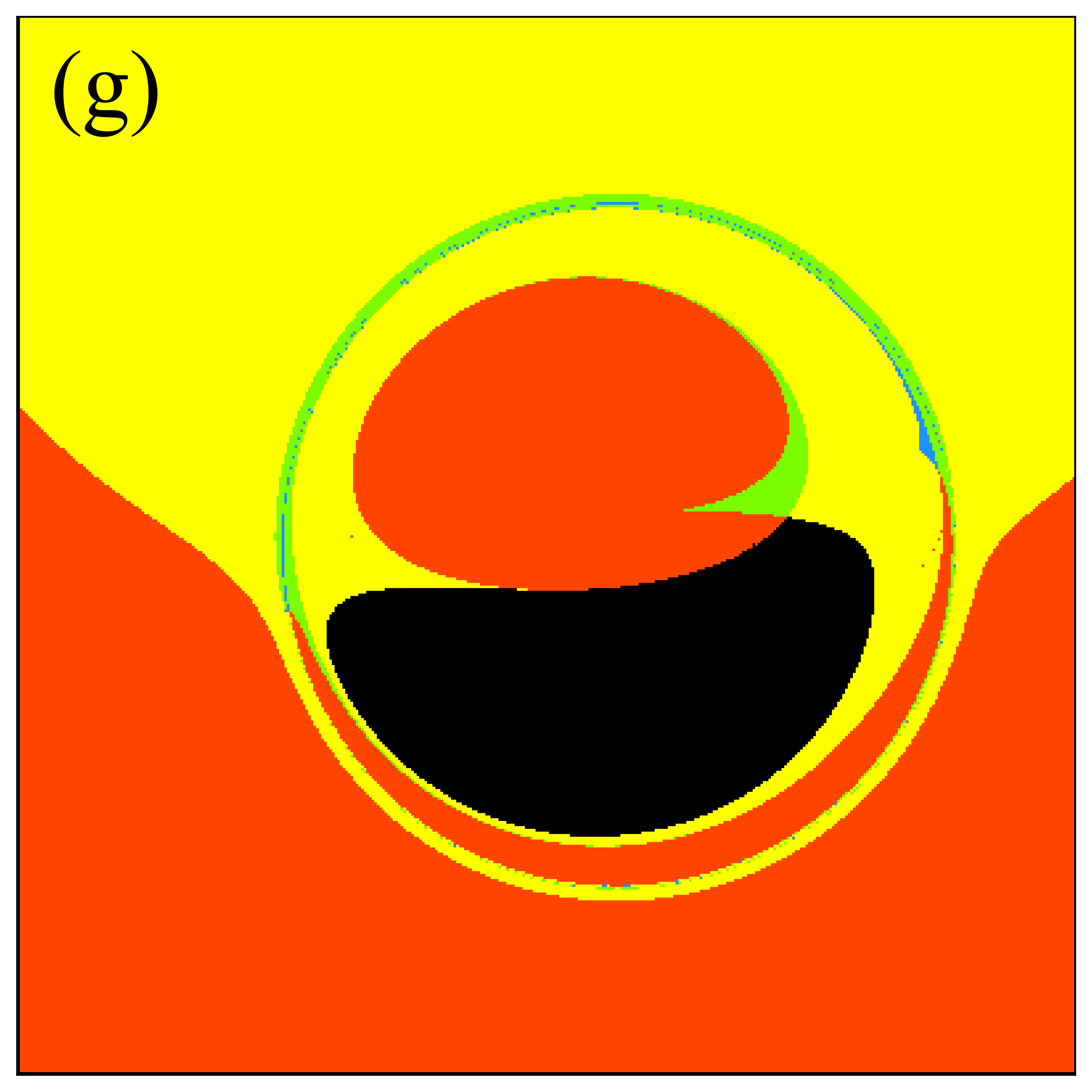}
\includegraphics[width=2.8cm]{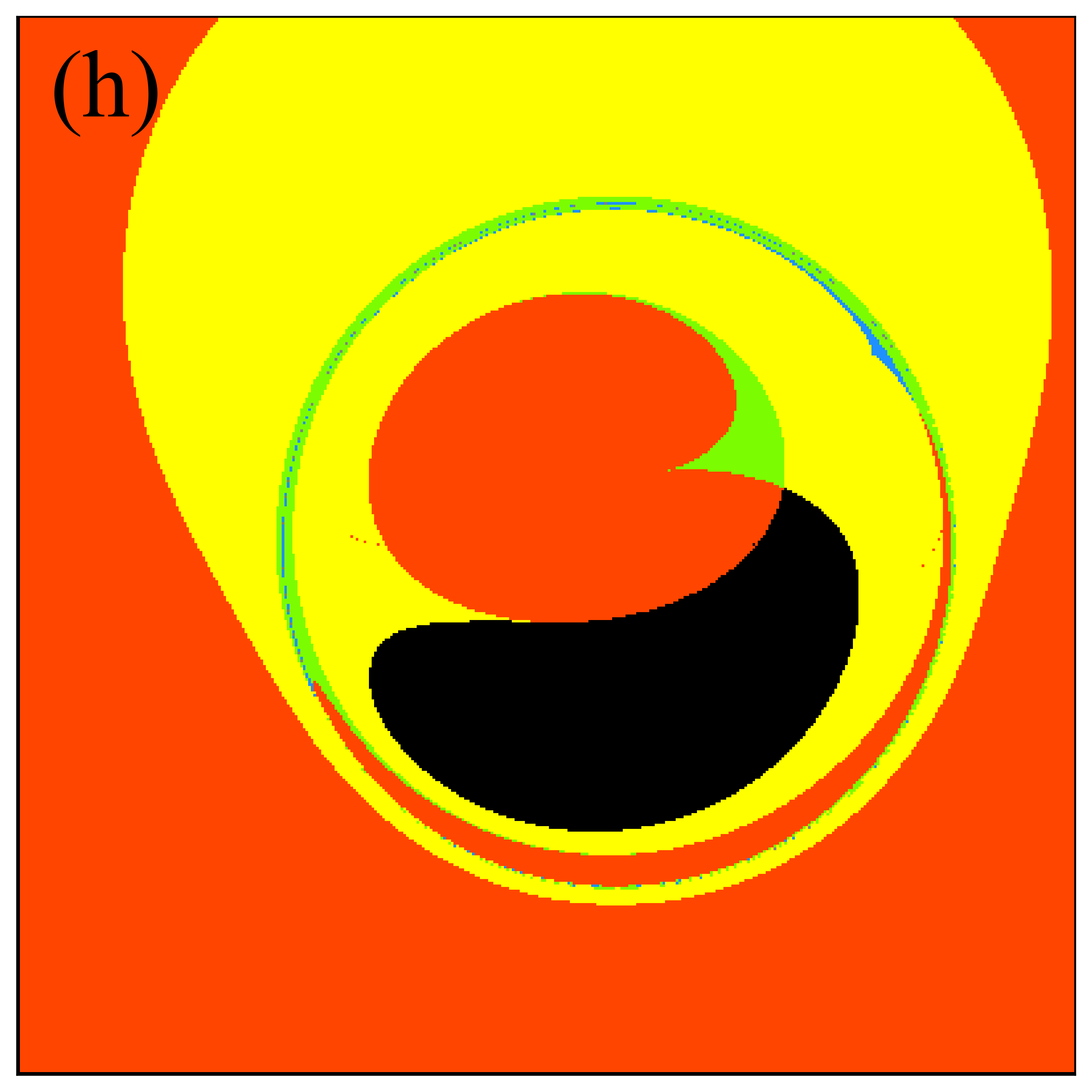}
\includegraphics[width=2.8cm]{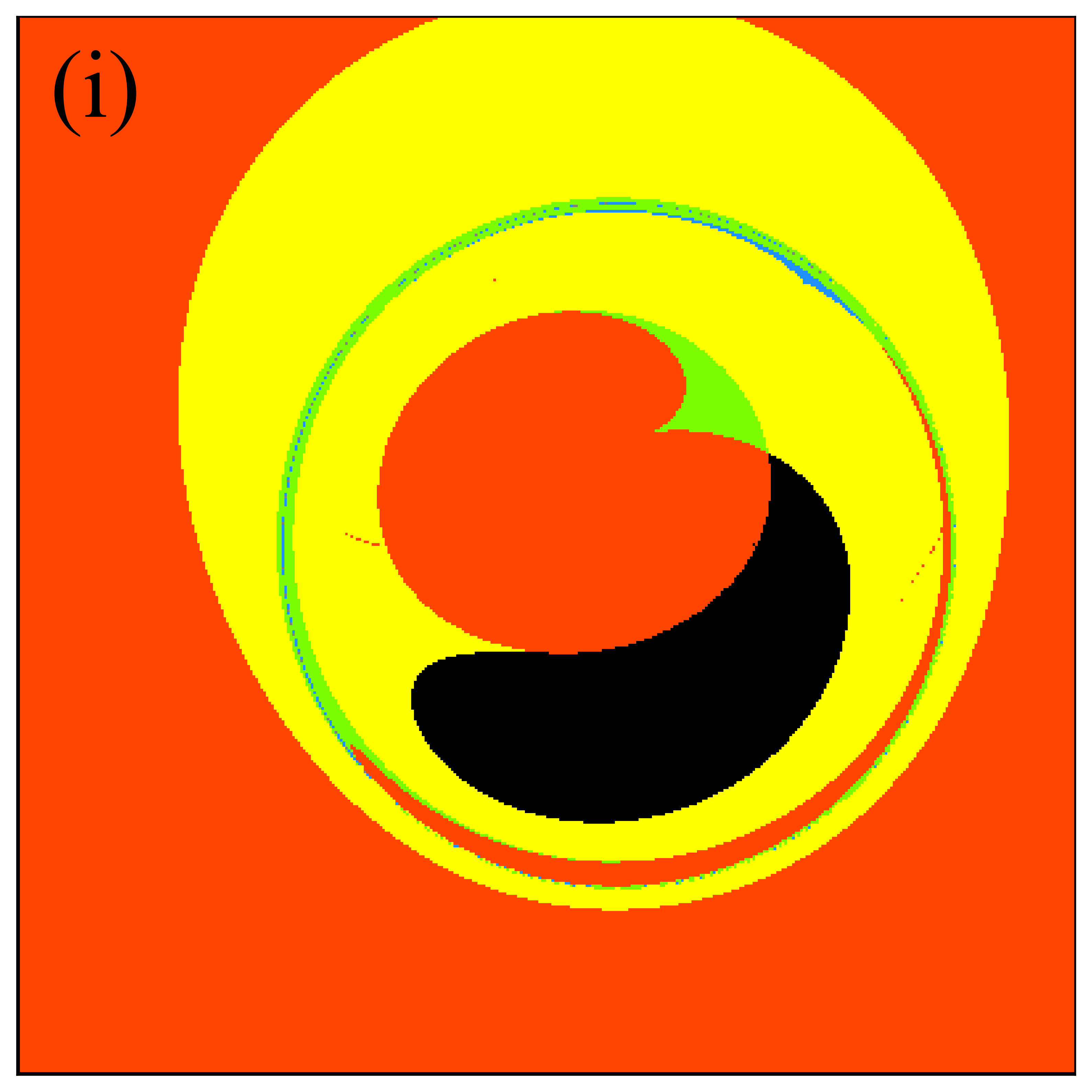}
\includegraphics[width=2.8cm]{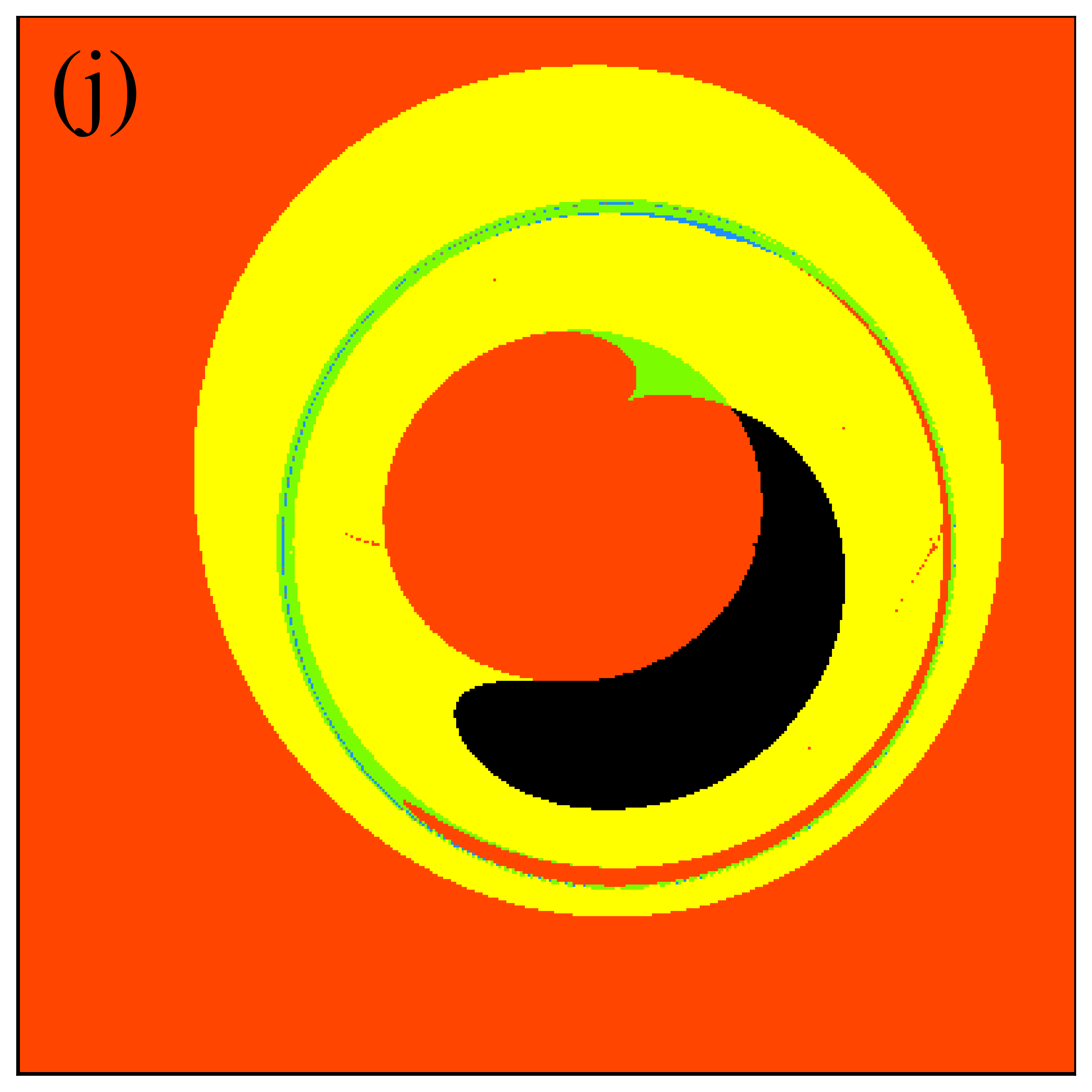}
\includegraphics[width=2.8cm]{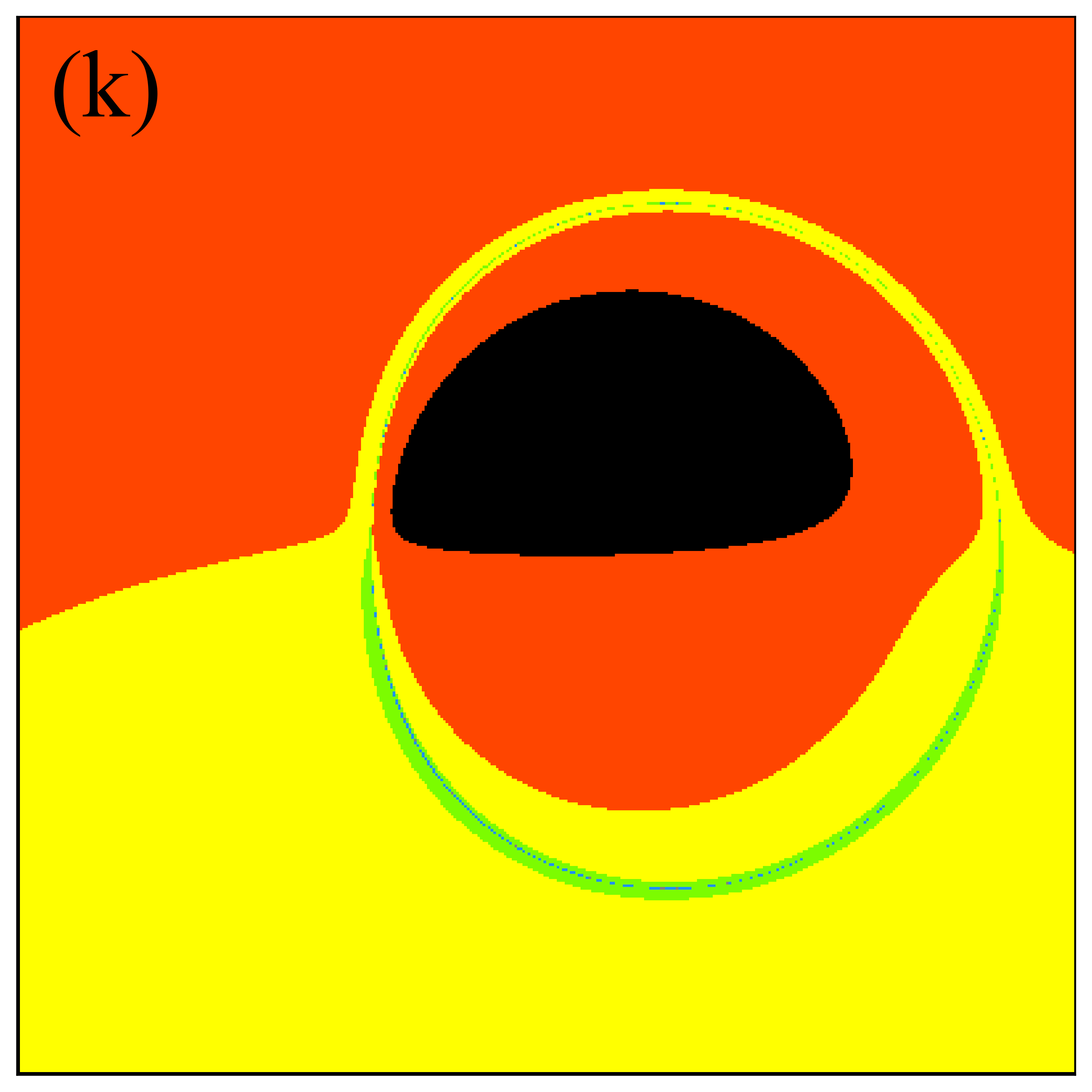}
\includegraphics[width=2.8cm]{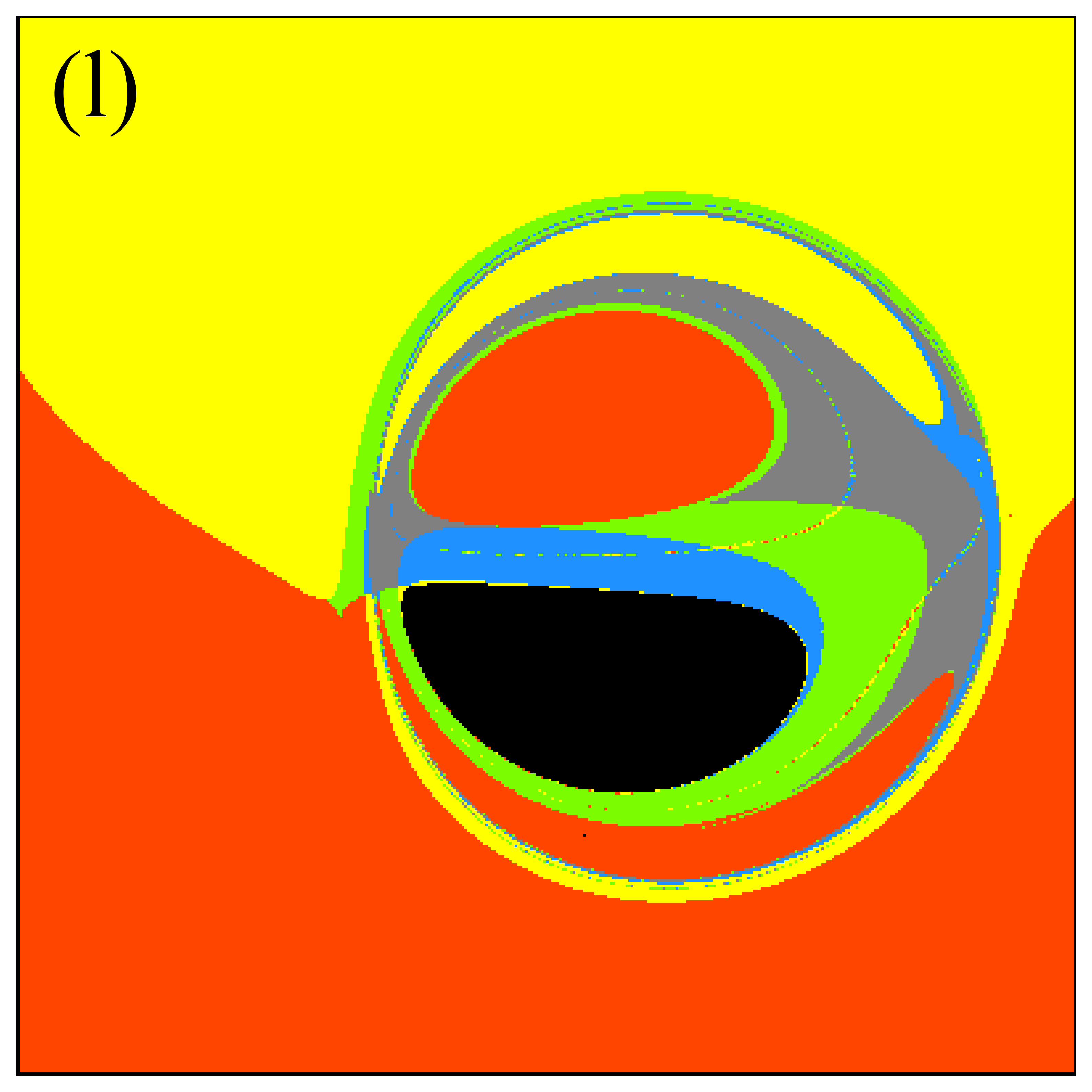}
\includegraphics[width=2.8cm]{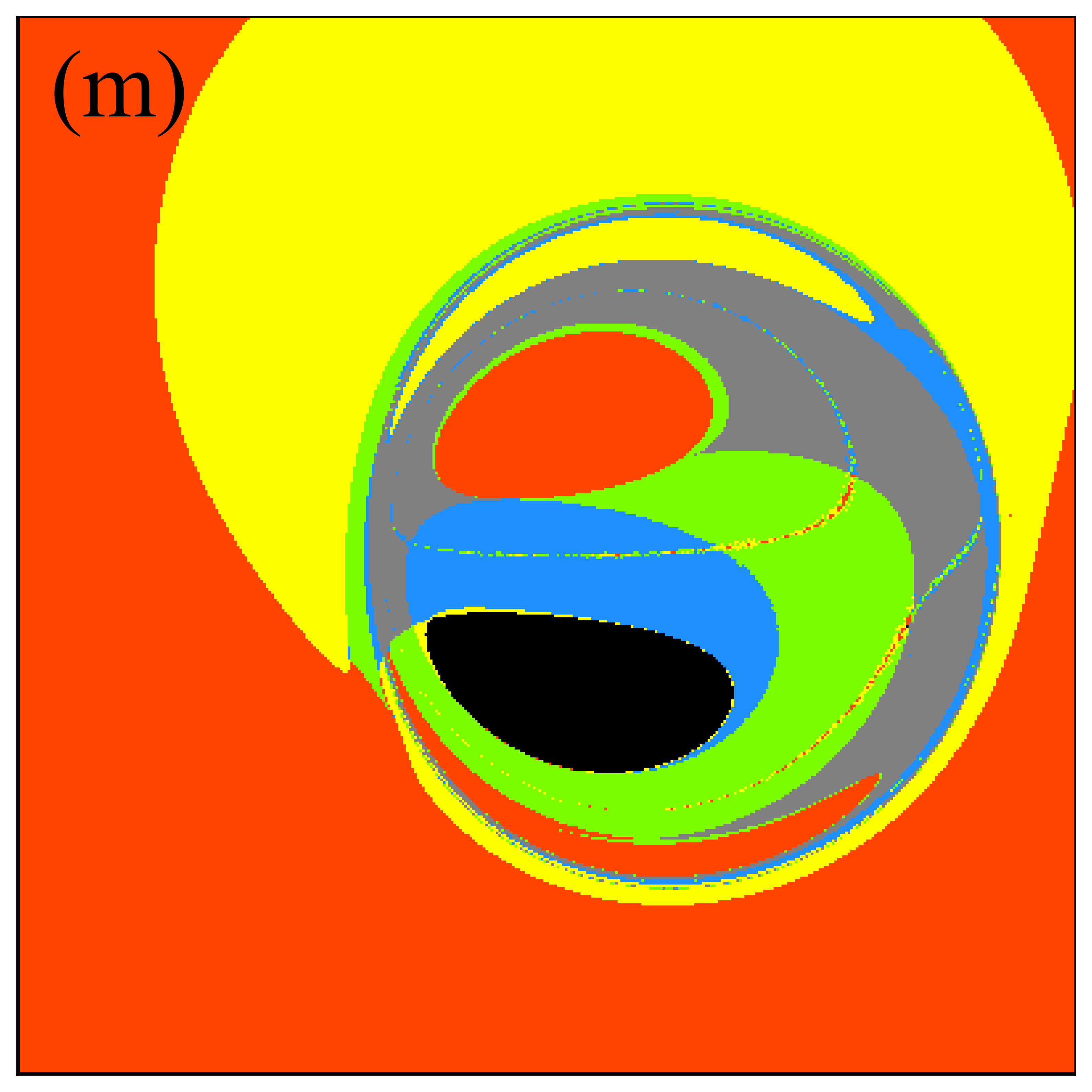}
\includegraphics[width=2.8cm]{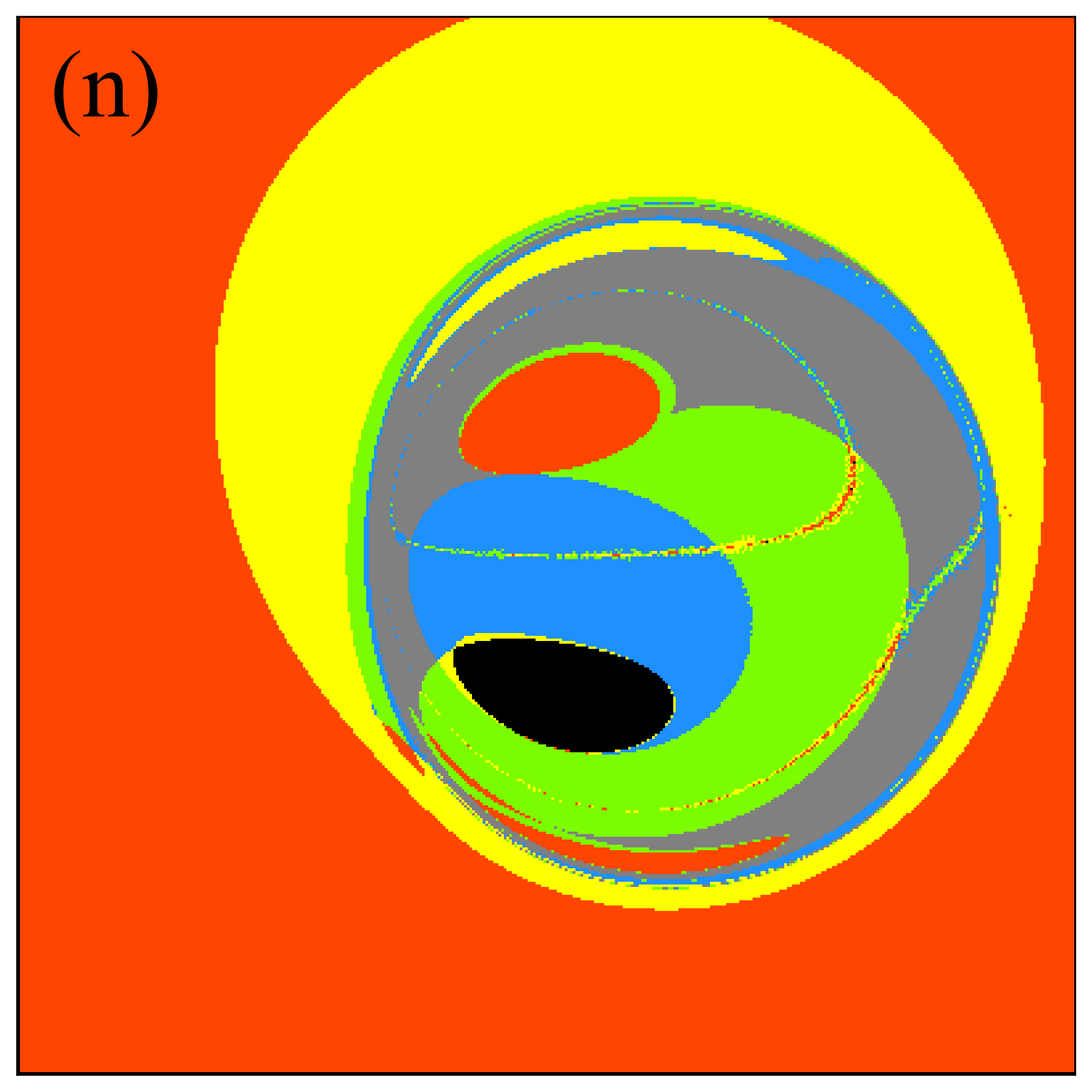}
\includegraphics[width=2.8cm]{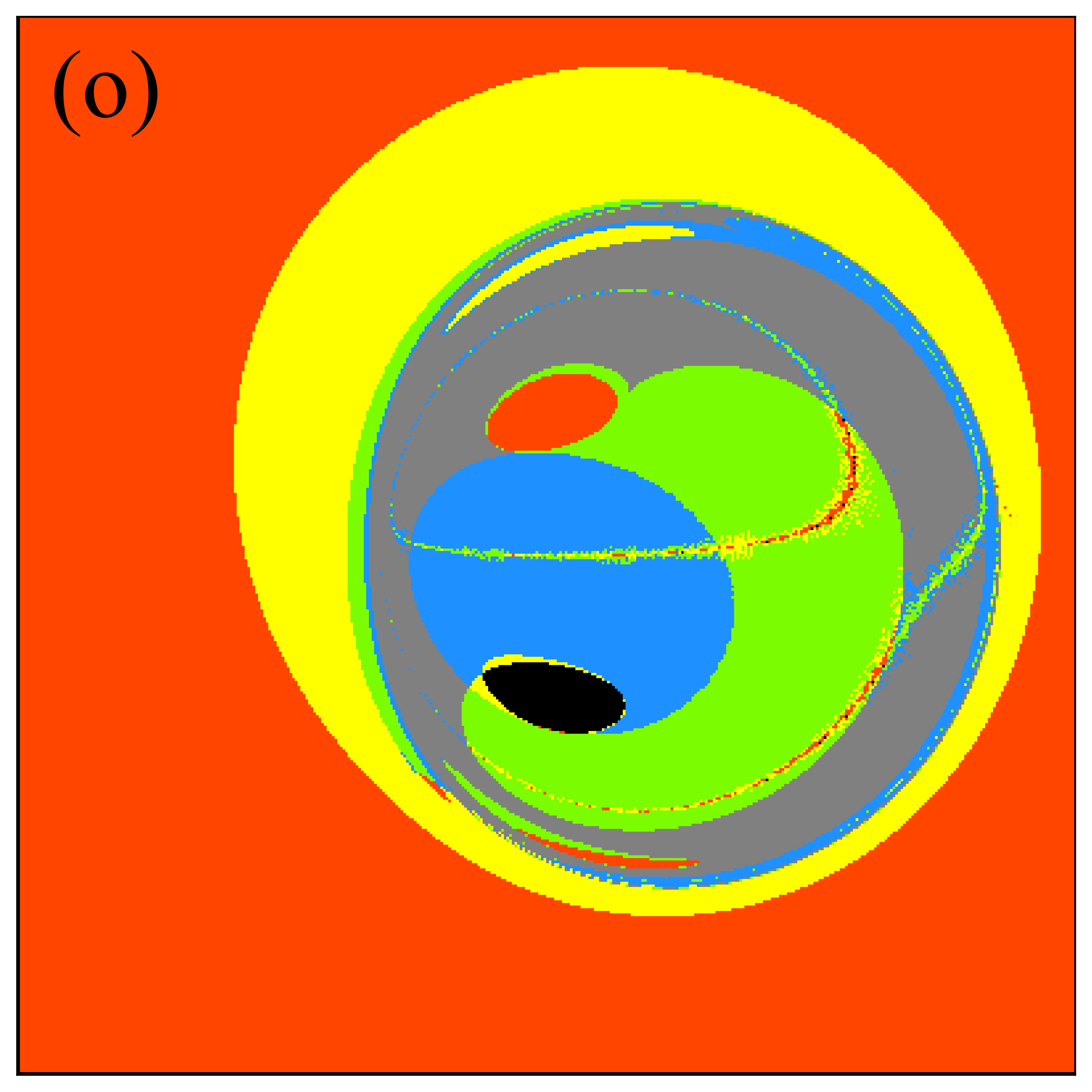}
\includegraphics[width=2.8cm]{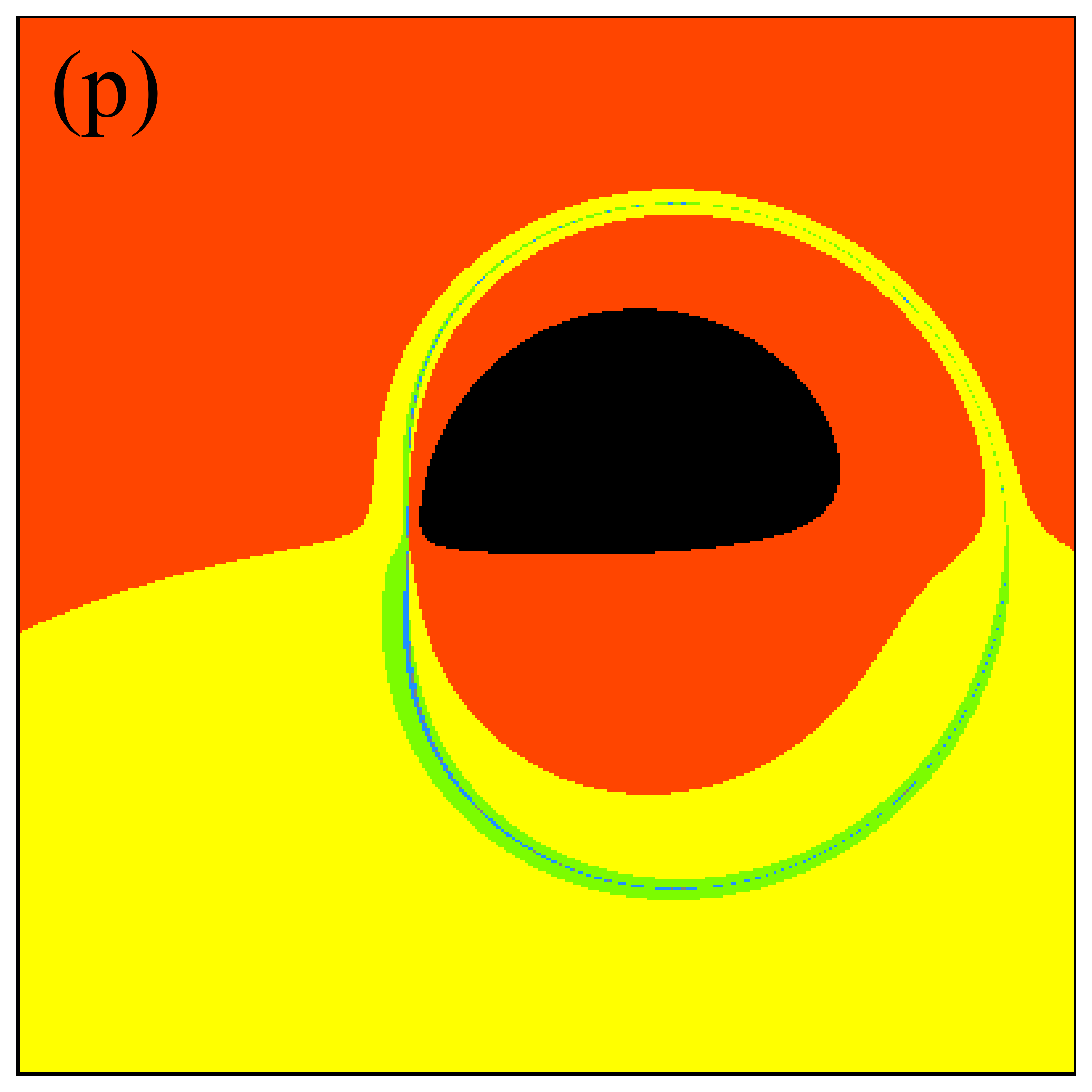}
\includegraphics[width=2.8cm]{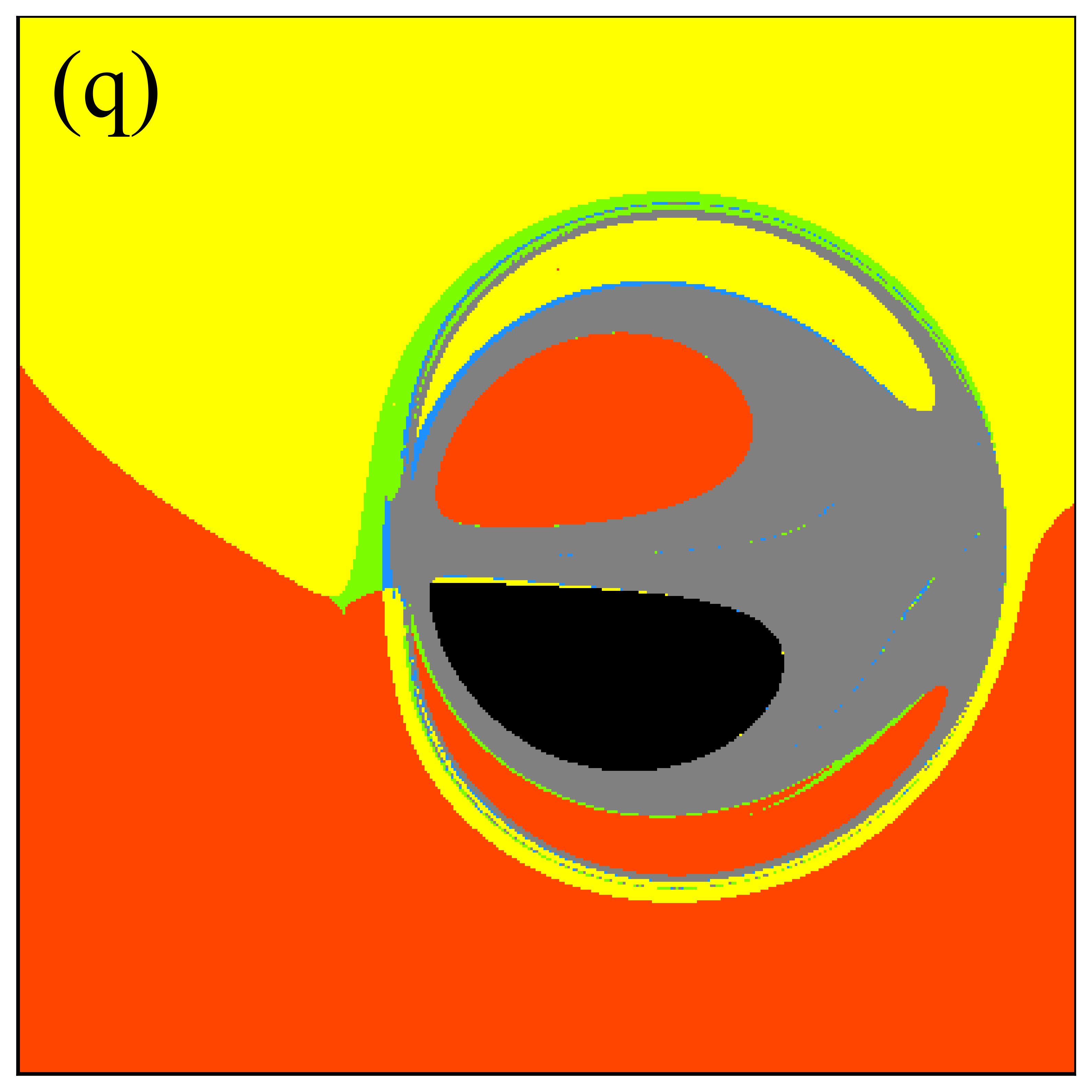}
\includegraphics[width=2.8cm]{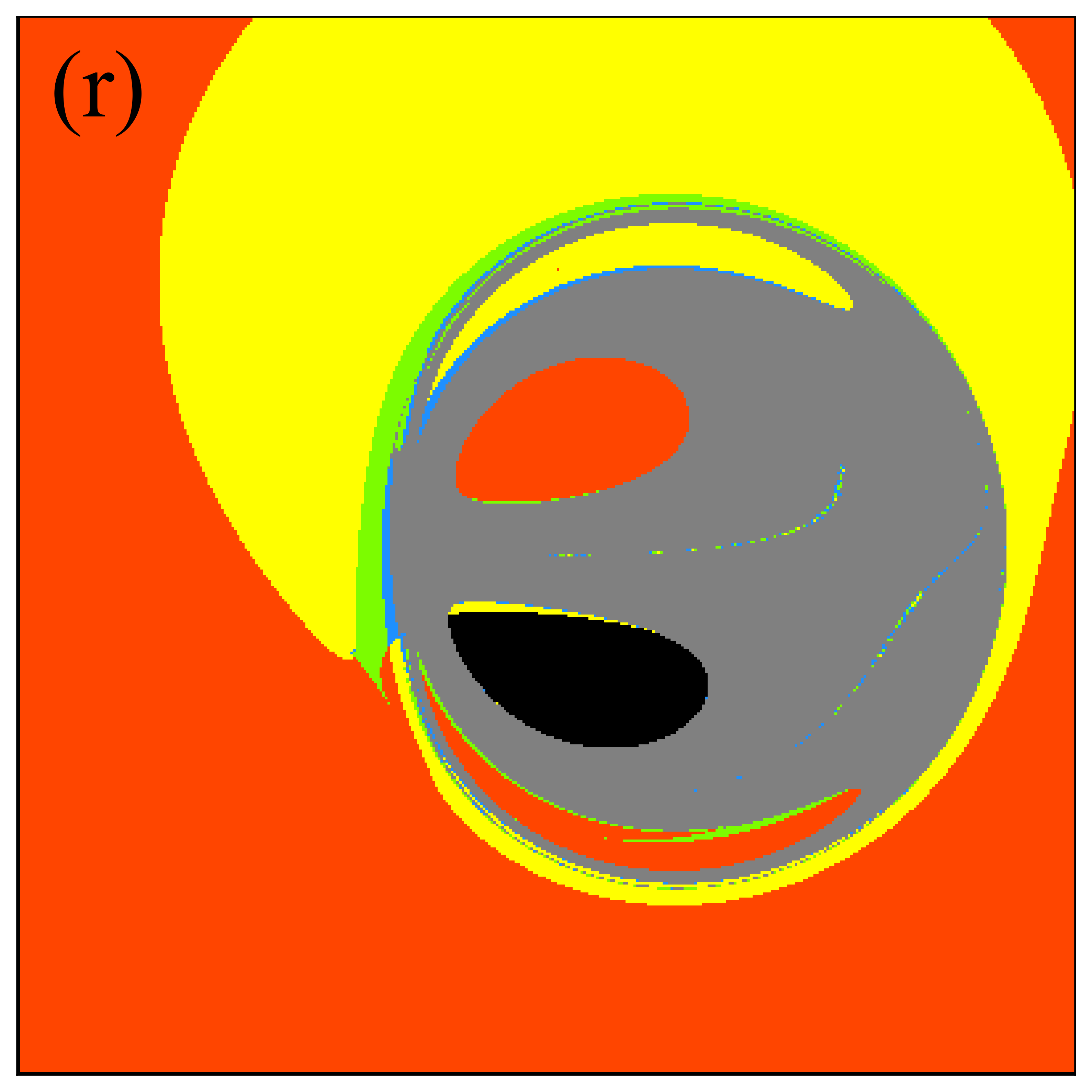}
\includegraphics[width=2.8cm]{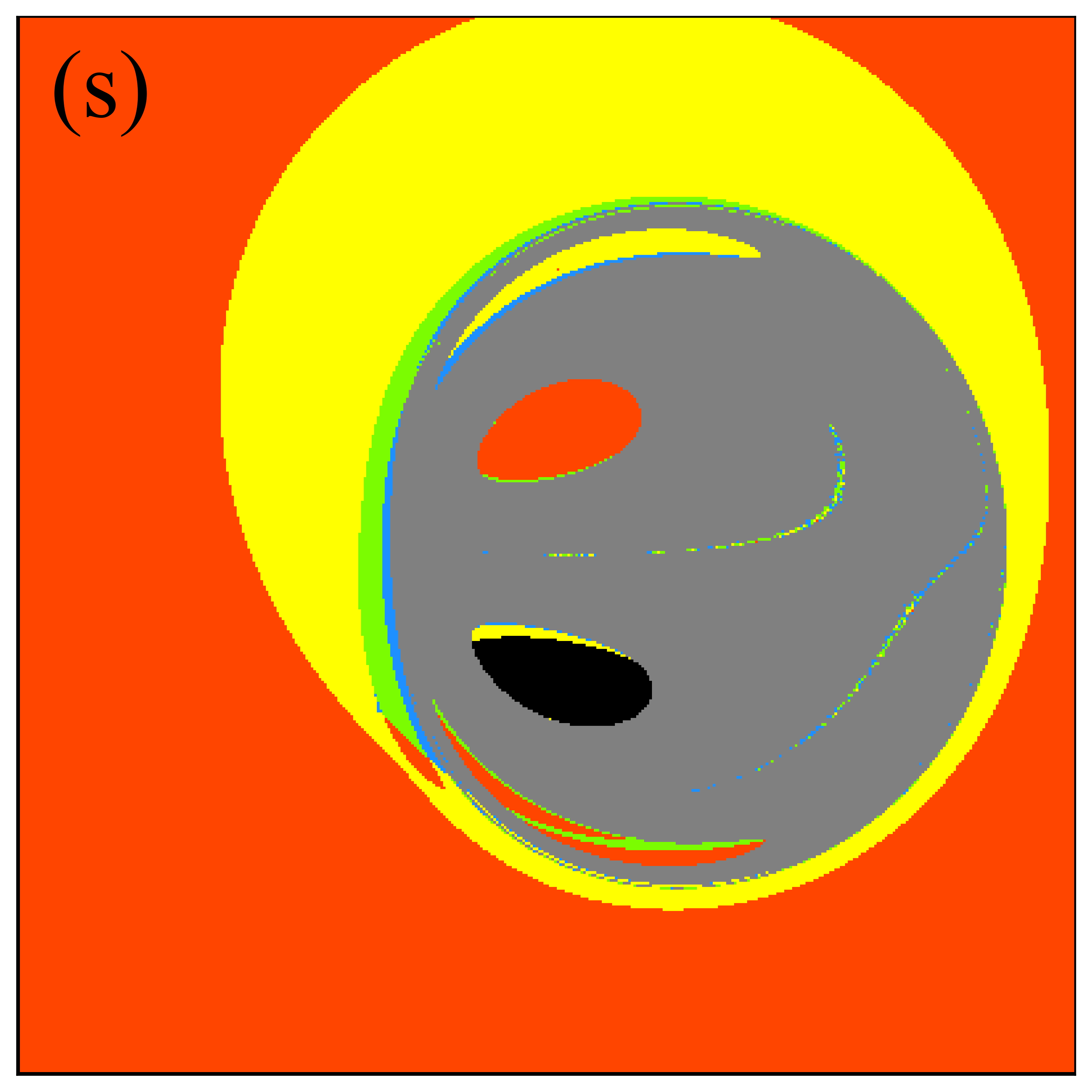}
\includegraphics[width=2.8cm]{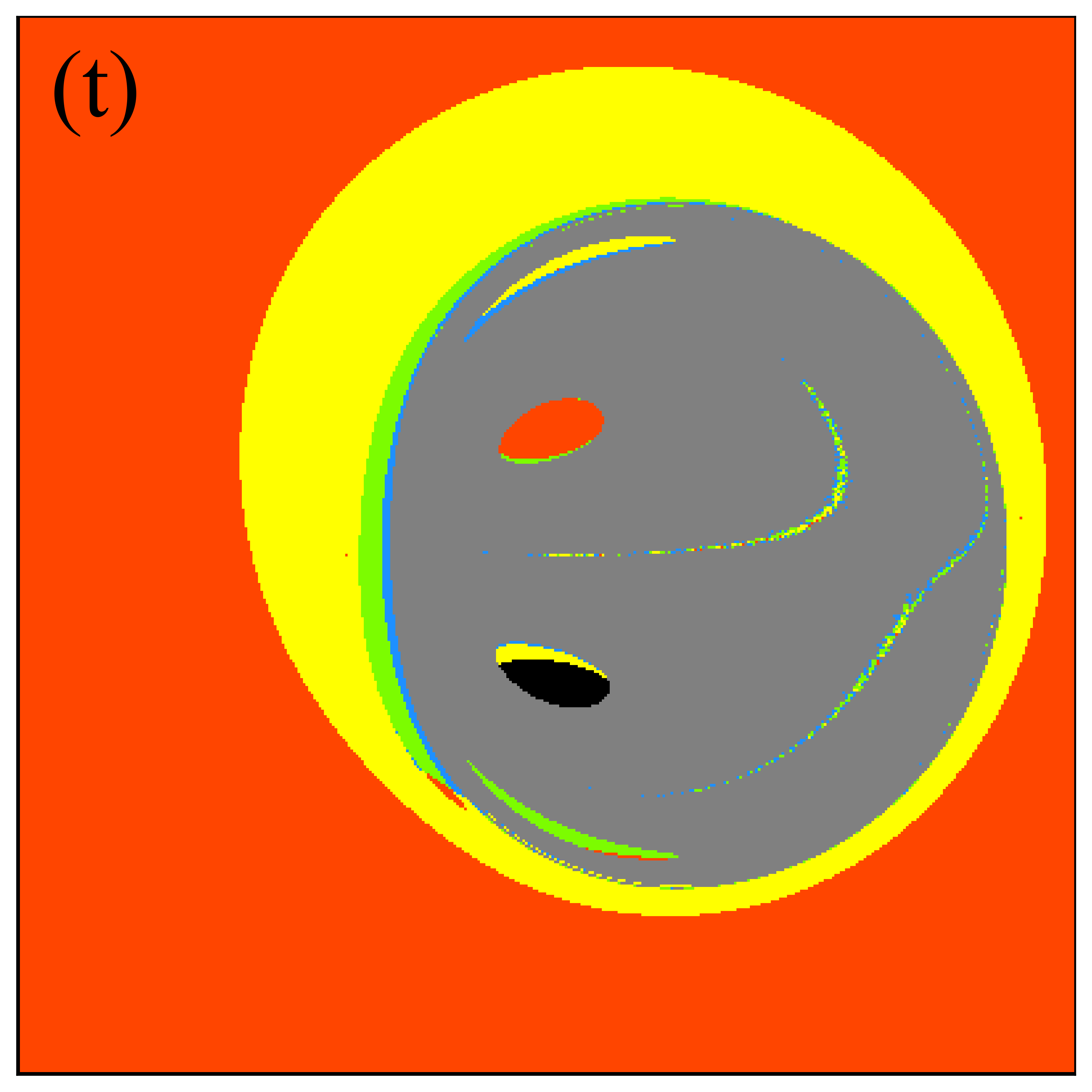}
\caption{Similar to figure 7, but for the observation angle of $85^{\circ}$.}}\label{fig8}
\end{figure*}
\subsection{$\Phi=90^{\circ}$ case}
In the case of an equatorial accretion disk, the observed black hole image is independent of the observer's azimuthal angle. Therefore, we focus only on the case of non-zero $\sigma$ in the following analysis.
\begin{figure*}
\center{
\includegraphics[width=3.7cm]{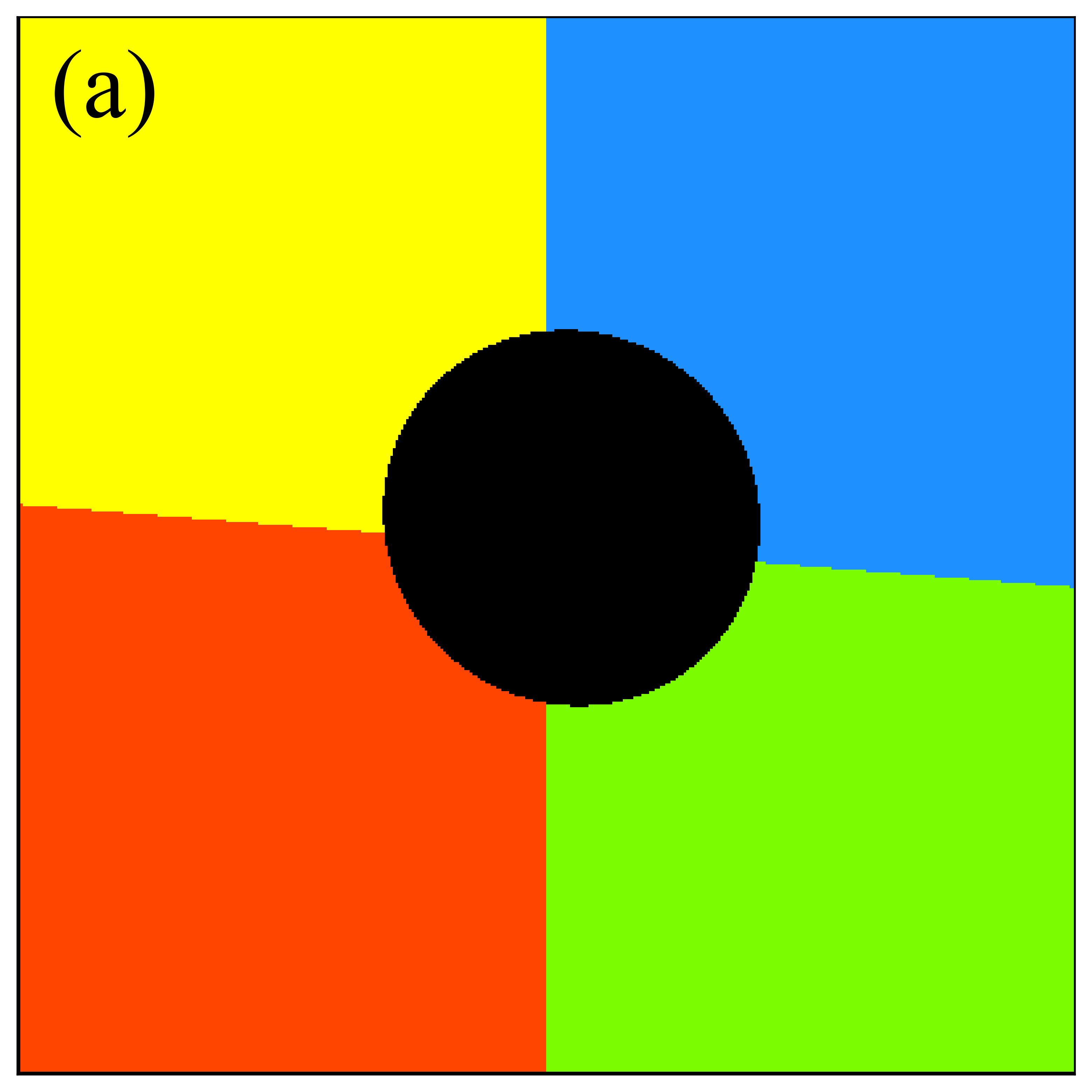}
\includegraphics[width=3.7cm]{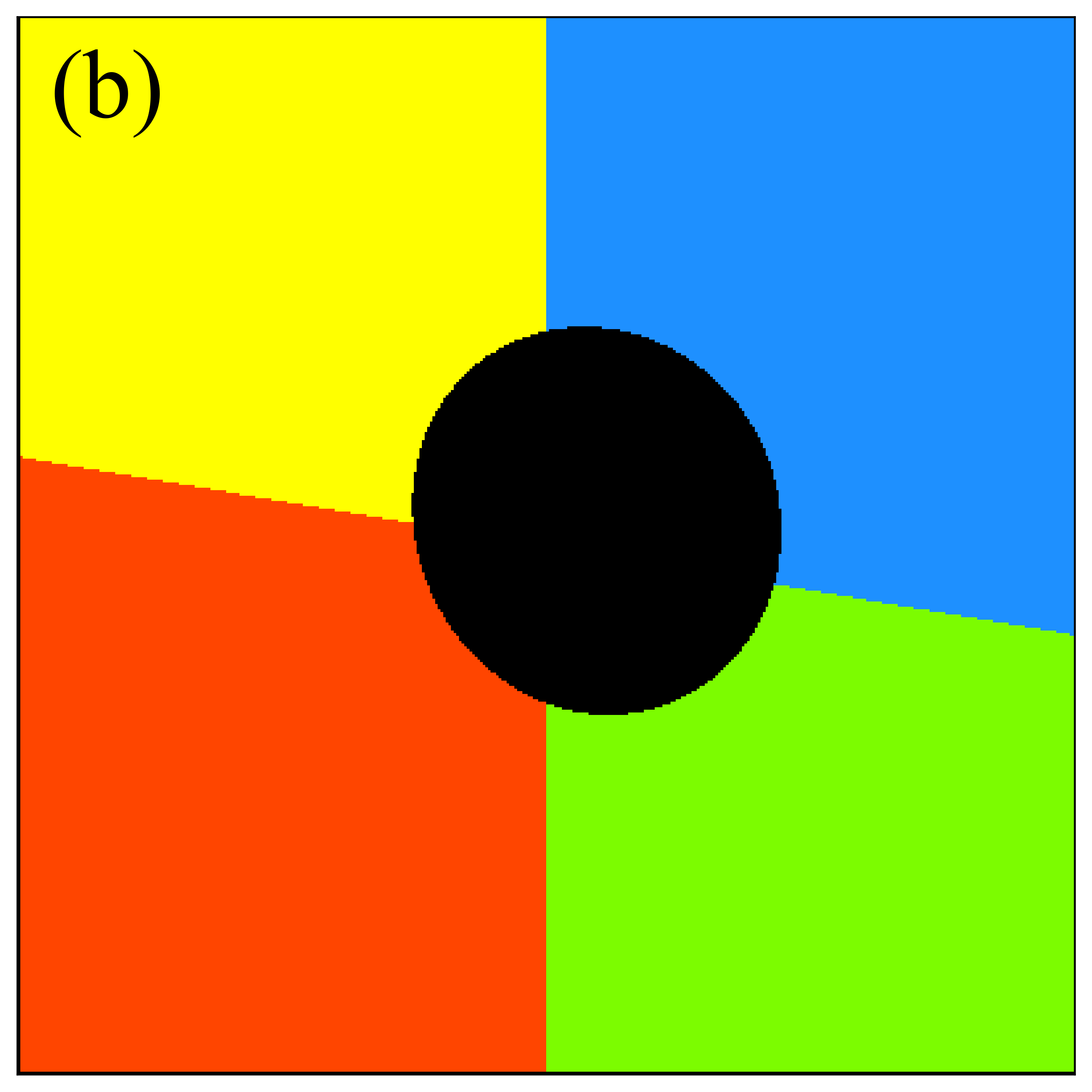}
\includegraphics[width=3.7cm]{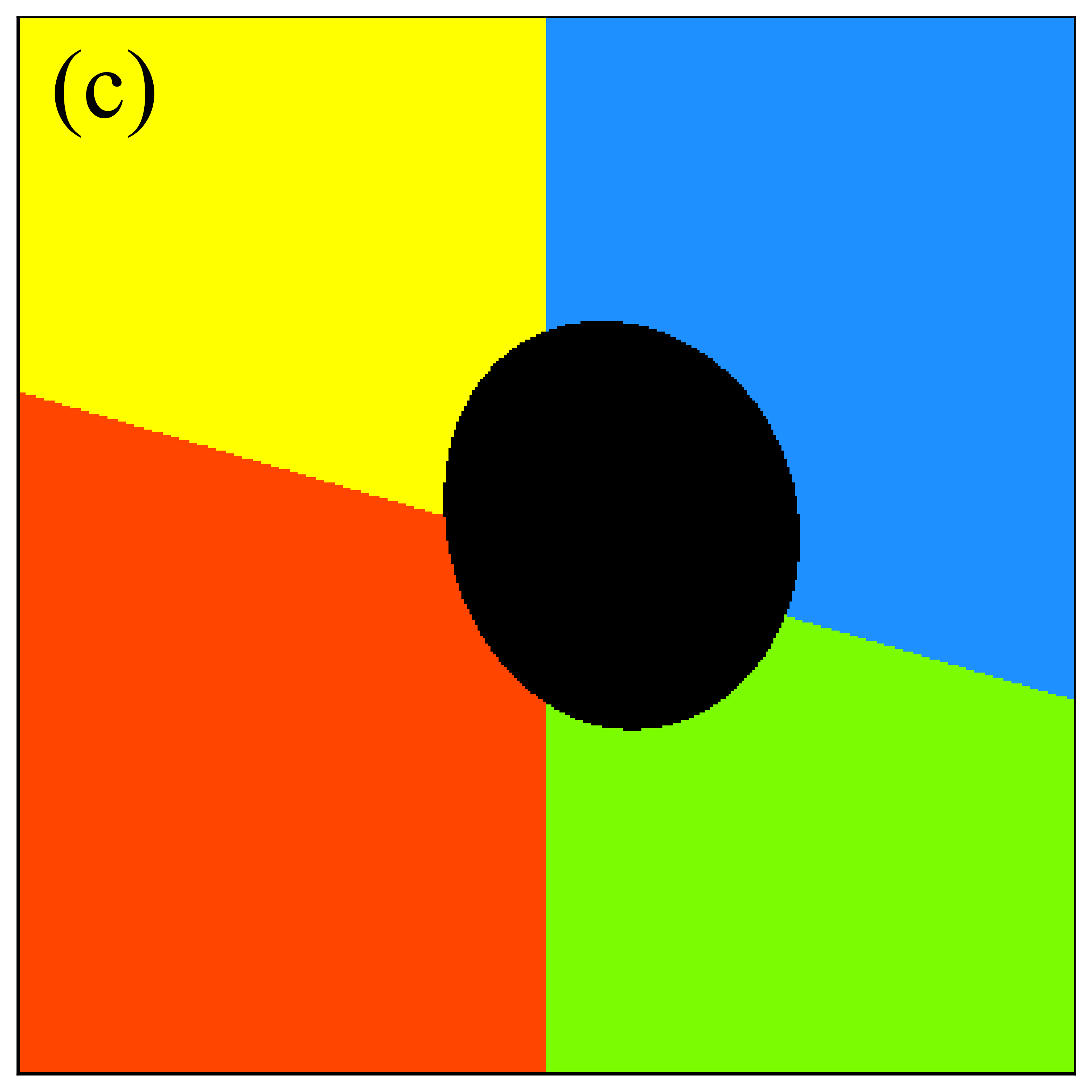}
\includegraphics[width=3.7cm]{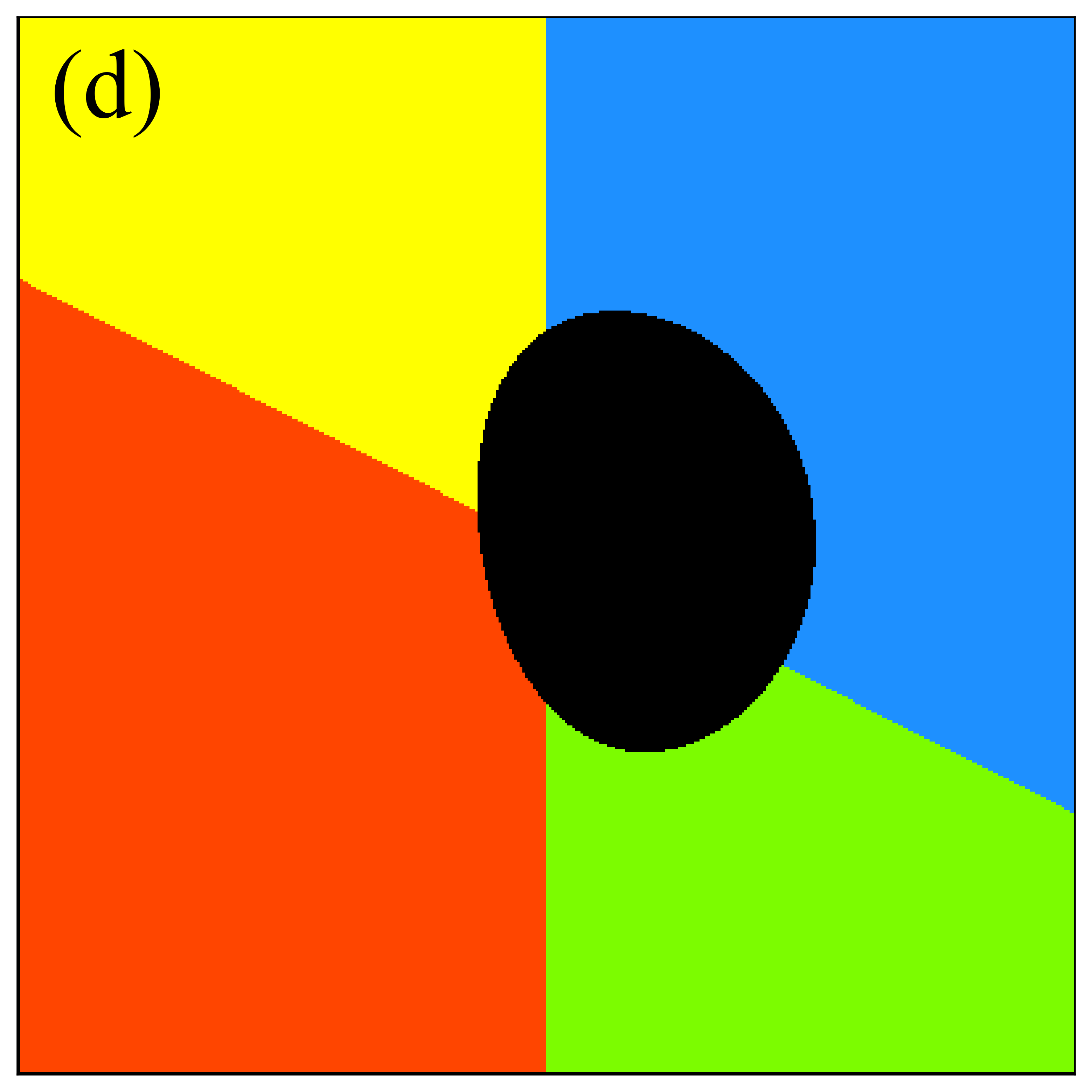}
\includegraphics[width=3.7cm]{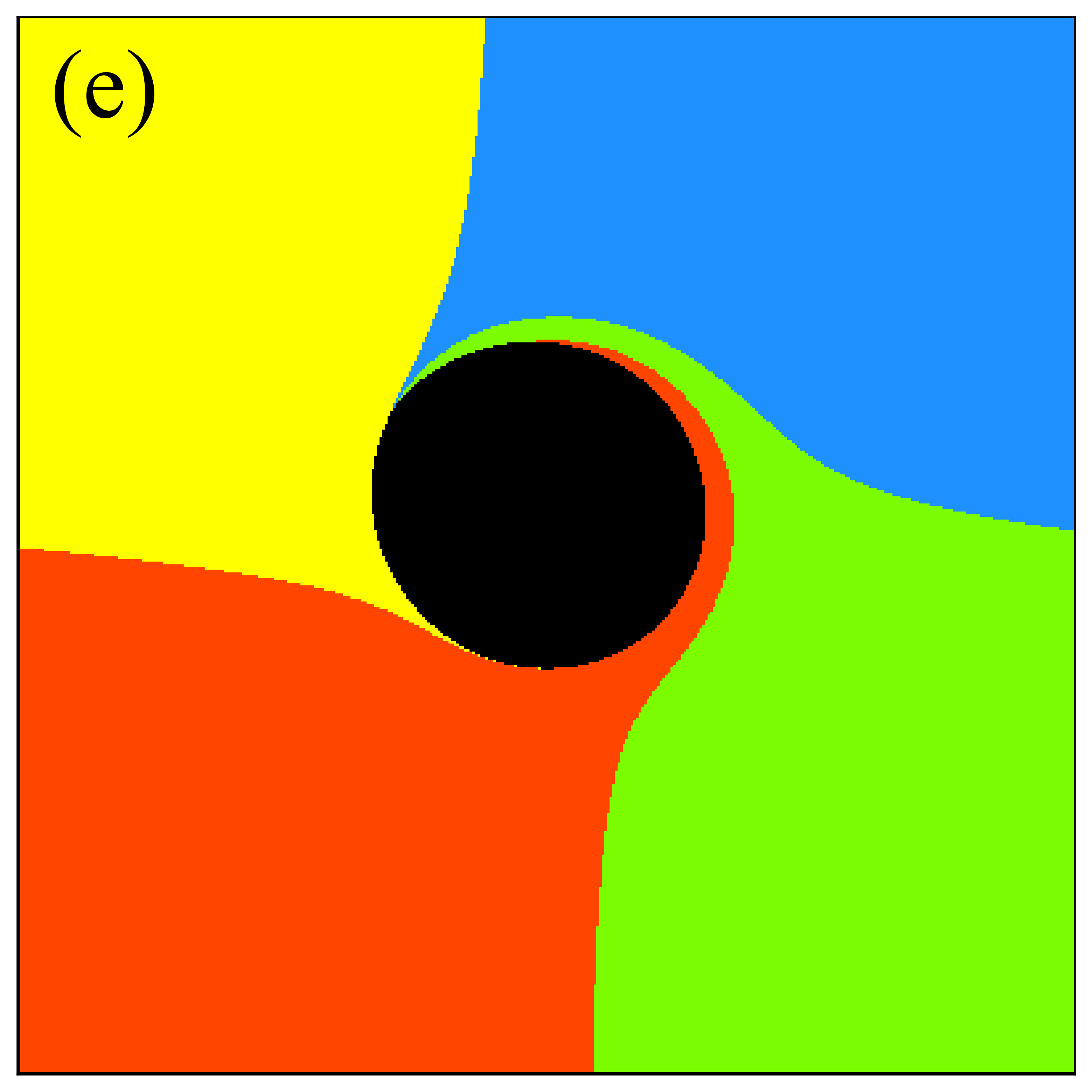}
\includegraphics[width=3.7cm]{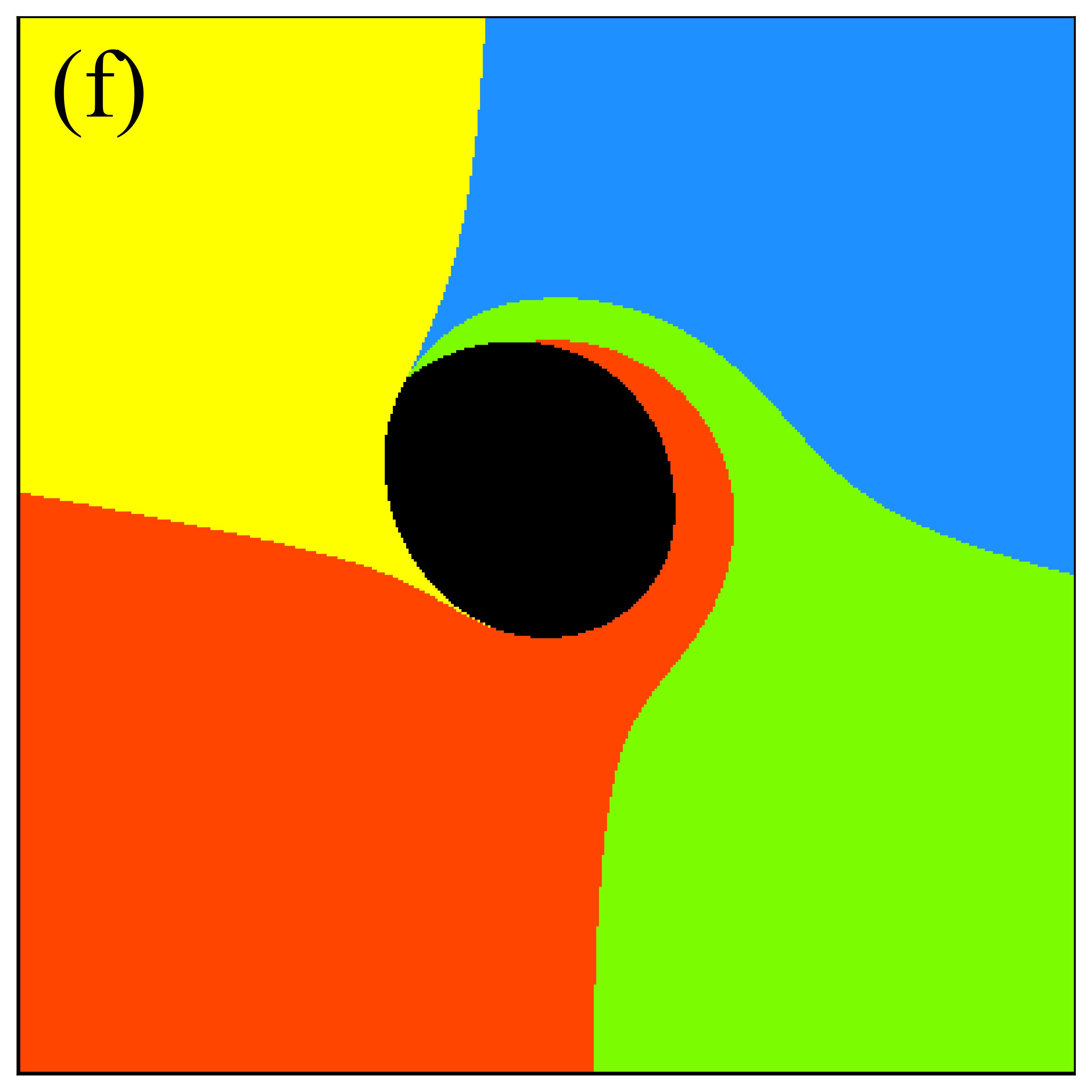}
\includegraphics[width=3.7cm]{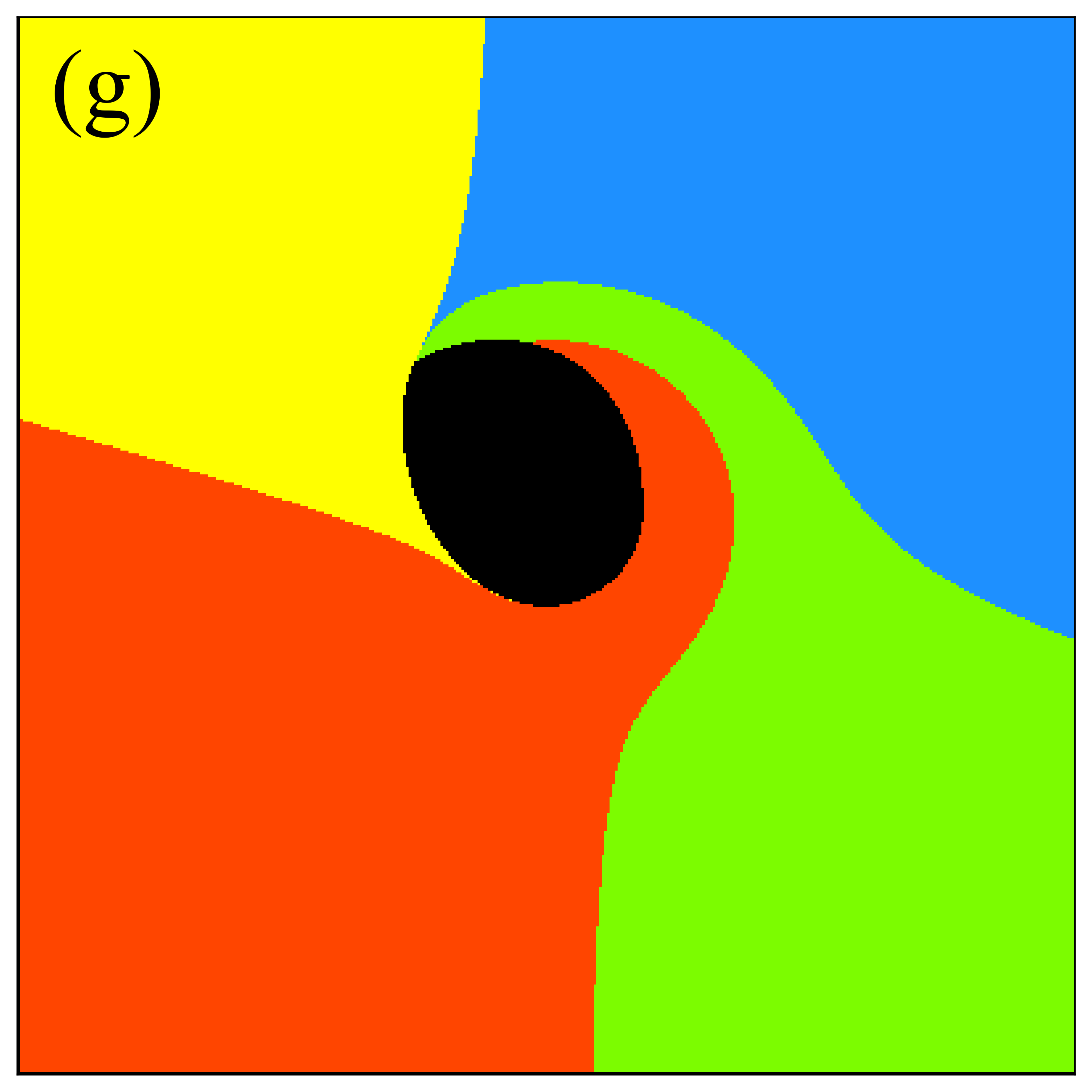}
\includegraphics[width=3.7cm]{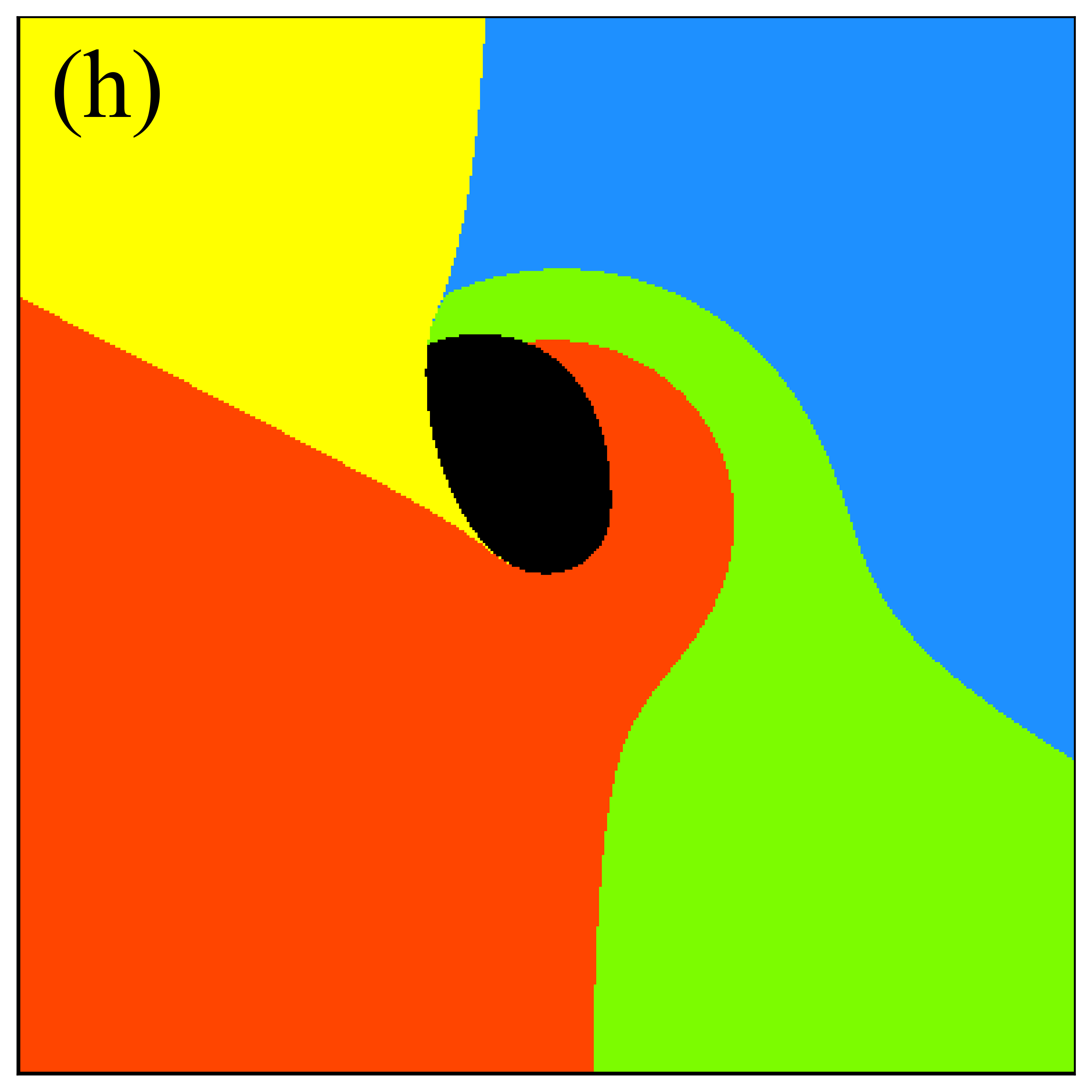}
\includegraphics[width=3.7cm]{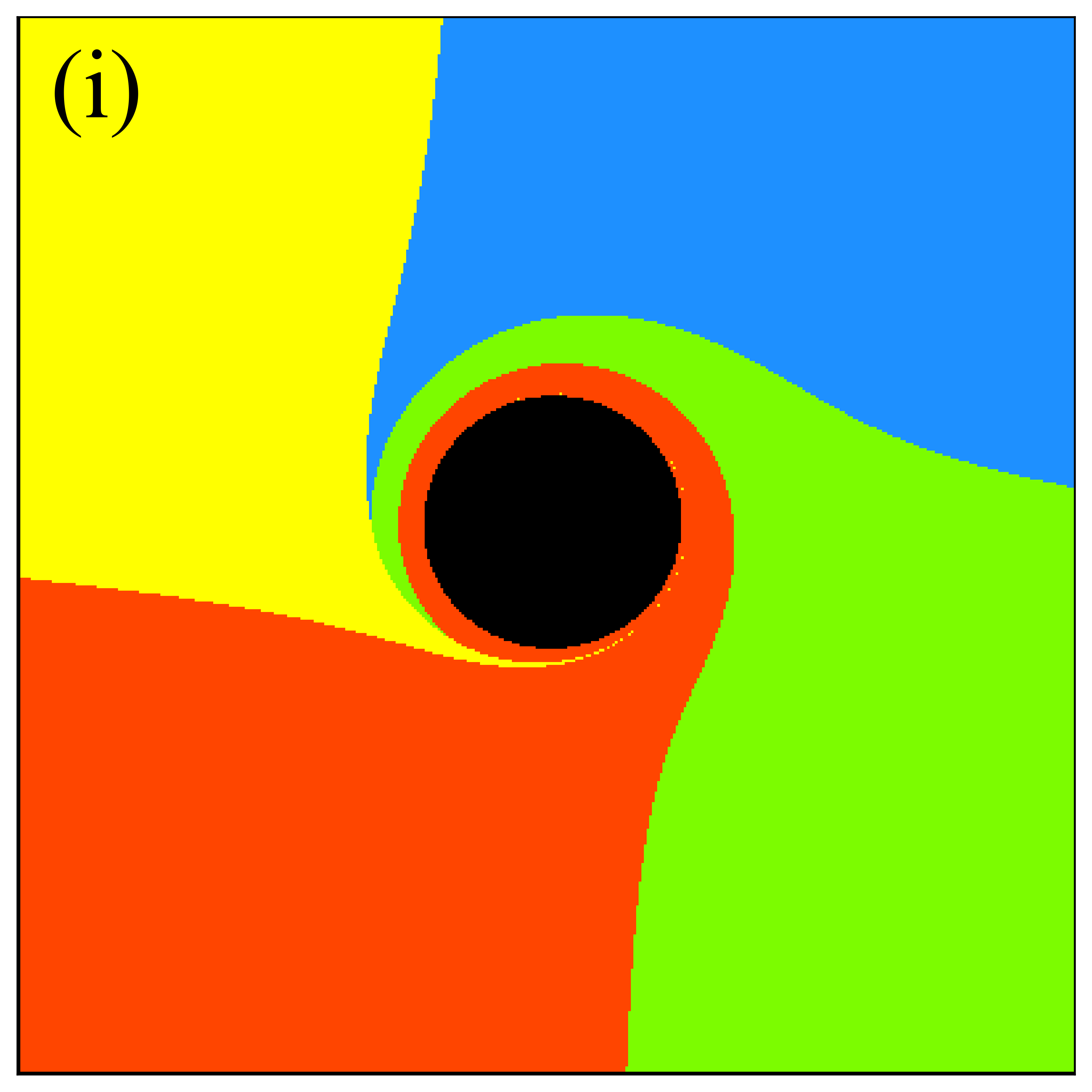}
\includegraphics[width=3.7cm]{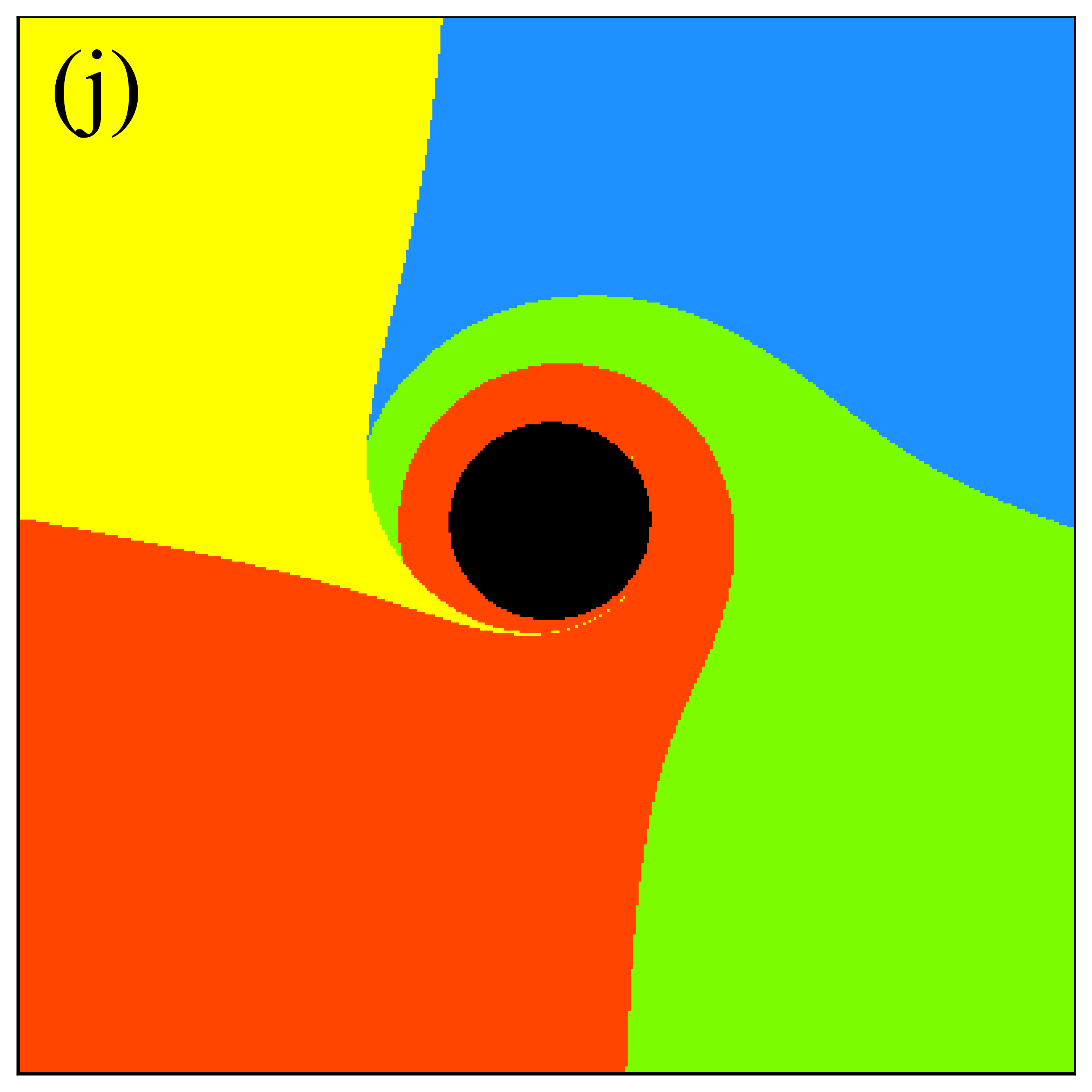}
\includegraphics[width=3.7cm]{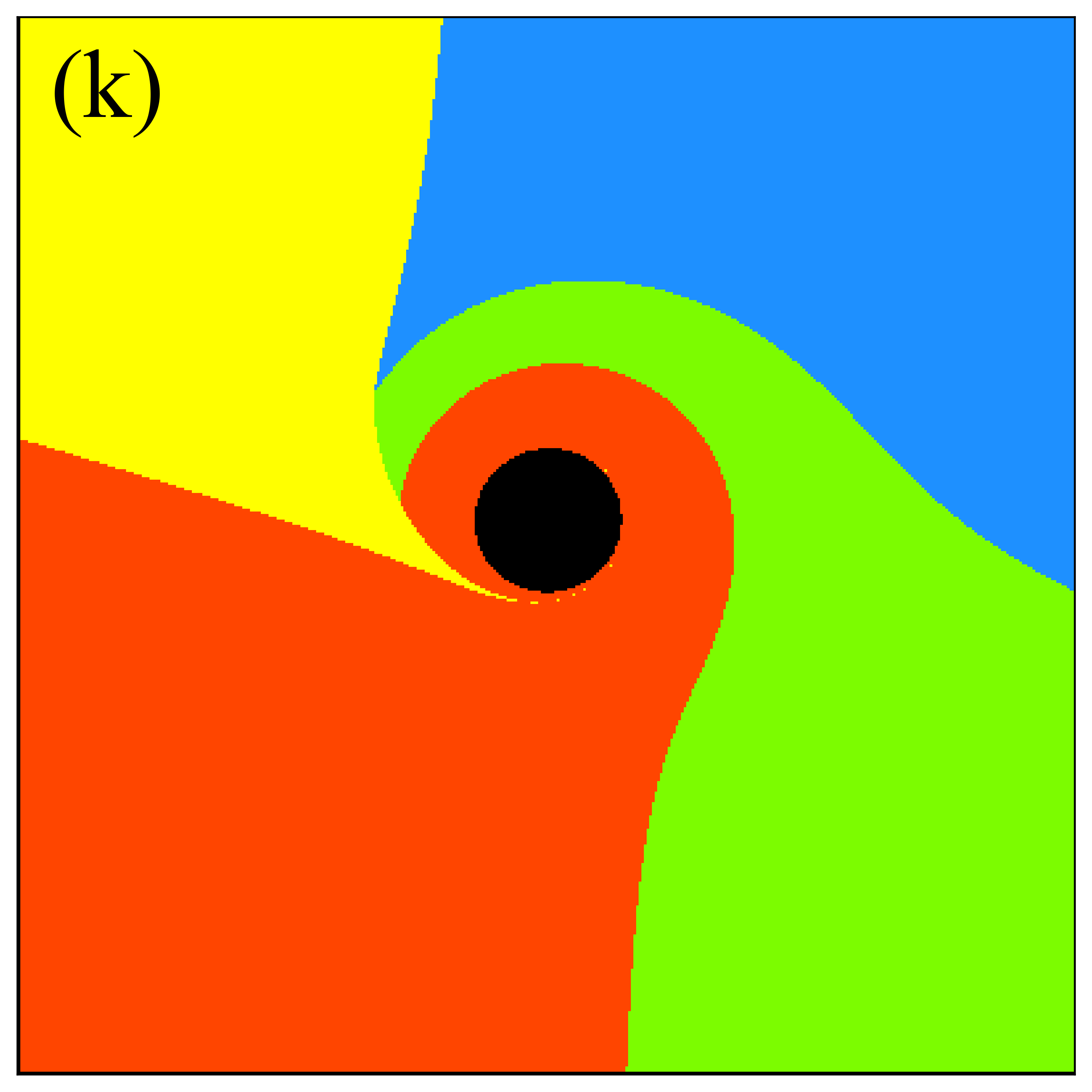}
\includegraphics[width=3.7cm]{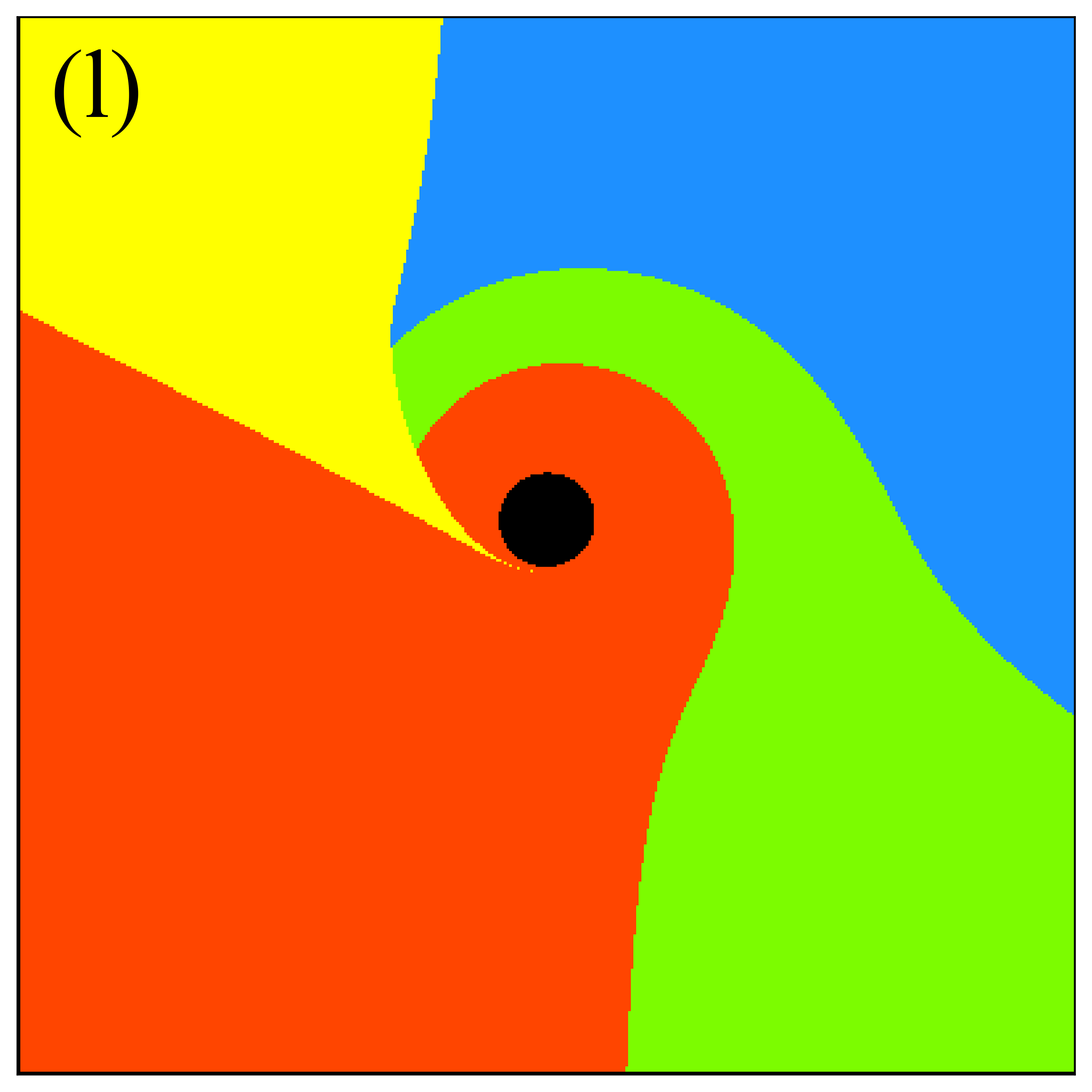}
\includegraphics[width=3.7cm]{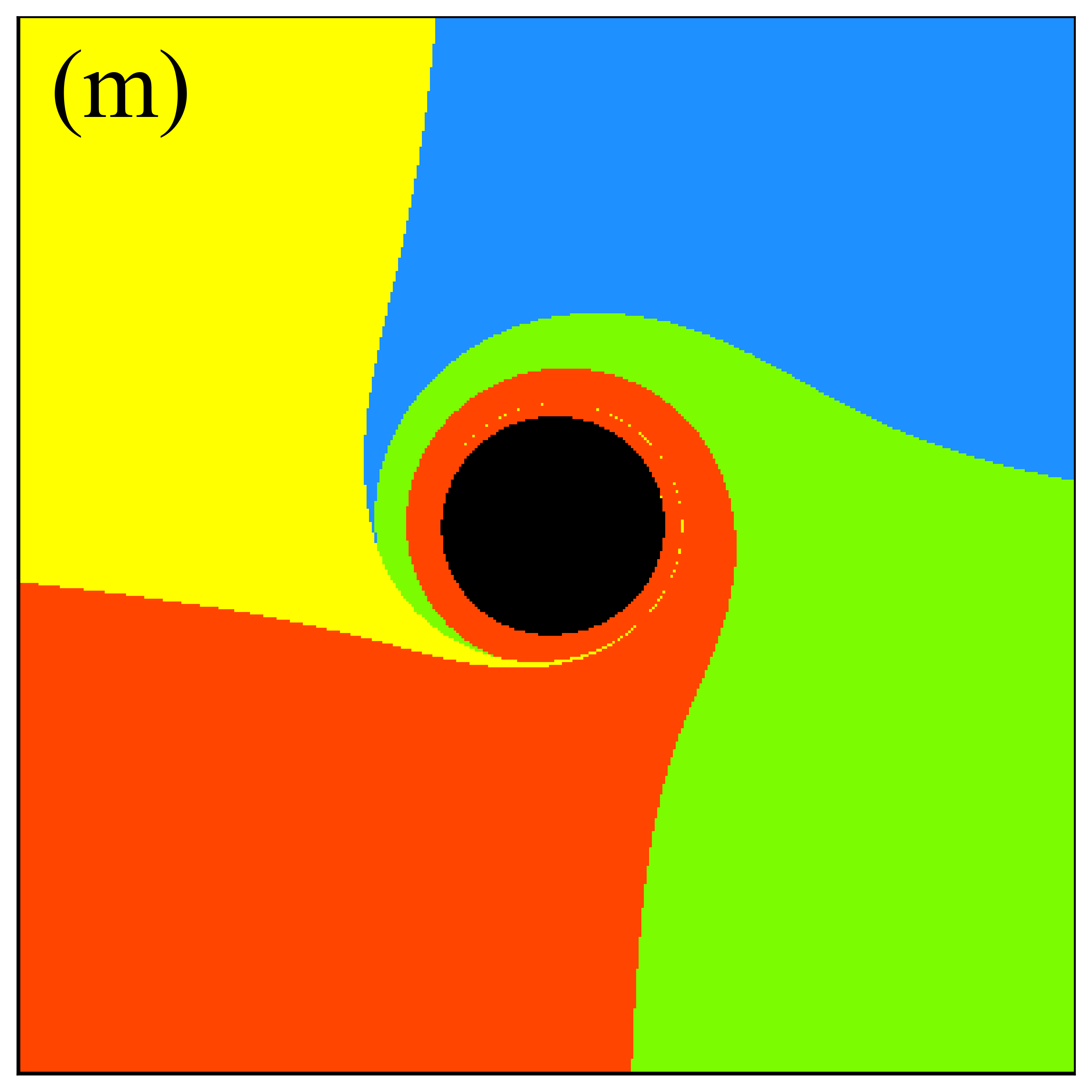}
\includegraphics[width=3.7cm]{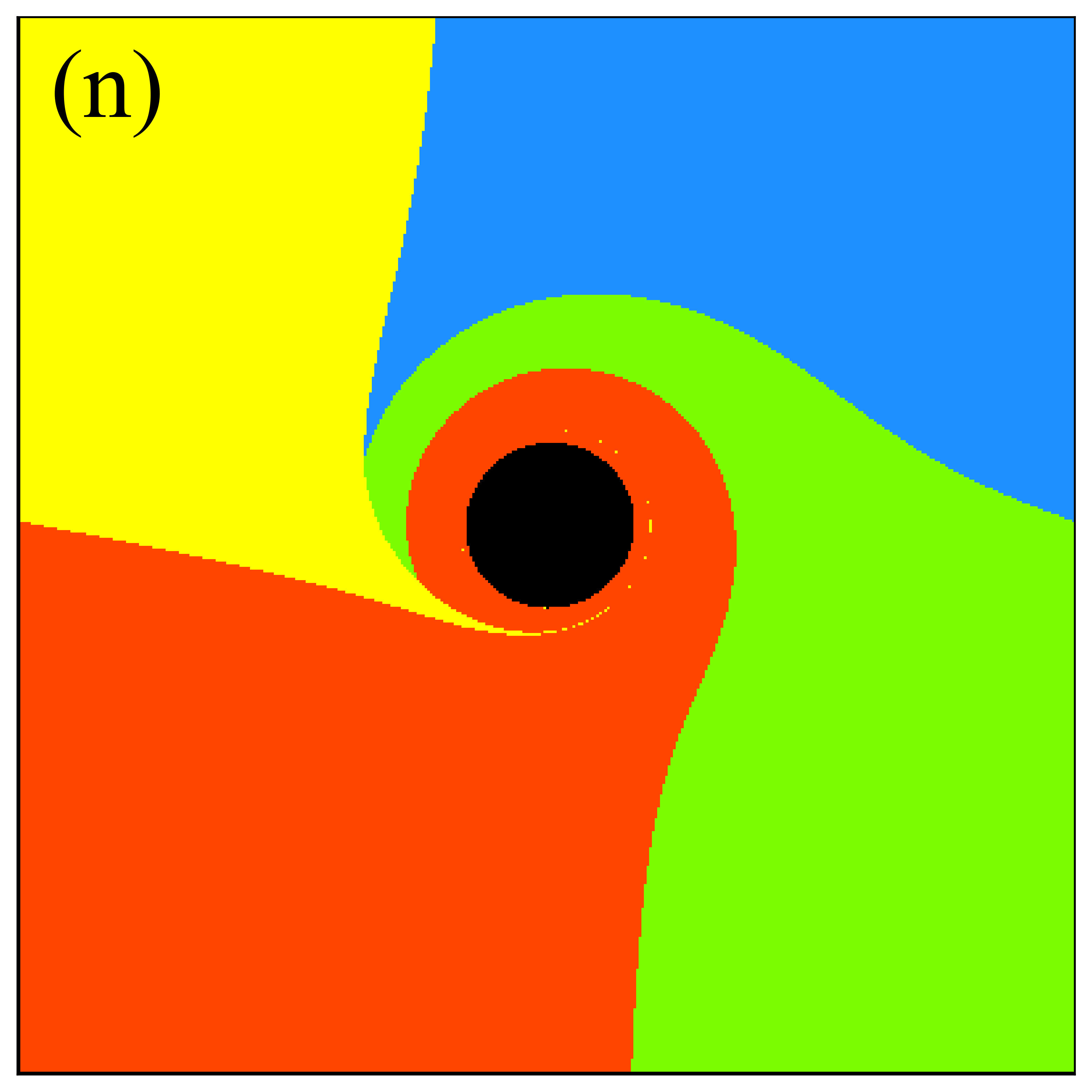}
\includegraphics[width=3.7cm]{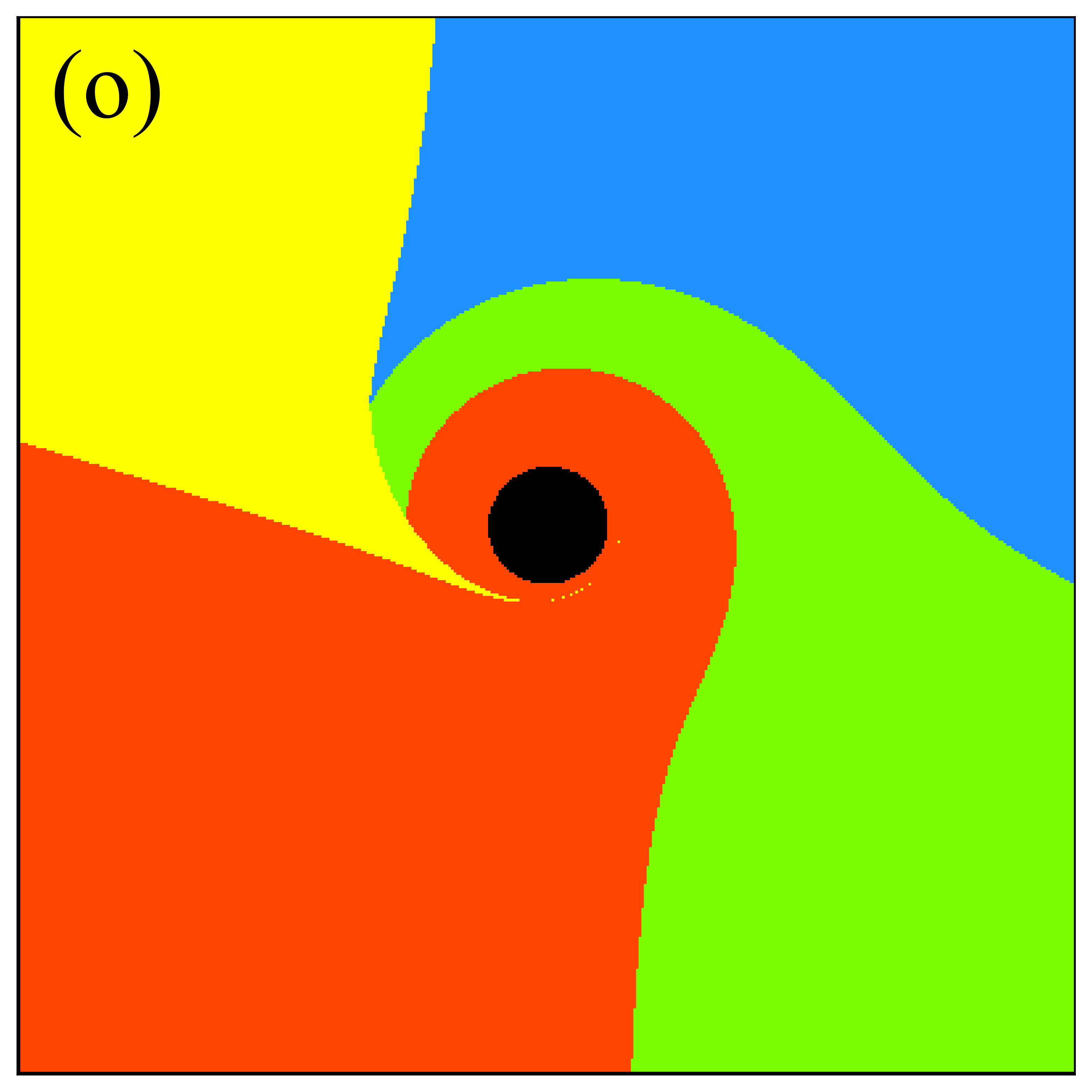}
\includegraphics[width=3.7cm]{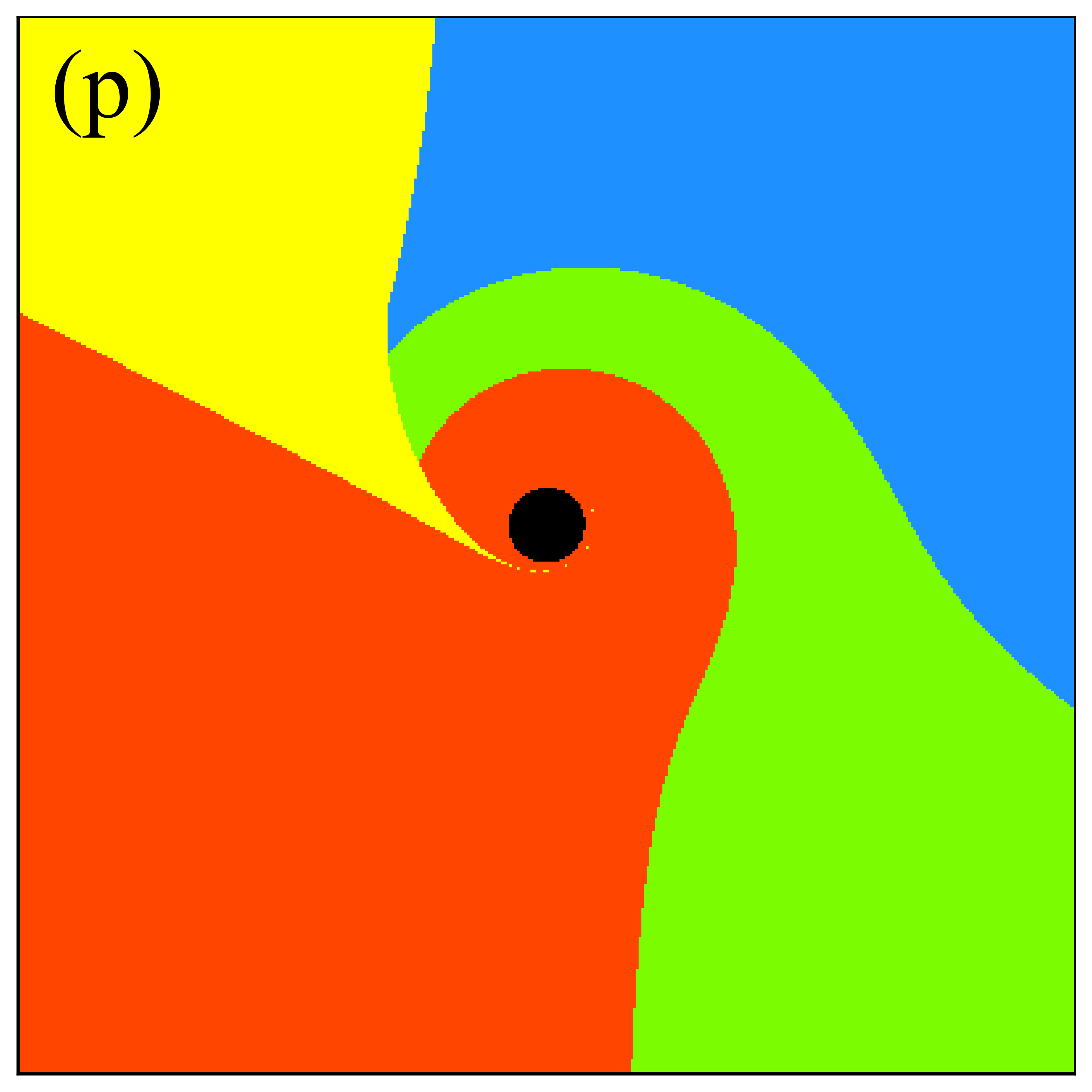}
\caption{Evolution of the inner shadow of a Kerr black hole with a tilted thin accretion disk with respect to disk inclination $\sigma$ and spin parameter $a$. From left to right, the values of disk inclination are $15^{\circ}$, $30^{\circ}$, $45^{\circ}$, and $60^{\circ}$; from top to bottom, the spin parameters are $0$, $0.54$, $0.94$, and $0.9985$. Here, the observation angle and azimuth are fixed at $17^{\circ}$ and $90^{\circ}$, respectively. The field of view and resolution are consistent with the settings in figure 2.}}\label{fig9}
\end{figure*}

We numerically simulated the inner shadow of a Kerr black hole with a tilted accretion disk under viewing angles of $17^{\circ}$, $50^{\circ}$, and $85^{\circ}$, with the results presented in figures 9-11, respectively. Similar to the case of $\Phi=0$, we observe that the tilted accretion disk can obscure the inner shadow, leading to a size significantly smaller than $S^{\textrm{min}}_{\Theta=0}$ in certain parameter spaces. Moreover, the introduction of the disk tilt $\sigma$ results in more complex inner shadow structures. Notably, we identify the appearance of dual shadows, as illustrated in figures 10(h), 11(d), and the second row in figure 11. This structure consists of a `primary shadow' located closer to the image center and a `secondary shadow' that is relatively distant but non-negligible. The shape and size of the secondary shadow, often eyebrow-shaped, are influenced by both the spin parameter and the inclination of the accretion disk, with the secondary shadow being more likely to appear for moderate spin values. It is worth mentioning that such dual-shadow structures are unlikely to occur in the equatorial disk scenario. Additionally, while dual-shadow structures can also emerge in binary black hole systems and in the single black hole in vacuum \cite{Nitta et al. (2011),Nitta et al. (2012),Bohn et al. (2015),Abdolrahimi et al. (2015)}, the dual-shadow structure presented in this paper is distinct from those previously reported. Hence, this structure has the potential to serve as a probe for tilted accretion disks.
\begin{figure*}
\center{
\includegraphics[width=3.7cm]{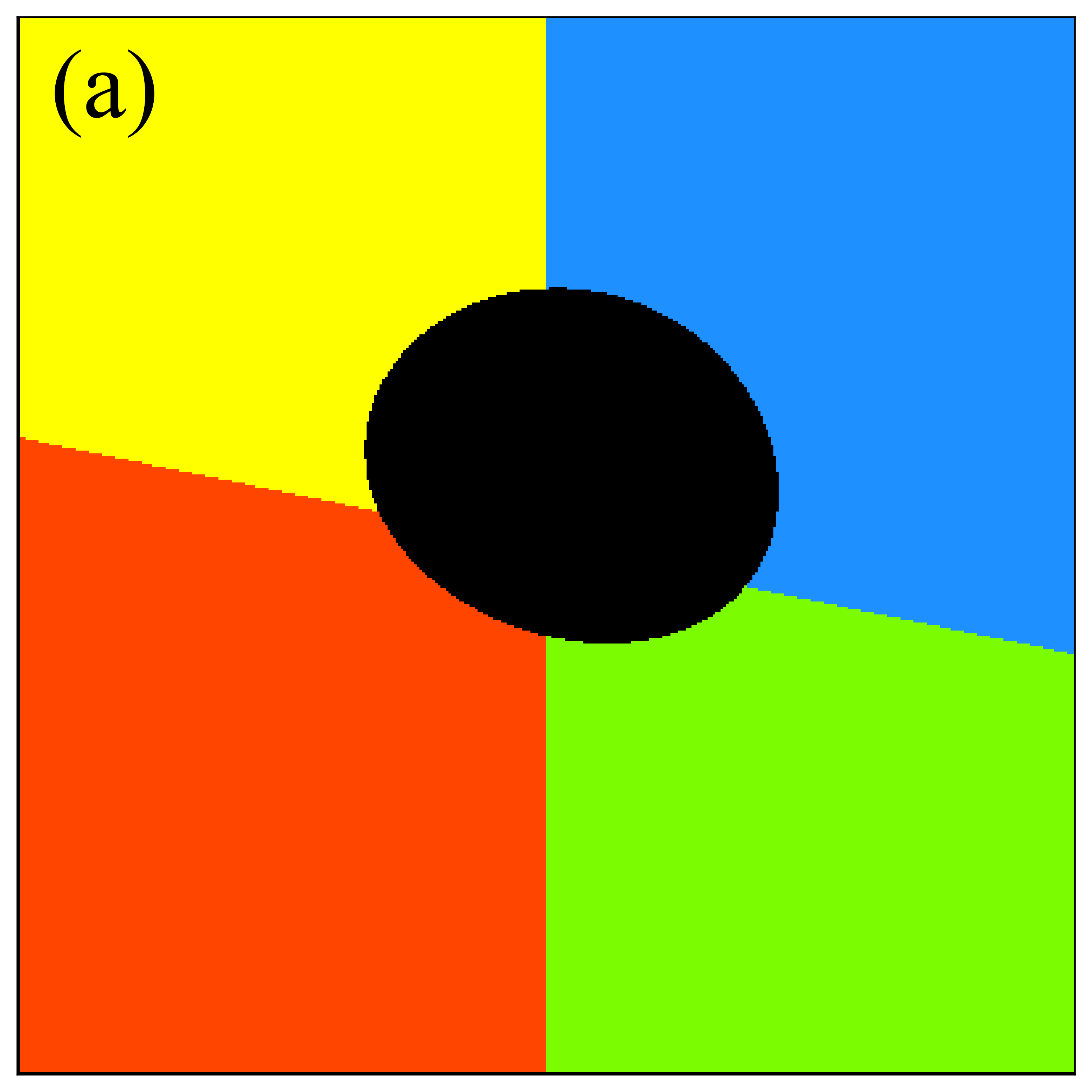}
\includegraphics[width=3.7cm]{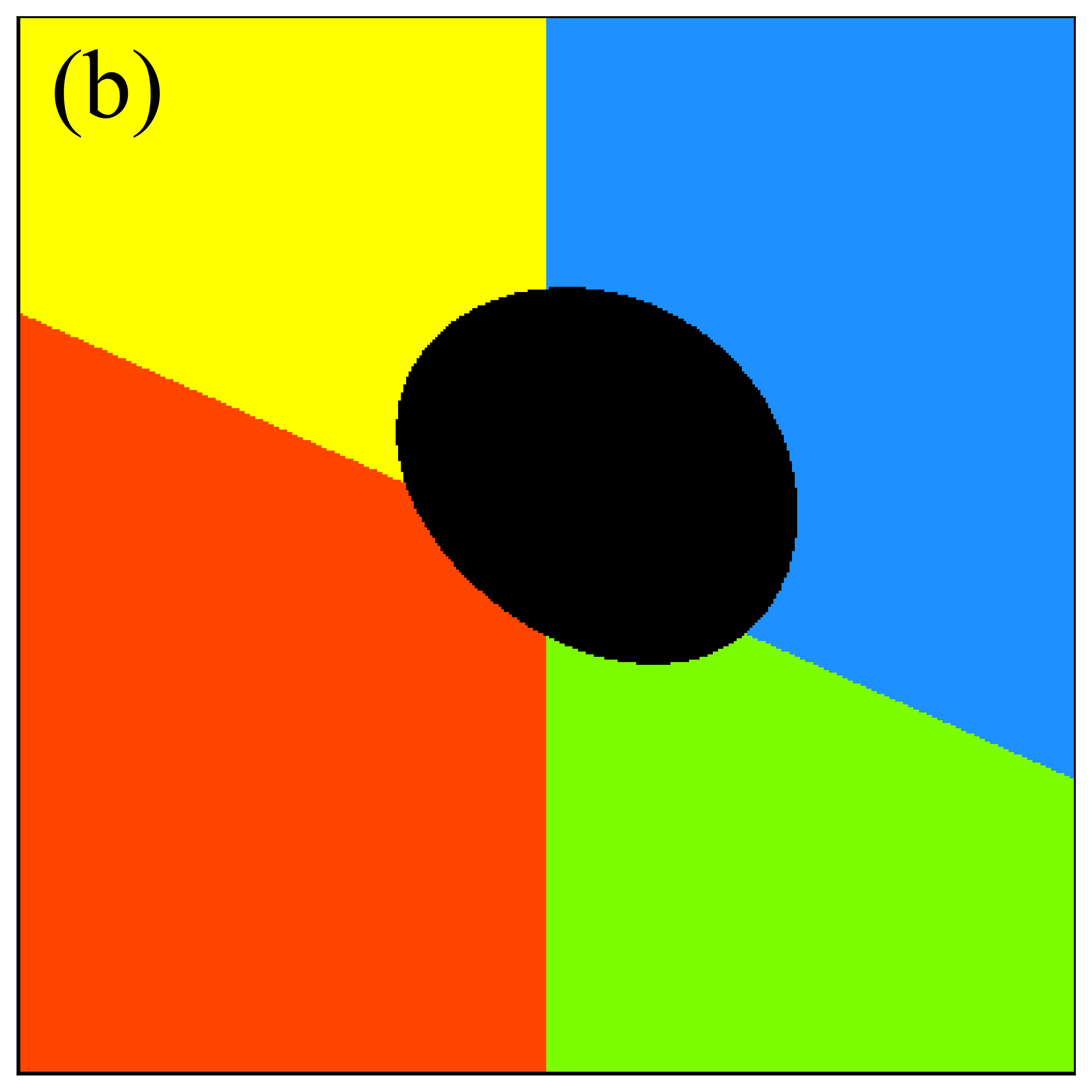}
\includegraphics[width=3.7cm]{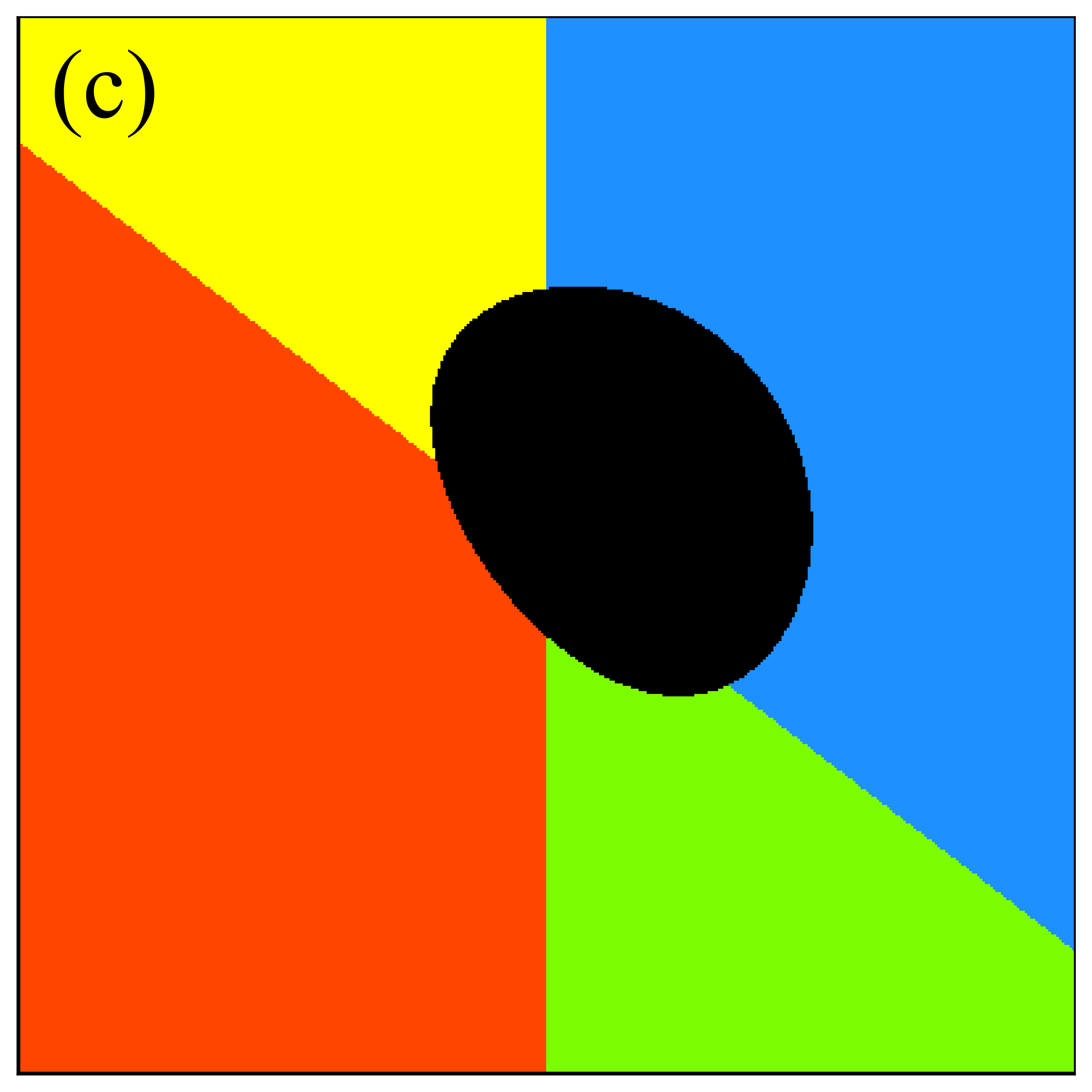}
\includegraphics[width=3.7cm]{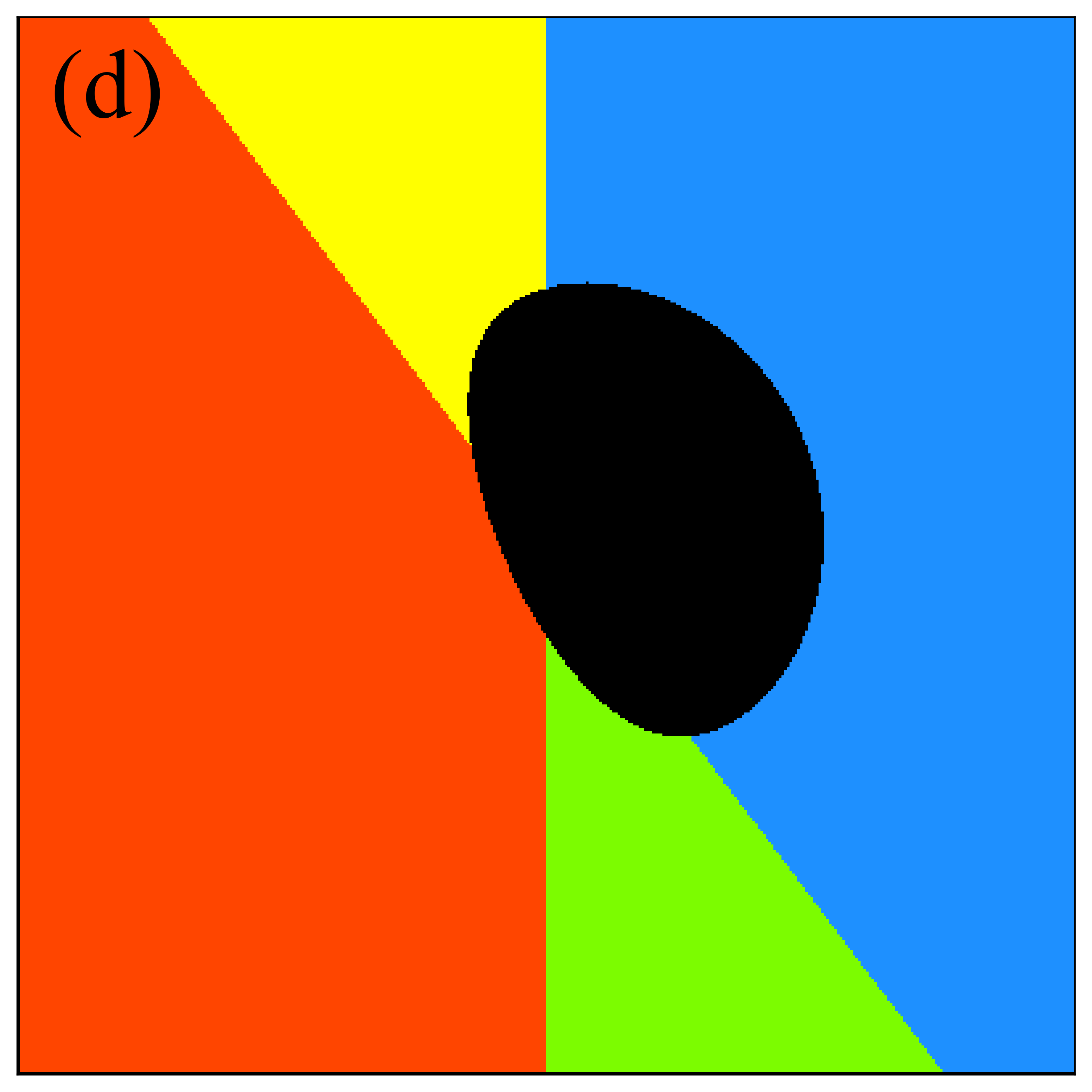}
\includegraphics[width=3.7cm]{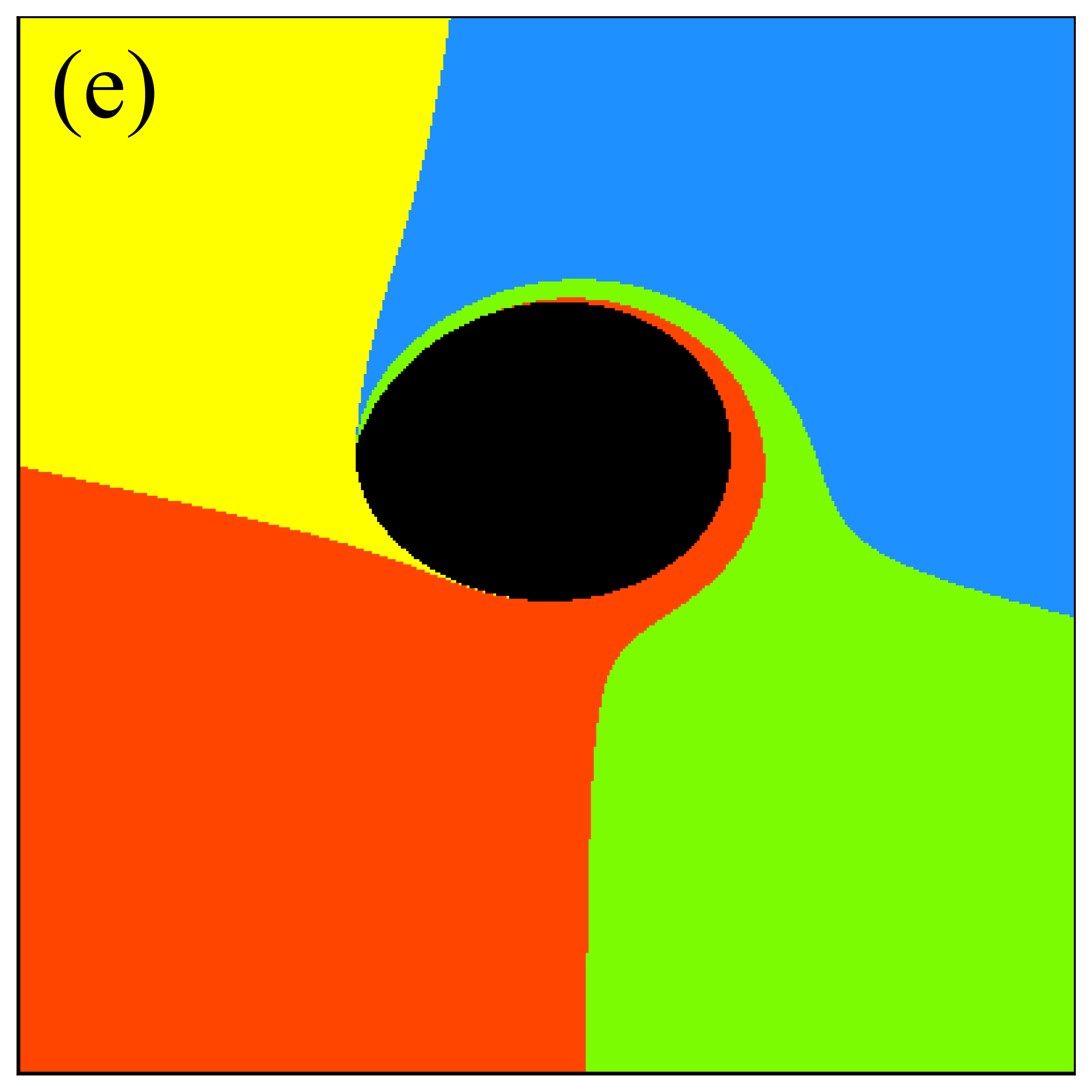}
\includegraphics[width=3.7cm]{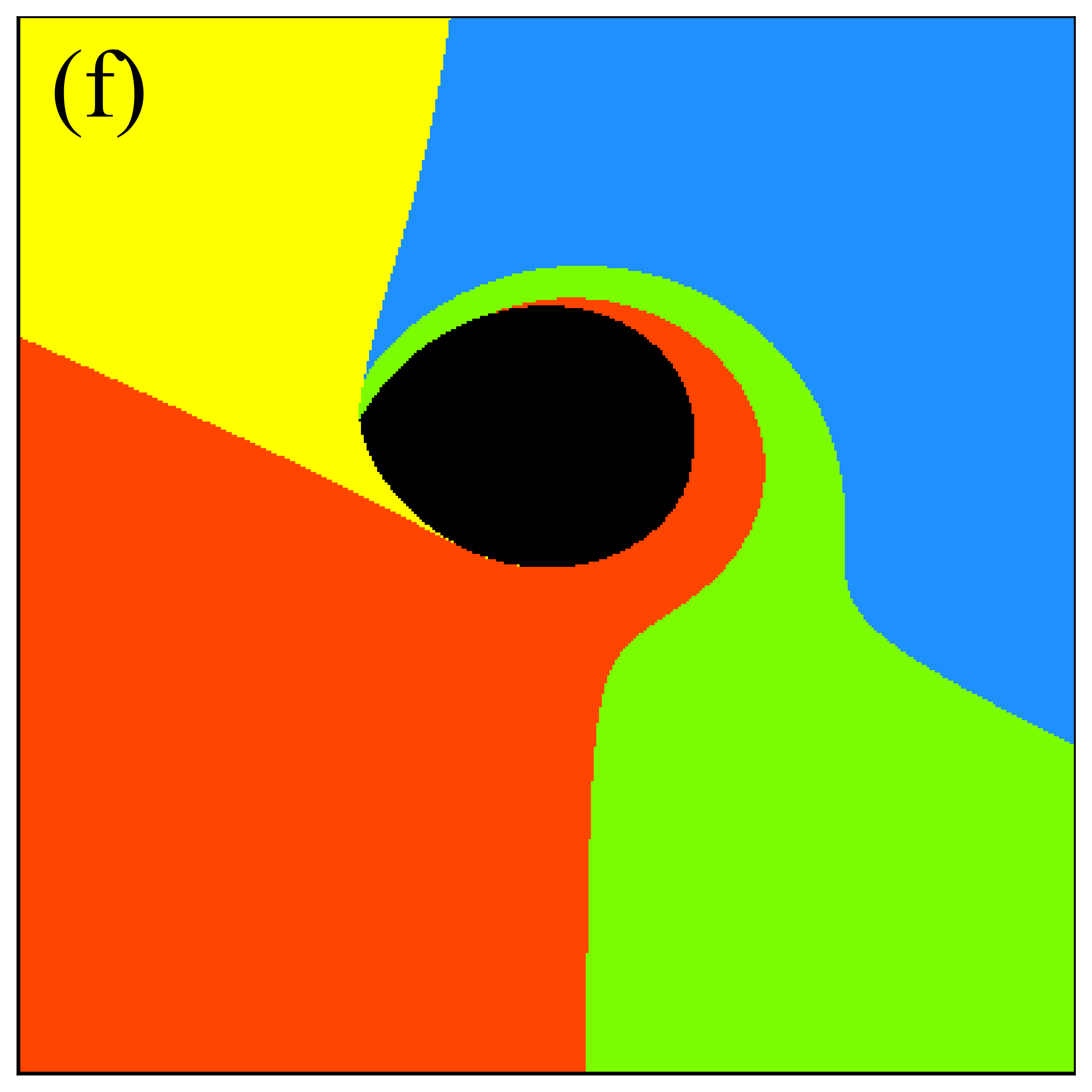}
\includegraphics[width=3.7cm]{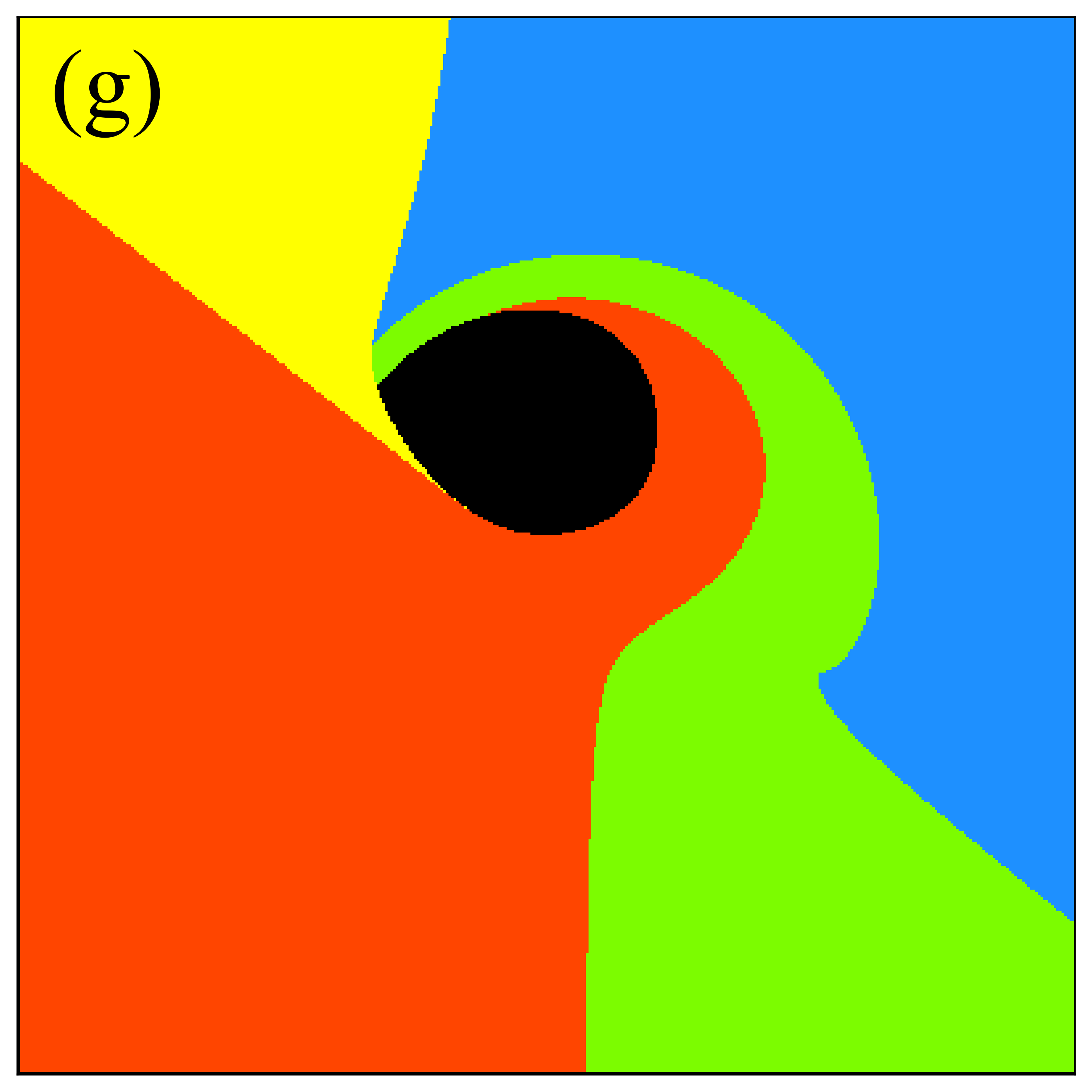}
\includegraphics[width=3.7cm]{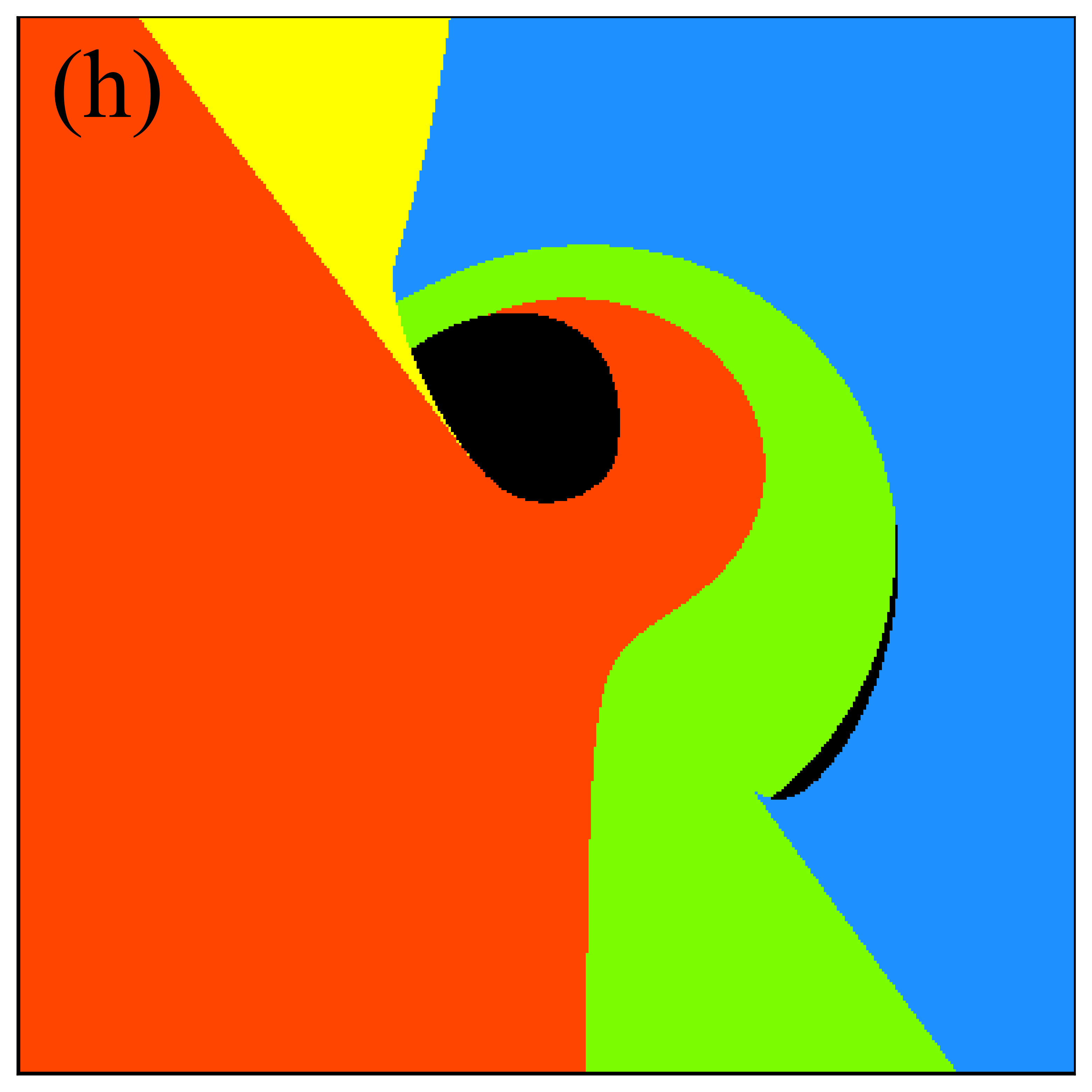}
\includegraphics[width=3.7cm]{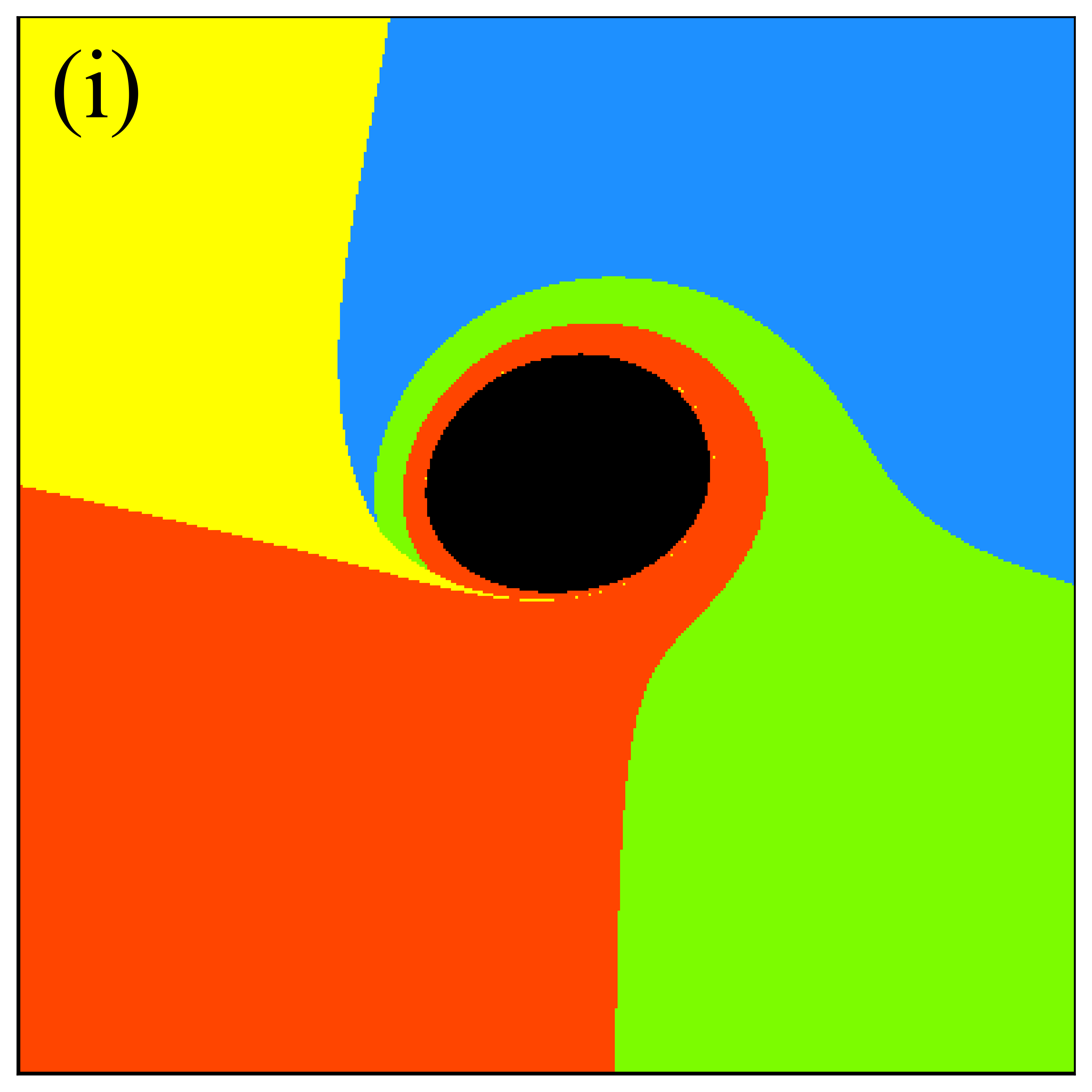}
\includegraphics[width=3.7cm]{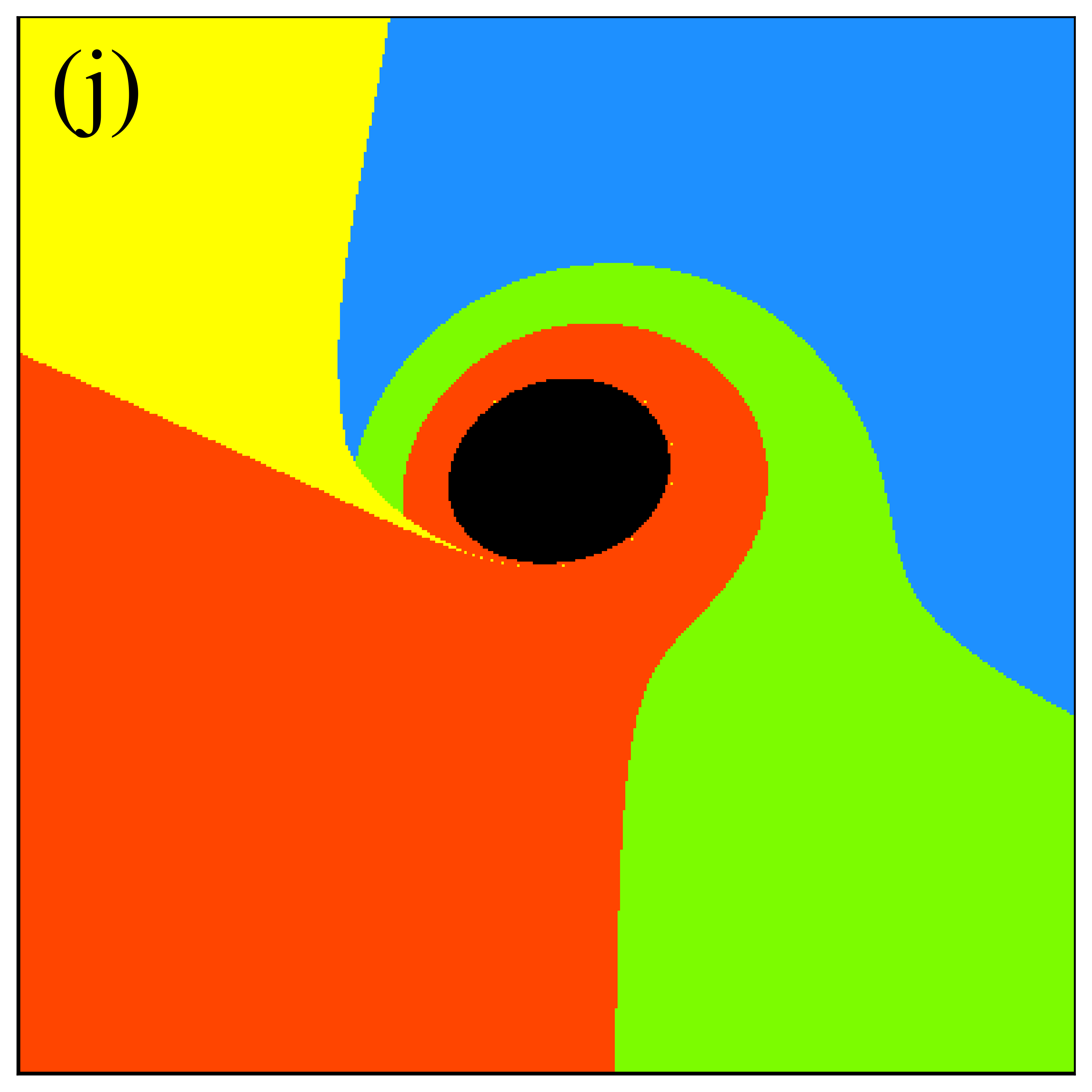}
\includegraphics[width=3.7cm]{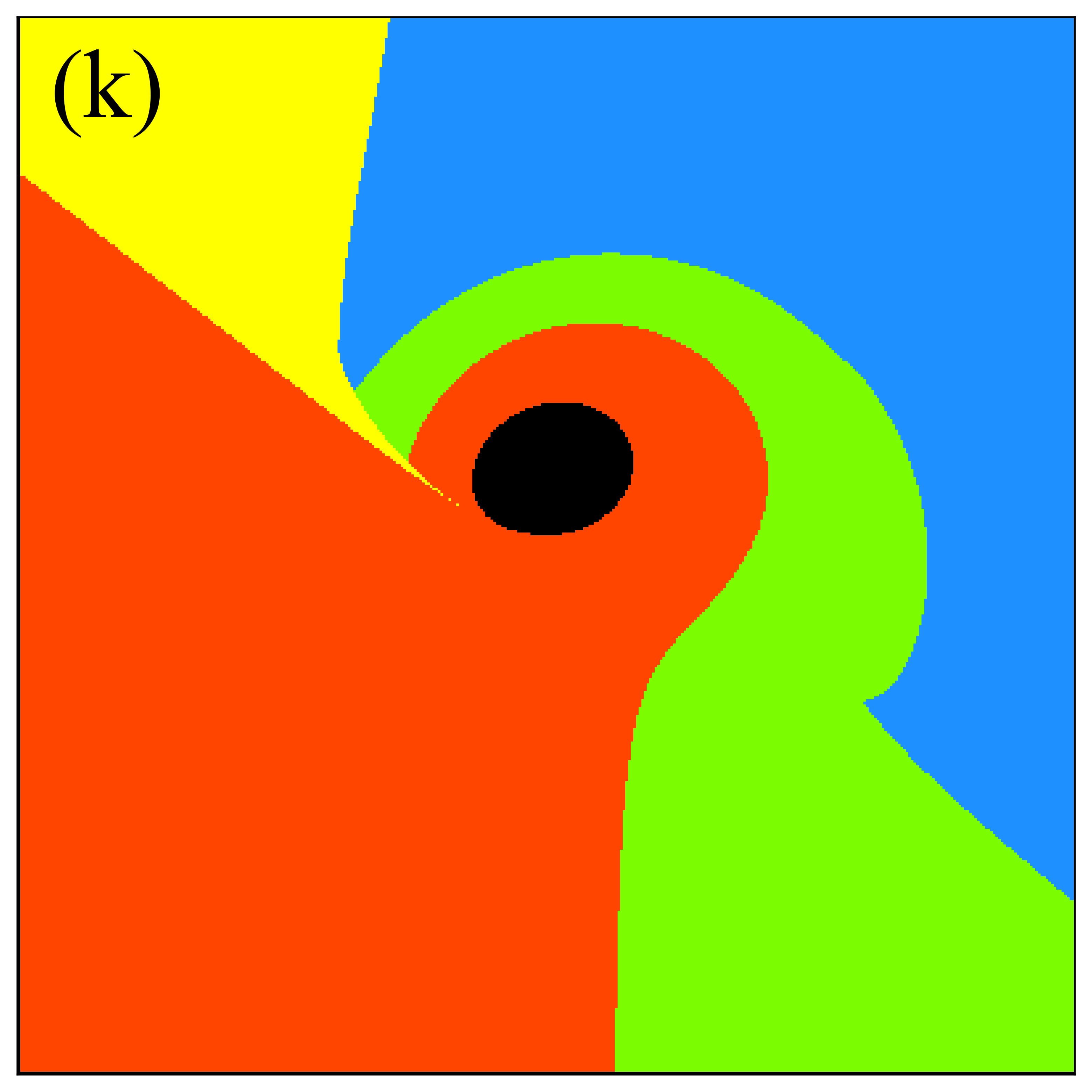}
\includegraphics[width=3.7cm]{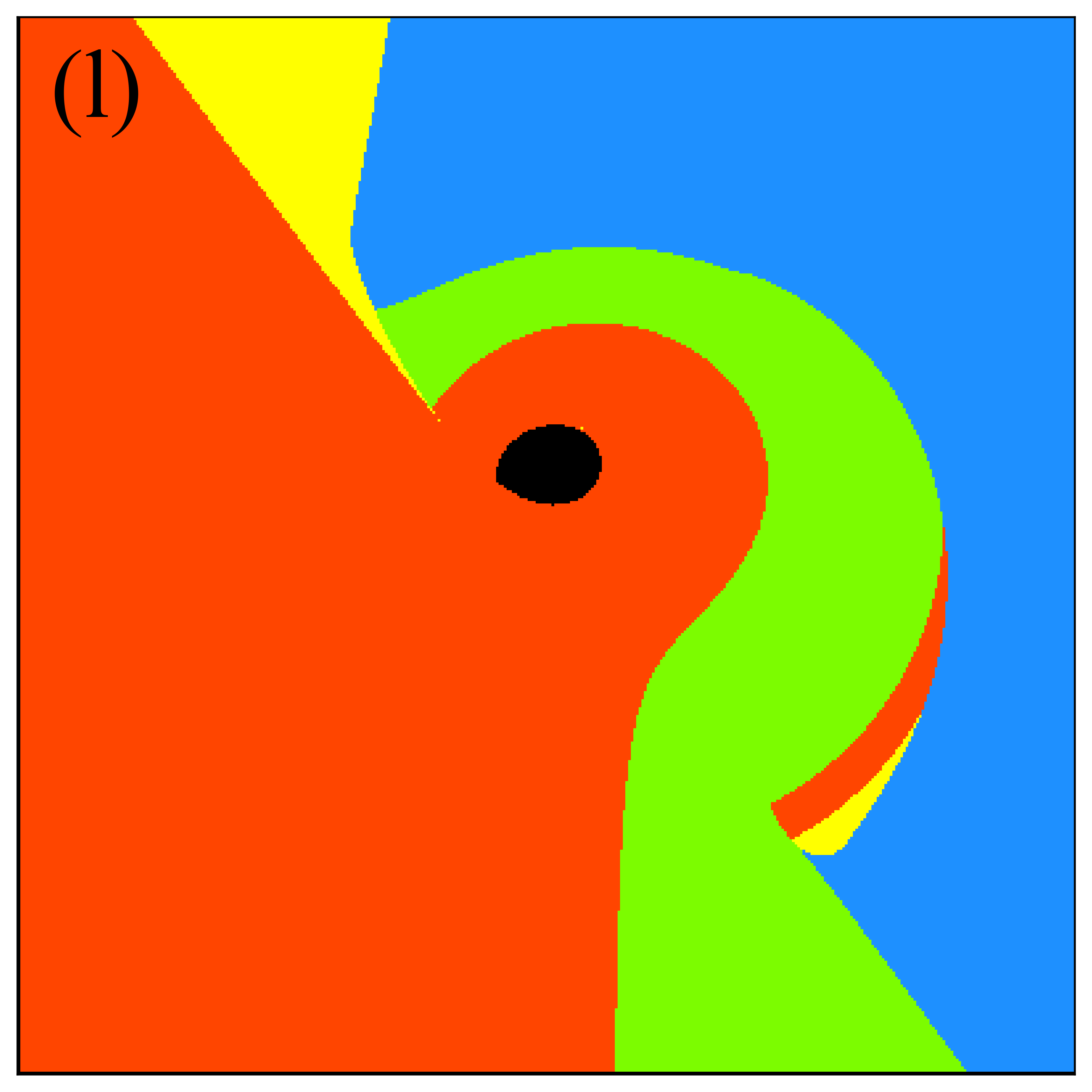}
\includegraphics[width=3.7cm]{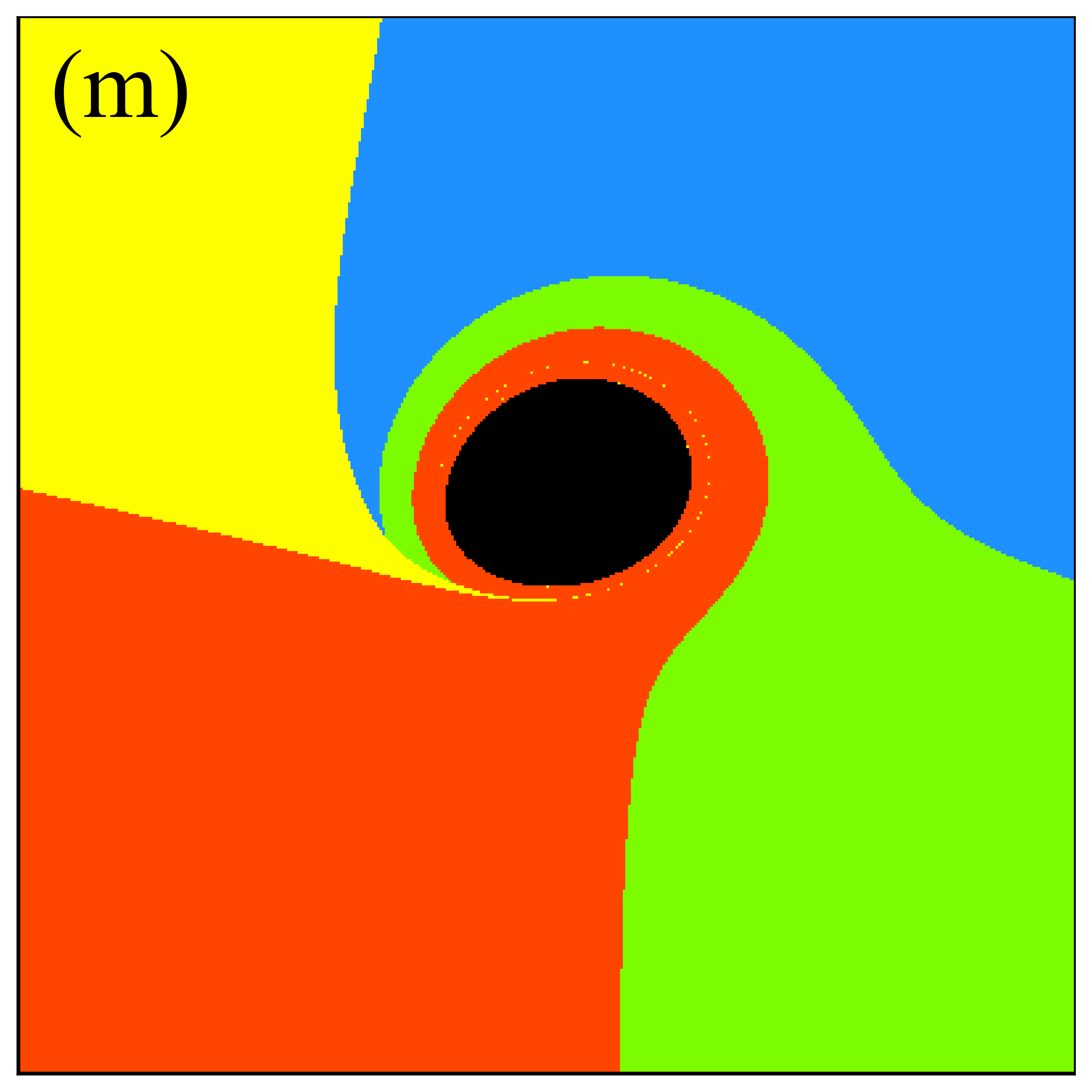}
\includegraphics[width=3.7cm]{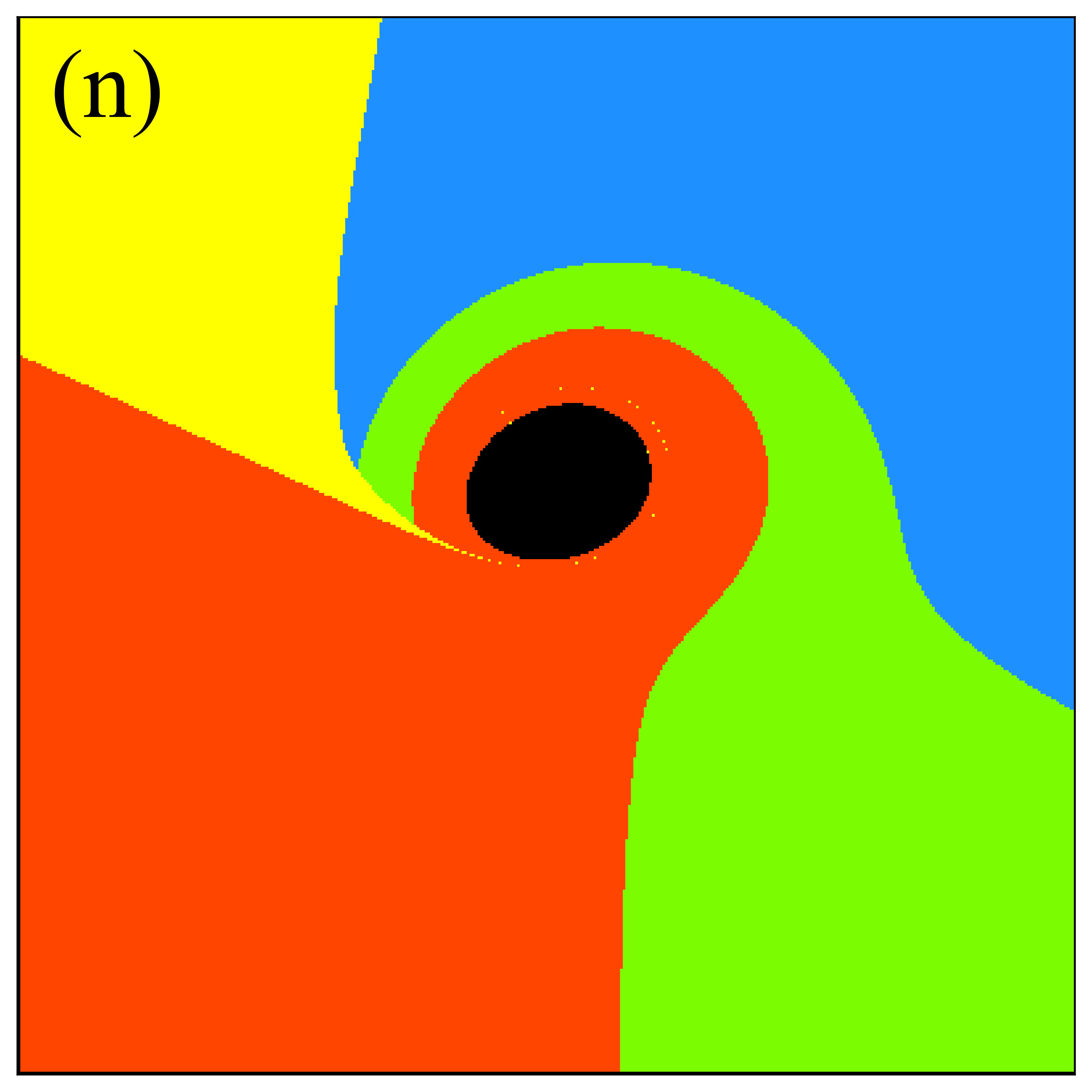}
\includegraphics[width=3.7cm]{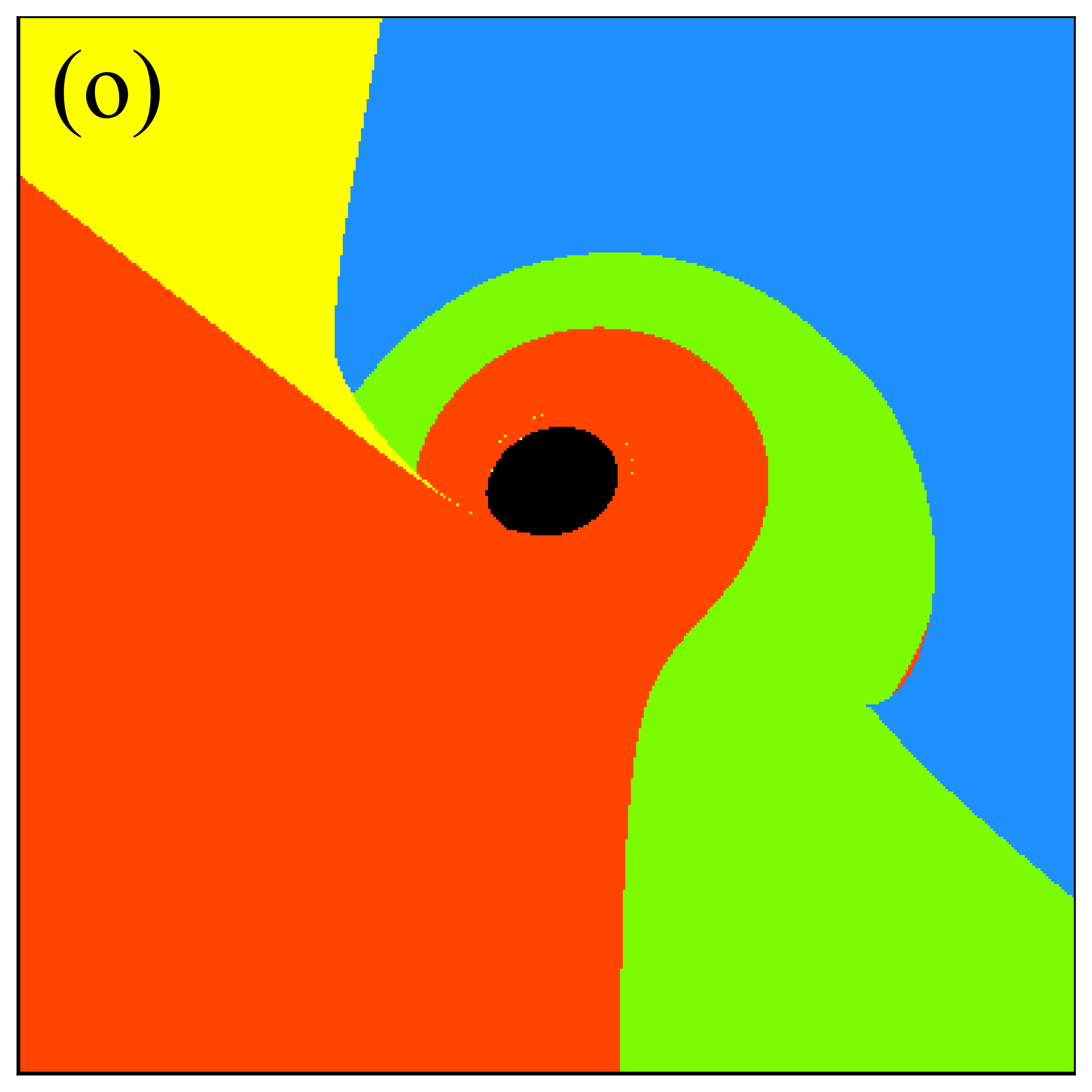}
\includegraphics[width=3.7cm]{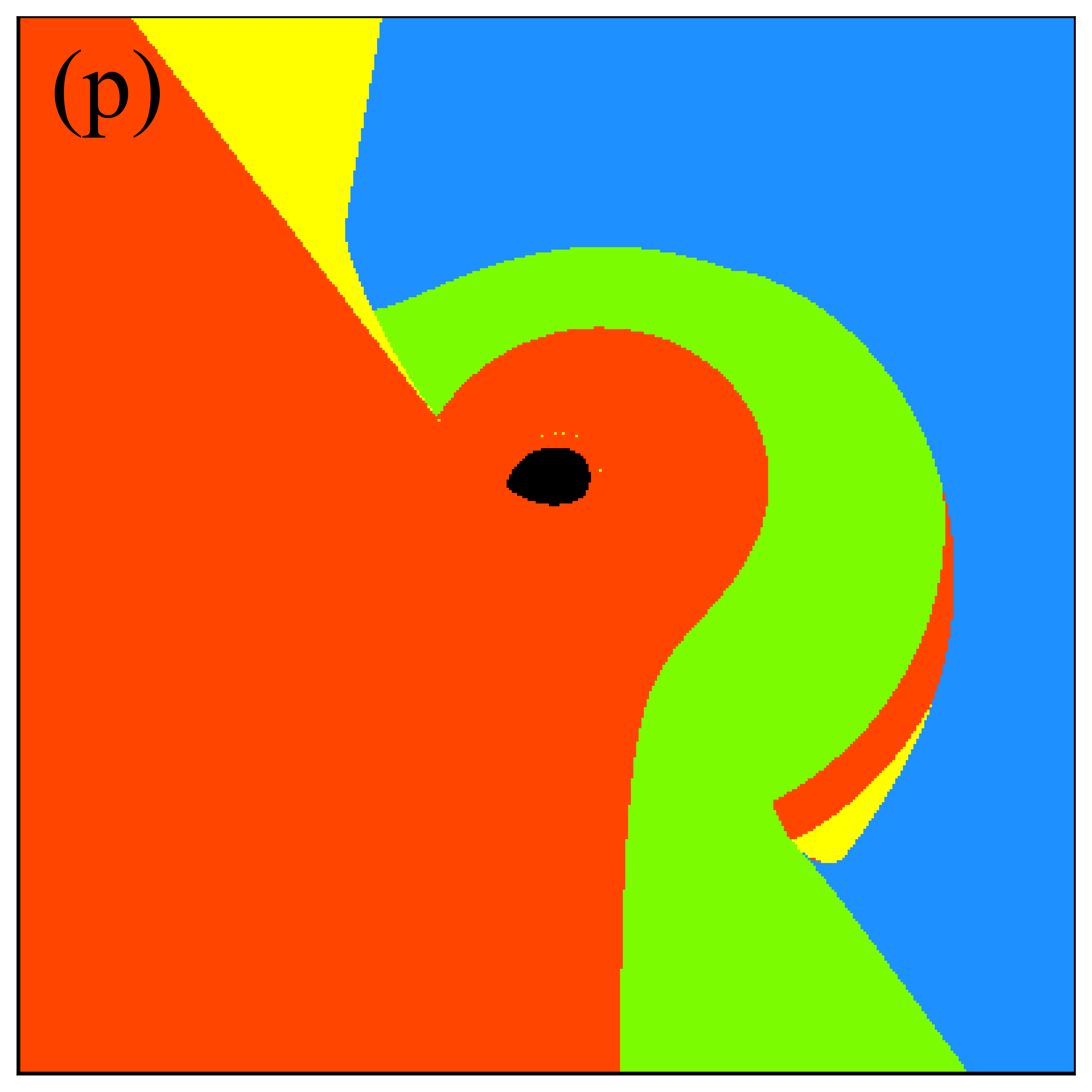}
\caption{Similar to figure 9, but for the observation angle of $50^{\circ}$.}}\label{fig10}
\end{figure*}
\begin{figure*}
\center{
\includegraphics[width=3.7cm]{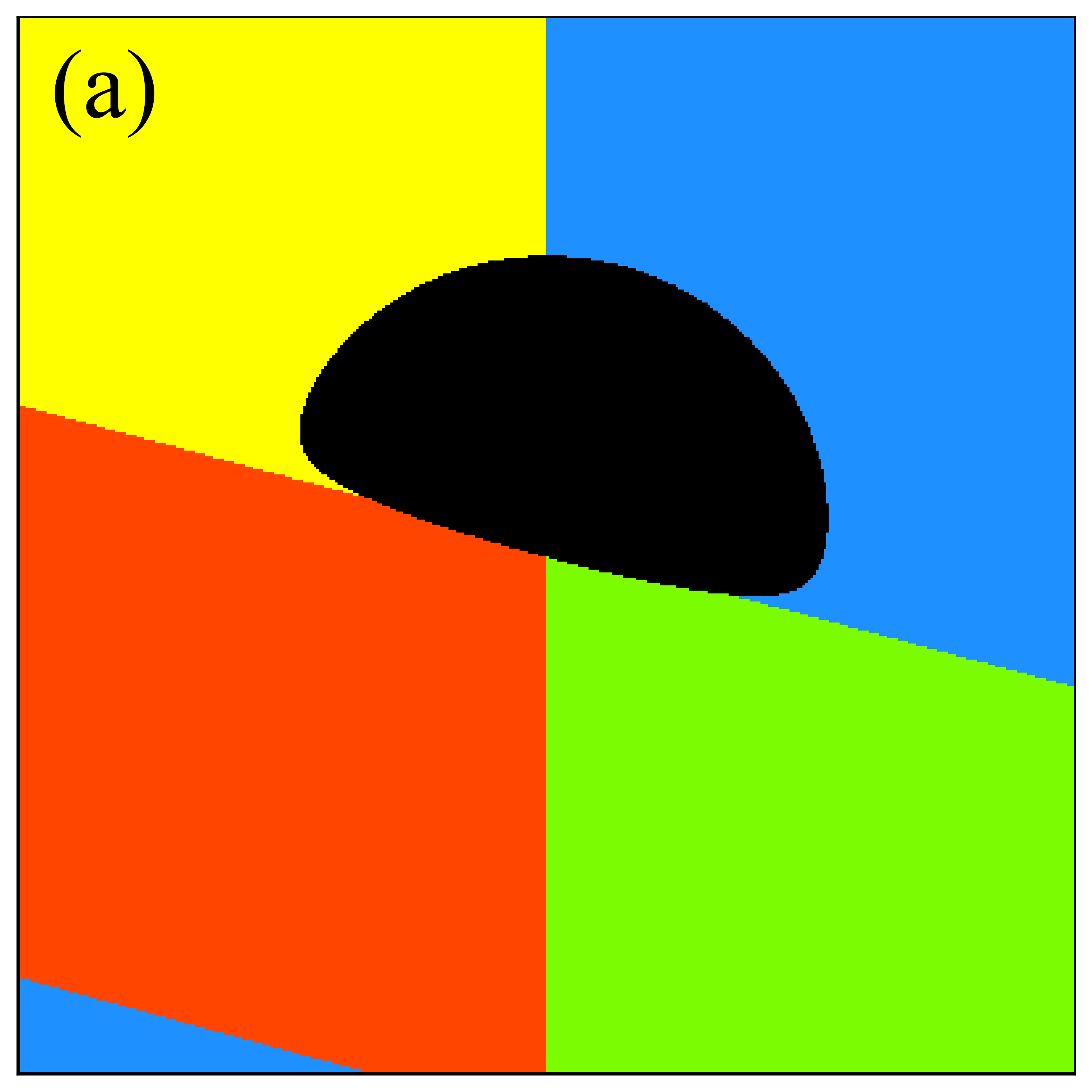}
\includegraphics[width=3.7cm]{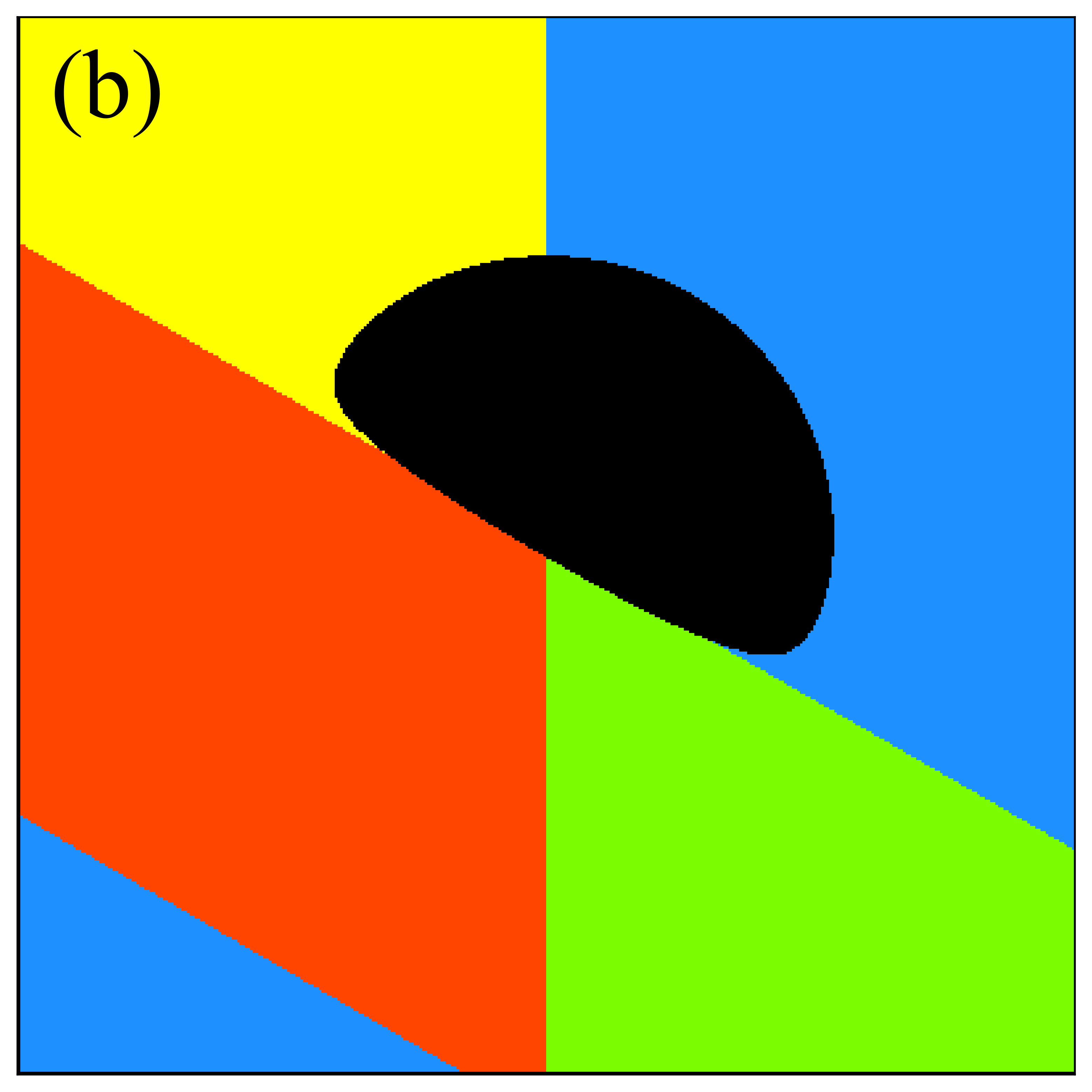}
\includegraphics[width=3.7cm]{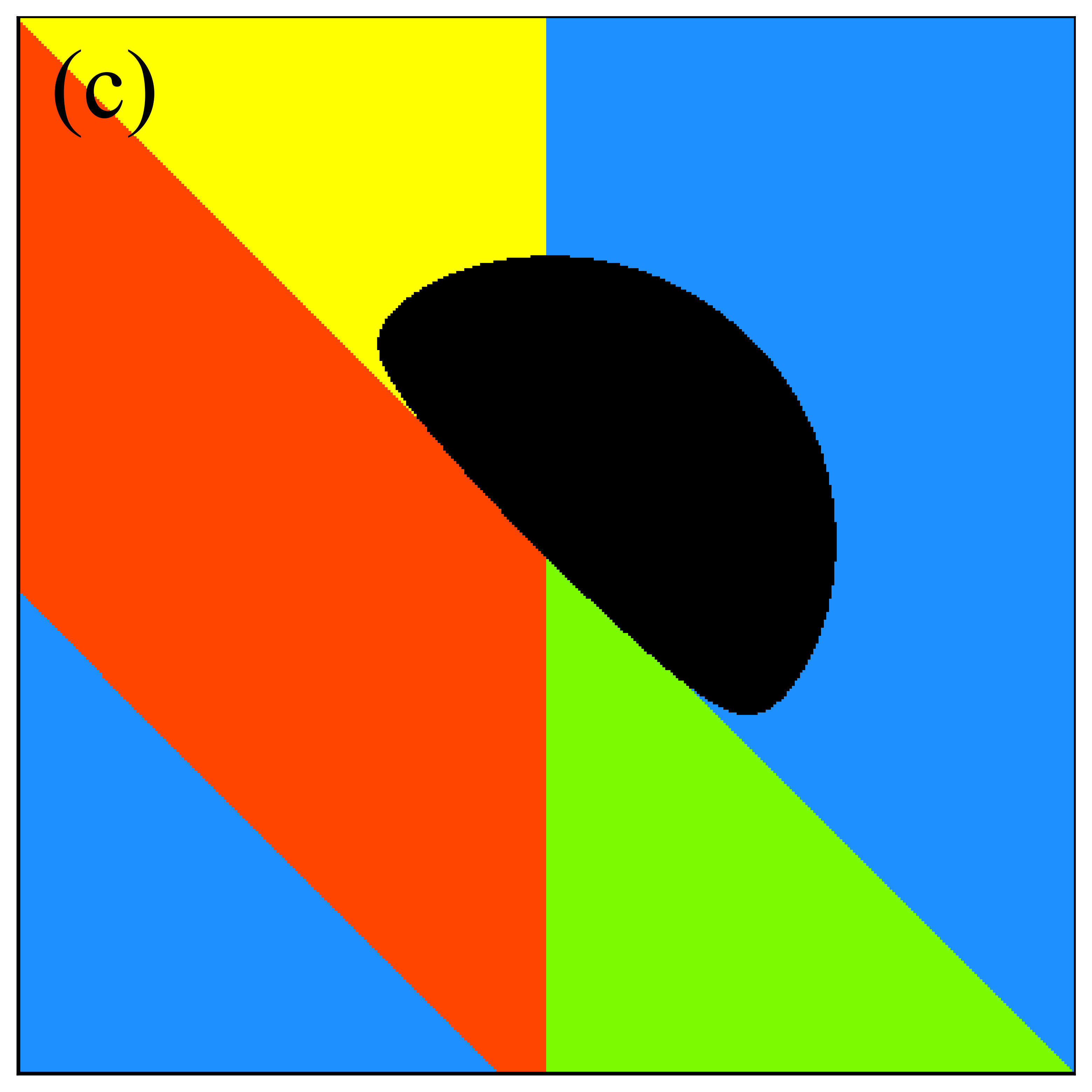}
\includegraphics[width=3.7cm]{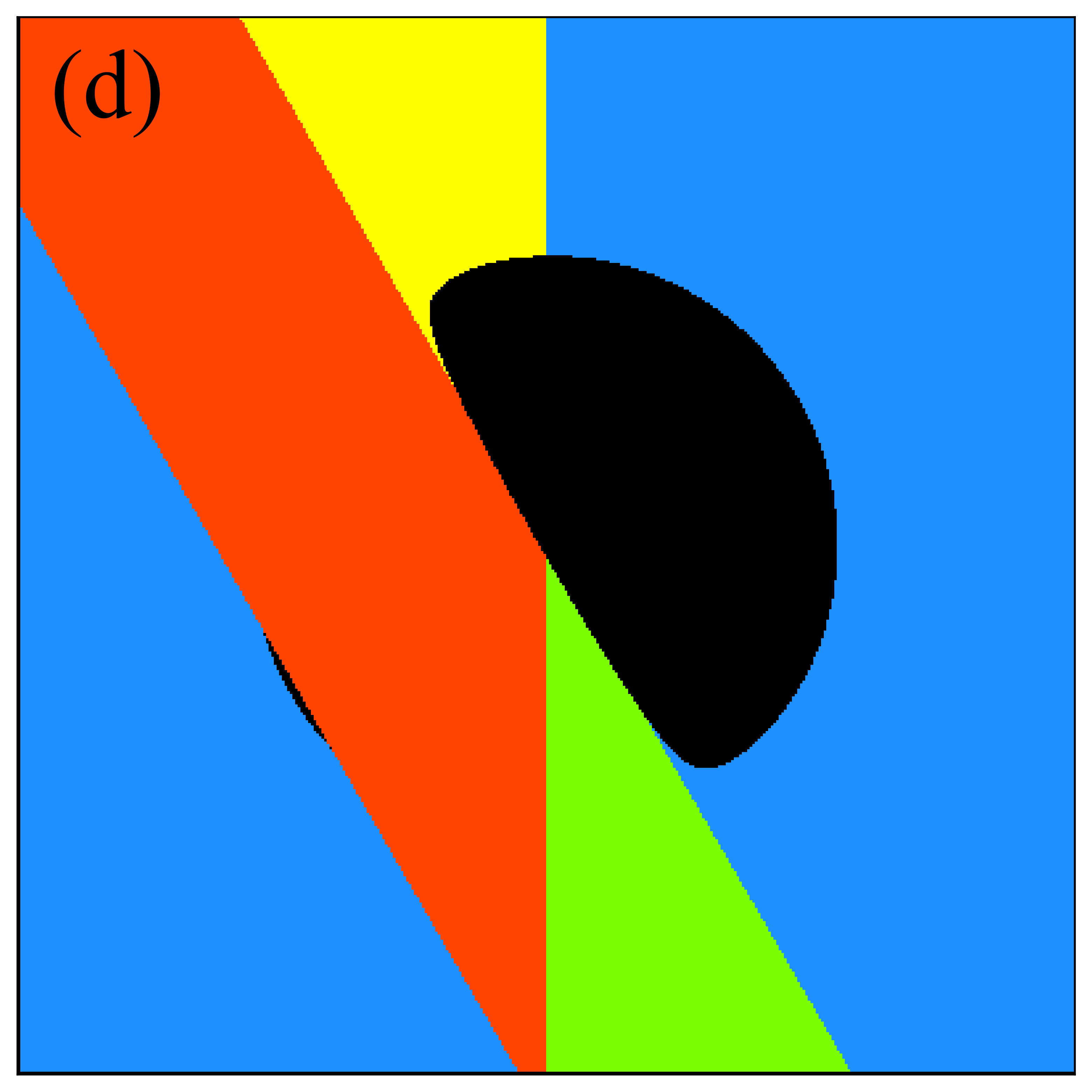}
\includegraphics[width=3.7cm]{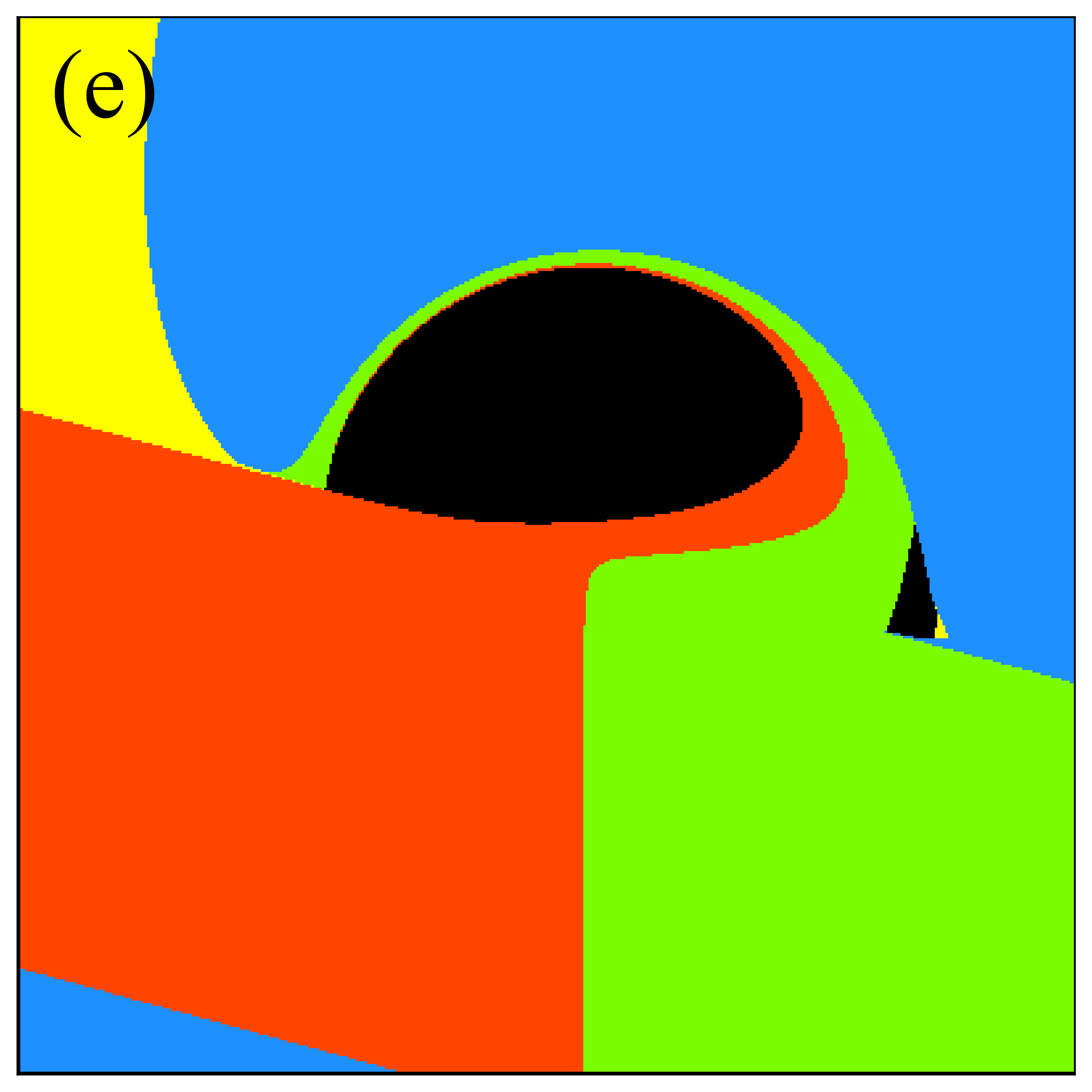}
\includegraphics[width=3.7cm]{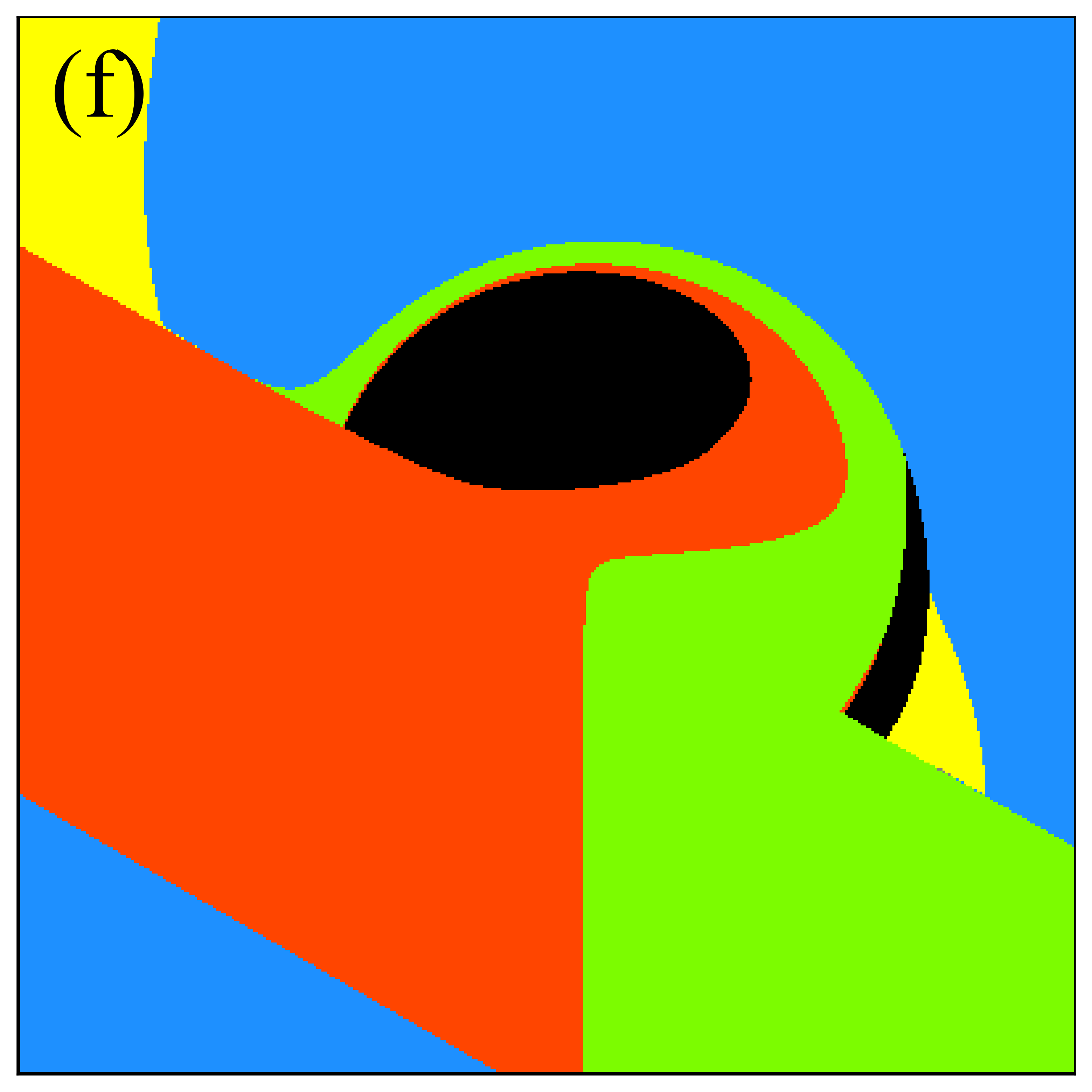}
\includegraphics[width=3.7cm]{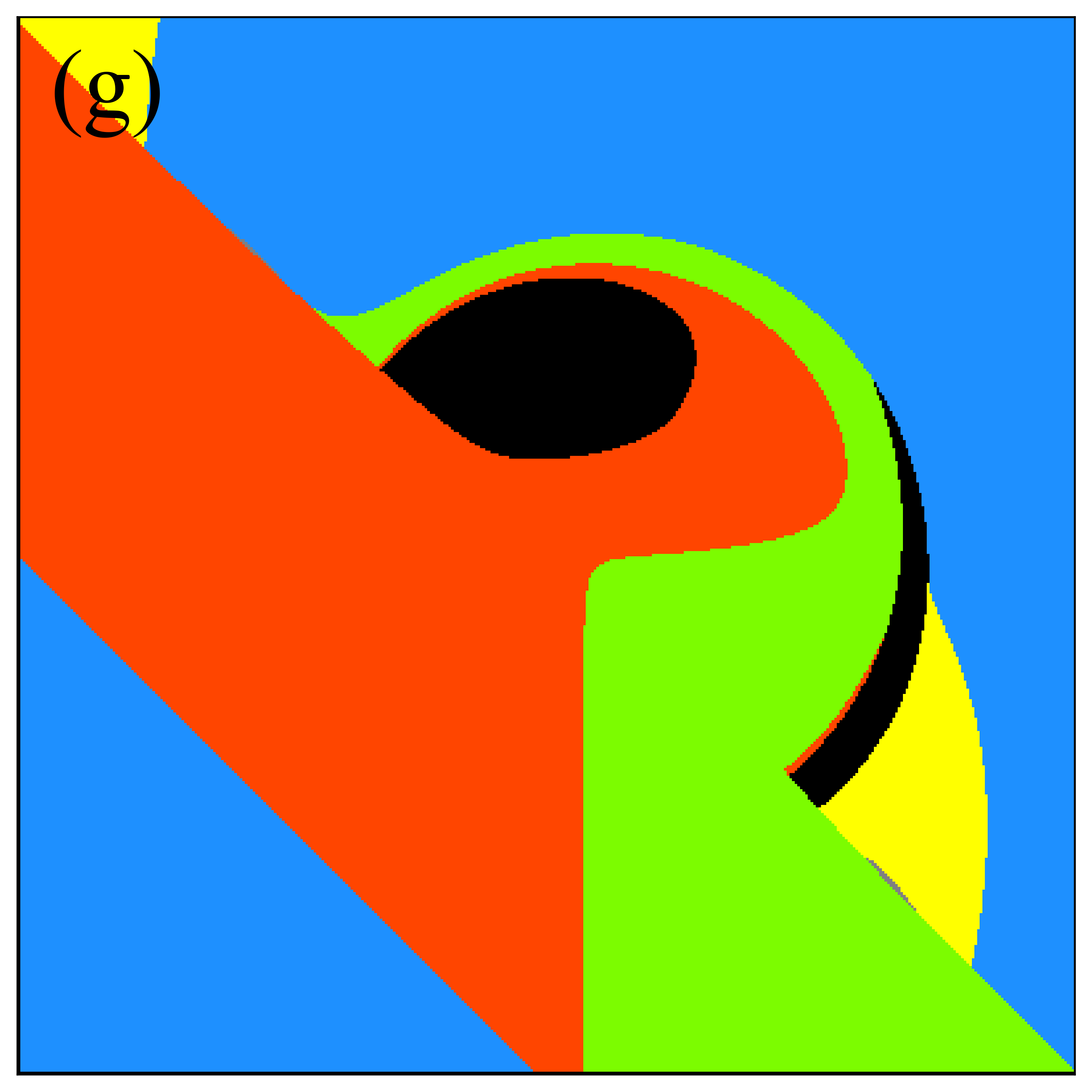}
\includegraphics[width=3.7cm]{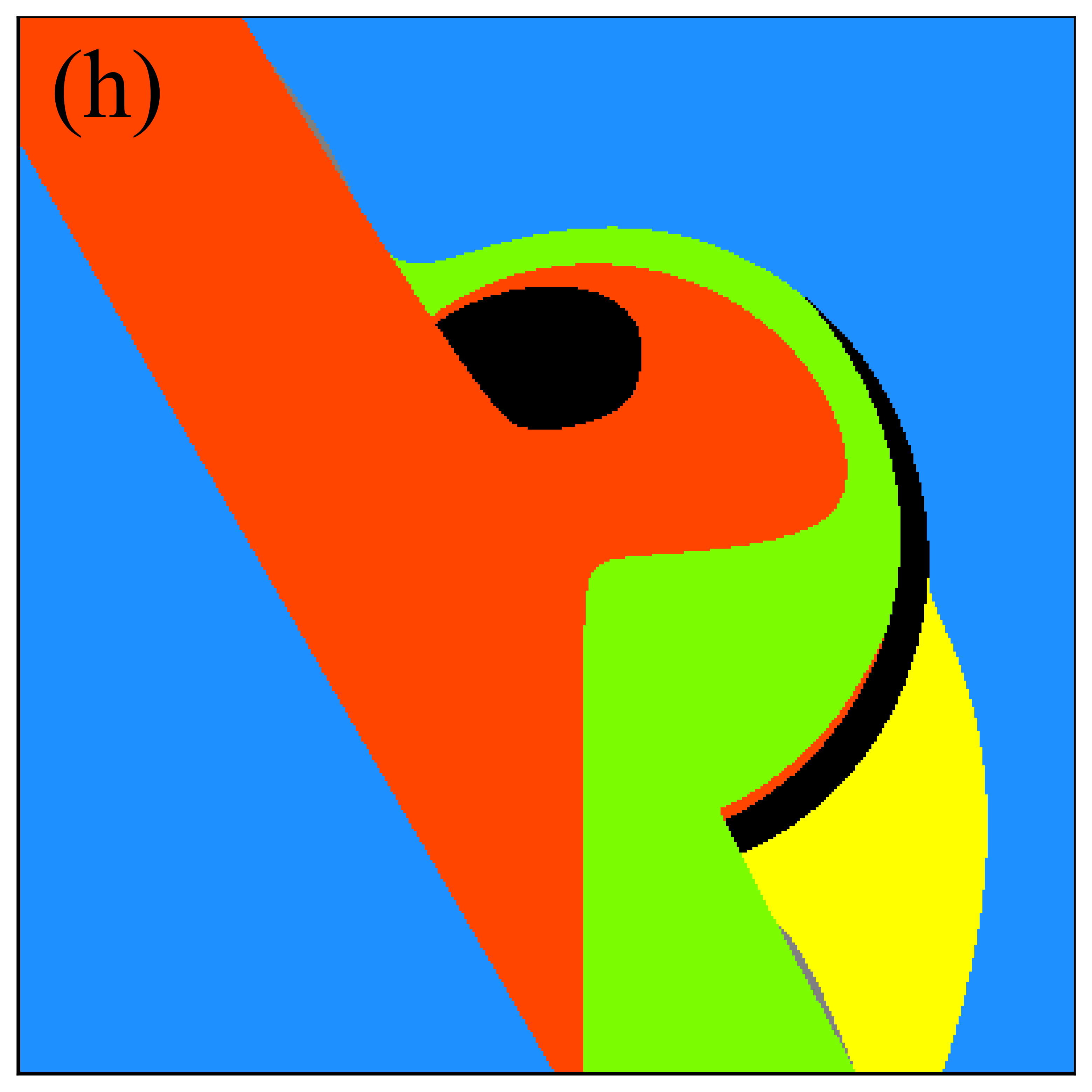}
\includegraphics[width=3.7cm]{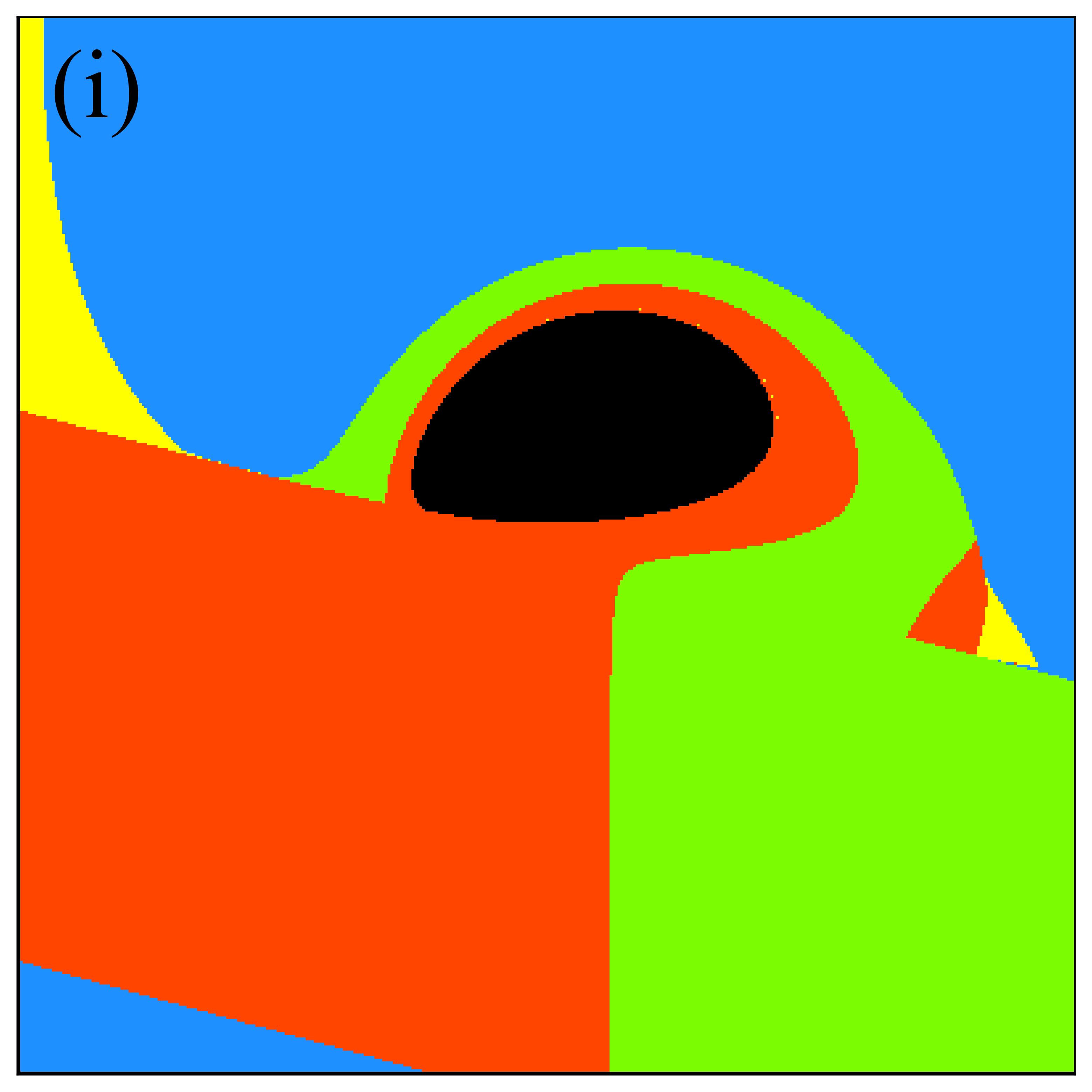}
\includegraphics[width=3.7cm]{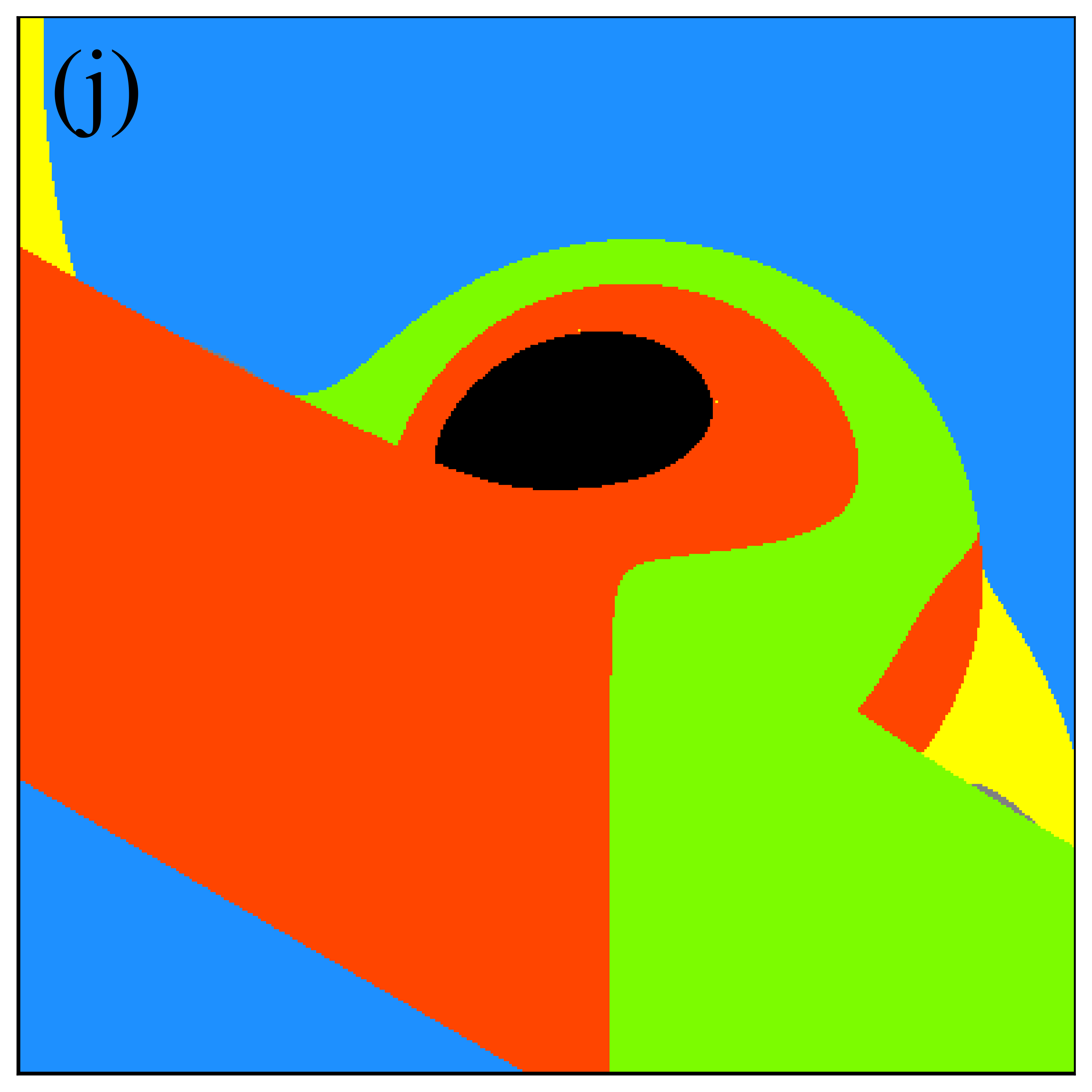}
\includegraphics[width=3.7cm]{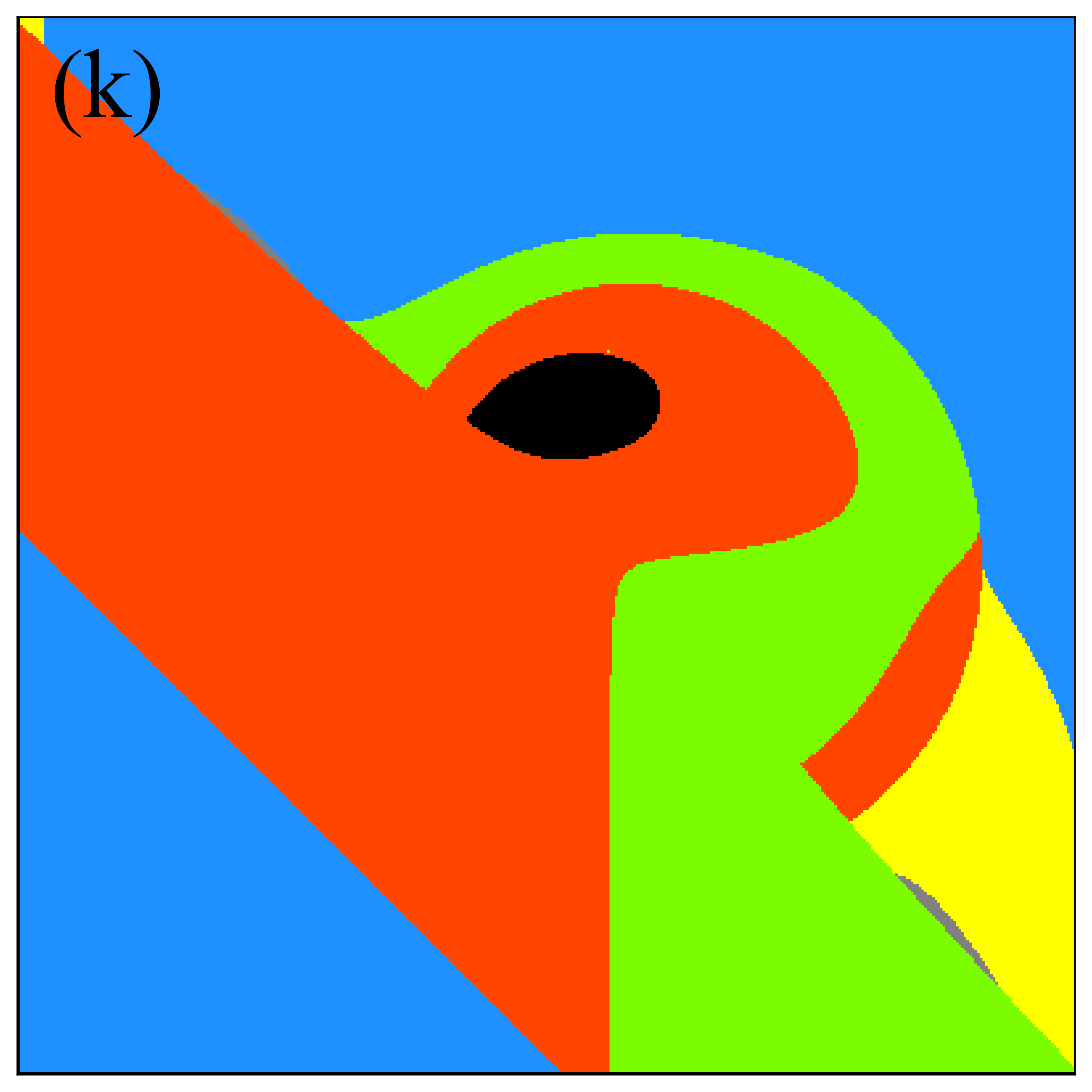}
\includegraphics[width=3.7cm]{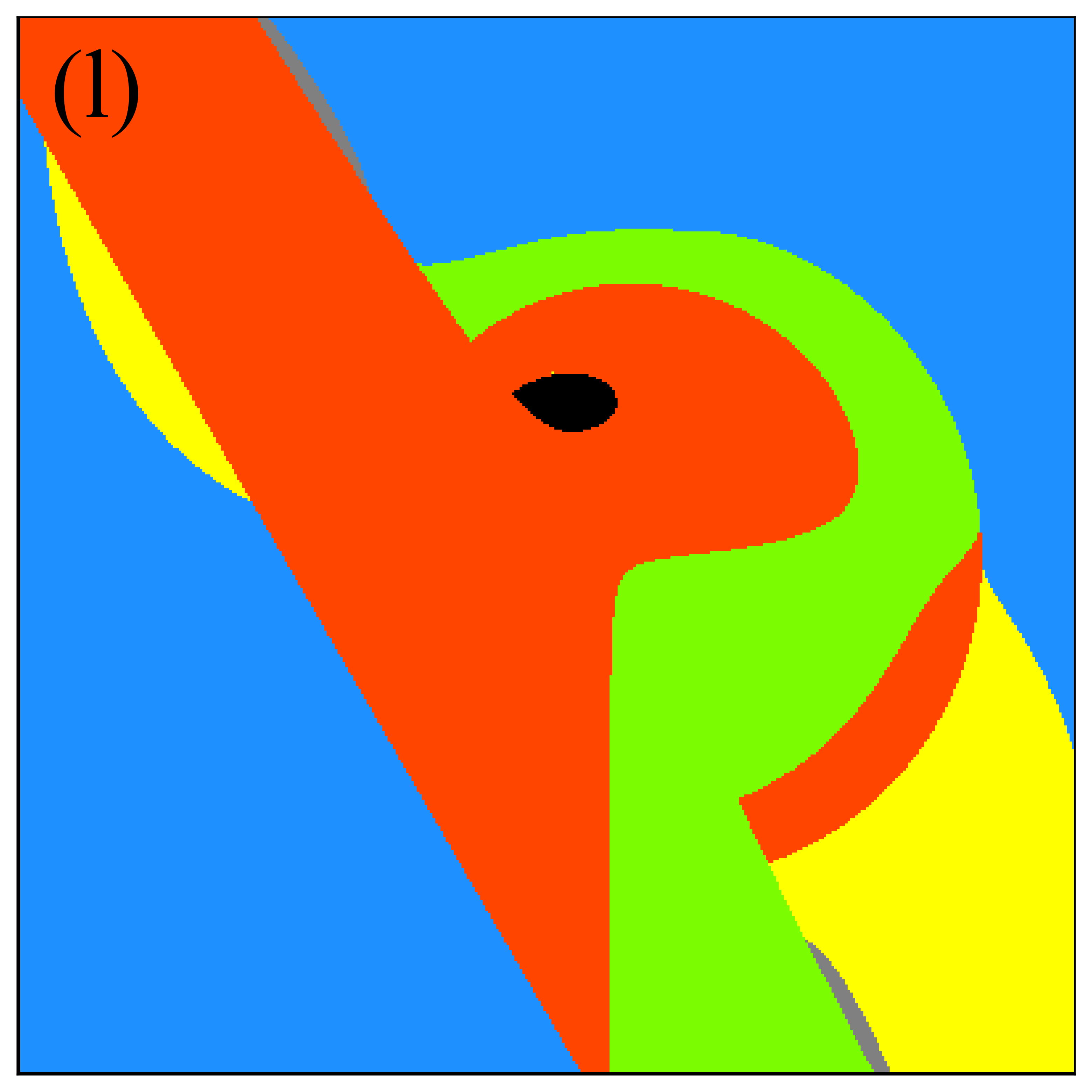}
\includegraphics[width=3.7cm]{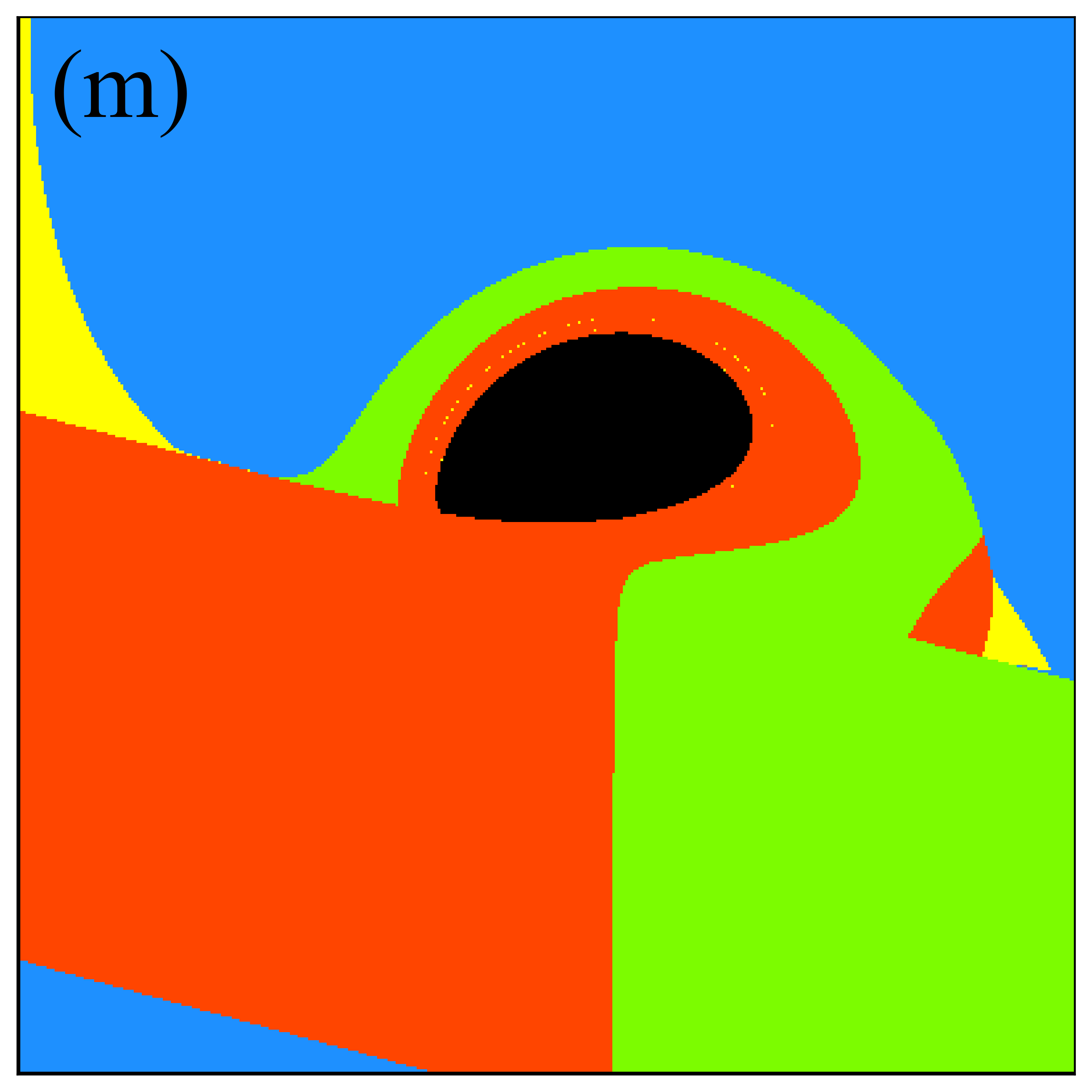}
\includegraphics[width=3.7cm]{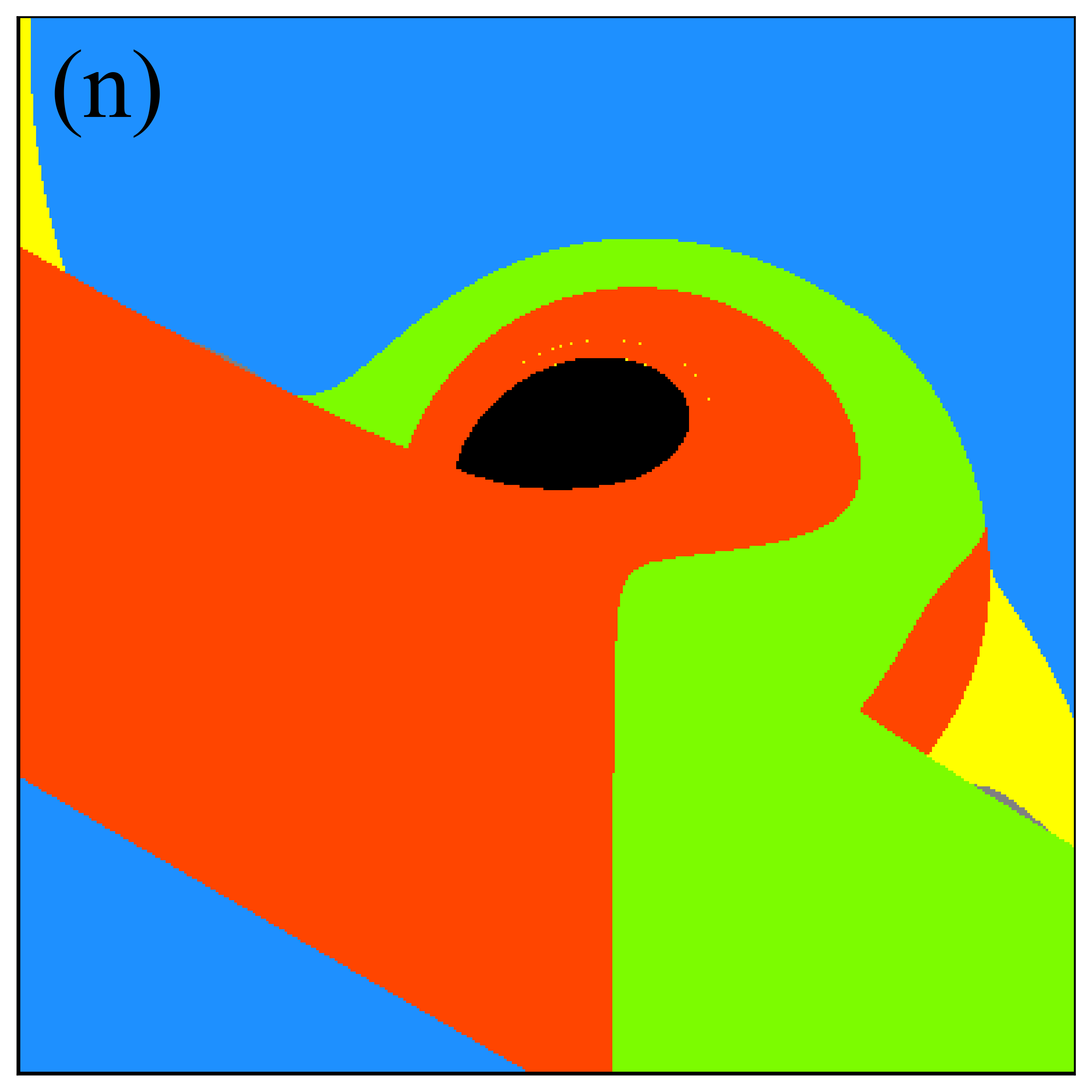}
\includegraphics[width=3.7cm]{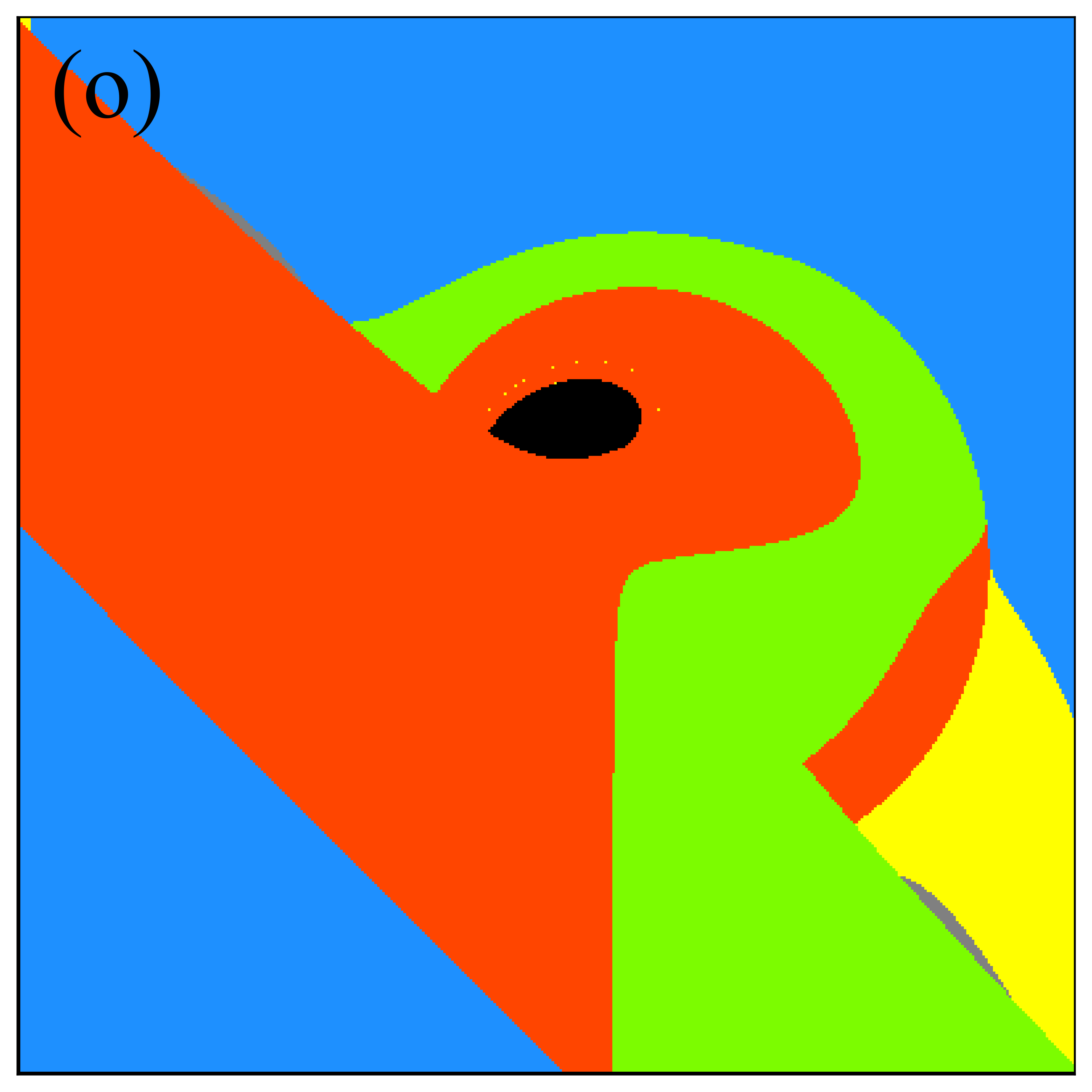}
\includegraphics[width=3.7cm]{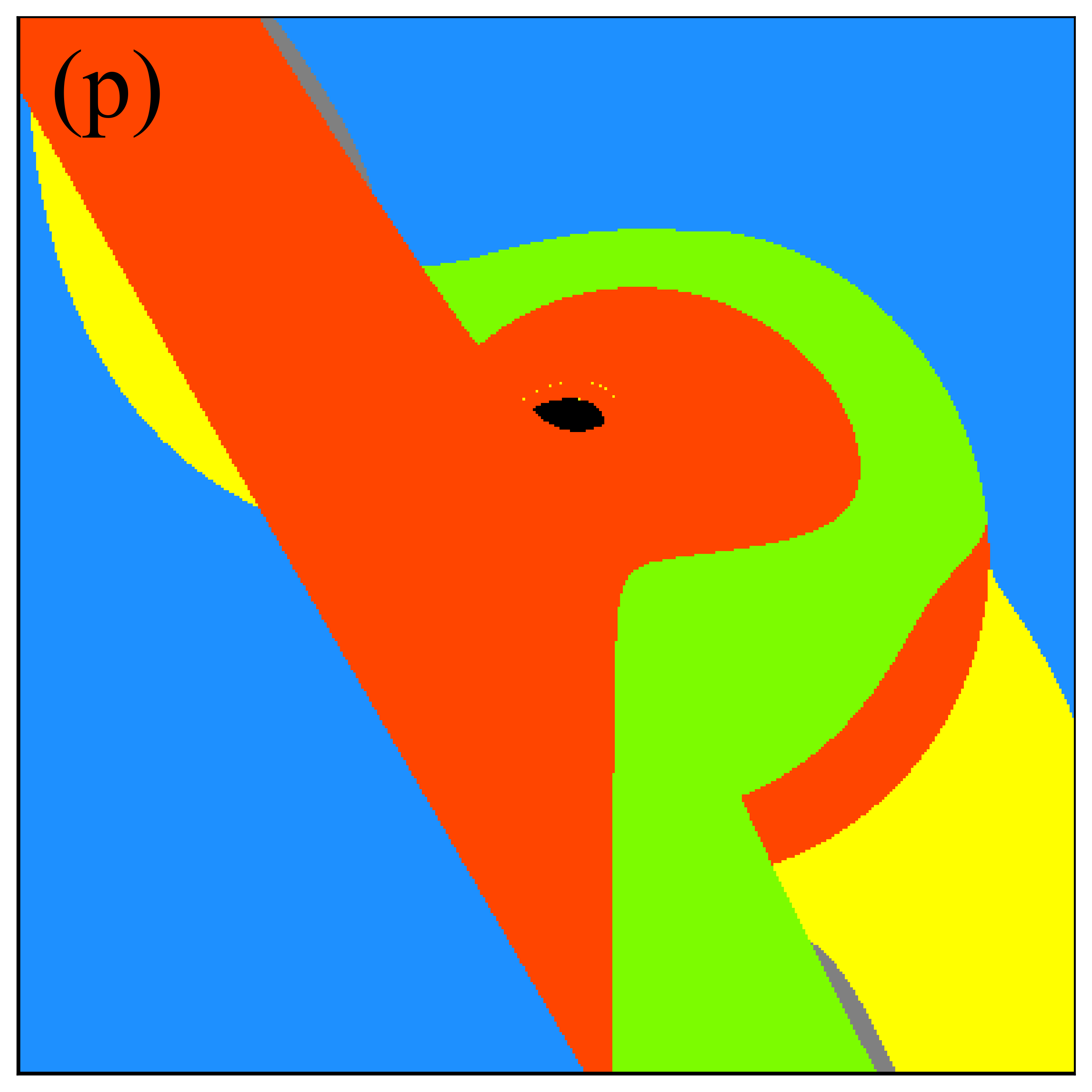}
\caption{Similar to figure 9, but for the observation angle of $85^{\circ}$.}}\label{fig11}
\end{figure*}
\section{Conclusions and Discussions}
An inner shadow emerges in the black hole image as the inner boundary of the equatorial accretion disk intersects the event horizon. This mechanism remains effective in the tilted accretion disk scenario. In this study, we employ a geometric approach to define a thin, tilted accretion disk and utilize a ray-tracing algorithm to simulate the inner shadow of a Kerr black hole. We find that in the case of an equatorial accretion disk, the morphology of the inner shadow of a Kerr black hole is shaped by both the viewing angle and the spin parameter, with its size increasing with the former and decreasing with the latter. Thus, when the viewing angle is $0^{\circ}$ and the spin parameter approaches $1$, the Kerr black hole surrounded by the equatorial accretion disk exhibits a minimum inner shadow size of approximately $S_{\textrm{min}} = 13.075$ M$^{2}$.

However, in the scenario of a tilted accretion disk, both the size and shape of the inner shadow of the Kerr black hole undergo significant transformations. We discover that the tilted accretion disk has a considerable capacity to obscure the inner shadow, resulting in novel shapes such as crescent, petal, and eyebrow configurations. Moreover, in most parameter spaces, the size of the inner shadow is less than $S_{\textrm{min}}$. This indicates that the inner shadow is highly sensitive to the accretion environment, making it unsuitable for constraining system parameters. Nevertheless, the inner shadow still has the potential to serve as a probe for tilted accretion disks and modified gravity theories. Specifically, when an inner shadow significantly smaller than expected is observed, it may indicate the presence of a tilted accretion disk or provide evidence supporting modified gravity.

Furthermore, in specific parameter spaces, we identify that a tilted accretion disk can give rise to a dual-shadow structure in the Kerr black hole. In this configuration, the primary shadow manifests as a petal shape, while the secondary shadow often takes on an elongated eyebrow form, distinguishing it from the multi-shadow structures associated with binary black holes \cite{Nitta et al. (2011),Nitta et al. (2012),Bohn et al. (2015)} and the single black hole in vacuum \cite{Abdolrahimi et al. (2015)}. Consequently, the dual-shadow identified in this study could also serve as an indicator of the existence of tilted accretion disks.

Moreover, it is crucial to emphasize that the tilted accretion disk assumed in this work serves purely as a geometrically defined light source, and the derived inner shadow remains theoretical in nature. From an astrophysical standpoint, the radiation from accreting matter near the black hole's event horizon is subject to both Doppler redshift and gravitational redshift. This may creates an extremely faint region around the inner shadow that is virtually indistinguishable from the shadow itself, resulting in an observed inner shadow size slightly larger than theoretical predictions. We will investigate this discrepancy in detail in future work.

\acknowledgments
The authors are very grateful to the referee for insightful comments and valuable suggestions. This research has been supported by the National Natural Science Foundation of China [Grant No. 12403081].

\end{document}